\documentclass[11pt]{article}
\usepackage{amsmath,amsthm,comment,amssymb}
\usepackage{makecell}
\usepackage{mathrsfs}
\usepackage{algorithm}
\usepackage{algorithmic}
\usepackage{amsfonts}
\usepackage{graphicx}
\usepackage{multirow}
\usepackage{booktabs}
\usepackage{bbm}
\usepackage{enumitem}
\usepackage[top=1in, bottom=1in, left=1in, right=1in]{geometry}
\usepackage{graphicx}
\usepackage{leftindex}
\usepackage{dsfont}
\usepackage{bm}
\usepackage{color}
\usepackage{xcolor}
\usepackage{subcaption}
\usepackage{caption}
\usepackage[colorlinks=true,citecolor=blue,linkcolor=red,urlcolor=blue]{hyperref}
\usepackage[round,authoryear]{natbib}
\usepackage{indentfirst}
\allowdisplaybreaks
\usepackage{appendix}
\usepackage{tikz}
\numberwithin{equation}{section}
\newtheorem{theorem}{Theorem}[section]

\newtheorem{lemma}{Lemma}[section]
\newtheorem{proposition}{Proposition}[section]
\theoremstyle{definition}

\newtheorem{remark}{Remark}[section]
\newtheorem{assumption}{\textbf{Assumption}}[section]

\newcommand{\var}{\mathrm{VaR}}
\newcommand{\covar}{\mathrm{CoVaR}}

\newcommand{\coes}{\mathrm{CoES}}
\newcommand{\bE}{\mathbb{E}}

\newcommand{\bP}{\mathbb{P}}
\newcommand{\bbR}{\mathbb{R}}

\usepackage{xr}
\begin{document}

\title{Estimations of Extreme CoVaR and CoES under Asymptotic Independence}
\author{
Qingzhao Zhong\footnote{\scriptsize School of Data Science, Fudan University, Email: {\color{blue}qzzhong22@m.fudan.edu.cn}}
}
\date{\today}
\maketitle

\begin{abstract}
The two popular systemic risk measures CoVaR (Conditional Value-at-Risk) and CoES (Conditional Expected Shortfall) have recently been receiving growing attention on applications in economics and finance. In this paper, we study the estimations of extreme CoVaR and CoES when the two random variables are asymptotic independent but positively associated. We propose two types of extrapolative approaches: the first relies on intermediate VaR and extrapolates it to extreme CoVaR/CoES via an adjustment factor; the second directly extrapolates the estimated intermediate CoVaR/CoES to the extreme tails. All estimators, including both intermediate and extreme ones, are shown to be asymptotically normal. Finally, we explore the empirical performances of our methods through conducting a series of Monte Carlo simulations and a real data analysis on S\&P500 Index with 12 constituent stock data.
\end{abstract}

\section{Introduction}\label{sec:intro}

Systemic risk refers to the widespread crisis of the entire financial system triggered by the distress of a single institution or event through interconnectedness and contagion mechanisms. In the wake of the global financial crisis and the COVID-19 pandemic, modeling and measuring systemic risk have risen to the key priorities in risk management. As extensions to the well-known Value-at-Risk (VaR) and Expected Shortfall (ES), the CoVaR and CoES have become two widely used approaches for measuring systemic risk, introduced by \cite{Tobias2016} and \cite{Huang2018}, respectively. In this paper, we proceed with CoVaR defined in \cite{Girardi2013} and \cite{Nolde2020} to facilitate nonparametric inferences, which conditions on the event that the system loss exceeds its VaR. Specifically, let $Y$ denote the loss variable of a system (or market) and $X$ the loss variable of interest. For a given quantile level $\tau \in (0,1)$, the $\covar_{X|Y}(\tau)$ is defined by,
\begin{equation}\label{eq:covar_X}
\bP(X \ge \covar_{X|Y}(\tau)|Y \ge \var_Y(\tau)) =  1 - \tau,
\end{equation}
where $\var_Y(\tau)$ denotes the VaR of $Y$, \emph{i.e.}, $\var_Y(\tau) := \inf\{ y \in \bbR: F_Y(y) \ge \tau \}$, with $F_Y$ being $Y's$ marginal distribution. Correspondingly, the definition of $\coes_{X|Y}(\tau)$ is given by,
\begin{equation}\label{eq:coes_X}
\coes_{X|Y}(\tau) := \frac{1}{1-\tau} \int_{\tau}^{1} \covar_{X|Y}(\alpha) \,d \alpha.
\end{equation}
The assumed continuity of $F_{X,Y}(\cdot,\cdot)$ implies that the $\coes_{X|Y}(\tau)$ coincides with the following conditional expectation
\begin{equation}\label{eq:coes_X_exp}
\coes_{X|Y}(\tau) = \bE[X|X \ge \covar_{X|Y}(\tau), Y \ge \var_Y(\tau)].
\end{equation}
A vast body of literature on CoVaR and CoES has emerged, with a primary focus on statistical inferences and their various applications in econometrics. Major directions include: statistical methodologies \citep{Francq2025,Dimitriadis2025,Huang2024}, backtesting \citep{Fissler2024}, portfolio selection \citep{Capponi2022,Fung2026}, and risk forecasting \citep{Girardi2013,Bernardi2019}, etc. 

Most of the aforementioned studies on CoVaR and CoES focus on fixed risk levels. In order to better characterize tail risk, it is both necessary and beneficial to model probability level $\tau$ as a sequence that depends on the sample size $n$ and approaches to 1 as $n$ increases, \emph{i.e.}, \emph{extreme risk}. In the bivariate extreme value theory, a classic approach involves assuming the tail dependence structure between $X$ and $Y$, \emph{i.e.}, tail copula is non-degenerate. For example, \cite{Cai2015} and \cite{Hou2021} respectively developed extrapolative estimators for the Marginal Expected Shortfall (MES, see \cite{Acharya2017}) and the Tail Gini functional at extreme levels within the framework of tail dependence. In the context of extreme CoVaR, two semi-parametric methods have been proposed by \cite{Nolde2020} and \cite{Nolde2022}. The first approximates the conditional probability in \eqref{eq:covar_X} via the tail of a skew-elliptical distribution, while the second assumes that the tail copula function follows a particular parametric family.

Although most research in bivariate extreme theory focuses on tail dependence, growing evidence suggests that asymptotic independence may exist in bivariate tail region across various applications, including spatial statistics \citep{Wadsworth2012,Le2018,Lalancette2021}, risk management \citep{Kulik2015,Das2018,Lalancette2020}. To measure the strength of the extremal dependence asymptotically, \cite{Ledford1996} and \cite{Ledford1997} proposed a more flexible framework via a coefficient of tail dependence, denoted as $\eta$. Namely, we assume that there exists an $\eta \in (0,1]$ such that the following limit exits and is positive for all $(x,y) \in (0,\infty)^2$,
\begin{equation}\label{eq:asy_ind}
\lim_{t \to \infty} t^{\frac{1}{\eta}} \bP \left( \overline{F}_X(X) < x/t, \overline{F}_Y(Y) < y/t  \right) =: C(x,y) \in (0,\infty).
\end{equation}
Let $C(x,y)=0$ for either $x=0$ or $y=0$. It is worthy noting that $C(x,y)$ is a homogeneous function of order $1/\eta$, \emph{i.e.}, $C(ax,ay) = a^{1/\eta}C(x,y)$ for any $a>0$. The coefficient $\eta$ describes the strength of tail dependence: $\eta = 1$ implies that $X$ and $Y$ are asymptotically dependent; if $\eta \in (1/2,1)$, we say that $X$ and $Y$ are asymptotically independent but positively associated; if $\eta \in (0,1/2)$, we say that $X$ and $Y$ are asymptotically independent but negatively associated. When $X$ and $Y$ are independent, it follows $\eta = 1/2$. We refer to \cite{Ledford1996}, \cite{Draisma2004} for more detailed interpretations of $\eta$. Under the framework of asymptotic independence \eqref{eq:asy_ind} with $\eta \in (1/2,1)$, \cite{Cai2020}, \cite{Zhang2024} and \cite{Wang2025} have re-estimated the MES, Expectile-based MES and Tail Gini functional at extreme levels, respectively, since all tail-dependence methods fall.

The main contribution of this paper lie in addressing the estimations for extreme CoVaR and CoES under asymptotic independence but positive association \emph{i.e.}, $\eta \in (1/2,1)$. To our best knowledge, there is still no literature considering the estimation issue of CoVaR and CoES within a framework of asymptotic independence. Our work is of great theoretical and practical implication to provide the powerful complement to the gaps in existing research. Due to the scarcity of tail observations, we develop two types of approaches to extrapolating CoVaR and CoES at an extreme level. More specifically, the first one is built upon the estimated intermediate VaR, with an extrapolation relationship that depends on an adjustment factor linking $\covar_{X|Y}(\tau)$ to the marginal $\var_X(\tau)$ via \eqref{eq:covar_var}, together with a procedure for estimating this adjustment factor at an intermediate level. The second approach involves the estimations of intermediate CoVaR and CoES, from which the extrapolation relationship is constructed, thereby avoiding the adjustment factor. Accounting for the intrinsic connection between CoVaR and CoES, we establish the joint bivariate asymptotic normalities of their extrapolative estimators.

The rest of this paper is organized as follows. In Section \ref{sec:meth}, we begin with describing the basic model setting, and then elaborate in detail on the two proposed classes of extrapolative estimations for CoVaR and CoES under asymptotic independence. We also study the asymptotic behaviors of all estimations therein. The empirical performance of our estimations is evaluated by a simulation study in Section \ref{sec:simulation}, and a real application in Section \ref{sec:realanslysis}. All technical proofs are collected in supplementary material.

\section{Methodologies}\label{sec:meth}

Recall that $(X,Y)$ is a two-dimensional random vector with joint distribution $F_{X,Y}$, and marginal distributions $F_X(\cdot) = F_{X,Y}(\cdot, \infty)$, $F_Y(\cdot) = F_{X,Y}(\infty,\cdot)$. Denote $\overline{F}_X(\cdot):=1-F_X(\cdot)$ and $\overline{F}_Y(\cdot):=1-F_Y(\cdot)$ as the (upper) tails of $F_X(\cdot)$ and $F_Y(\cdot)$, respectively. Suppose the bivariate samples $\{(X_i,Y_i)\}_{i=1}^n$ are independent copies of $(X,Y)$. Let $k = k(n)$ be an intermediate sequence of integers, such that $k \to \infty$ and $k/n \to 0$ as $n \to \infty$. Accordingly, $1 - k/n$ can serve as an \emph{intermediate} level. Let $\tau'_n$ be an \emph{extreme} level, satisfying $\tau'_n \to 1$ and $n(1-\tau'_n) \to c \in [0, \infty)$. Denote $A^{\top}$ as the transpose of a matrix or vector $A$. Denote $U_1 = (1/\overline{F}_X)^{\leftarrow}$ as the \emph{tail quantile} function with ``$\leftarrow$" denoting the left-continuous inverse.

We formulate our basic framework through Assumption \ref{ass:basic_ass} (a) and (b), which specify first-order conditions on the tail distribution of $X$ and the asymptotic independence between $X$ and $Y$.
\begin{assumption}\label{ass:basic_ass}
~
\begin{itemize}
\item[(a)] There exists a $\gamma_1 > 0$ such that, for $x > 0$, the following limit exists,
\begin{equation}\label{eq:rv_UX}
\lim_{t \to \infty} \frac{U_1(tx)}{U_1(t)} = x^{\gamma_1}.
\end{equation}
\item[(b)] There exists a $\eta \in (1/2,1)$ such that for all $(x,y) \in (0,\infty)^2$, the following limit exits,
\begin{equation*}
\lim_{t \to \infty} t^{\frac{1}{\eta}} \bP \left( \overline{F}_X(X) < x/t, \overline{F}_Y(Y) < y/t  \right) =: C(x,y) \in (0,\infty).
\end{equation*}
\item[(c)] The function $C(x,y)$ is differentiable at each coordinate with partial derivatives
\begin{equation*}
C_1(x,y) = \frac{\partial}{\partial x}C(x,y), \text{~and~} C_2(x,y) = \frac{\partial}{\partial y}C(x,y).
\end{equation*}
The partial derivatives $C_i(x,y)$ are continuous with respect to $x$ in the neighborhood of $(0,1)$ and $C_i(0,1) > 0$ with $i = 1,2$
\end{itemize}
\end{assumption}

Assumption \ref{ass:basic_ass} (a) assumes that $X$ follows a heavy-tail distribution, characterized by regular variation \eqref{eq:rv_UX}, where $\gamma_1$ is the \emph{extreme value index} (EVI), see \cite{dehaan2006}. This implies that $\overline{F}_X$ is also regularly varying with index $-1/\gamma_1$, \emph{i.e.}, $\lim_{t \to \infty}\frac{\overline{F}_X(tx)}{\overline{F}_X(t)} = x^{-1/\gamma_1}$.
Secondly, we impose $\eta \in (1/2,1)$ in \eqref{eq:asy_ind} throughout this paper. This range of $\eta$ models tail asymptotic independence alongside a positive association between $X$ and $Y$, a feature that makes the framework particularly well-suited for modeling financial data. 
The smoothness condition imposed in Assumption \ref{ass:basic_ass} (c) on function $C(\cdot,\cdot)$ and its derivatives ensures the key limit relationships presented in Proposition \ref{pro:coes_covar} and Lemma \ref{lem:lim_xi_xistar} below.

\begin{proposition}\label{pro:coes_covar}
Suppose that Assumption \ref{ass:basic_ass} holds, we have that, for $\gamma_1 \in (0,1)$,
\begin{equation}\label{eq:coes_covar}
\lim_{\tau \uparrow 1} \frac{\coes_{X|Y}(\tau)}{\covar_{X|Y}(\tau)} = \frac{1}{1-\gamma_1}.
\end{equation}
\end{proposition}

The restriction $\gamma_1 \in (0,1)$ guarantees both the existence of the limit and a finite expectation $\mathbb{E}[X]$. The resulting limit relationship \eqref{eq:coes_covar} between CoVaR and CoES as $\tau \uparrow 1$ is fundamental to our procedure for extrapolating CoES from CoVaR.

\subsection{Estimations based on intermediate VaR}\label{sec:extra_VaR}

It is the goal of this paper to estimate CoVaR and CoES at an extreme level $\tau'_n$ under asymptotic independence specified in Assumption \ref{ass:basic_ass} (b). To this end, we consider two types of extrapolative approaches: one based on an estimated intermediate VaR and the other based on an estimated intermediate CoVaR/CoES. This section introduces the first method, while the second will be discussed in Section \ref{sec:extra_CRES}.

For the VaR-based method, \cite{Nolde2022} proposes an adjustment factor $\xi_\tau$ associated with $\tau$, via
\begin{equation}\label{eq:covar_var}
\covar_{X|Y}(\tau) = \var_X(1-(1-\tau)\xi_\tau).
\end{equation}
This implies that $\covar_{X|Y}(\tau)$ reduces to an unconditional quantile evaluated at the adjusted level $1 - (1-\tau)\xi_\tau$ and an extrapolation procedure governed by $\xi_\tau$ is thereby induced. Under mild distributional conditions, \cite{Zhong2025} developed the sufficient and necessary condition for the existence and uniqueness of $\xi_\tau$, \emph{i.e.}, 
\begin{equation}\label{eq:equ_con}
\bP(X \ge \var_X(\tau), Y \ge \var_Y(\tau)) > (1-\tau)^2,
\end{equation}
holds on a tail region $\tau \in (\tau_0,1)$ for some threshold $\tau_0$.

\begin{remark}
First, condition \eqref{eq:equ_con} is generally regarded as a weaker condition than the positive quadrant dependence (PQD) used in \cite{Nolde2022} and it restricts $\tau$ to a tail region, rather than $(0,1)$; second, as shown in \cite{Zhong2025}, the existence and uniqueness of $\xi_\tau$ does not rely on the specific structure of tail dependence, hence making it compatible with asymptotic independence framework as well; third, $\xi_\tau$ indeed admits a closed solution,
\begin{equation*}
\xi_\tau: = \frac{\bP(X \ge \covar_{X|Y}(\tau))}{\bP(X \ge \covar_{X|Y}(\tau) | Y \ge \var_Y(\tau))}.
\end{equation*}
\end{remark}

For brevity, we omit the detailed discussions on the existence and uniqueness of $\xi_\tau$ as it is identical to that of \cite{Zhong2025} and just assume it holds by default. We also observe that, the quantile level $1-(1-\tau)\xi_\tau$ is higher than $\tau$ as $\xi_\tau \in (0,1)$ deduced by \eqref{eq:equ_con}. Therefore, combining the limit condition \eqref{eq:asy_ind} with \eqref{eq:covar_X} and \eqref{eq:covar_var} implies that,
\begin{equation}\label{eq:limrela}
\begin{split}
 \left( 1-\tau \right)^{2-\frac{1}{\eta}} & = \left( 1-\tau \right)^{-\frac{1}{\eta}} \bP (X \ge \covar_{X|Y}(\tau), Y \ge \var_Y(\tau)) \\
  & = \left( 1-\tau \right)^{-\frac{1}{\eta}} \bP ( \overline{F}_X(X) \leq (1-\tau)\xi_\tau, \overline{F}_F(Y) \leq 1-\tau).
\end{split}
\end{equation}
By letting $\tau \uparrow 1$, \eqref{eq:asy_ind} and \eqref{eq:limrela} suggest a possibility of approximating the true adjustment factor $\xi_\tau$ with an approximation, denoted as $\xi_\tau^*$, which is defined implicitly via
\begin{equation}\label{eq:xi_star}
C(\xi_\tau^*,1) = \left( 1-\tau \right)^{2-\frac{1}{\eta}}.
\end{equation}
The assumed continuity of maps: $x \mapsto C(x,1)$ implies that $\xi_\tau^* \to 0$ as $\tau \uparrow 1$. We can further derive the limit relationship between $\xi_\tau$ and $\xi^*_\tau$ by imposing a second-order condition on \eqref{eq:asy_ind} (see Assumption \ref{ass:reg_cond} (b)). For $t > 0$, define
\begin{equation}\label{eq:c_t}
C_t(x,y) = t^{\frac{1}{\eta}} \bP(\overline{F}_X(X) < x/t,\overline{F}_Y(Y) < y/t ),
\end{equation}
which indicates $\lim_{t \to \infty} C_t(x,y) = C(x,y)$ for all $(x,y) \in (0,\infty)^2$ by \eqref{eq:asy_ind}.

\begin{lemma}\label{lem:lim_xi_xistar}
Suppose that Assumption \ref{ass:basic_ass} (c) and Assumption \ref{ass:reg_cond} (b) hold, we have that,
\begin{equation*}
\lim_{\tau \uparrow 1}\frac{(1-\tau)^{2-\frac{1}{\eta}}}{\xi_\tau^*} = C_1(0,1), \quad \text{and} \quad \lim_{\tau \uparrow 1}\frac{\xi_\tau}{\xi_\tau^*} = 1.
\end{equation*}
\end{lemma}

This lemma indicates that $\xi_\tau \to 0$ with the same speed as $(1-\tau)^{2-\frac{1}{\eta}}$.

\begin{remark}
Notably, a loose constraint $\alpha < \frac{1}{\eta} - 2$ is sufficient to ensure Lemma \ref{lem:lim_xi_xistar} holds. However, in order to study the asymptotic behaviors, we need to impose an additional condition on the growth rate of the intermediate sequence $k$, namely $k = O(n^{\iota})$ with $\iota \in \left( \frac{2}{3}, 1 - \frac{1}{1 - 2\alpha} \right)$ as $n \to \infty$. It is straightforward to verify that only if $\alpha < -1$ can ensure $\frac{2}{3} < 1 - \frac{1}{1 - 2 \alpha}$. Therefore, we formulate Assumption \ref{ass:reg_cond} (b) with $\alpha < - 1$ (by noting that $-1 < \frac{1}{\eta} - 2$) for the sake of condition consistency.
\end{remark}

Next, we consider the VaR-based extrapolative estimation for $\covar_{X|Y}(\tau'_n)$. Using \eqref{eq:covar_var} and regular variation \eqref{eq:rv_UX}, we have that,
\begin{equation*}
\frac{\covar_{X|Y}(\tau'_n)}{\var_X(1-k/n)} = \frac{\var_X(1-(1-\tau'_n)\xi_{\tau'_n})}{\var_X(1-k/n)} \sim \left( \frac{k}{n(1-\tau'_n)} \right)^{\gamma_1} \xi_{\tau'_n}^{-\gamma_1},
\end{equation*}
which implies that,
\begin{equation}\label{eq:ex_covar1}
\covar_{X|Y}(\tau'_n) \sim \left( \frac{k}{n(1-\tau'_n)} \right)^{\gamma_1} \xi_{\tau'_n}^{-\gamma_1} \var_X(1-k/n).
\end{equation}
Moreover, by Lemma \ref{lem:lim_xi_xistar}, we can also extrapolate $\xi_{\tau'_n}$ from an intermediate $\xi_{1-k/n}$,
\begin{equation}\label{eq:extra_xi}
\xi_{\tau'_n} \sim \left( \frac{n(1-\tau'_n)}{k} \right)^{2-\frac{1}{\eta}} \xi_{1-k/n}.
\end{equation}
Then, combining \eqref{eq:ex_covar1} and \eqref{eq:extra_xi} yields,
\begin{equation}\label{eq:extra_covar1}
\covar_{X|Y}(\tau'_n) \sim \left( \frac{k}{n(1-\tau'_n)} \right)^{\gamma_1\left( 3 - \frac{1}{\eta} \right)} \xi_{1-k/n}^{-\gamma_1} \var_X(1-k/n).
\end{equation}

From \eqref{eq:extra_covar1}, there involves four quantities, $\var_X(1-k/n)$, $\gamma_1$, $\eta$, and $\xi_{1-k/n}$, supposed to estimate. Denote $X_{1,n} \leq X_{2,n} \leq \cdots \leq X_{n,n}$ are the order statistics of $\{ X_1,...,X_n\}$. First, we estimate $\var_X(1-k/n)$ by using the $(n-k)$-th order statistic,
\begin{equation}\label{eq:est_var_int}
\widehat{\var}_X(1-k/n) := X_{n-k,n}.
\end{equation}
Let $k_1$ and $k_2$ be two intermediate sequences, satisfying, $k_j=k_j(n) \to \infty$, $k_j/n \to 0$, as $n \to \infty$, for $j=1,2$. We then estimate $\gamma_1$ by the Hill estimator (\cite{Hill1975})
\begin{equation}\label{eq:hill_est}
\hat{\gamma}_1 = \frac{1}{k_1} \sum_{i=1}^{k_1} \log X_{n-i+1,n} - \log X_{n-k_1,n},
\end{equation}
and estimate $\eta$ by the method proposed in \cite{Draisma2004},
\begin{equation}\label{eq:est_eta}
\hat{\eta} = \frac{1}{k_2} \sum_{i=1}^{k_2} \log T_{n-i+1,n}^{(n)} - \log T_{n-k_2,n}^{(n)},
\end{equation}
where $T_{1,n}^{(n)} \leq T_{2,n}^{(n)} \leq \cdots \leq T_{n,n}^{(n)}$ are the order statistics of $\{T^{(n)}_1, T^{(n)}_2,...,T^{(n)}_n\}$ with $T_i^{(n)}= \frac{n+1}{n+1-R_i^X} \wedge \frac{n+1}{n+1-R_i^Y}$ and $R_i^X$, $R_i^Y$ denoting the ranks of $X_i$, $Y_i$ in their respective samples. The associating asymptotic behaviours of $\hat\gamma_1$ and $\hat{\eta}$ are summarized in Theorem 3.2.5 and Theorem 7.6.1 in \cite{dehaan2006}.

In addition, according to \eqref{eq:limrela}, $\xi_{1-k/n}$ can be estimated by,
\begin{equation}\label{eq:est_xi_int}
\hat{\xi}_{1-k/n} := \inf \left\{ \xi \in (0,1),~ \widehat{C}_{n/k} (\xi,1) \ge (k/n)^{2 - \frac{1}{\hat\eta}} \right\},
\end{equation}
where
\begin{equation}\label{eq:nonp_c}
\widehat{C}_{n/k} (x,y) = \left( \frac{n}{k}  \right)^{\frac{1}{\hat{\eta}}} \frac{1}{n} \sum_{i=1}^{n} I \left( 1 - \widehat{F}_X(X_i) \leq \frac{kx}{n}, 1-\widehat{F}_Y(Y_i) \leq \frac{ky}{n} \right),
\end{equation}
severs as an empirical version of the joint probability on the right-hand side of \eqref{eq:limrela} by embedding the well-estimated $\hat{\eta}$ and the marginal empirical distributions $\widehat{F}_X(\cdot) = \frac{1}{n} \sum_{i=1}^{n} I(X_i \leq \cdot)$, $\widehat{F}_Y(\cdot) = \frac{1}{n} \sum_{i=1}^{n} I(Y_i \leq \cdot)$. As $\widehat{C}_{n/k} (x,y)$ is non-decreasing on $x$ given any $y$, $\hat{\xi}_{1-k/n}$ can be regarded as the generalized inverse function of $\widehat{C}_{n/k} (\xi,1)$ at an estimated level $(k/n)^{2 - \frac{1}{\hat\eta}}$.

To driven the asymptotic properties, we present the following regularity conditions used in this paper.

\begin{assumption}\label{ass:reg_cond}
~
\begin{itemize}
\item[(a)] There exists $\gamma_1 > 0$ such that
\begin{equation*}
\sup_{x > 1} \left| x^{-\gamma_1} \frac{U_1(tx)}{U_1(t)} - 1 \right| = O(A_1(t)),
\end{equation*}
where $A_1$ is an auxiliary function that is regularly varying with index $\rho_1 \leq 0$ and satisfies $\lim_{t \to \infty}A_1(t) = 0$.
\item[(b)] There exists $\beta_1 > \gamma_1$ and $\alpha < -1$ such that
\begin{equation*}
\sup_{0 < x \leq 1, 1/2 \leq y \leq 2} |C_t(x,y) - C(x,y)| x^{-\beta_1} = O(t^{\alpha}).
\end{equation*}
\item[(c)] The following integral is finite,
\begin{equation}\label{eq:infint}
\left|\int_{0}^{1} C(x,1) \,dx^{-\gamma_1} \right| < \infty.
\end{equation}
\item[(d)] As $n \to \infty$, $k=O(n^{\iota})$ with $\iota \in \left( \frac{2}{3}, 1 - \frac{1}{1-2\alpha} \right)$.
\item[(e)] Let $d_n = k/(n(1-\tau'_n))$. As $n \to \infty$, $\frac{k}{n} \log d_n \to 0$ when $n(1-\tau'_n) \to 0$.
\item[(f)] As $n \to \infty$, the two intermediate sequences $k_j$ satisfy $k_j = O(k)$ for $j = 1,2$. 
\end{itemize}
\end{assumption}

Note that Assumption \ref{ass:reg_cond} (a) and (b) are second-order strengthenings quantifying the rates of convergence in \eqref{eq:rv_UX} and \eqref{eq:asy_ind}, respectively. Assumption \ref{ass:reg_cond} (c) is a technical condition imposing some integrality on the function $C(\cdot,\cdot)$. We further require Assumption \ref{ass:reg_cond} (d) - (e) on the rates of the intermediate sequence $k$ and extreme level $\tau'_n$. The choice of $k_1$ and $k_2$ might be different from $k$. To ensure a uniform rate of convergence throughout the extrapolations, $k_1$ and $k_2$ should be the same order as $k$. Thus, Assumption \ref{ass:reg_cond} (f) implies that,
\begin{equation}\label{eq:cons_eta_gam}
\sqrt{k}(\hat\gamma_1 - \gamma_1) = O_{\bP}(1), \quad \text{and} \quad \sqrt{k}(\hat{\eta} - \eta) = O_{\bP}(1).
\end{equation}

The following proposition presents the asymptotic property for $\hat{\xi}_{1-k/n}$ characterized by a centered Gaussian process, $W(\cdot)$, with covariance structure,
\begin{equation*}
\bE[W(x)W(y)] = x \wedge y.
\end{equation*}

\begin{proposition}\label{prop:asy_xi}
Suppose that Assumption \ref{ass:basic_ass} (c) and Assumption \ref{ass:reg_cond} (b), (d), (f) hold with $\eta \in \left( \frac{7+\sqrt{17}}{16},1 \right)$. We have that, as $n \to \infty$,
\begin{equation}\label{eq:asy_xi}
\frac{k^{3/2}}{n} \left( \frac{\hat{\xi}_{1-k/n}}{\xi_{1-k/n}} - 1 \right) \xrightarrow{d} \left( \frac{1}{\eta} - 2 \right) C_2(0,1)W(1).
\end{equation}
\end{proposition}

Therefore, by recalling the extrapolative expression given in \eqref{eq:extra_covar1}, we can estimate $\covar_{X|Y}(\tau'_n)$ by plugging in the estimators $\widehat{\var}_X(1-k/n)$, $\hat{\gamma}_1$, $\hat{\eta}$ and $\hat{\xi}_{1-k/n}$,
\begin{equation}\label{eq:extra_CR1}
\widetilde{\covar}_{X|Y}^{(1)}(\tau'_n) = \left( \frac{k}{n(1-\tau'_n)} \right)^{\hat\gamma_1\left( 3 - \frac{1}{\hat\eta} \right)} \hat{\xi}_{1-k/n}^{-\hat\gamma_1} \widehat{\var}_X(1-k/n).
\end{equation}
Then, combining \eqref{eq:extra_CR1} and \eqref{eq:coes_covar}, we can extrapolate the $\coes_{X|Y}(\tau'_n)$ by
\begin{equation}\label{eq:extra_CES1}
\begin{split}
\widetilde{\coes}_{X|Y}^{(1)}(\tau'_n) & = \frac{1}{1-\hat{\gamma}_1} \widetilde{\covar}_{X|Y}^{(1)}(\tau'_n) \\
& = \frac{1}{1-\hat{\gamma}_1} \left( \frac{k}{n(1-\tau'_n)} \right)^{\hat\gamma_1\left( 3 - \frac{1}{\hat\eta} \right)} \hat{\xi}_{1-k/n}^{-\hat\gamma_1} \widehat{\var}_X(1-k/n).
\end{split}
\end{equation}

Under the second-order conditions specified in Assumption \ref{ass:reg_cond} (a) and (b), we establish the joint asymptotic properties of $\widetilde{\covar}_{X|Y}^{(1)}(\tau'_n)$ and $\widetilde{\coes}_{X|Y}^{(1)}(\tau'_n)$ in the following theorem.
\begin{theorem}\label{th:asy_CRCES1}
Suppose that Assumption \ref{ass:basic_ass} (c) and Assumption \ref{ass:reg_cond} (a), (b), (d), (e), (f) hold with $\eta \in \left(  \frac{7 + \sqrt{17}}{16},1\right)$. We have that, as $n \to \infty$,
\begin{equation}\label{eq:asy_CRCES1}
\frac{k^{3/2}}{n}\left( \frac{\widetilde{\covar}_{X|Y}^{(1)}(\tau'_n)}{\covar_{X|Y}(\tau'_n)} - 1, \frac{\widetilde{\coes}_{X|Y}^{(1)}(\tau'_n)}{\coes_{X|Y}(\tau'_n)} -1 \right)^{\top} \xrightarrow{d} (1,1)^{\top} \gamma_1 \left( 2 - \frac{1}{\eta}  \right) C_2(0,1) W(1).
\end{equation}
\end{theorem}

\subsection{Estimations based on intermediate CoVaR and CoES}\label{sec:extra_CRES}

Now, we turn our attention to the second approach of extrapolation for $\covar_{X|Y}(\tau'_n)$ and $\coes_{X|Y}(\tau'_n)$, namely, the intermediate-CoVaR/CoES approach. This approach necessitates the estimators of $\covar_{X|Y}(1-k/n)$ and $\coes_{X|Y}(1-k/n)$, denoted as $\widehat{\covar}_{X|Y}(1-k/n)$ and $\widehat{\coes}_{X|Y}(1-k/n)$, respectively. First, according to \eqref{eq:covar_X}, $\widehat{\covar}_{X|Y}(1-k/n)$ can be defined by
\begin{equation}\label{eq:est_CR_int}
\widehat{\covar}_{X|Y}(1-k/n) := \sup \left\{ s \in (0,\infty), ~ P_n(s) \ge (k/n)^2  \right\},
\end{equation}
where
\begin{equation*}
P_n(s) := \frac{1}{n} \sum_{i=1}^{n} I(X_i \ge s, Y_i \ge \widehat{\var}_Y(1-k/n)),
\end{equation*}
and $\widehat{\var}_Y(1-k/n) := Y_{n-k,n}$ corresponding to the $(n-k)$-th order statistic of $Y_1,...,Y_n$. Then, \eqref{eq:est_CR_int} indicates that $\widehat{\covar}_{X|Y}(1-k/n)$ can be treated as the generalized inverse function of $P_n(\cdot)$ at the quantile level $1-k/n$. Furthermore, according to the expected form of CoES given in \eqref{eq:coes_X_exp}, $\widehat{\coes}_{X|Y}(1-k/n)$ can be defined by
\begin{equation}\label{eq:est_CES_int}
\widehat{\coes}_{X|Y}(1-k/n) := \frac{n}{k^2} \sum_{i=1}^{n} X_i I\left( X_i \ge \widehat{\covar}_{X|Y}(1-k/n), Y_i \ge \widehat{\var}_Y(1-k/n) \right).
\end{equation}

The following proposition establishes the joint asymptotic normality for $\widehat{\covar}_{X|Y}(1-k/n)$ and $\widehat{\coes}_{X|Y}(1-k/n)$.

\begin{proposition}\label{pro:asy_CRCES_int}
Suppose that Assumption \ref{ass:basic_ass} (c) and Assumption \ref{ass:reg_cond} (b), (c), (d), (f) hold with $\eta \in \left( \frac{7+\sqrt{17}}{16} ,1\right)$. Then, we have that, for any $0 < \gamma_1 < \frac{1}{4} \left( 3 - \frac{1}{\eta} \right)$, as $n \to \infty$,
\begin{equation}\label{eq:asy_CRCES_int}
\frac{k^{3/2}}{n}\left( \frac{\widehat{\covar}_{X|Y}(1-k/n)}{\covar_{X|Y}(1-k/n)} - 1, \frac{\widehat{\coes}_{X|Y}(1-k/n)}{\coes_{X|Y}(1-k/n)} - 1 \right)^{\top} \xrightarrow{d} (1,2)^{\top} \gamma_1 \left(2 - \frac{1}{\eta}\right) C_2(0,1) W(1).
\end{equation}
\end{proposition}

Note that the narrower constraint $0 < \gamma_1 < \frac{1}{4} \left( 3 - \frac{1}{\eta} \right) < \frac{1}{2}$ is imposed to ensure the the asymptotic behavior of $\widehat{\coes}_{X|Y}(1-k/n)$. Then, using \eqref{eq:rv_UX}, \eqref{eq:coes_covar}, and \eqref{eq:extra_xi} once again, we obtain the asymptotic relationship between $\covar_{X|Y}(\tau'_n)$ and $\covar_{X|Y}(1-k/n)$,
\begin{equation*}
\frac{\covar_{X|Y}(\tau'_n)}{\covar_{X|Y}(1-k/n)} = \frac{\var_X(1-(1-\tau'_n)\xi_{\tau'_n})}{\var_X(1-k\xi_{1-k/n}/n)} \sim \left( \frac{k}{n(1-\tau'_n)} \right)^{\gamma_1 \left( 3 - \frac{1}{\eta} \right)},
\end{equation*}
suggesting the second intermediate-CoVaR-based extrapolations for $\covar_{X|Y}(\tau'_n)$ and $\coes_{X|Y}(\tau'_n)$ given by,
\begin{equation}\label{eq:extra_CR2}
\widetilde{\covar}^{(2)}_{X|Y}(\tau'_n) = \left( \frac{k}{n(1-\tau'_n)} \right)^{\hat\gamma_1 \left( 3 - \frac{1}{\hat\eta} \right)} \widehat{\covar}_{X|Y}(1-k/n),
\end{equation}
and
\begin{equation}\label{eq:extra_CES2}
\begin{split}
\widetilde{\coes}^{(2)}_{X|Y}(\tau'_n) & = \frac{1}{1-\hat{\gamma}_1} \widetilde{\covar}^{(2)}_{X|Y}(\tau'_n) \\
& = \frac{1}{1-\hat{\gamma}_1} \left( \frac{k}{n(1-\tau'_n)} \right)^{\hat\gamma_1 \left( 3 - \frac{1}{\hat\eta} \right)} \widehat{\covar}_{X|Y}(1-k/n).
\end{split}
\end{equation}

Finally, the intermediate-CoES-based extrapolation for $\coes_{X|Y}(\tau'_n)$ is given by
\begin{equation}\label{eq:extra_CES3}
\widetilde{\coes}^{(3)}_{X|Y}(\tau'_n) = \left( \frac{k}{n(1-\tau'_n)} \right)^{\hat\gamma_1 \left( 3 - \frac{1}{\hat\eta} \right)} \widehat{\coes}_{X|Y}(1-k/n),
\end{equation}
which follows the asymptotic relationship,
\begin{align*}
\frac{\coes_{X|Y}(\tau'_n)}{\coes_{X|Y}(1-k/n)} & = \frac{\coes_{X|Y}(\tau'_n)}{\covar_{X|Y}(\tau'_n)} \frac{\covar_{X|Y}(1-k/n)}{\coes_{X|Y}(1-k/n)} \frac{\covar_{X|Y}(\tau'_n)}{\covar_{X|Y}(1-k/n)} \\
& \sim \left( \frac{k}{n(1-\tau'_n)} \right)^{\gamma_1\left( 3 - \frac{1}{\eta} \right)}.
\end{align*}

The following theorem establishes the joint asymptotic normalities of $\widetilde{\coes}^{(2)}_{X|Y}(\tau'_n)$/$\widetilde{\coes}^{(3)}_{X|Y}(\tau'_n)$ and $\widetilde{\covar}_{X|Y}^{(2)}(\tau'_n)$. For the result concerning $\widetilde{\coes}^{(3)}_{X|Y}(\tau'_n)$, which depends on $\widehat{\covar}_{X|Y}(1-k/n)$, the EVI $\gamma_1$ should be restricted to the interval $\left( 0, \frac{1}{4} \left( 3 - \frac{1}{\eta} \right) \right)$ as well.

\begin{theorem}\label{th:asy_CRCES23}
Suppose that Assumption \ref{ass:basic_ass} (c) and Assumption \ref{ass:reg_cond} hold with $\eta \in \left( \frac{7+\sqrt{17}}{16} ,1\right)$. Then, we have that, for any $\gamma_1 \in (0,1)$, as $n \to \infty$,
\begin{equation}\label{eq:asy_CRCES2}
\frac{k^{3/2}}{n}\left( \frac{\widetilde{\covar}_{X|Y}^{(2)}(\tau'_n)}{\covar_{X|Y}(\tau'_n)} - 1, \frac{\widetilde{\coes}_{X|Y}^{(2)}(\tau'_n)}{\coes_{X|Y}(\tau'_n)} -1 \right)^{\top} \xrightarrow{d} (1,1)^{\top} \gamma_1 \left(2 - \frac{1}{\eta} \right) C_2(0,1) W(1),
\end{equation}
and moreover, for any $\gamma_1 \in \left( 0, \frac{1}{4} \left( 3 - \frac{1}{\eta} \right) \right)$,
\begin{equation}\label{eq:asy_CRCES3}
\frac{k^{3/2}}{n}\left( \frac{\widetilde{\covar}^{(2)}_{X|Y}(\tau'_n)}{\covar_{X|Y}(\tau'_n)} - 1, \frac{\widetilde{\coes}^{(3)}_{X|Y}(\tau'_n)}{\coes_{X|Y}(\tau'_n)} - 1 \right)^{\top} \xrightarrow{d} (1,2)^{\top} \gamma_1 \left(2 - \frac{1}{\eta}\right) C_2(0,1) W(1).
\end{equation}
\end{theorem}

\section{Simulation Study}\label{sec:simulation}

In this section, we implement some Monte Carlo simulations to investigate the finite sample performance of our extrapolative methods. We consider the following bivariate models:
\begin{itemize}
  \item \textbf{Model 1.} Define
  \begin{equation*}
    (X,Y) = (Z_1,Z_2),
  \end{equation*}
  where $(Z_1, Z_2)$ has identical Pareto marginal distributions $\Psi_a(x) = 1 - x^{-a}$ with common parameter $a > 0$ and a dependence structure given by Marshall-Olkin survival copula, defined by $\overline{C}_{MO}(u,v) = uv \min \{ u^{-a_1},v^{-a_2} \}$ for all $u,v \in (0,1)^2$ and some $a_1, a_2 \in (0,1)$. For this model, we have that, $\gamma_1 = 1/a$, $\eta = \frac{1}{2 - a_2}$, $A_1(t) = 0$, and
  \begin{equation*}
  C(x,y) = 
  \begin{cases}
    x^{1-a_1}y, & \mbox{if } a_1 < a_2, \\
    xy \max \{x,y \}^{-a_1}, & \mbox{if } a_1 = a_2, \\
    xy^{1-a_2}, & \mbox{if } a_1 > a_2.
  \end{cases}
  \end{equation*}
  In this simulation, we set $a = 3$, $a_1 = 5/6$, $a_2 = 2/3$, implying $\gamma_1 = 1/3$, $\eta = 3/4$ and $C(x,y)=xy^{1/3}$. We refer to see Example 4 of \cite{Das2018} for more details about this model.
  \item \textbf{Model 2.} Continue using \textbf{Model 1.} with parameters $a = 3$, $a_1 = a_2 = 7/10$, implying $\gamma_1 = 1/3$, $\eta = 10/13$ and $C(x,y) = xy \max \{x,y \}^{-7/10}$. 
  \item \textbf{Model 3.} Define
  \begin{equation*}
    (X,Y) = B(Z_1,Z_3) + (1-B)(Z_2,Z_2),
  \end{equation*}
  where $Z_1, Z_2, Z_2$ are independent Pareto random variables with parameters $a, b, a$, respectively, and $B$ is a Bernoulli(1/2) random variable independent of $Z_i's$. For this model, we have  $\gamma_1 = 1/a$, $\eta = a/b$, $A_1(t) = 0$, and $C(x,y)=2^{b/a - 1}(x \wedge y)^{b/a}$. In this simulation, we set $a= 3$, $b = 4$, implying $\gamma_1 = 1/3$, $\eta = 3/4$, and $C(x,y)=2^{1/3}(x \wedge y)^{4/3}$. This model is also used in \cite{Cai2020} and \cite{Wang2025}.
\end{itemize}

Note that, in the third model, $C_i(0,1) = 0$ holds for both $i = 1,2$, which slightly departs from Assumption \ref{ass:basic_ass} (c). This mainly has two implications. First, when $C_1(0,1)=0$, the first limit in Lemma \ref{lem:lim_xi_xistar} vanishes, so that the convergence $\xi_\tau^* \to 0$ as $\tau \uparrow 1$ can no longer be inferred. Second, when $C_2(0,1)=0$, the asymptotic normality of all estimators is no longer ensured, although consistency remains valid. Overall, the condition $C_1(0,1)>0$ plays a more crucial role for the validity of our estimations. In the following simulation, we compare the proposed extrapolations:
\begin{itemize}
  \item CoVaR-I: estimator $\widetilde{\covar}_{X|Y}^{(1)}(\tau'_n)$, given in \eqref{eq:extra_CR1};
  \item CoVaR-II: estimator $\widetilde{\covar}_{X|Y}^{(2)}(\tau'_n)$, given in \eqref{eq:extra_CR2};
  \item CoES-I: estimator $\widetilde{\coes}_{X|Y}^{(1)}(\tau'_n)$, given in \eqref{eq:extra_CES1};
  \item CoES-II: estimator $\widetilde{\coes}_{X|Y}^{(2)}(\tau'_n)$, given in \eqref{eq:extra_CES2};
  \item CoES-III: estimator $\widetilde{\coes}_{X|Y}^{(3)}(\tau'_n)$, given in \eqref{eq:extra_CES3}.
\end{itemize}
For all simulated data, we set sample size $n = 500, 1000, 2000$, and 5000, extremes levels $\tau'_n = 0.99, 0.999$, and repeat the simulation $N = 1000$ times. To evaluate the performance of all estimators, we calculate the \emph{Mean Square Relative Error} (MSRE) based on $N$ replications, defined as,
\begin{equation*}
{\rm MSRE} = \frac{1}{N} \sum_{i=1}^{N} \left( \frac{\hat{\theta}_n^{(i)}}{\theta} -1\right)^2,
\end{equation*}
where $\hat{\theta}_n^{(i)}$ is the estimator we are interested in given the simulated data of the $i$-th replication, and $\theta$ is the true value.

Another delicate problem in the extreme value theory is the proper choice of three hyperparameters $k, k_1$ and $k_2$, \emph{that is}, the number of tail observations used in the estimations of $\covar_{X|Y}(1-k/n)$ (or $\coes_{X|Y}(1-k/n)$), $\gamma_1$, and $\eta$. Following the principle of computational simplicity, we set $k_1 = k_2$ and examine the sensitivity of the MSREs with respect to the values of $k$ and $k_1$. Specifically, we take the estimator $\widetilde{\covar}_{X|Y}^{(1)}(\tau'_n)$ as an example, consider a range of values for $k$ and $k_1$ according to different sample size, and compute the MSRE for each pair $(k,k_1)$ based on $N=100$ replications, then select the pair $(k,k_1)$ that minimizes the MSRE. We observe that larger values of $k$ and $k_1$ will result in a lower MSRE's value. Table \ref{tab:msre} summarizes the selected values of $k$ and $k_1$ across the three models at $\tau'_n = 0.99, 0.999$.

Moreover, Table \ref{tab:msre} also reports the specific MSRE values for the three models under different risk levels and sample sizes. It can be directly observed that all extrapolative estimators exhibit satisfactory MSRE values across the three models, which decrease as the sample size increases and the risk level decreases. In more detail, the performance of the two CoVaR extrapolations, namely, $\widetilde{\covar}_{X|Y}^{(i)}(\tau'_n)$ ($i=1,2$) is similar, as evidenced by only marginal differences in their MSRE across varying sample sizes and risk levels. As a result, this close performance is also mirrored in the first two extrapolations of CoES, namely, $\widetilde{\coes}_{X|Y}^{(i)}(\tau'_n)$ ($i=1,2$). This is because their expressions all take the form of $\widetilde{\covar}_{X|Y}^{(i)}(\tau'_n)$ with $i = 1, 2$ multiplied by a $\hat{\gamma}_1$-dependent term, which implies that the disparities in their empirical performances are entirely determined by $\widetilde{\covar}_{X|Y}^{(i)}(\tau'_n)$ ($i=1,2$). However, the third extrapolation $\widetilde{\coes}_{X|Y}^{(3)}(\tau'_n)$ is observed to yield little larger MSRE values across most scenarios, which depends on the intermediate estimation $\widehat{\coes}_{X|Y}(1-k/n)$.

\begin{figure}[htbp]
\centering
\begin{minipage}[b]{0.4\textwidth}
\includegraphics[width=\textwidth,height = 0.2\textheight]{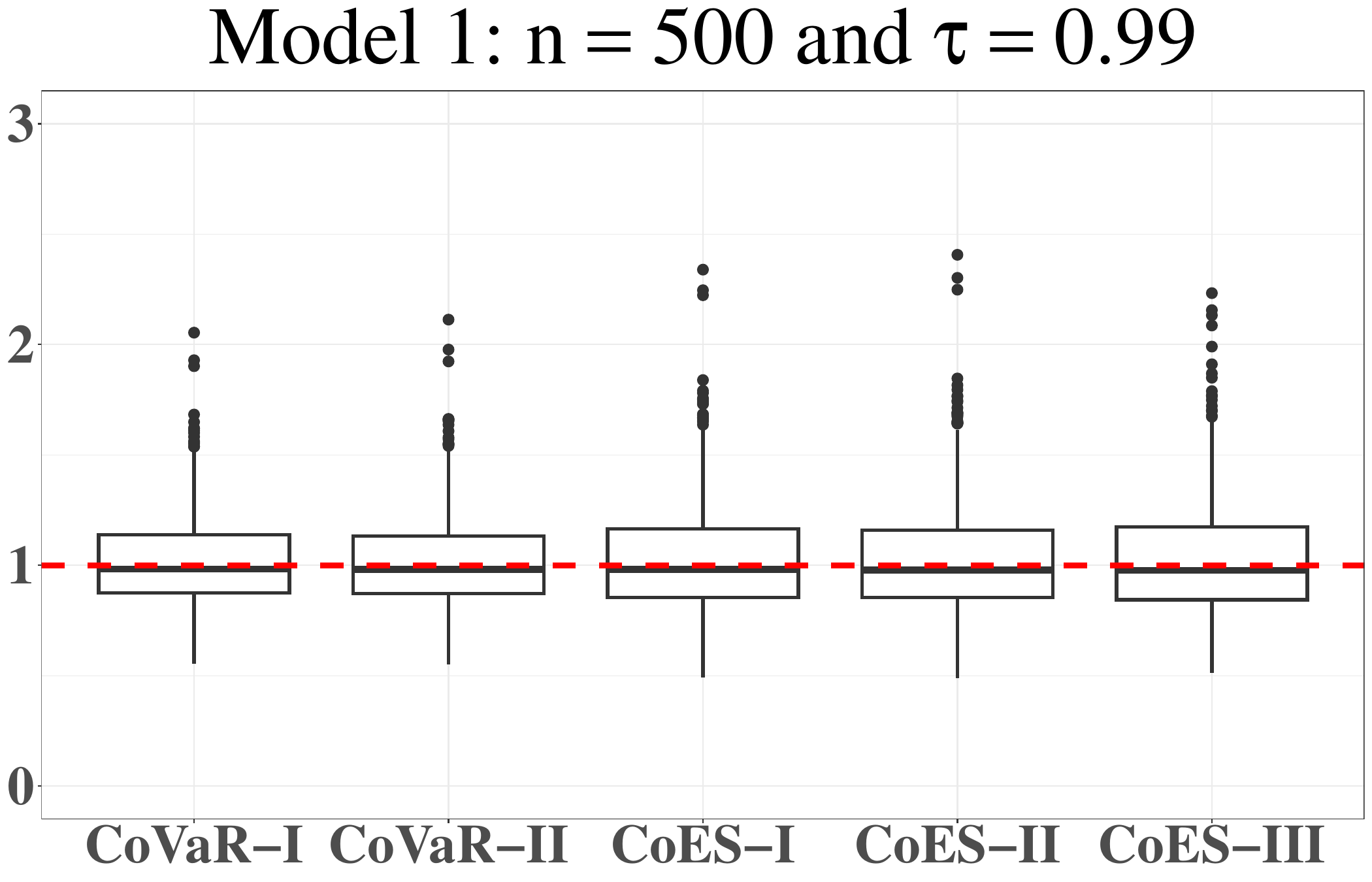}
\end{minipage}
\hspace{0.05\textwidth}
\begin{minipage}[b]{0.4\textwidth}
\includegraphics[width=\textwidth,height = 0.2\textheight]{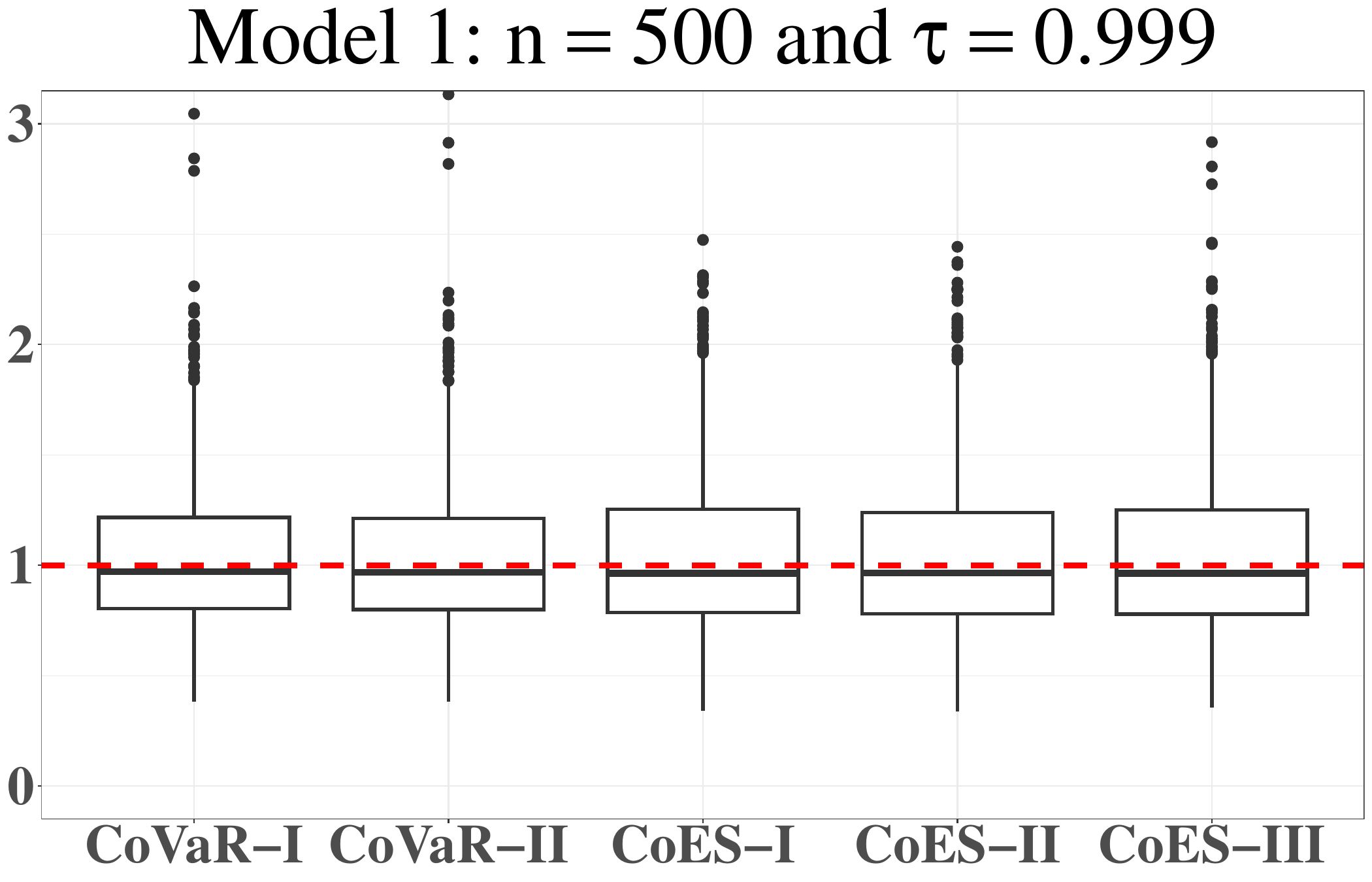}
\end{minipage}
\\[10pt]
\begin{minipage}[b]{0.4\textwidth}
\includegraphics[width=\textwidth,height = 0.2\textheight]{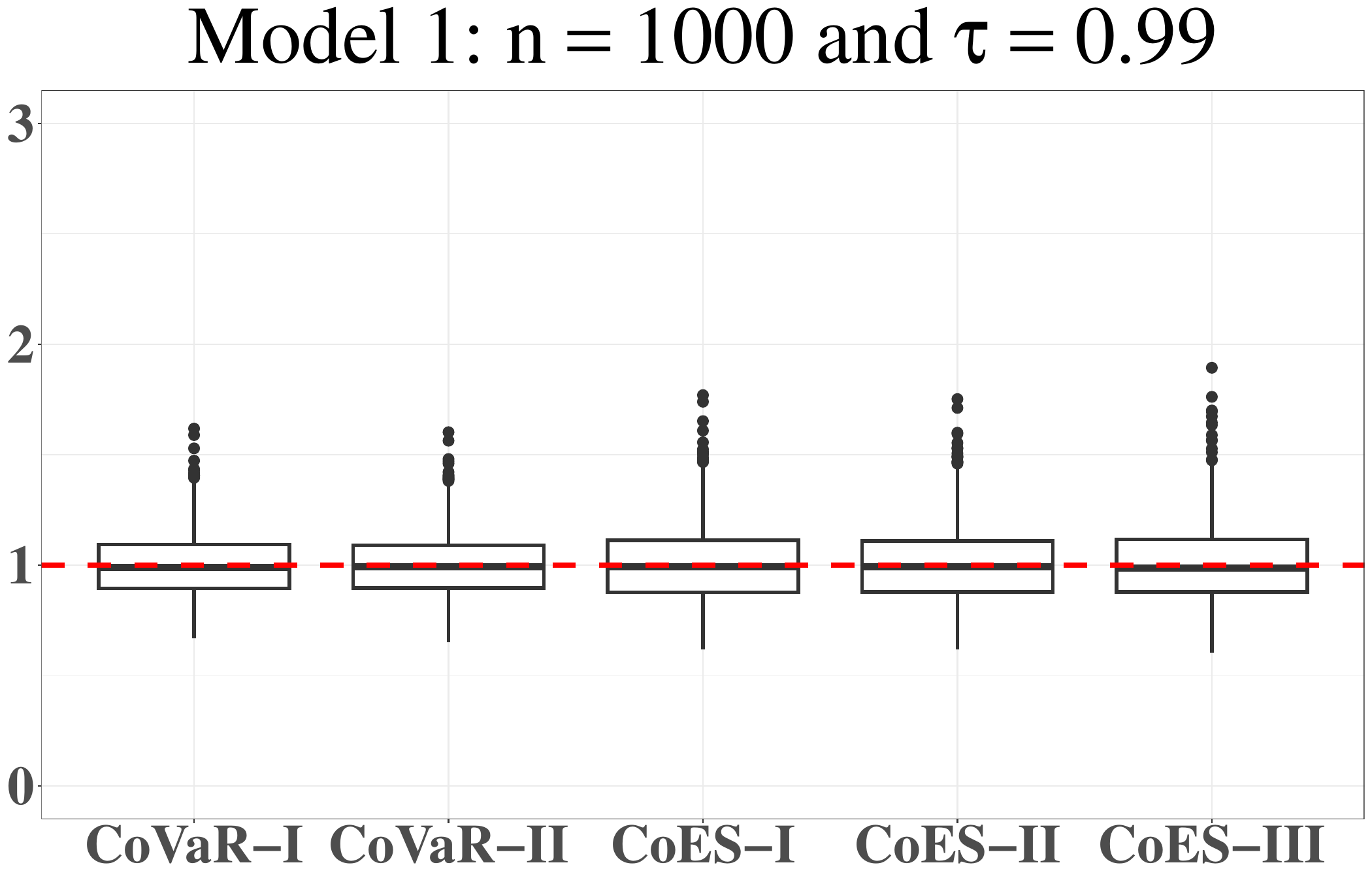}
\end{minipage}
\hspace{0.05\textwidth}
\begin{minipage}[b]{0.4\textwidth}
\includegraphics[width=\textwidth,height = 0.2\textheight]{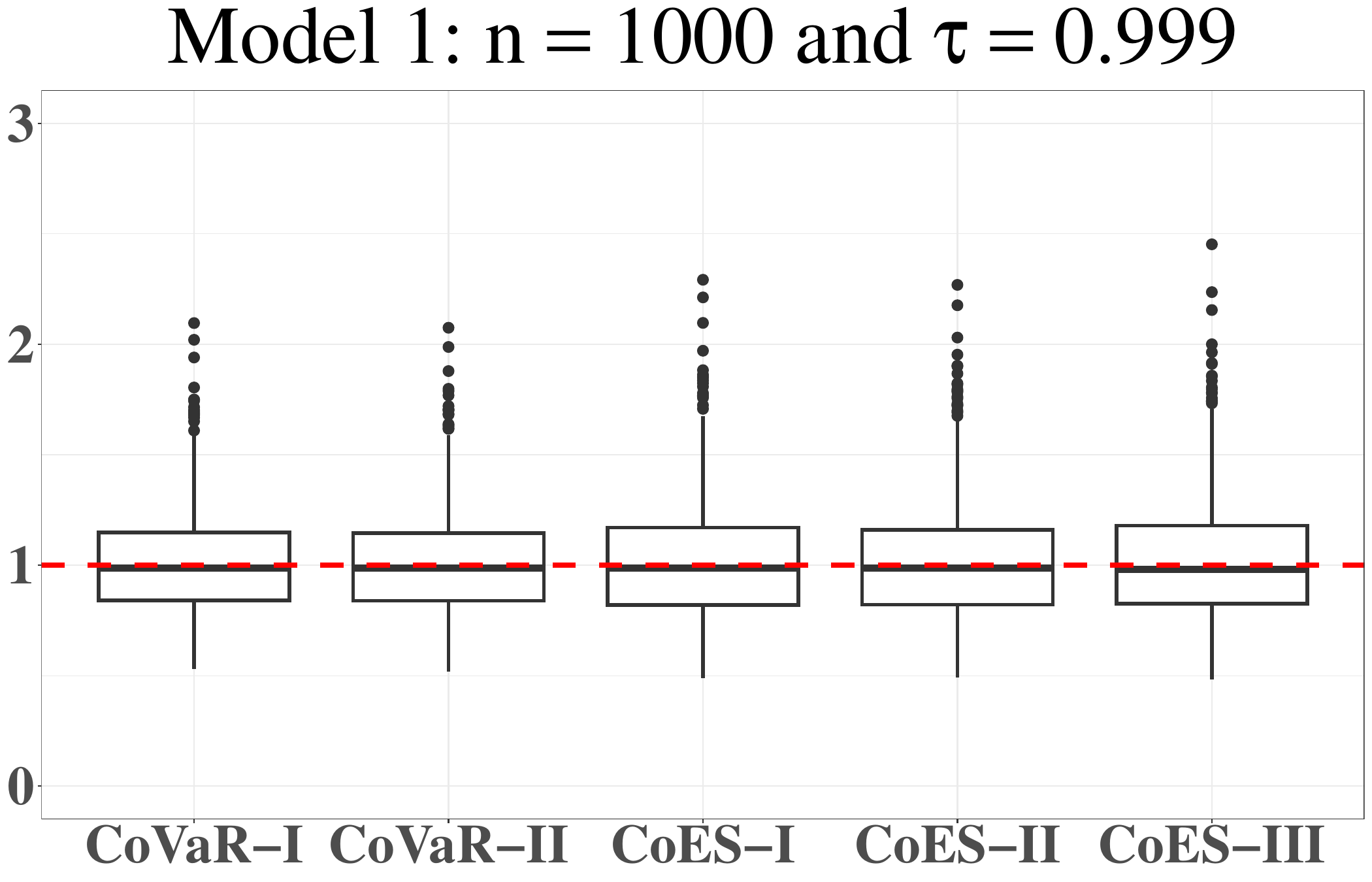}
\end{minipage}
\\[10pt]
\begin{minipage}[b]{0.4\textwidth}
\includegraphics[width=\textwidth,height = 0.2\textheight]{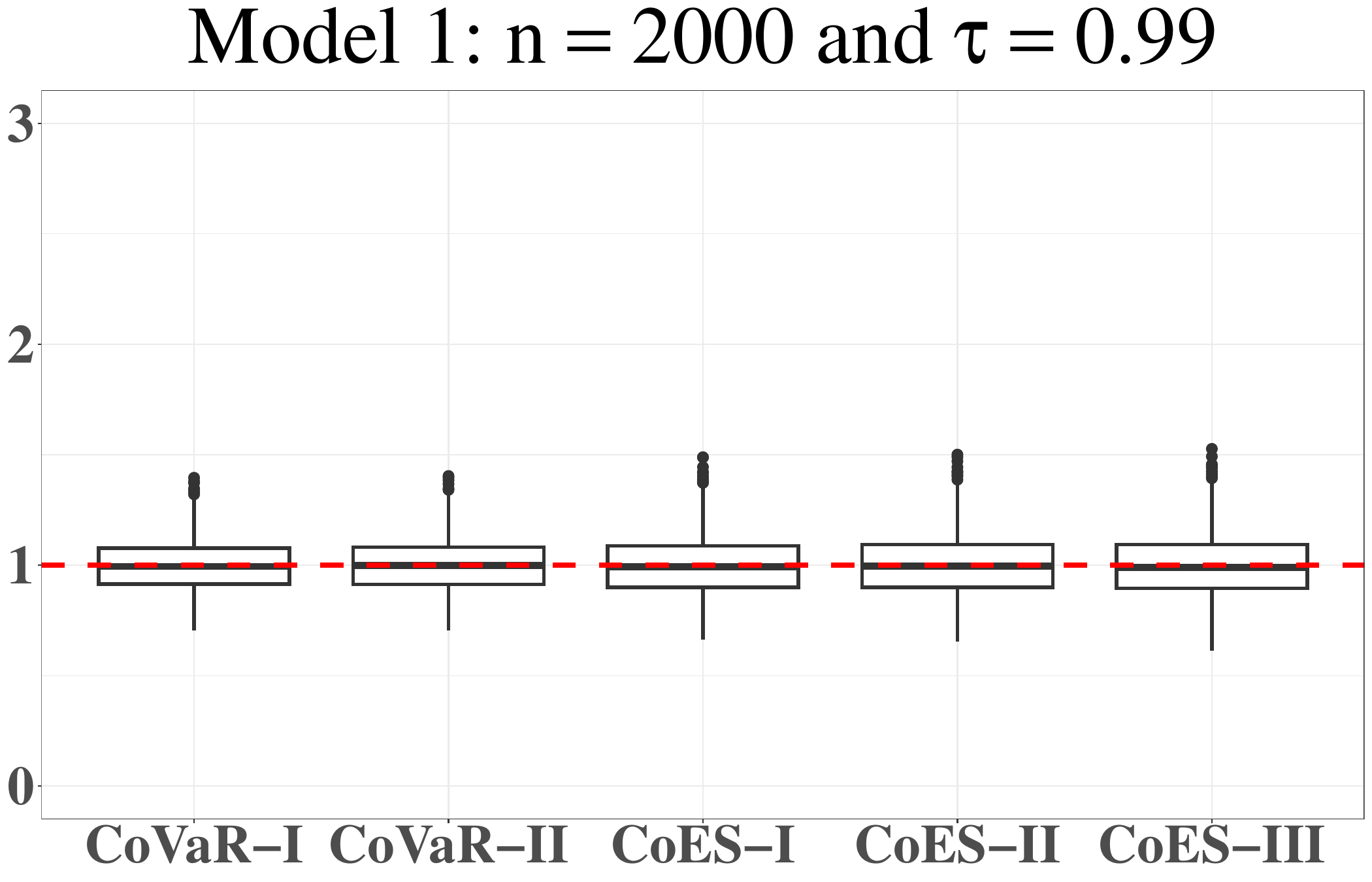}
\end{minipage}
\hspace{0.05\textwidth}
\begin{minipage}[b]{0.4\textwidth}
\includegraphics[width=\textwidth,height = 0.2\textheight]{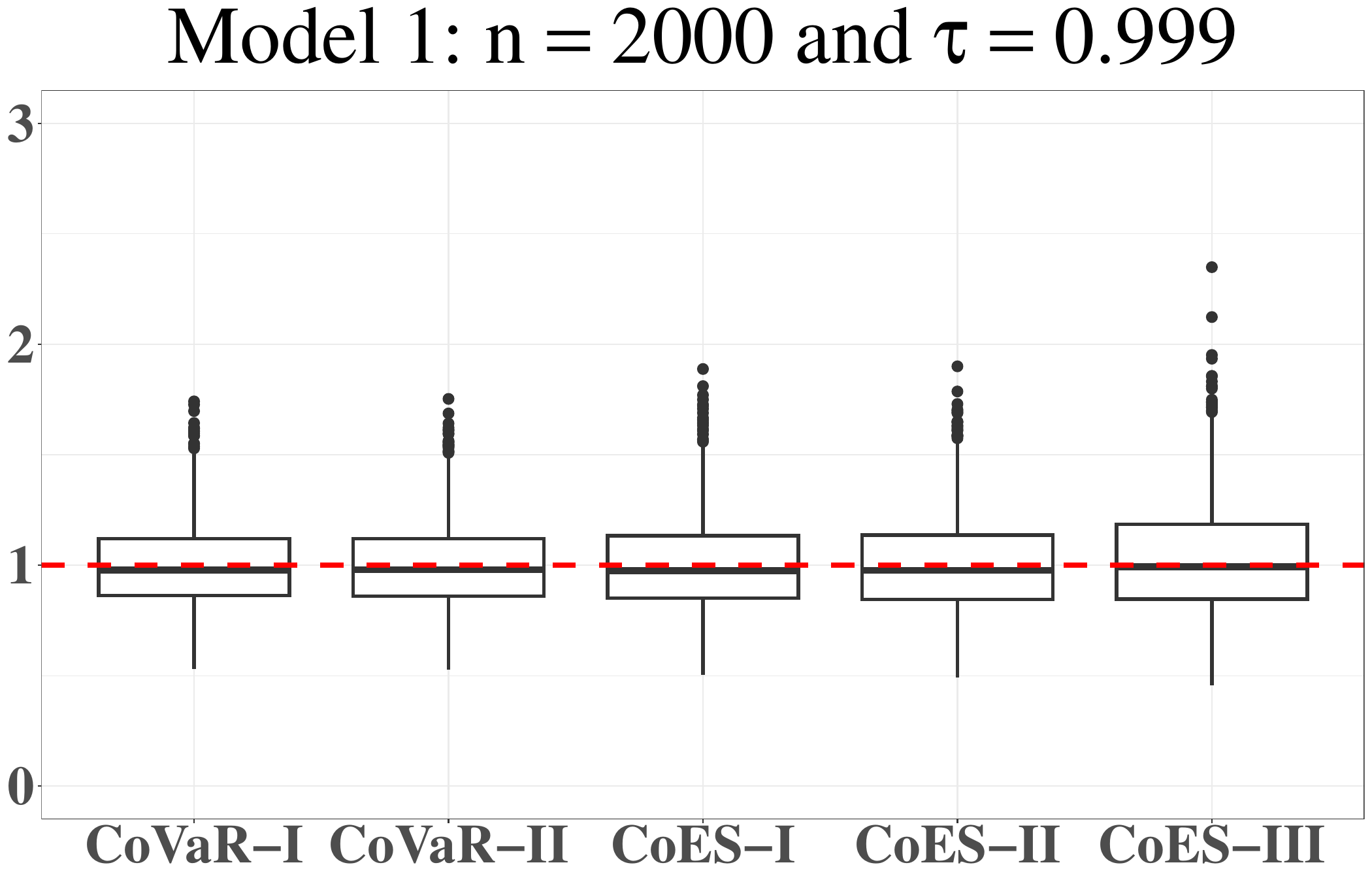}
\end{minipage}
\\[10pt]
\begin{minipage}[b]{0.4\textwidth}
\includegraphics[width=\textwidth,height = 0.2\textheight]{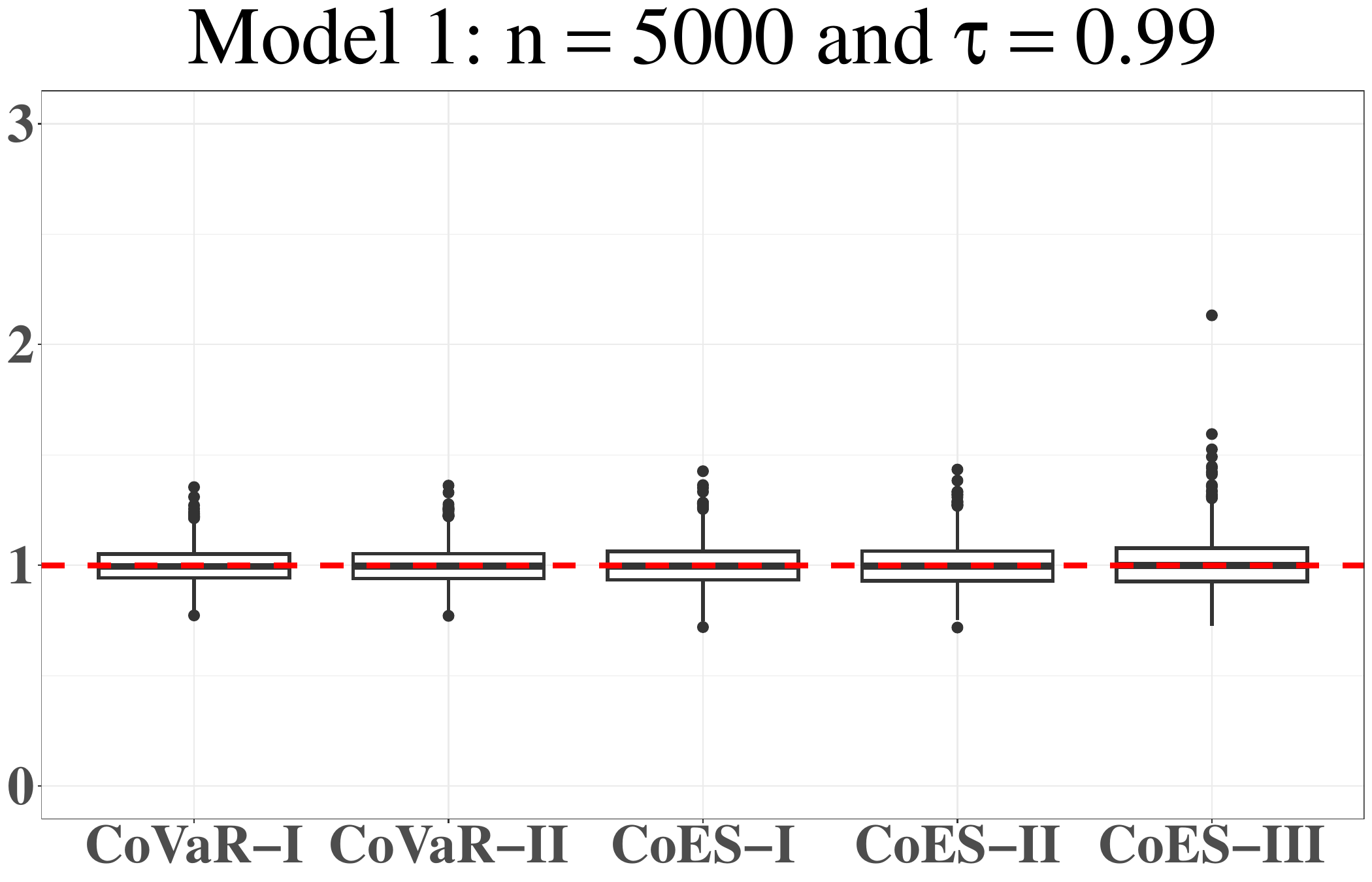}
\end{minipage}
\hspace{0.05\textwidth}
\begin{minipage}[b]{0.4\textwidth}
\includegraphics[width=\textwidth,height = 0.2\textheight]{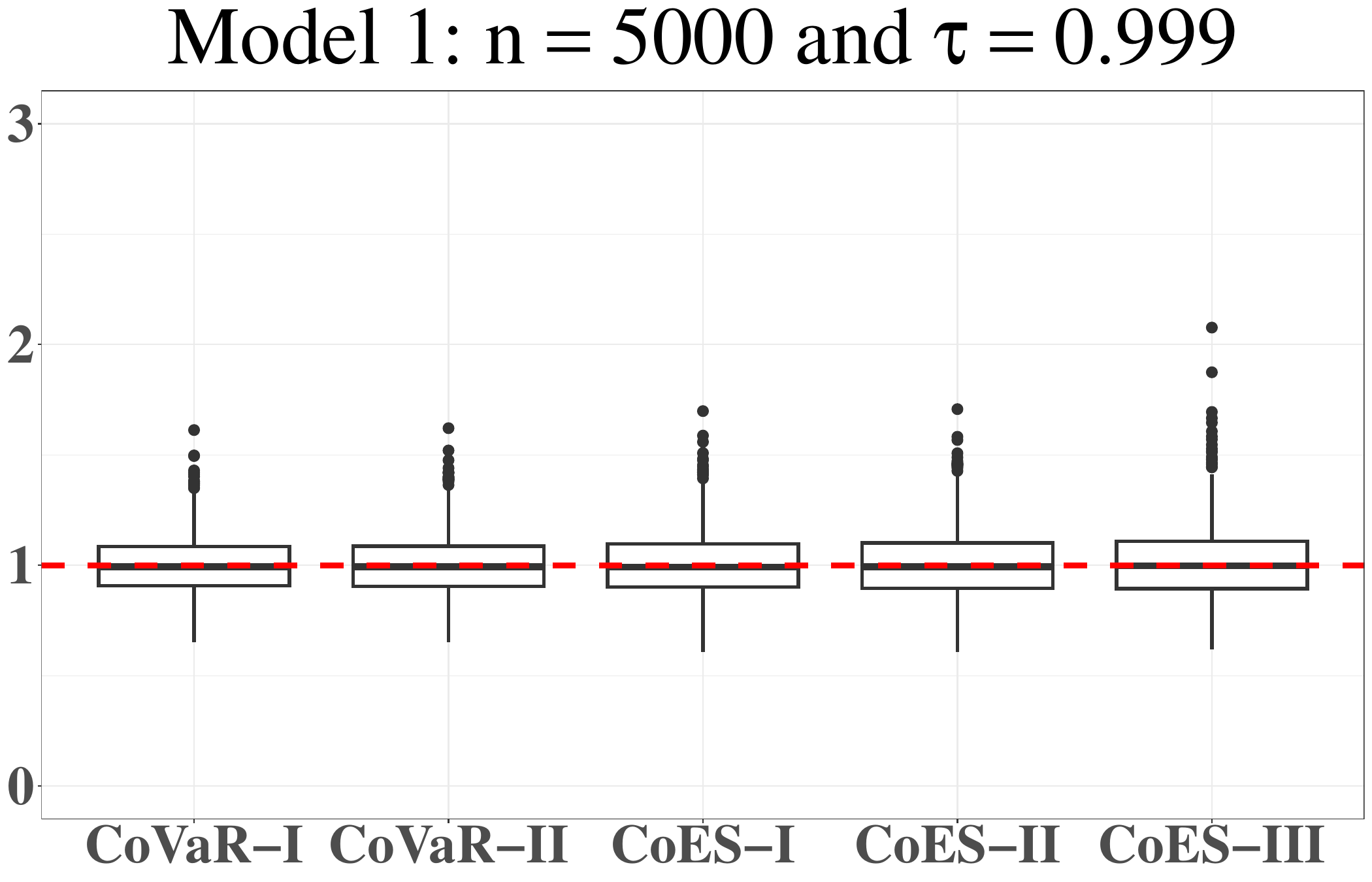}
\end{minipage}
\caption{Boxplots of the ratios between $\widetilde{\covar}_{X|Y}^{(i)}(\tau'_n)$ ($i=1,2$), $\widetilde{\coes}_{X|Y}^{(i)}(\tau'_n)$ ($i=1,2,3$) and their true values for \textbf{Model 1} with $n \in \{500, 1000, 2000, 5000\}$ and $\tau'_n \in \{0.99, 0.999\}$.}
\label{fig:model1}
\end{figure}

\begin{figure}[htbp]
\centering
\begin{minipage}[b]{0.4\textwidth}
\includegraphics[width=\textwidth,height = 0.2\textheight]{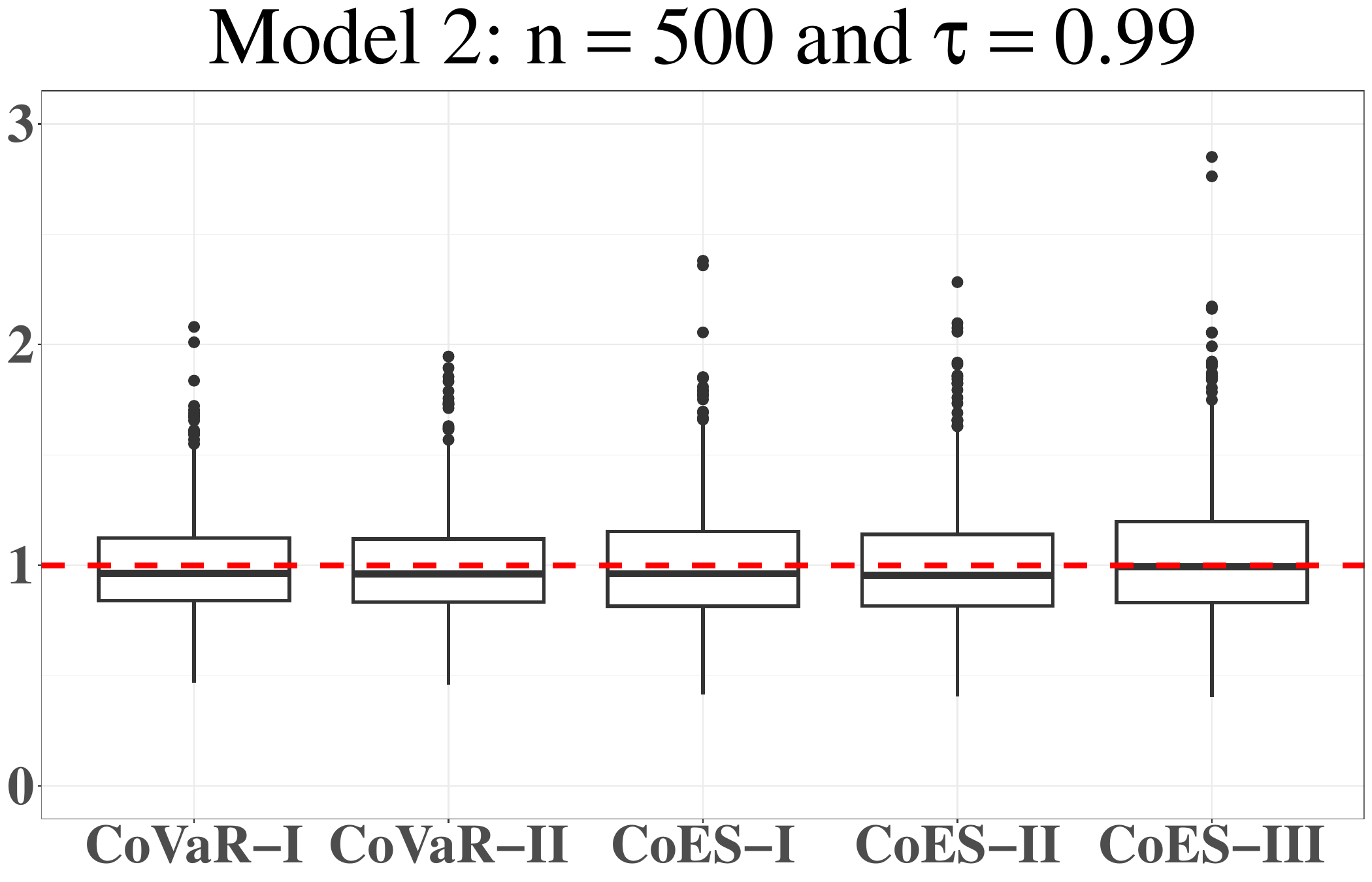}
\end{minipage}
\hspace{0.05\textwidth}
\begin{minipage}[b]{0.4\textwidth}
\includegraphics[width=\textwidth,height = 0.2\textheight]{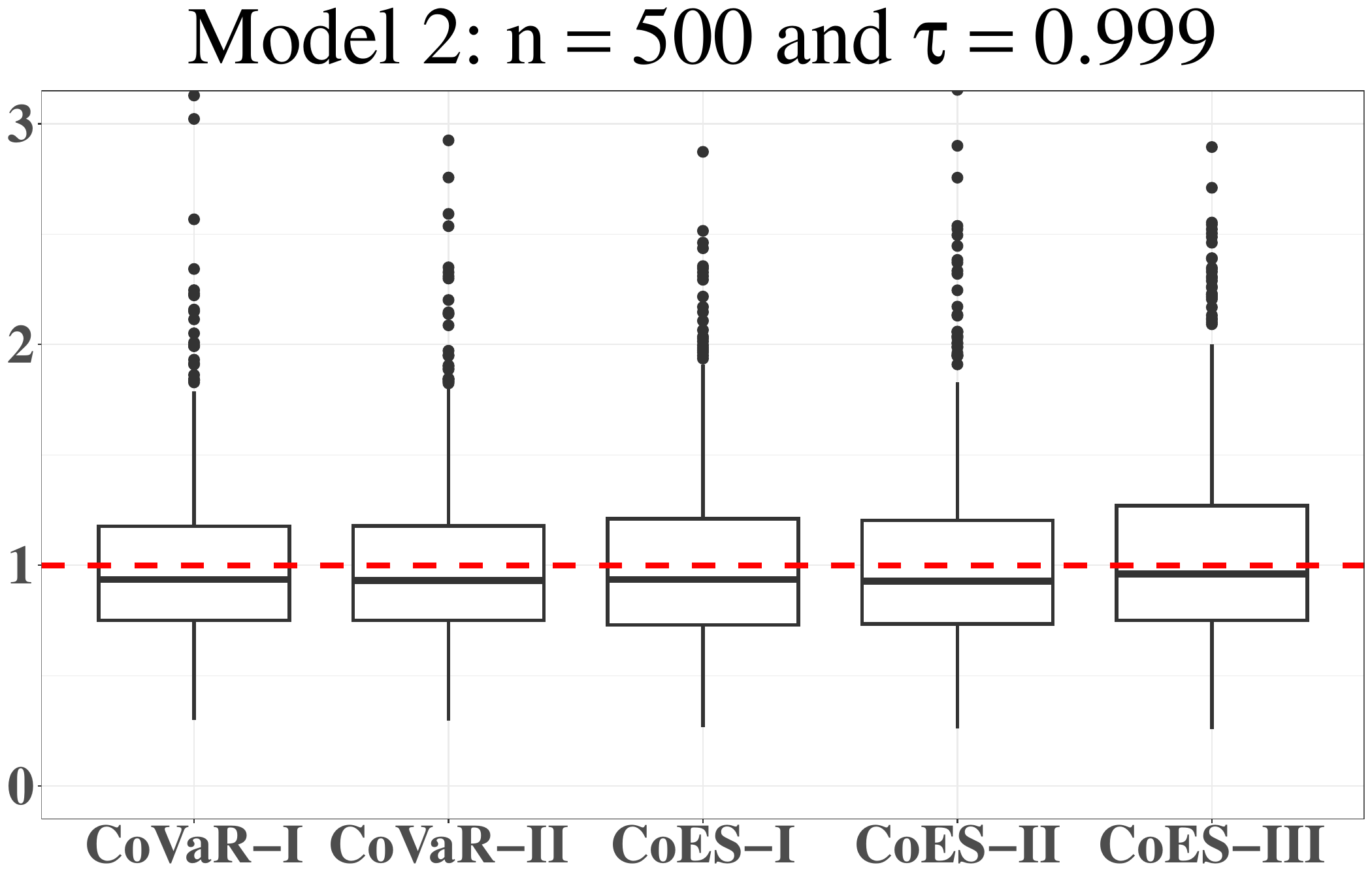}
\end{minipage}
\\[10pt]
\begin{minipage}[b]{0.4\textwidth}
\includegraphics[width=\textwidth,height = 0.2\textheight]{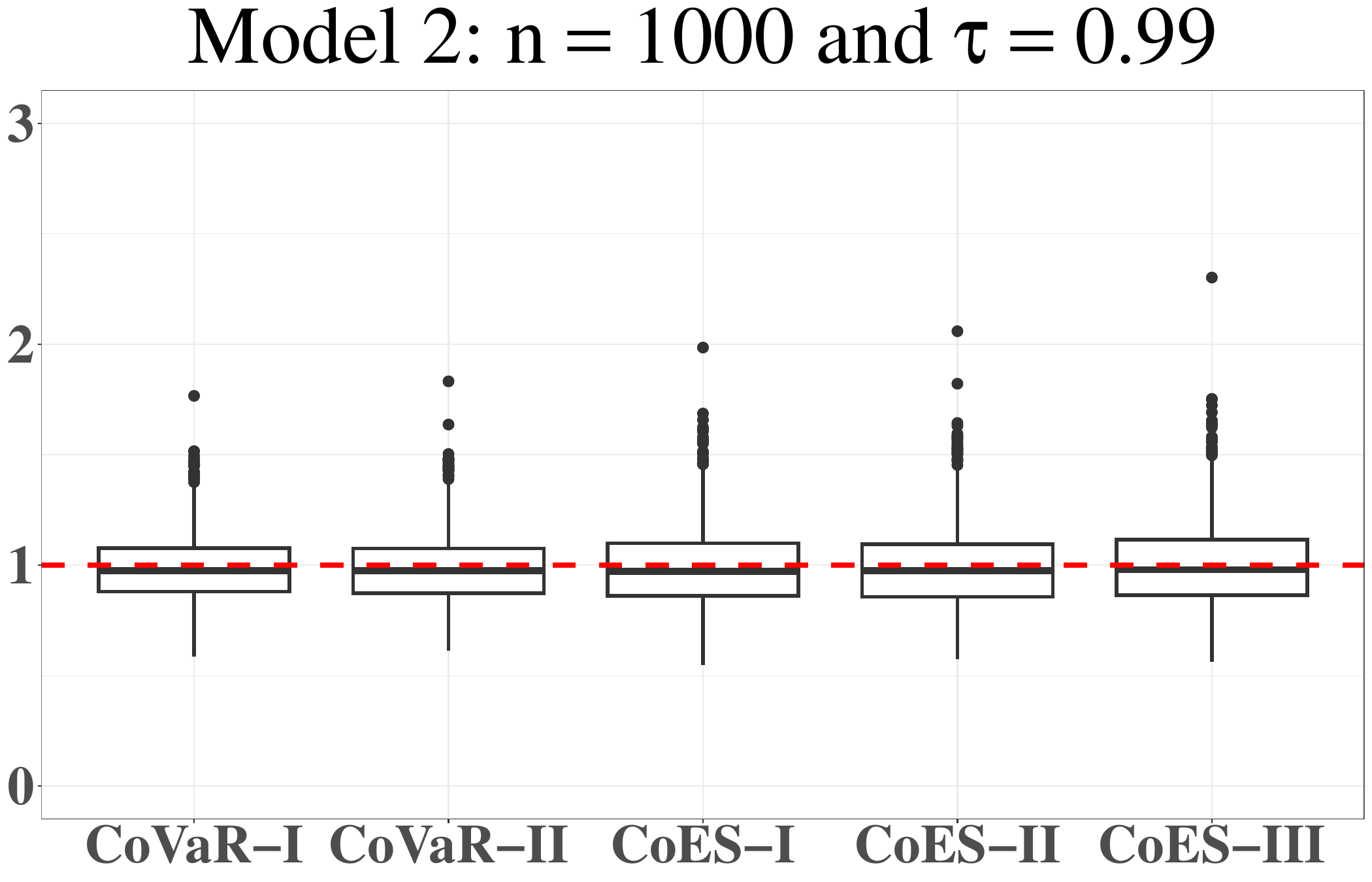}
\end{minipage}
\hspace{0.05\textwidth}
\begin{minipage}[b]{0.4\textwidth}
\includegraphics[width=\textwidth,height = 0.2\textheight]{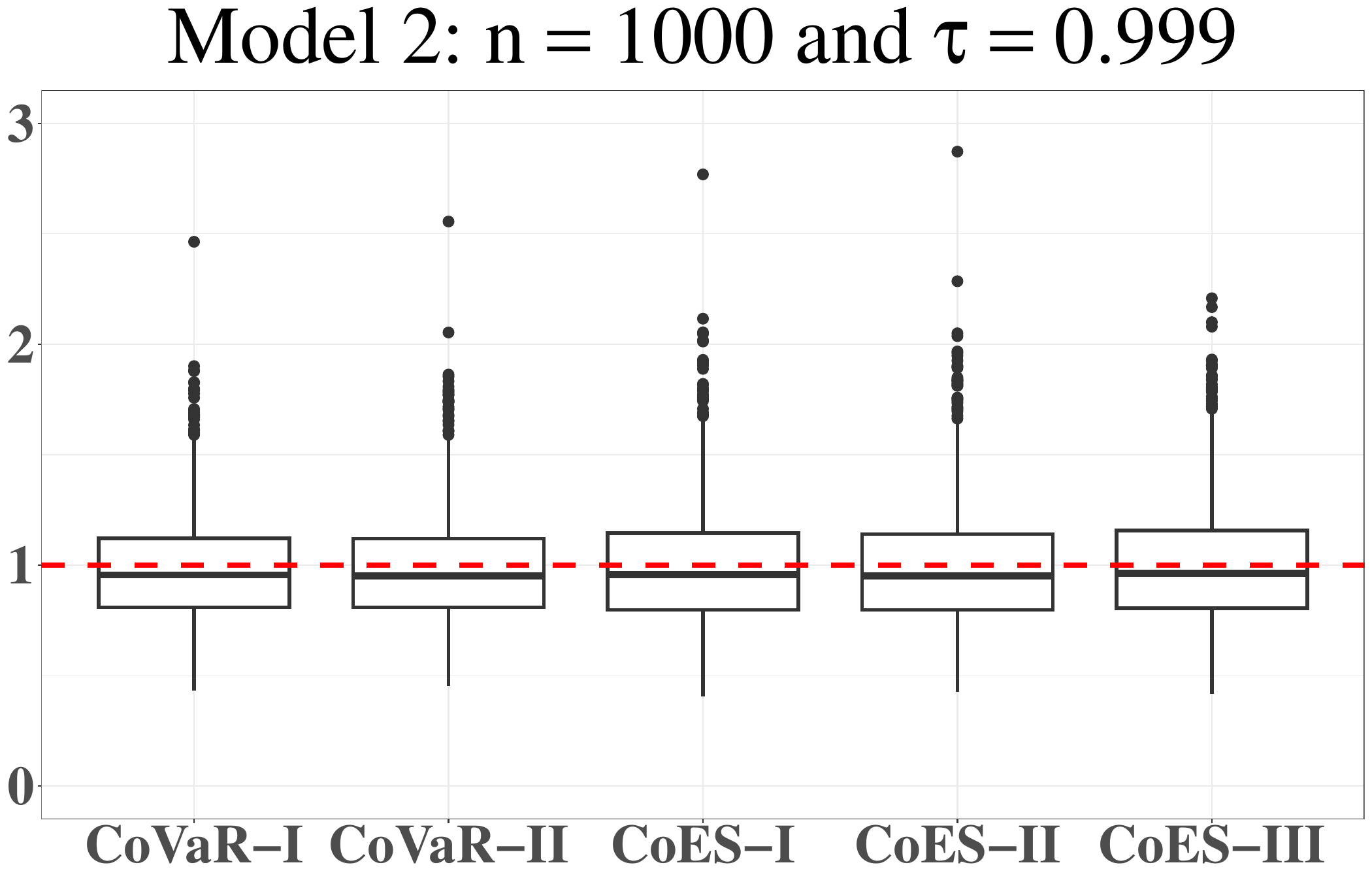}
\end{minipage}
\\[10pt]
\begin{minipage}[b]{0.4\textwidth}
\includegraphics[width=\textwidth,height = 0.2\textheight]{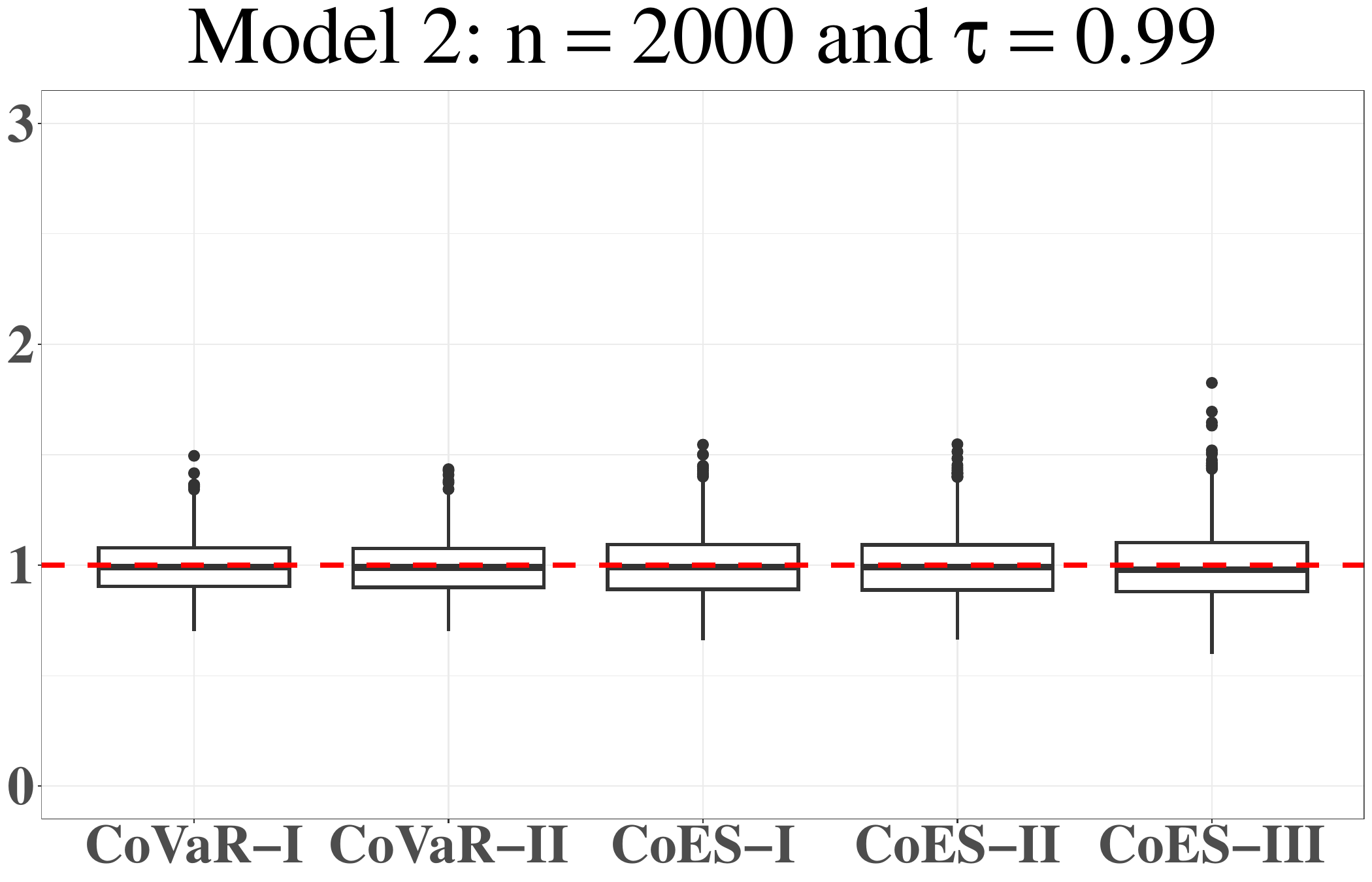}
\end{minipage}
\hspace{0.05\textwidth}
\begin{minipage}[b]{0.4\textwidth}
\includegraphics[width=\textwidth,height = 0.2\textheight]{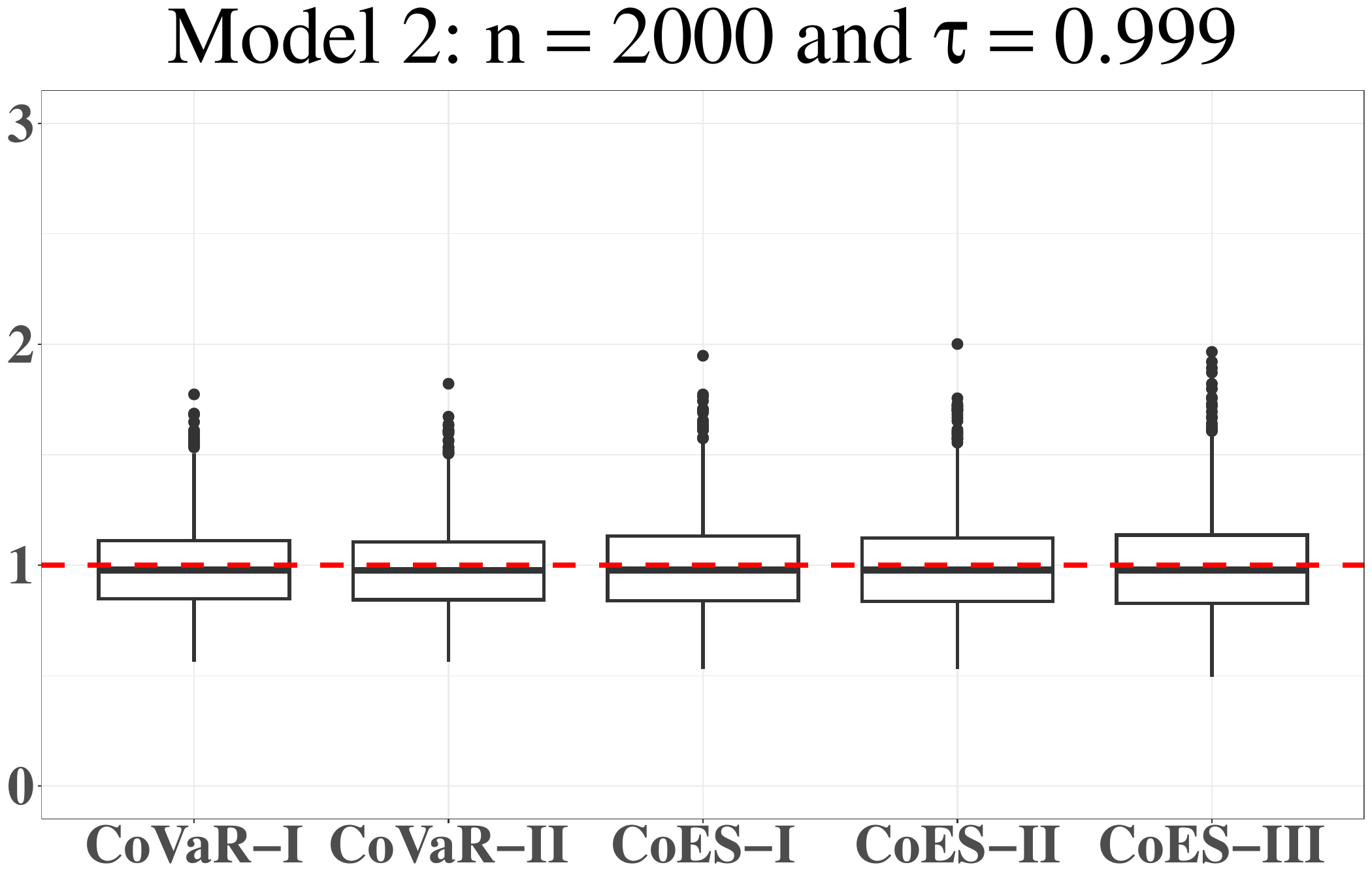}
\end{minipage}
\\[10pt]
\begin{minipage}[b]{0.4\textwidth}
\includegraphics[width=\textwidth,height = 0.2\textheight]{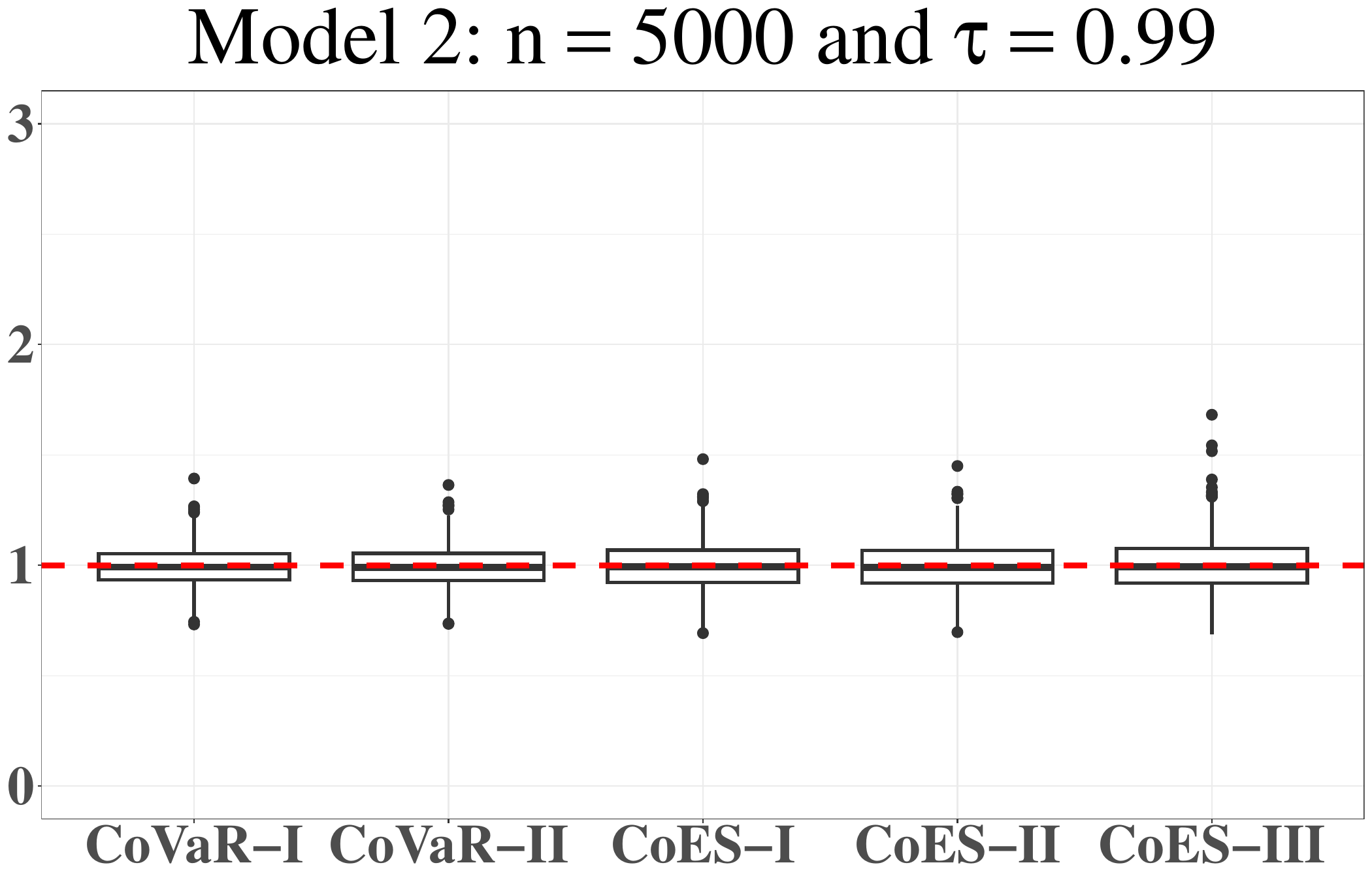}
\end{minipage}
\hspace{0.05\textwidth}
\begin{minipage}[b]{0.4\textwidth}
\includegraphics[width=\textwidth,height = 0.2\textheight]{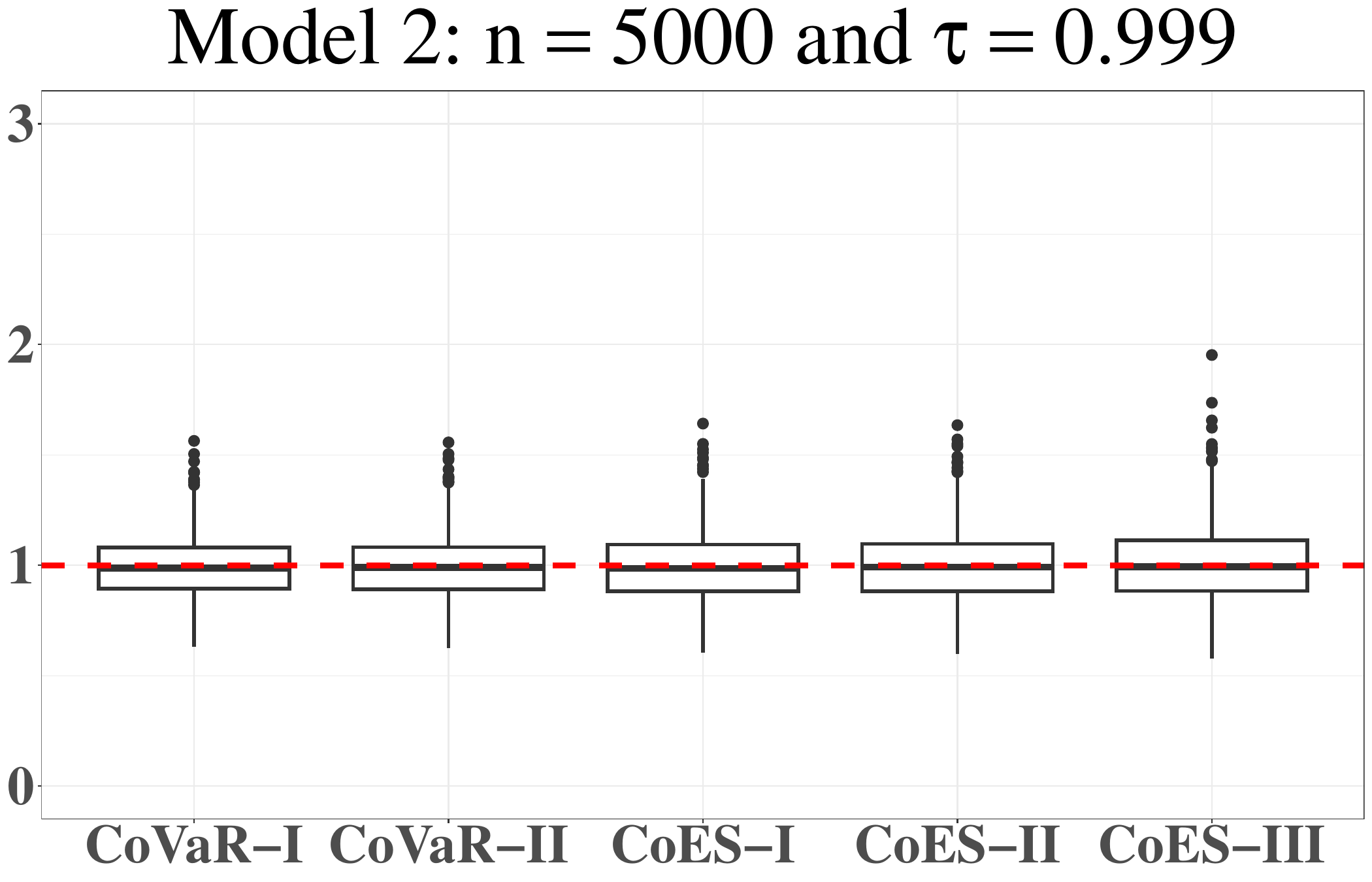}
\end{minipage}
\caption{Boxplots of the ratios between $\widetilde{\covar}_{X|Y}^{(i)}(\tau'_n)$ ($i=1,2$), $\widetilde{\coes}_{X|Y}^{(i)}(\tau'_n)$ ($i=1,2,3$) and their true values for \textbf{Model 2} with $n \in \{500, 1000, 2000, 5000\}$ and $\tau'_n \in \{0.99, 0.999\}$.}
\label{fig:model2}
\end{figure}

\begin{figure}[htbp]
\centering
\begin{minipage}[b]{0.4\textwidth}
\includegraphics[width=\textwidth,height = 0.2\textheight]{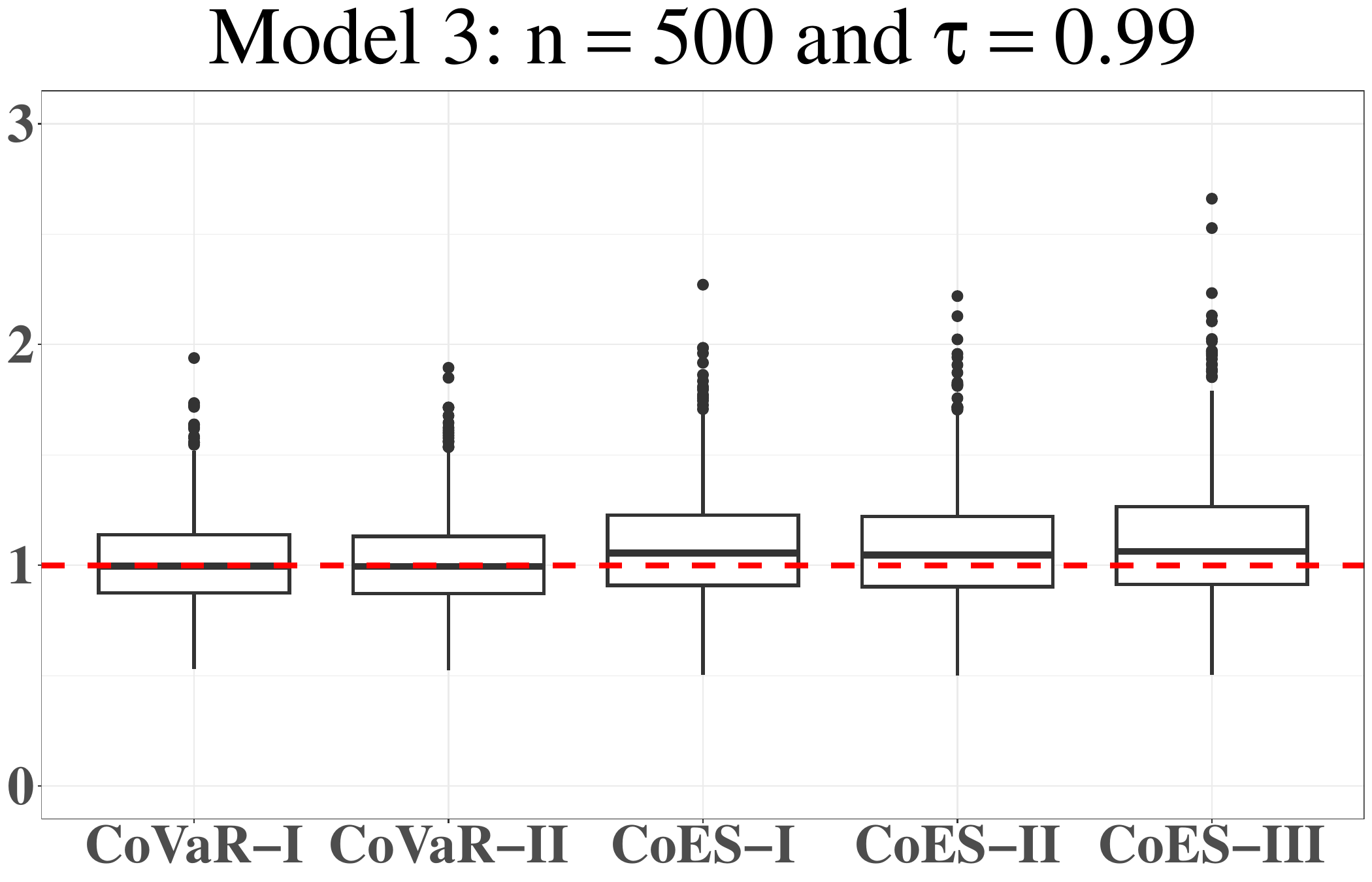}
\end{minipage}
\hspace{0.05\textwidth}
\begin{minipage}[b]{0.4\textwidth}
\includegraphics[width=\textwidth,height = 0.2\textheight]{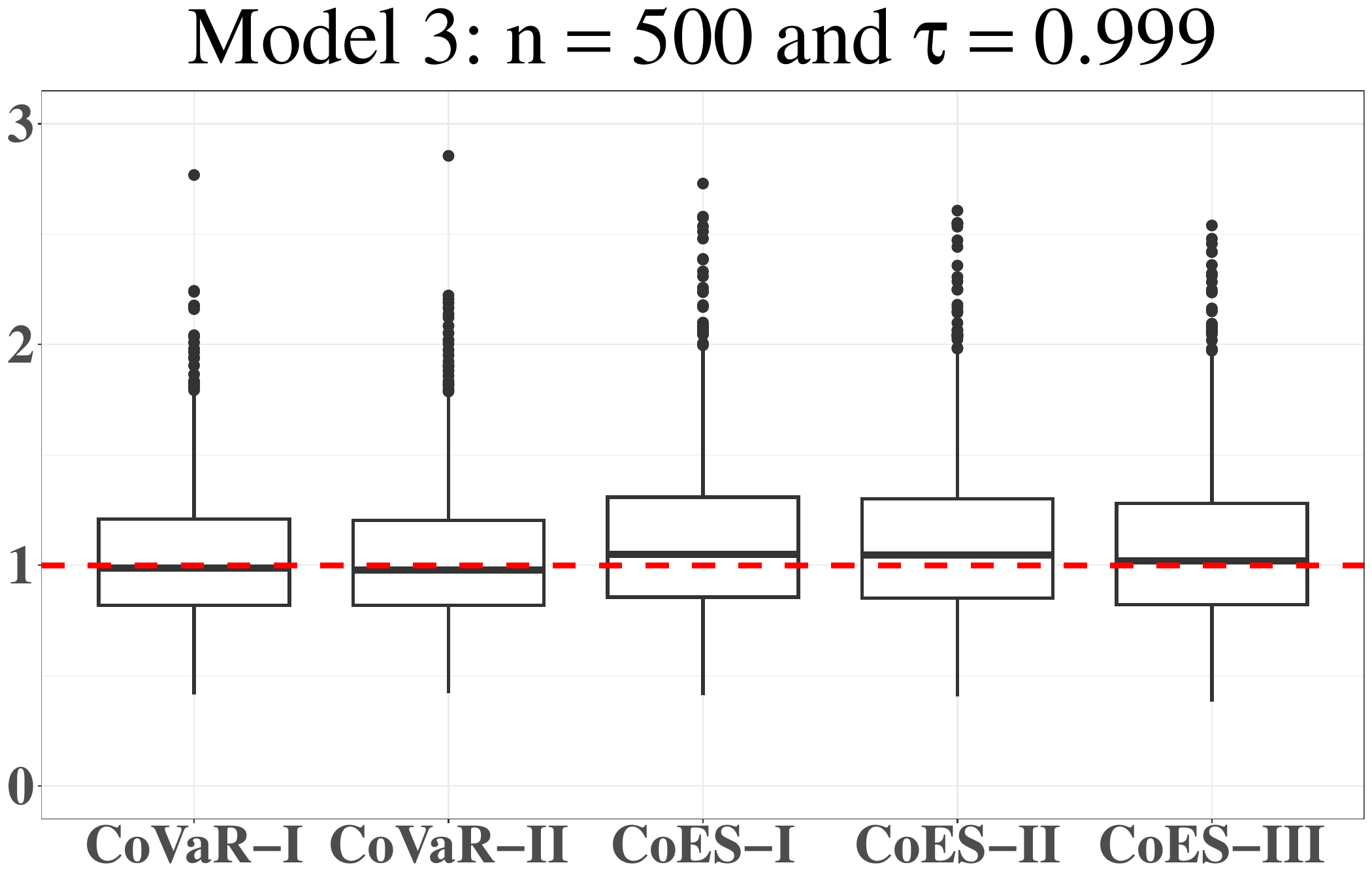}
\end{minipage}
\\[10pt]
\begin{minipage}[b]{0.4\textwidth}
\includegraphics[width=\textwidth,height = 0.2\textheight]{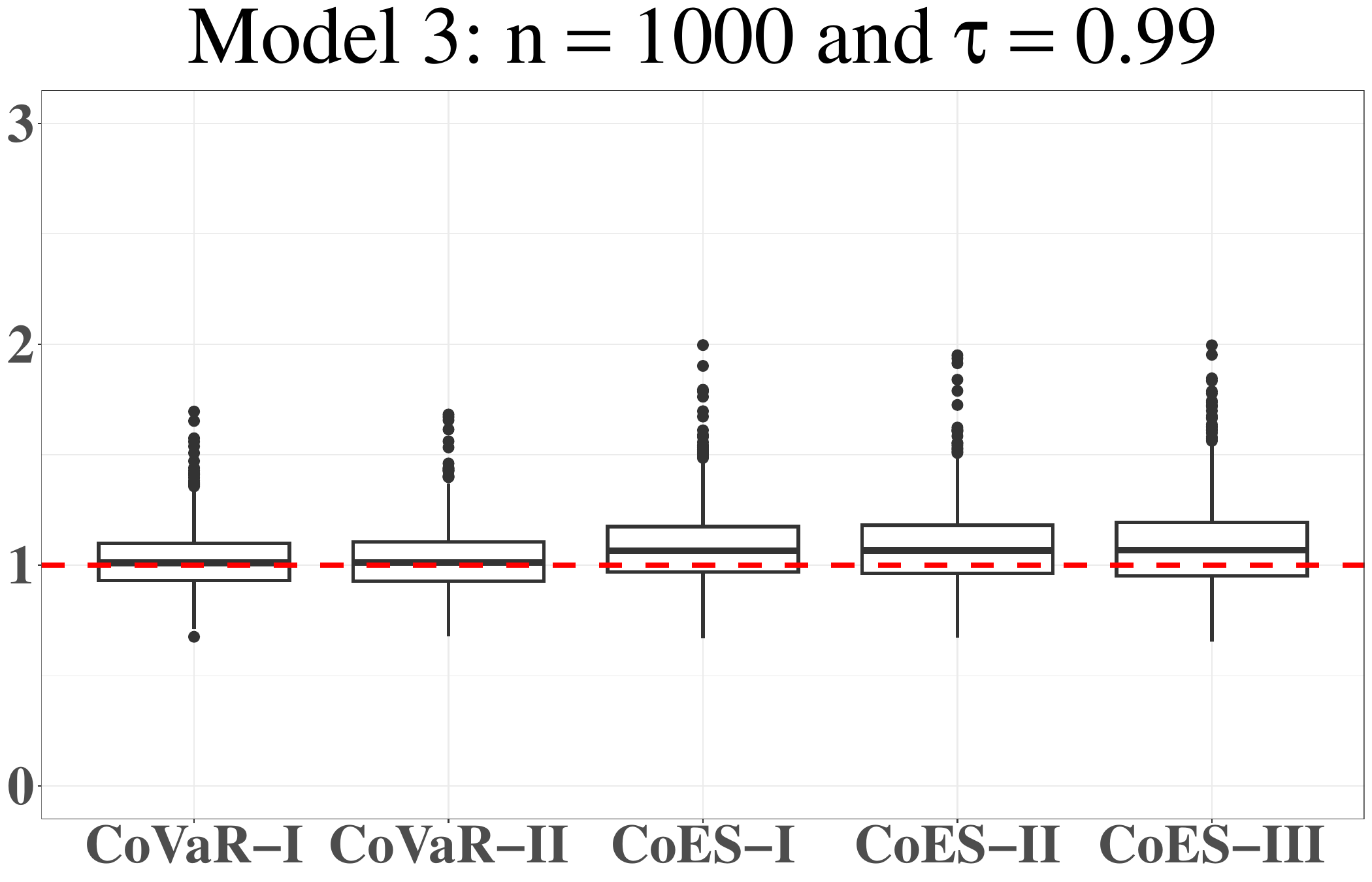}
\end{minipage}
\hspace{0.05\textwidth}
\begin{minipage}[b]{0.4\textwidth}
\includegraphics[width=\textwidth,height = 0.2\textheight]{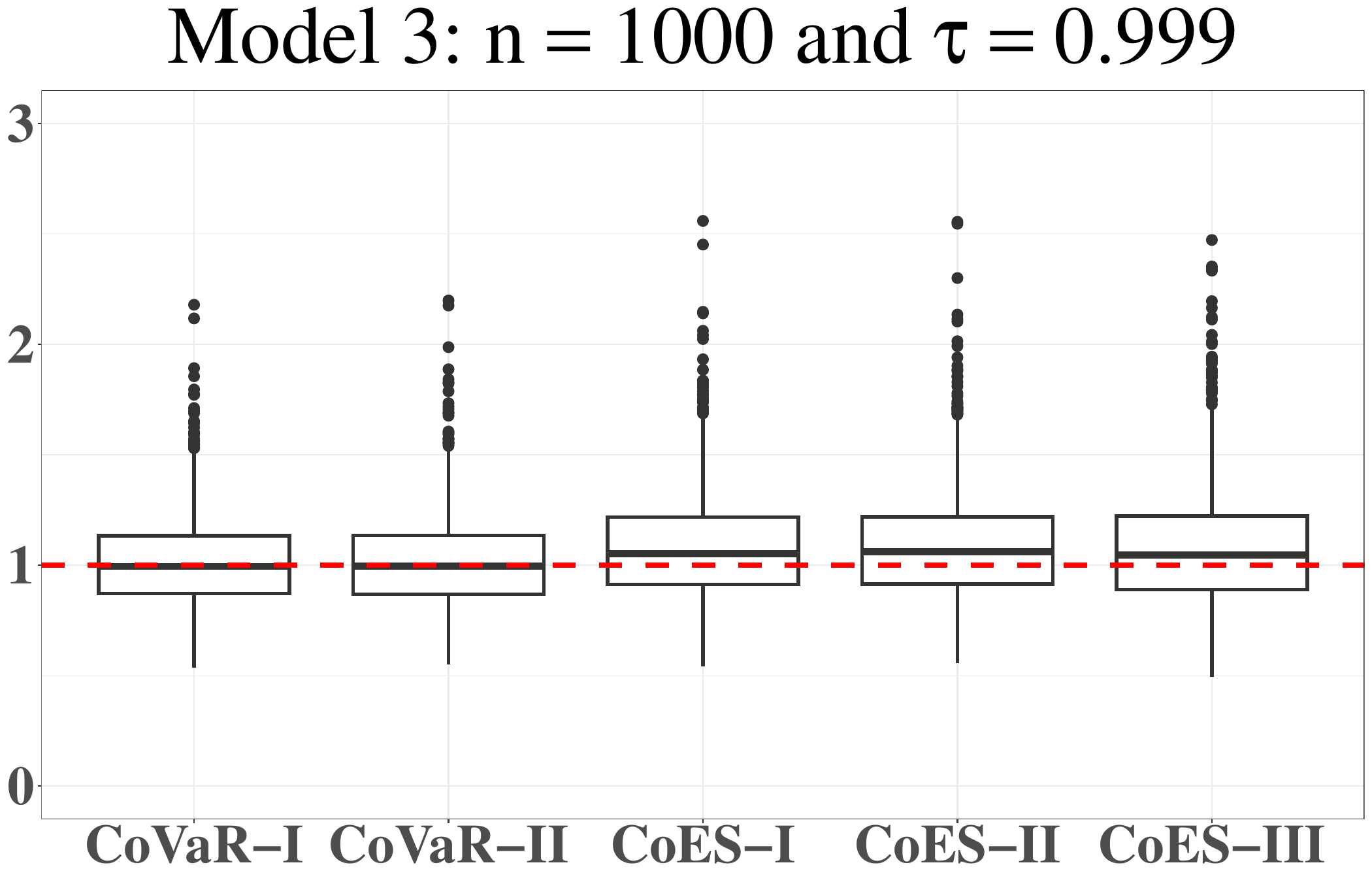}
\end{minipage}
\\[10pt]
\begin{minipage}[b]{0.4\textwidth}
\includegraphics[width=\textwidth,height = 0.2\textheight]{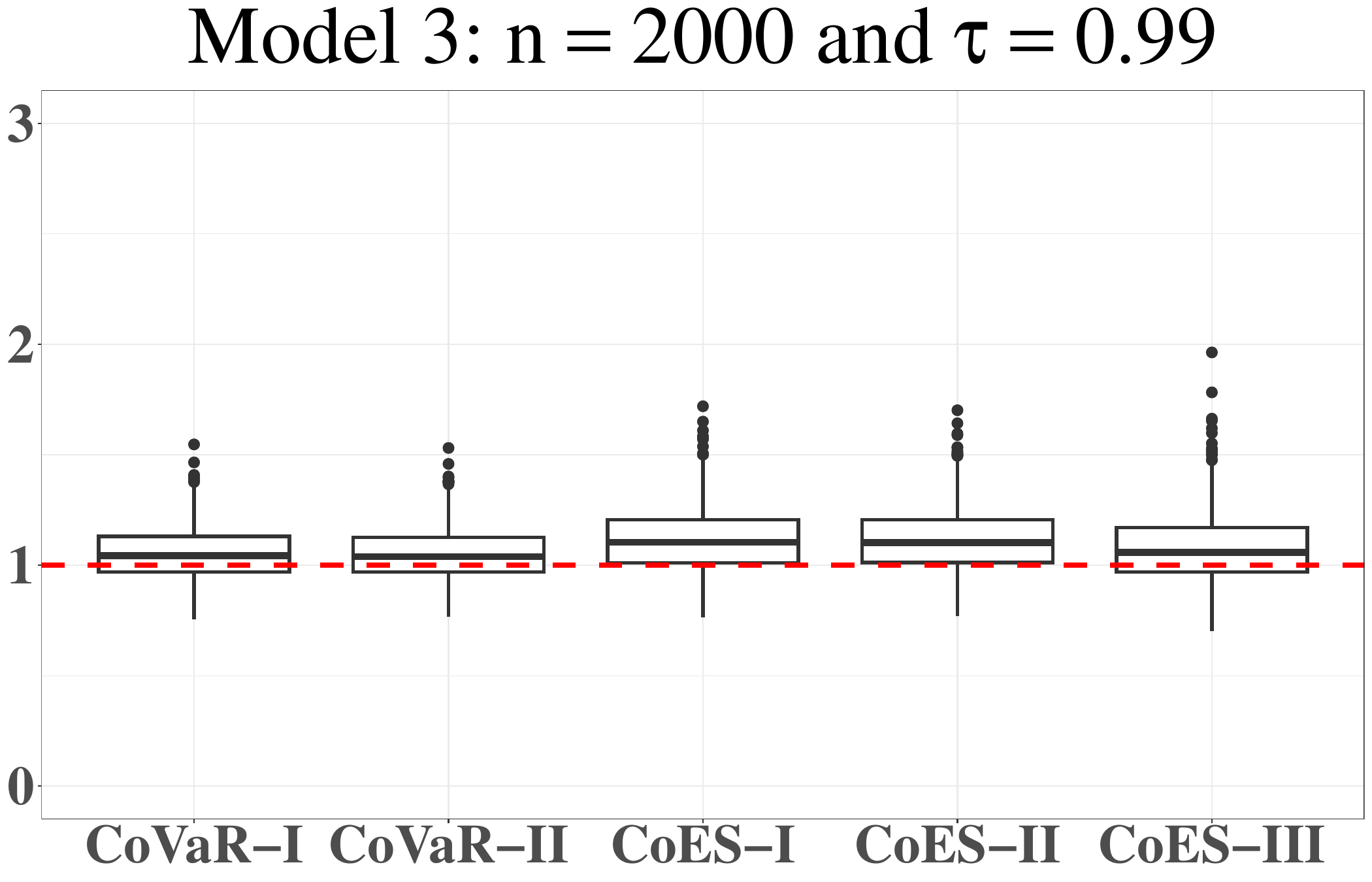}
\end{minipage}
\hspace{0.05\textwidth}
\begin{minipage}[b]{0.4\textwidth}
\includegraphics[width=\textwidth,height = 0.2\textheight]{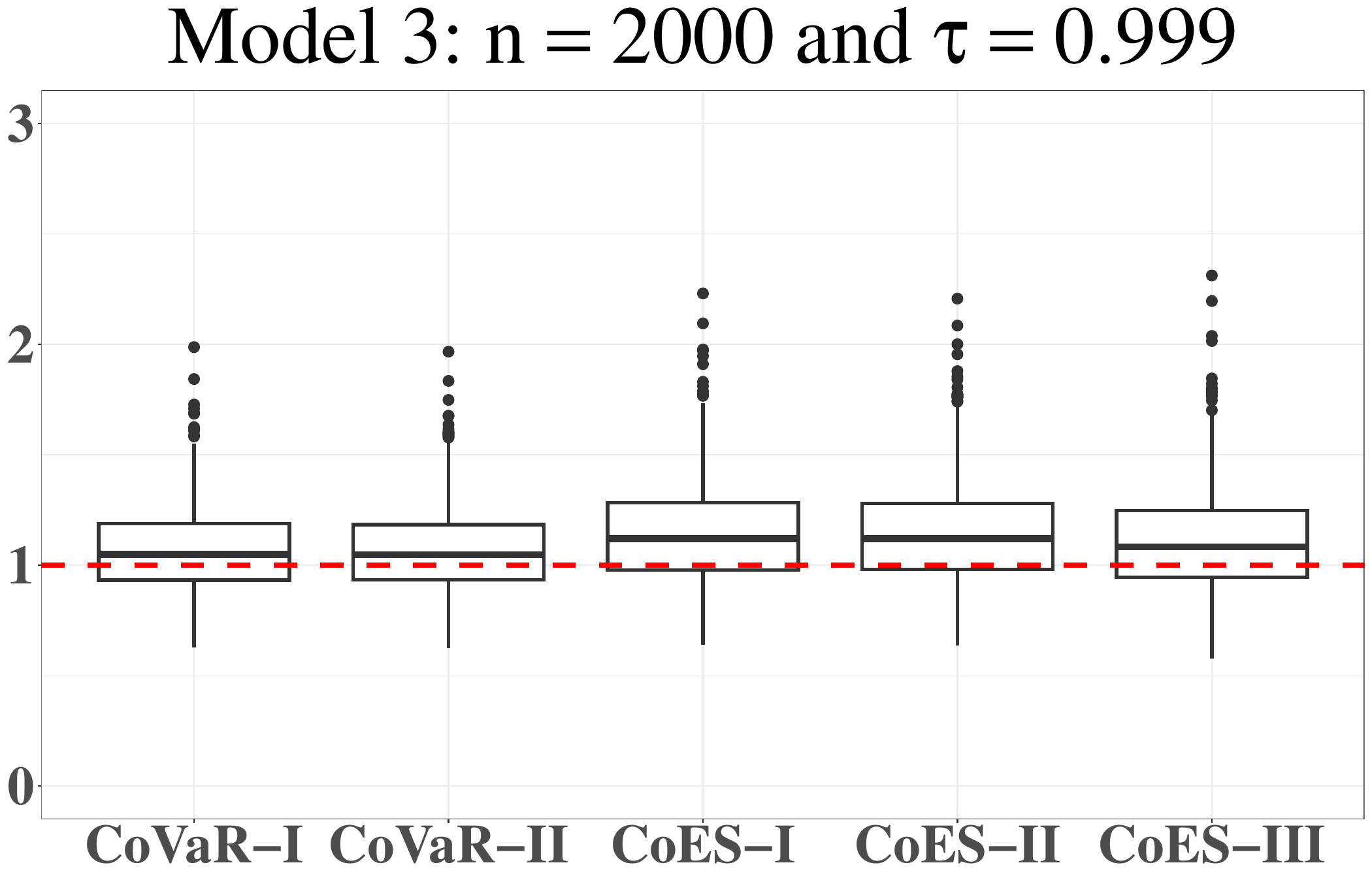}
\end{minipage}
\\[10pt]
\begin{minipage}[b]{0.4\textwidth}
\includegraphics[width=\textwidth,height = 0.2\textheight]{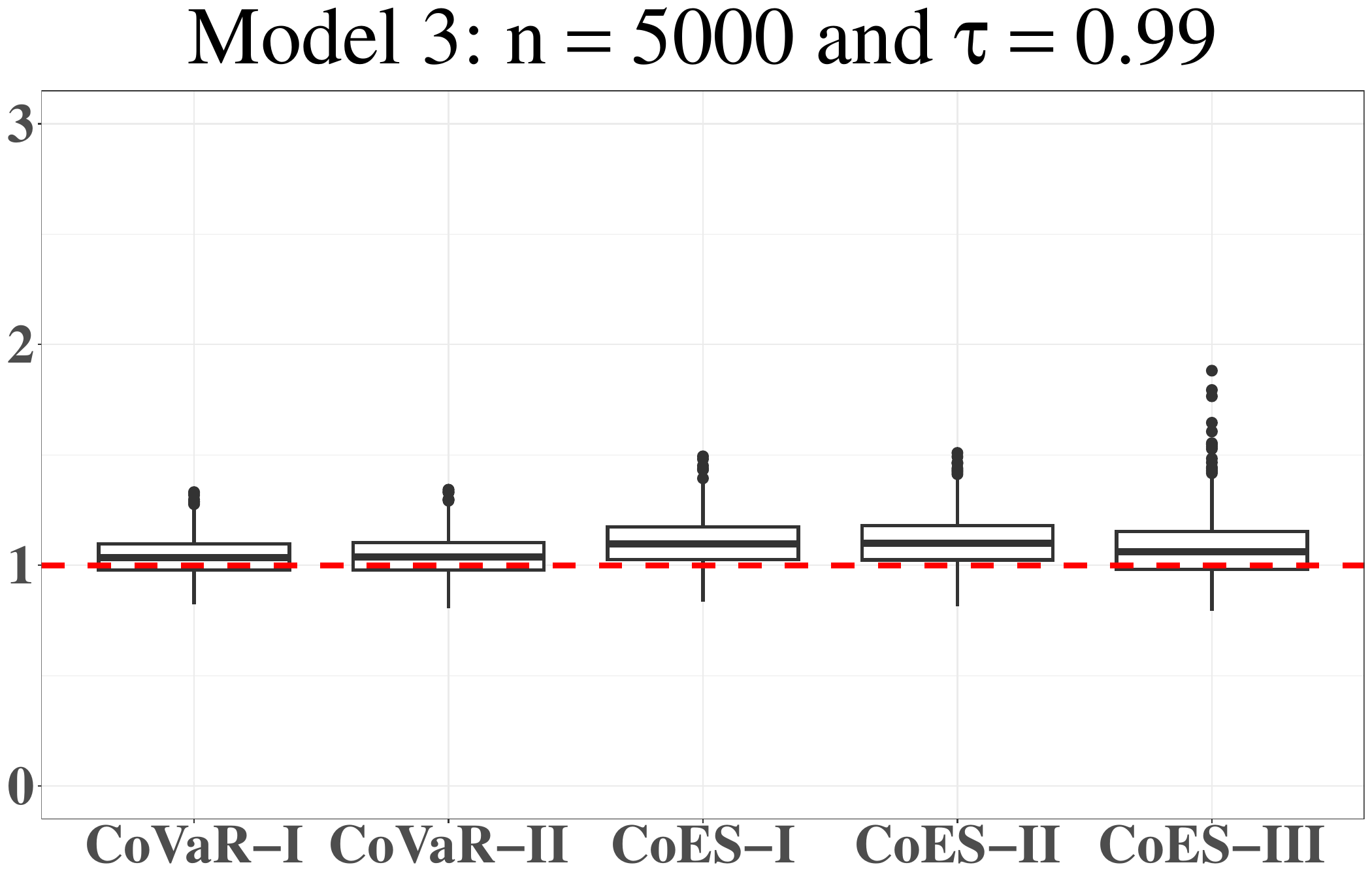}
\end{minipage}
\hspace{0.05\textwidth}
\begin{minipage}[b]{0.4\textwidth}
\includegraphics[width=\textwidth,height = 0.2\textheight]{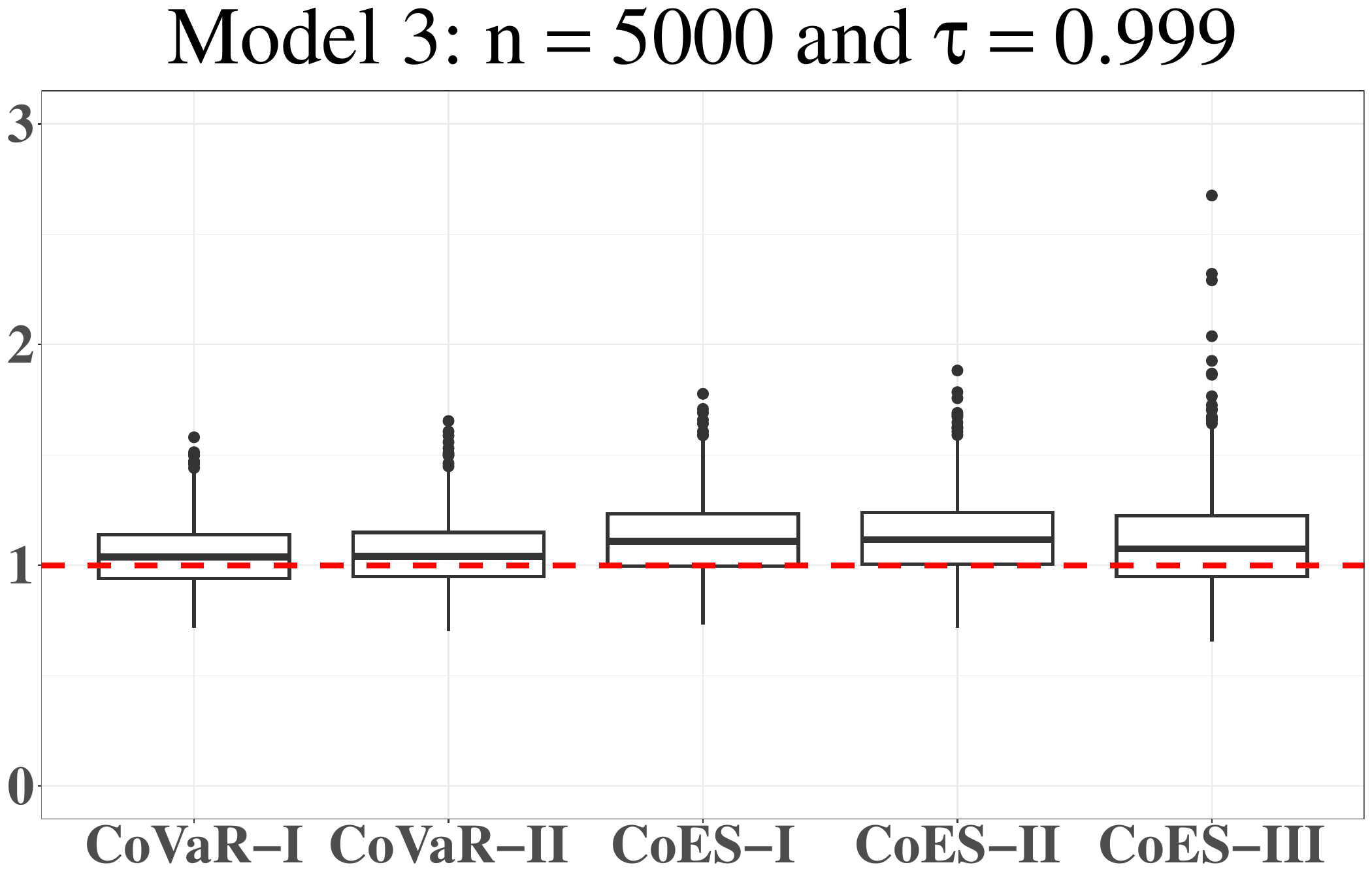}
\end{minipage}
\caption{Boxplots of the ratios between $\widetilde{\covar}_{X|Y}^{(i)}(\tau'_n)$ ($i=1,2$), $\widetilde{\coes}_{X|Y}^{(i)}(\tau'_n)$ ($i=1,2,3$) and their true values for \textbf{Model 3} with $n \in \{500, 1000, 2000, 5000\}$ and $\tau'_n \in \{0.99, 0.999\}$.}
\label{fig:model3}
\end{figure}

We further present a series of boxplots of the ratios between the estimations $\widetilde{\covar}_{X|Y}^{(i)}(\tau'_n)$, $\widetilde{\coes}_{X|Y}^{(i)}(\tau'_n)$ and the true values $\covar_{X|Y}(\tau'_n)$, $\coes_{X|Y}(\tau'_n)$ in Figures \ref{fig:model1} - \ref{fig:model3}, which reflect an intuitive comparison of consistency among these methods. From a global perspective, all extrapolative methods demonstrate a more concentrated pattern as the sample size grows, characterized by lower variation. Furthermore, the behavior of our proposed extrapolations remains highly consistent and stable. A somewhat surprising finding is that the extrapolations for Model 3 empirically outperform the first two models in certain scenarios, particularly with small samples. However, it shows a less robust profile with larger sample sizes, yielding higher MSRE values and more upward skewed boxplots than its predecessors, which possibly stems from the fact that Assumption \ref{ass:basic_ass} (c) is not satisfied for Model 3.

\begin{table}
\centering
\caption{The MSREs of $\widetilde{\covar}_{X|Y}^{(i)}(\tau'_n)$ ($i=1,2$) and $\widetilde{\coes}_{X|Y}^{(i)}(\tau'_n)$ ($i=1,2,3$) for \textbf{Models 1-3} with $\tau'_n = 0.99$ and $0.999$.}
\label{tab:msre}
\setlength{\tabcolsep}{10pt}
\begin{tabular}{@{}cccccccc@{}}
\hline\hline
\addlinespace[1.5pt]
& $\tau'_n$ & $(k,k_1)$ & CoVaR-I & CoVaR-II & CoES-I & CoES-II & CoES-III \\[1.5pt]
\hline
& \multicolumn{7}{c}{$n=500$} \\[1.5pt]
\hline
\textbf{Model 1} & 0.99 & $(137,143)$ & 0.04118 & 0.04117 & 0.06155 & 0.06146 & 0.06520 \\[1.5pt]
                 & 0.999& $(137,143)$ & 0.11382 & 0.11341 & 0.15278 & 0.15231 & 0.15578 \\[1.5pt]
\textbf{Model 2} & 0.99 & $(84,150)$  & 0.04611 & 0.04674 & 0.06538 & 0.06568 & 0.09129 \\[1.5pt]
                 & 0.999& $(84,150)$  & 0.12010 & 0.11969 & 0.15608 & 0.15492 & 0.19471 \\[1.5pt]
\textbf{Model 3} & 0.99 & $(71,137)$  & 0.03838 & 0.03907 & 0.06532 & 0.06592 & 0.08949 \\[1.5pt]
                 & 0.999& $(124,150)$ & 0.09767 & 0.09747 & 0.15209 & 0.15099 & 0.14587 \\[1.5pt]
\hline
& \multicolumn{7}{c}{$n=1000$} \\[1.5pt]
\hline
\textbf{Model 1} & 0.99 & $(287,274)$ & 0.02043 & 0.02017 & 0.02937 & 0.02899 & 0.03273 \\[1.5pt]
                 & 0.999& $(287,274)$ & 0.05298 & 0.05227 & 0.06803 & 0.06714 & 0.07246 \\[1.5pt]
\textbf{Model 2} & 0.99 & $(182,274)$ & 0.02491 & 0.02537 & 0.03507 & 0.03552 & 0.04136 \\[1.5pt]
                 & 0.999& $(182,274)$ & 0.06311 & 0.06375 & 0.08059 & 0.08129 & 0.08826 \\[1.5pt]
\textbf{Model 3} & 0.99 & $(155,287)$ & 0.01945 & 0.02026 & 0.03578 & 0.03694 & 0.04587 \\[1.5pt]
                 & 0.999& $(116,300)$ & 0.04618 & 0.04839 & 0.07214 & 0.07534 & 0.08829 \\[1.5pt]
\hline
& \multicolumn{7}{c}{$n=2000$} \\[1.5pt]
\hline
\textbf{Model 1} & 0.99 & $(384,400)$ & 0.01386 & 0.01423 & 0.01987 & 0.02028 & 0.02251 \\[1.5pt]
                 & 0.999& $(195,400)$ & 0.03878 & 0.03908 & 0.04873 & 0.04896 & 0.06612 \\[1.5pt]
\textbf{Model 2} & 0.99 & $(274,384)$ & 0.01596 & 0.01629 & 0.02237 & 0.02269 & 0.02830 \\[1.5pt]
                 & 0.999& $(305,384)$ & 0.04026 & 0.04087 & 0.05087 & 0.05149 & 0.05639 \\[1.5pt]
\textbf{Model 3} & 0.99 & $(274,384)$ & 0.01650 & 0.01635 & 0.03475 & 0.03425 & 0.03193 \\[1.5pt]
                 & 0.999& $(274,384)$ & 0.04263 & 0.04225 & 0.07509 & 0.07424 & 0.06861 \\[1.5pt]
\hline
& \multicolumn{7}{c}{$n=5000$} \\[1.5pt]
\hline
\textbf{Model 1} & 0.99 & $(671,724)$ & 0.00708 & 0.00727 & 0.01015 & 0.01034 & 0.01572 \\[1.5pt]
                 & 0.999& $(671,724)$ & 0.01886 & 0.01908 & 0.02396 & 0.02418 & 0.03072 \\[1.5pt]
\textbf{Model 2} & 0.99 & $(697,697)$ & 0.00755 & 0.00778 & 0.01074 & 0.01098 & 0.01377 \\[1.5pt]
                 & 0.999& $(592,750)$ & 0.01906 & 0.01940 & 0.02391 & 0.02425 & 0.02933 \\[1.5pt]
\textbf{Model 3} & 0.99 & $(382,724)$ & 0.00947 & 0.01020 & 0.02247 & 0.02368 & 0.02445 \\[1.5pt]
                 & 0.999& $(250,750)$ & 0.02325 & 0.02569 & 0.04380 & 0.04815 & 0.05687 \\[1.5pt]
\hline\hline
\end{tabular}
\end{table}

\section{Real Data Analysis}\label{sec:realanslysis}

\begin{figure}[htbp]
\centering
\begin{minipage}[b]{0.32\textwidth}
\includegraphics[width=\textwidth,height = 0.13\textheight]{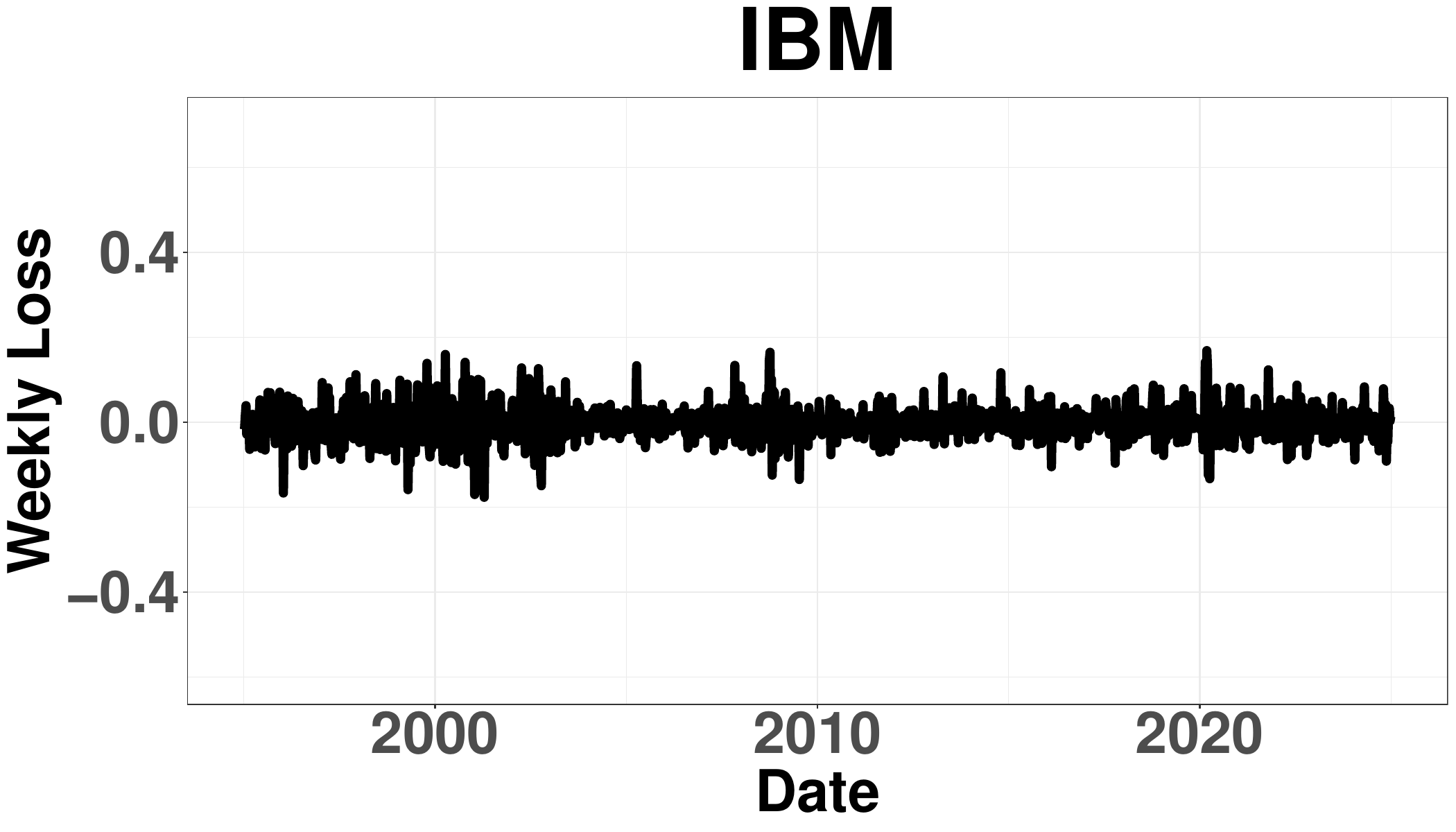}
\end{minipage}
\begin{minipage}[b]{0.32\textwidth}
\includegraphics[width=\textwidth,height = 0.13\textheight]{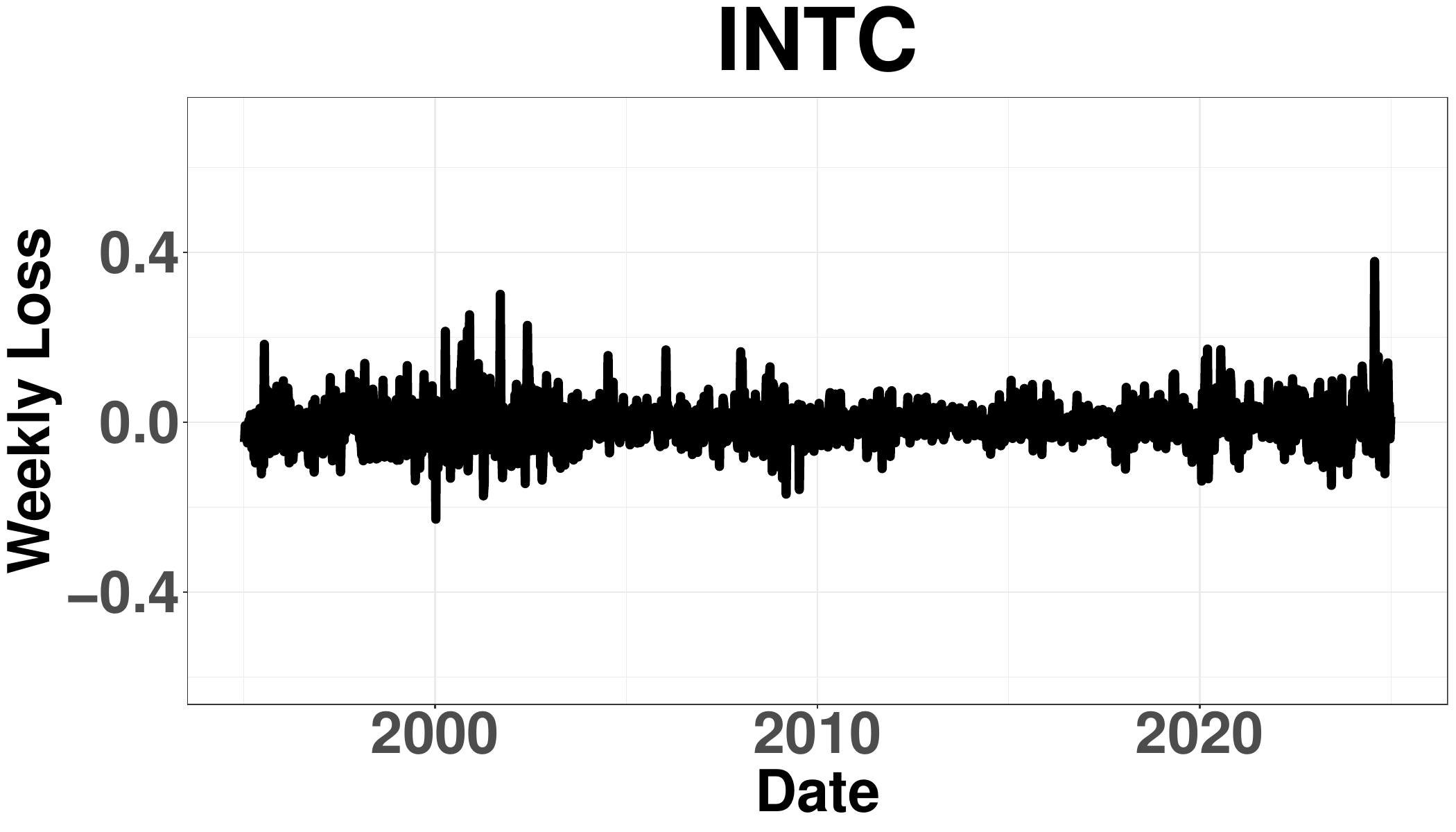}
\end{minipage}
\begin{minipage}[b]{0.32\textwidth}
\includegraphics[width=\textwidth,height = 0.13\textheight]{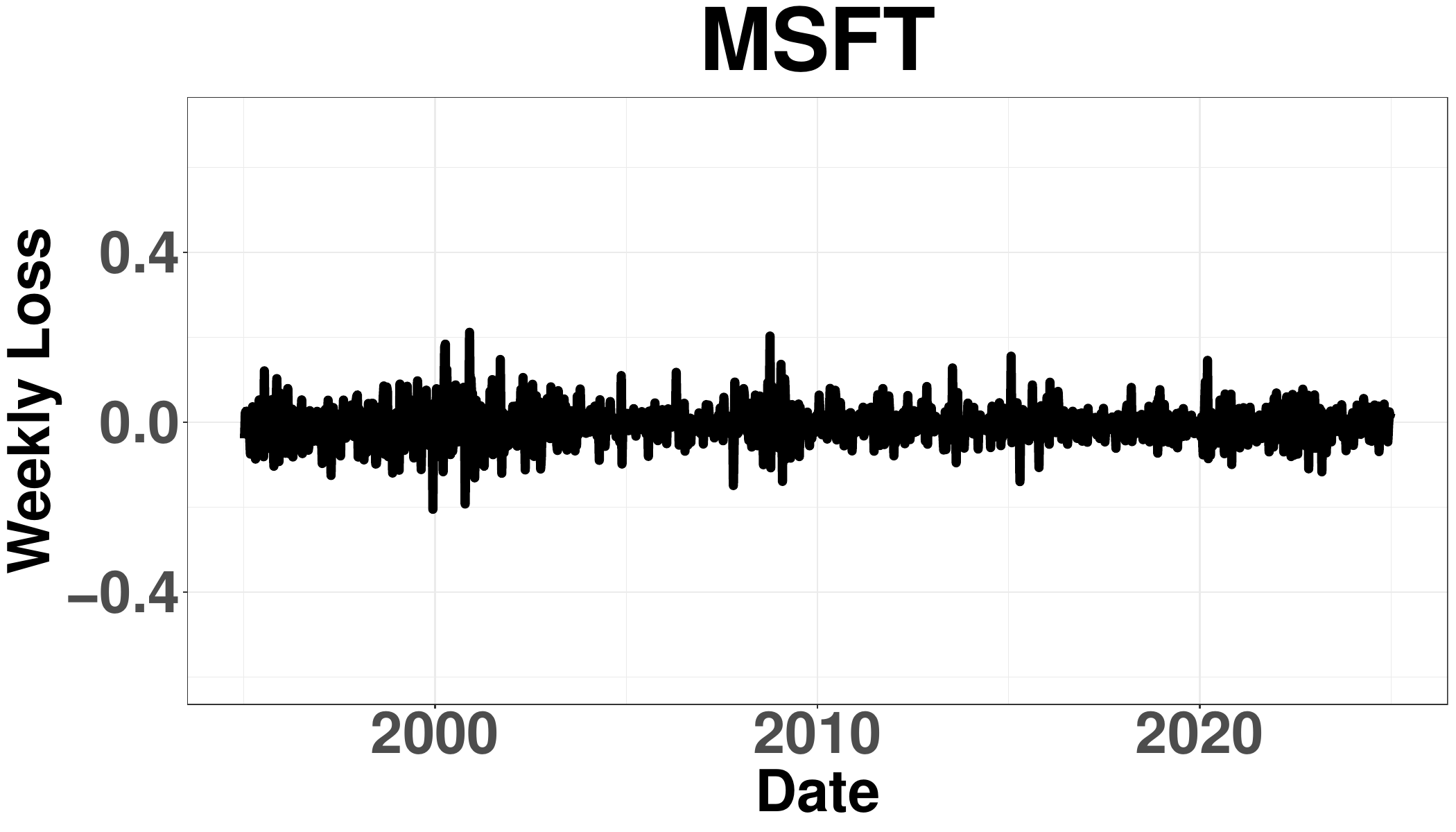}
\end{minipage}
\\
\begin{minipage}[b]{0.32\textwidth}
\includegraphics[width=\textwidth,height = 0.13\textheight]{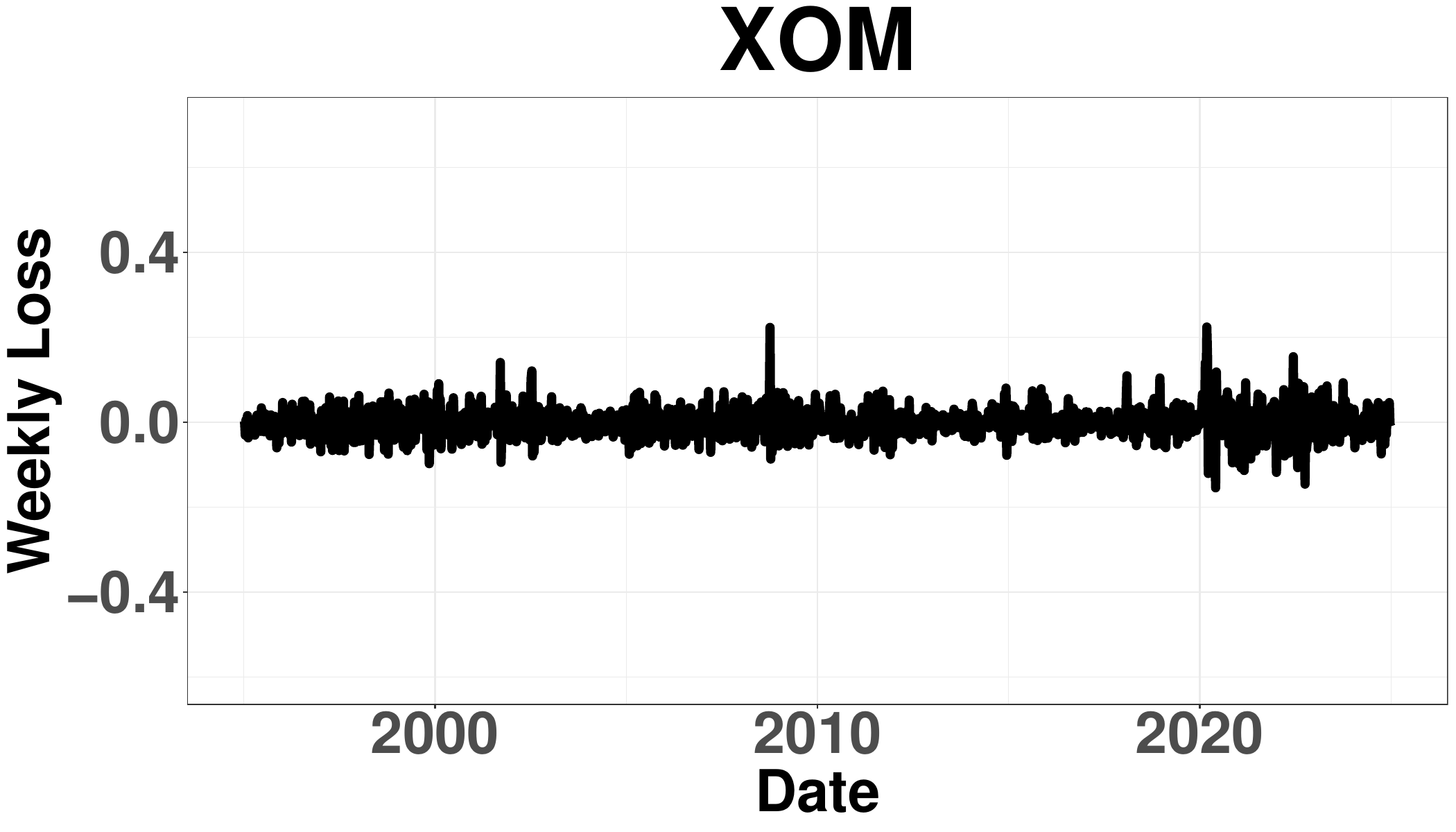}
\end{minipage}
\begin{minipage}[b]{0.32\textwidth}
\includegraphics[width=\textwidth,height = 0.13\textheight]{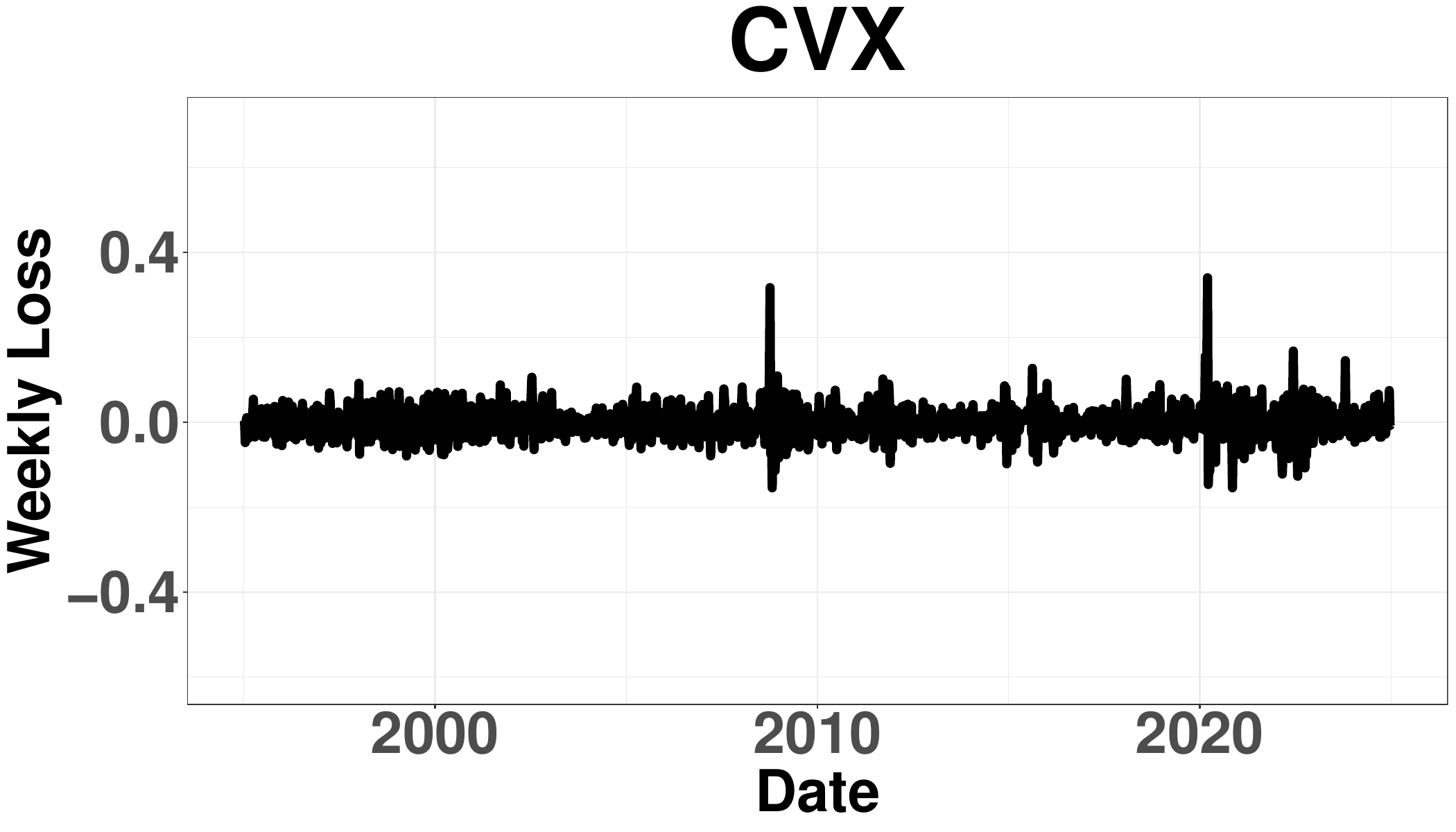}
\end{minipage}
\begin{minipage}[b]{0.32\textwidth}
\includegraphics[width=\textwidth,height = 0.13\textheight]{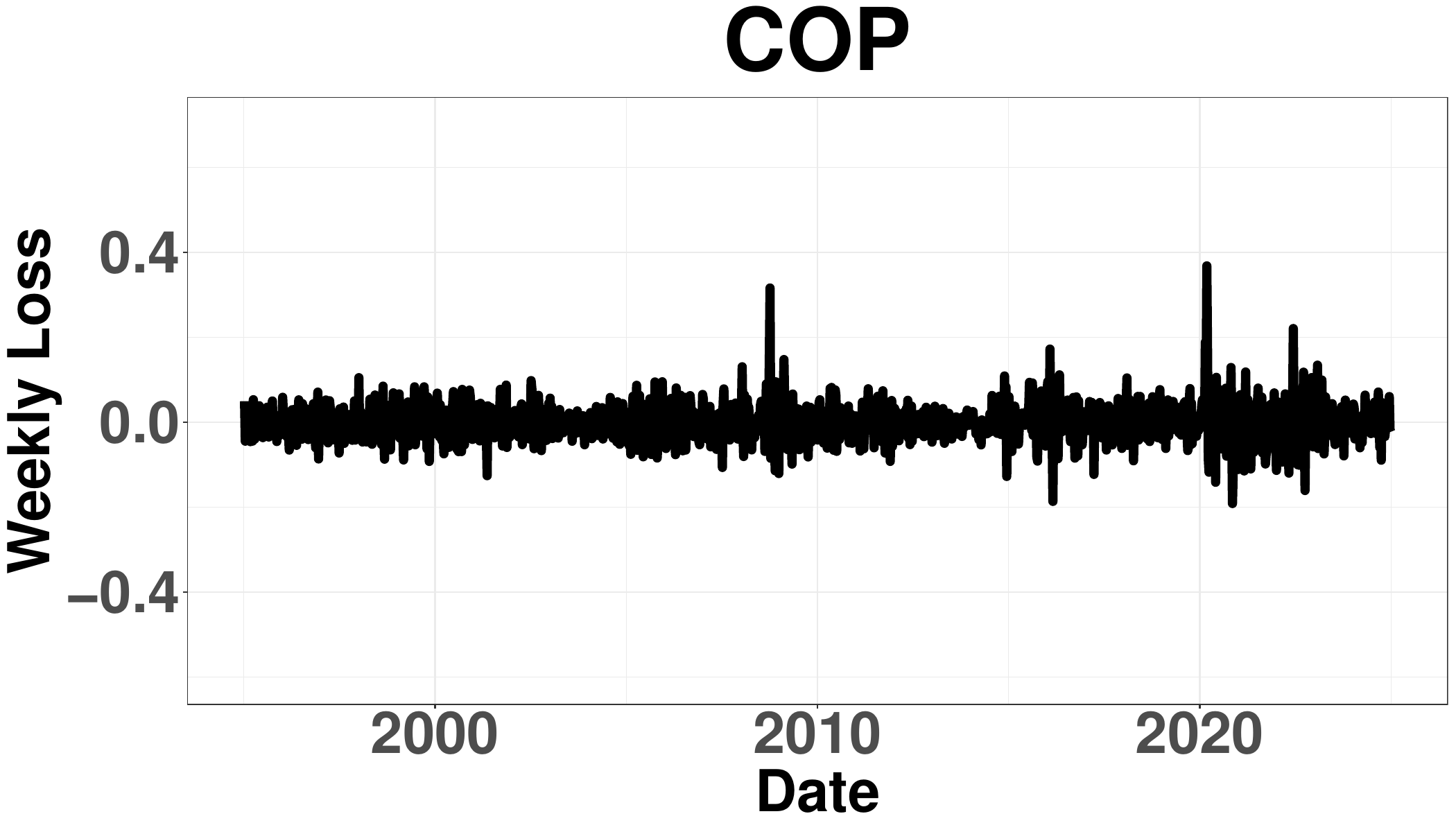}
\end{minipage}
\\
\begin{minipage}[b]{0.32\textwidth}
\includegraphics[width=\textwidth,height = 0.13\textheight]{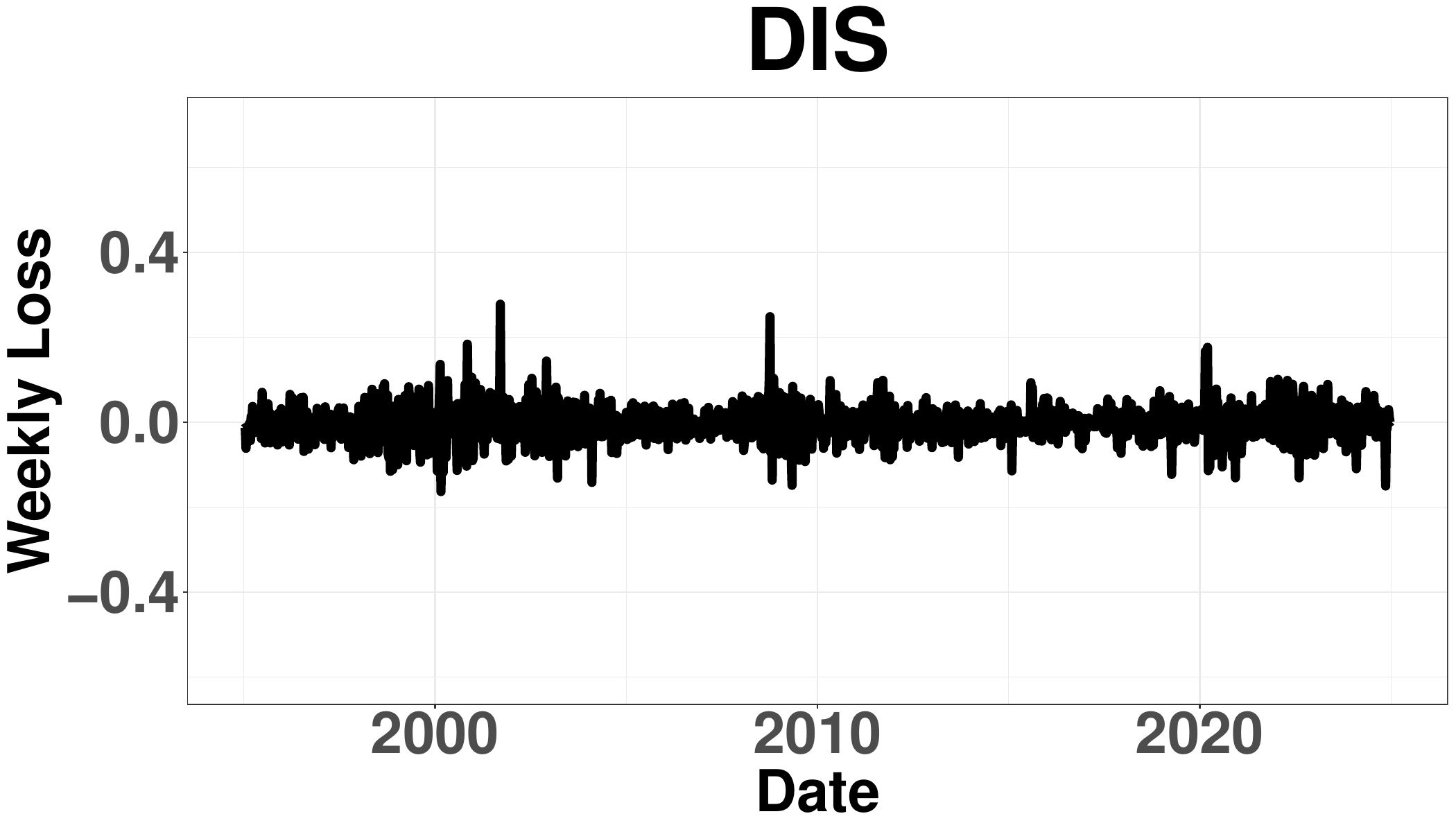}
\end{minipage}
\begin{minipage}[b]{0.32\textwidth}
\includegraphics[width=\textwidth,height = 0.13\textheight]{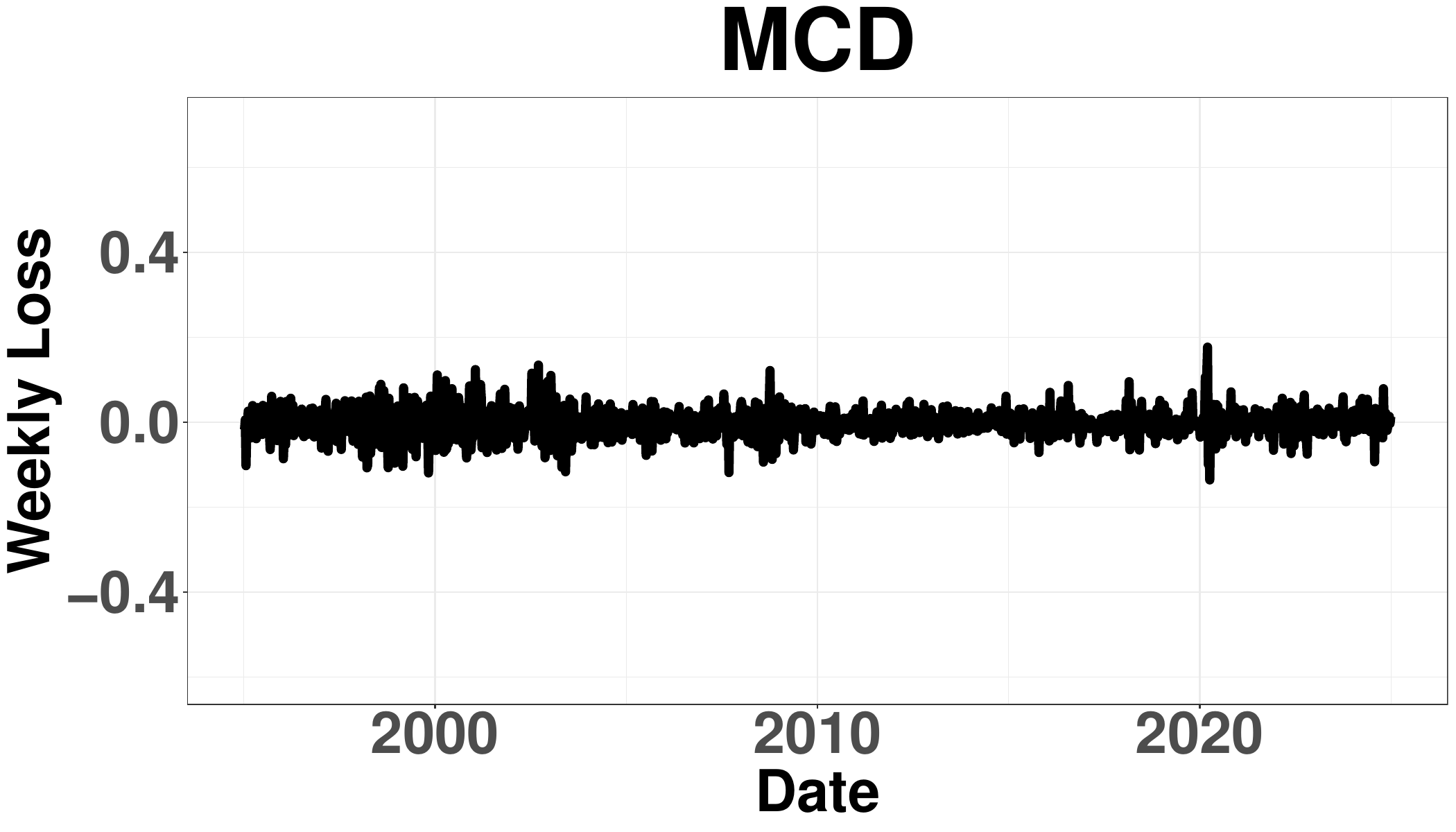}
\end{minipage}
\begin{minipage}[b]{0.32\textwidth}
\includegraphics[width=\textwidth,height = 0.13\textheight]{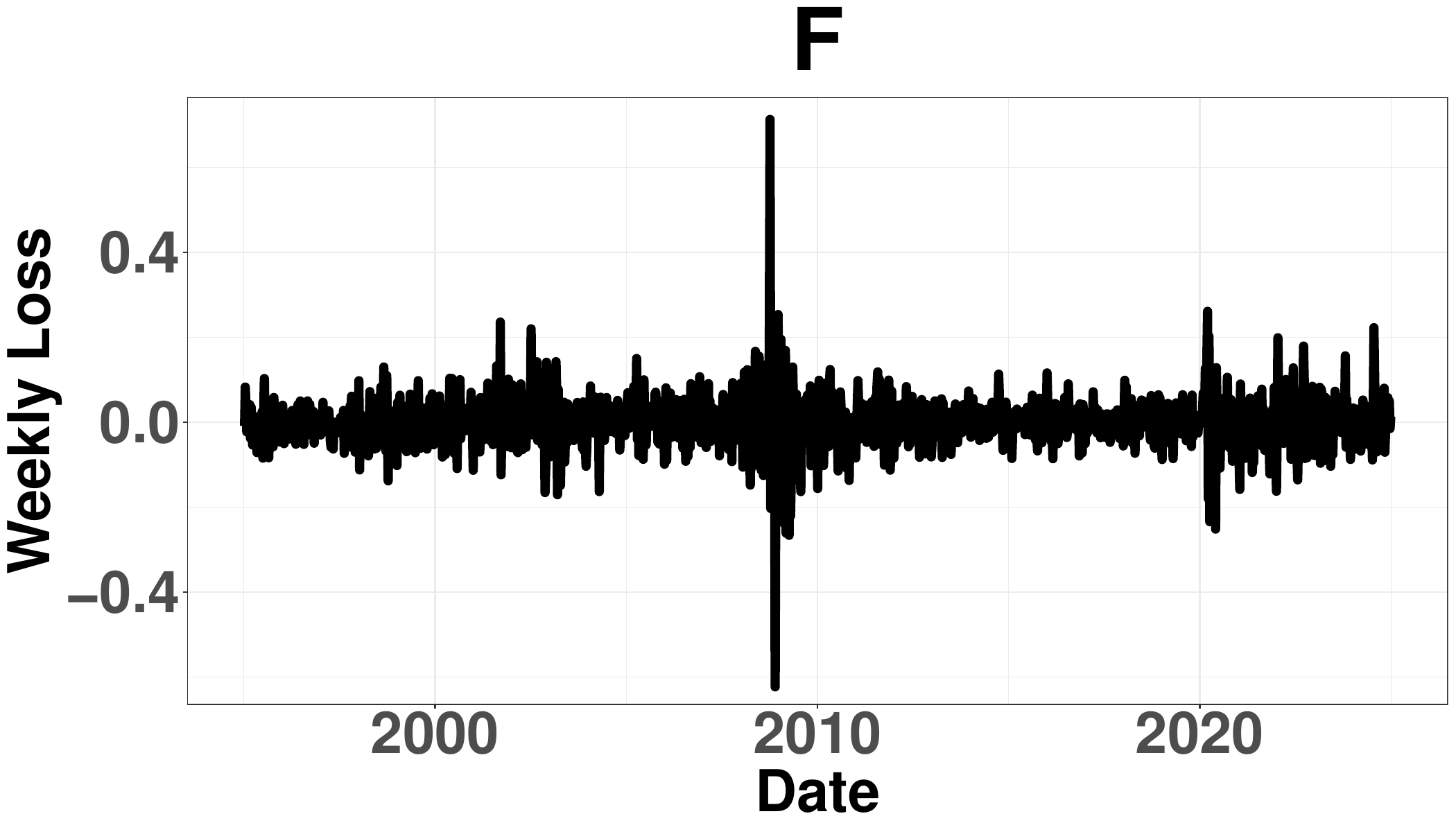}
\end{minipage}
\\
\begin{minipage}[b]{0.32\textwidth}
\includegraphics[width=\textwidth,height = 0.13\textheight]{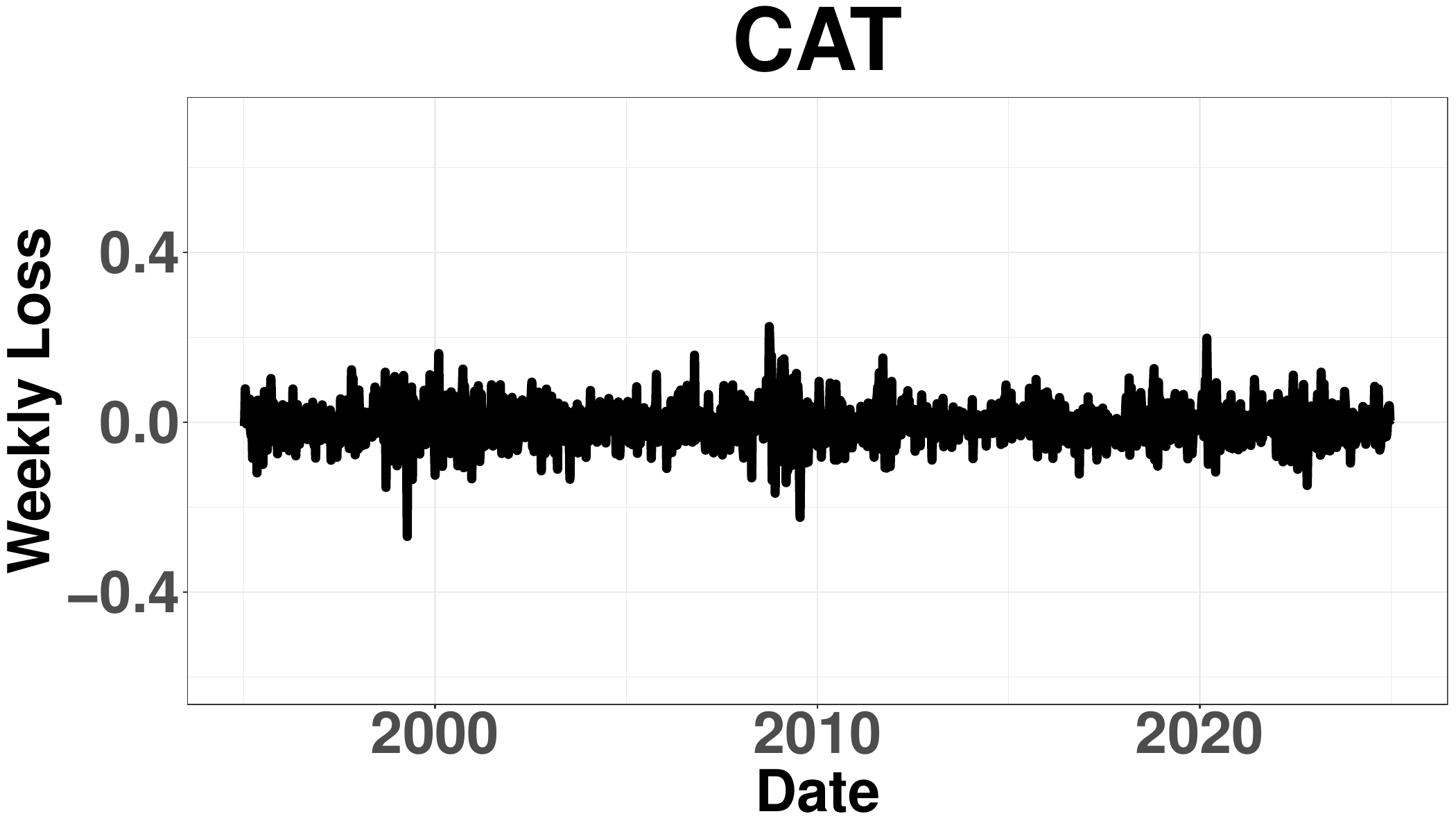}
\end{minipage}
\begin{minipage}[b]{0.32\textwidth}
\includegraphics[width=\textwidth,height = 0.13\textheight]{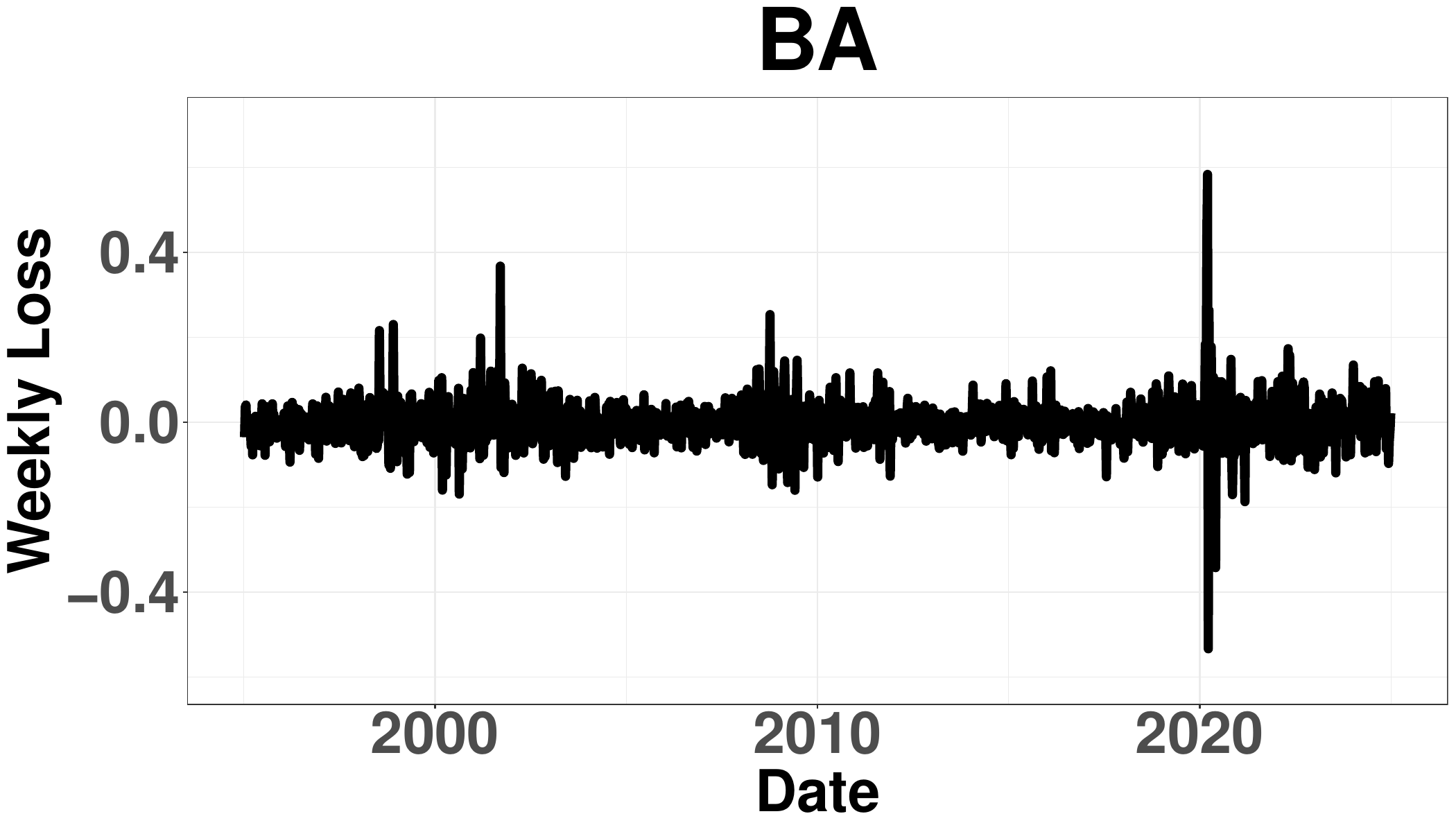}
\end{minipage}
\begin{minipage}[b]{0.32\textwidth}
\includegraphics[width=\textwidth,height = 0.13\textheight]{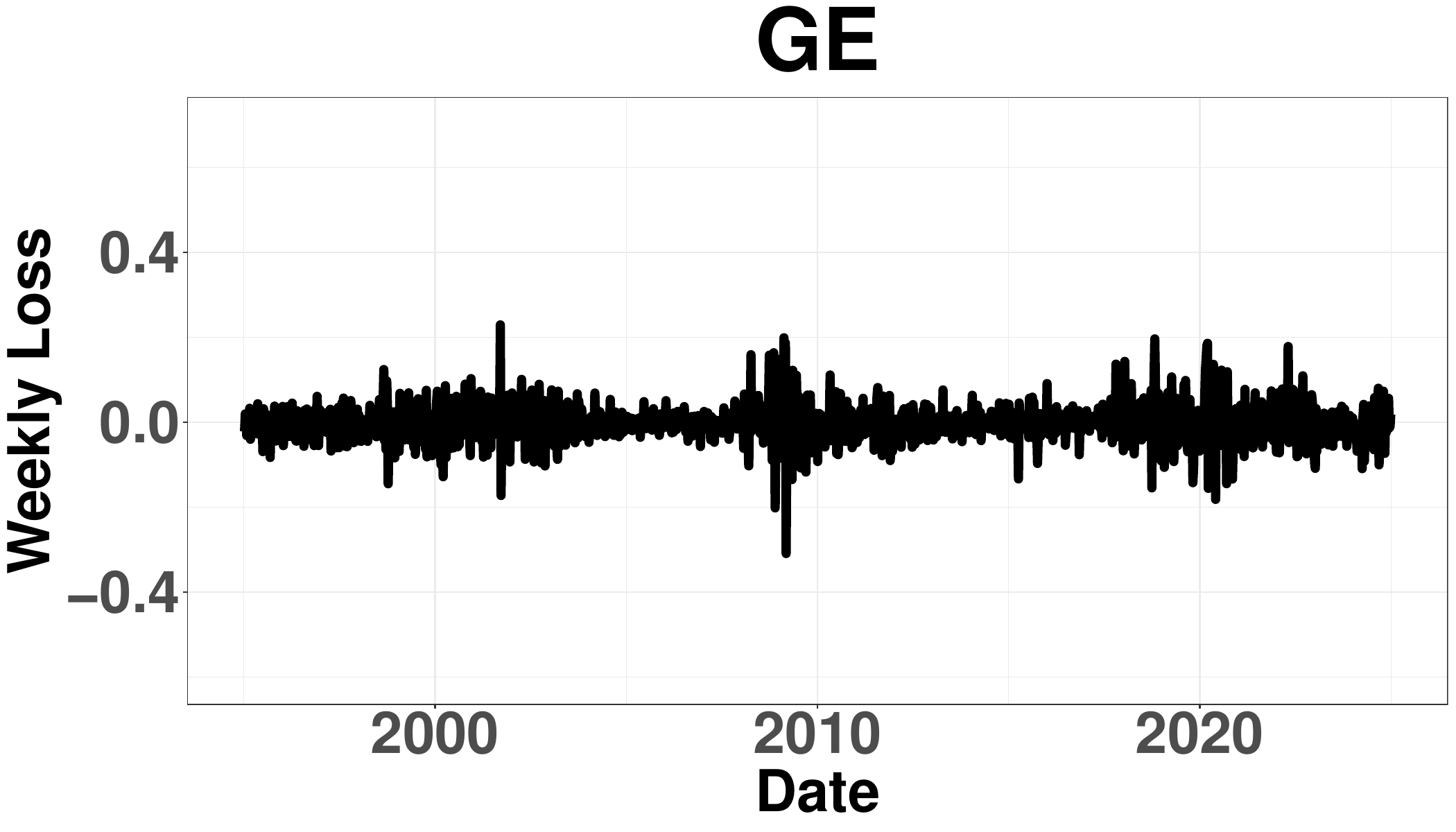}
\end{minipage}
\caption{Time series plots of weekly losses for the 12 constituent stocks.}
\label{fig:TSplots}
\end{figure}

In this section, we further examine the finite-sample performance of the proposed extrapolative methods by conducting a  real-world analysis on stock loss data. In our analysis, we consider 12 constituent stocks from 4 major sectors of the S\&P 500 Index, namely,
\begin{itemize}
  \item IT sector: IBM (International Business Machines), INTC (Intel), MSFT (Microsoft);
  \item Energy sector: XOM (Exxon Mobil), CVX (Chevron), COP (ConocoPhillips);
  \item Consumer Discretionary sector: DIS (The Walt Disney), MCD (McDonald's), F (Ford Motor);
  \item Industrials sector: CAT (Caterpillar), BA (The Boeing), GE (GE Aerospace).
\end{itemize} 
We treat the S\&P500 Index as the system variable ($Y$) and the 12 firms as the individual variables $(X_i, i=1,...,12)$. As the theories for statistical methodologies are derived for independent and identically distributed samples, we utilize the weekly historical adjusted closing prices of their stocks to reduce the potential serial dependence. The data spans from January 1, 1995, to December 29, 2024, consisting of 1566 trading records. The weekly losses were calculated as negative log returns. Figure \ref{fig:TSplots} shows the time series plots of weekly losses for the 12 stocks, illustrating pronounced volatility clustering, particularly during the global financial crisis and the COVID-19 pandemic.

\begin{figure}[htbp]
\centering
\begin{minipage}[b]{0.32\textwidth}
\includegraphics[width=\textwidth,height = 0.15\textheight]{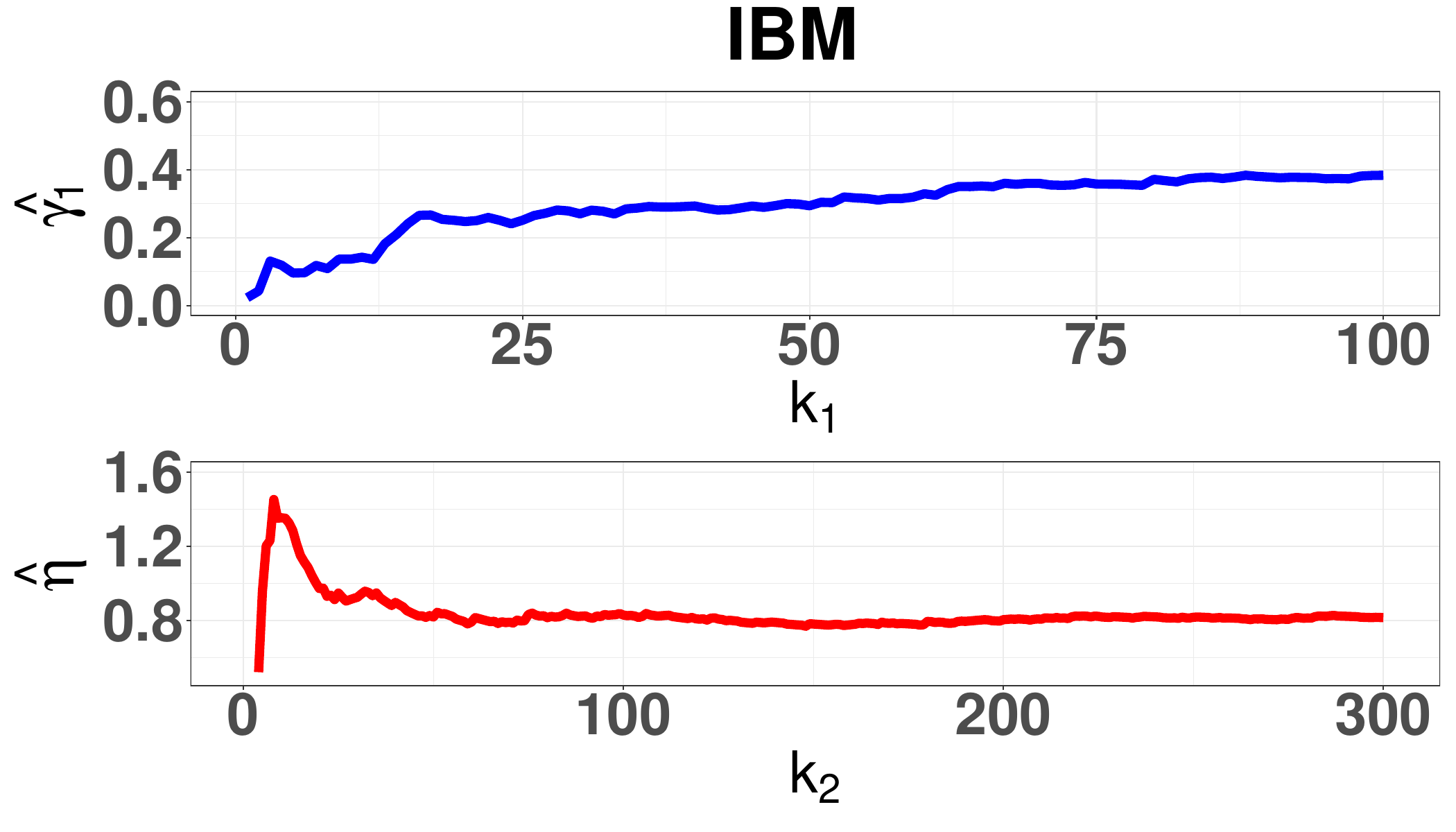}
\end{minipage}
\begin{minipage}[b]{0.32\textwidth}
\includegraphics[width=\textwidth,height = 0.15\textheight]{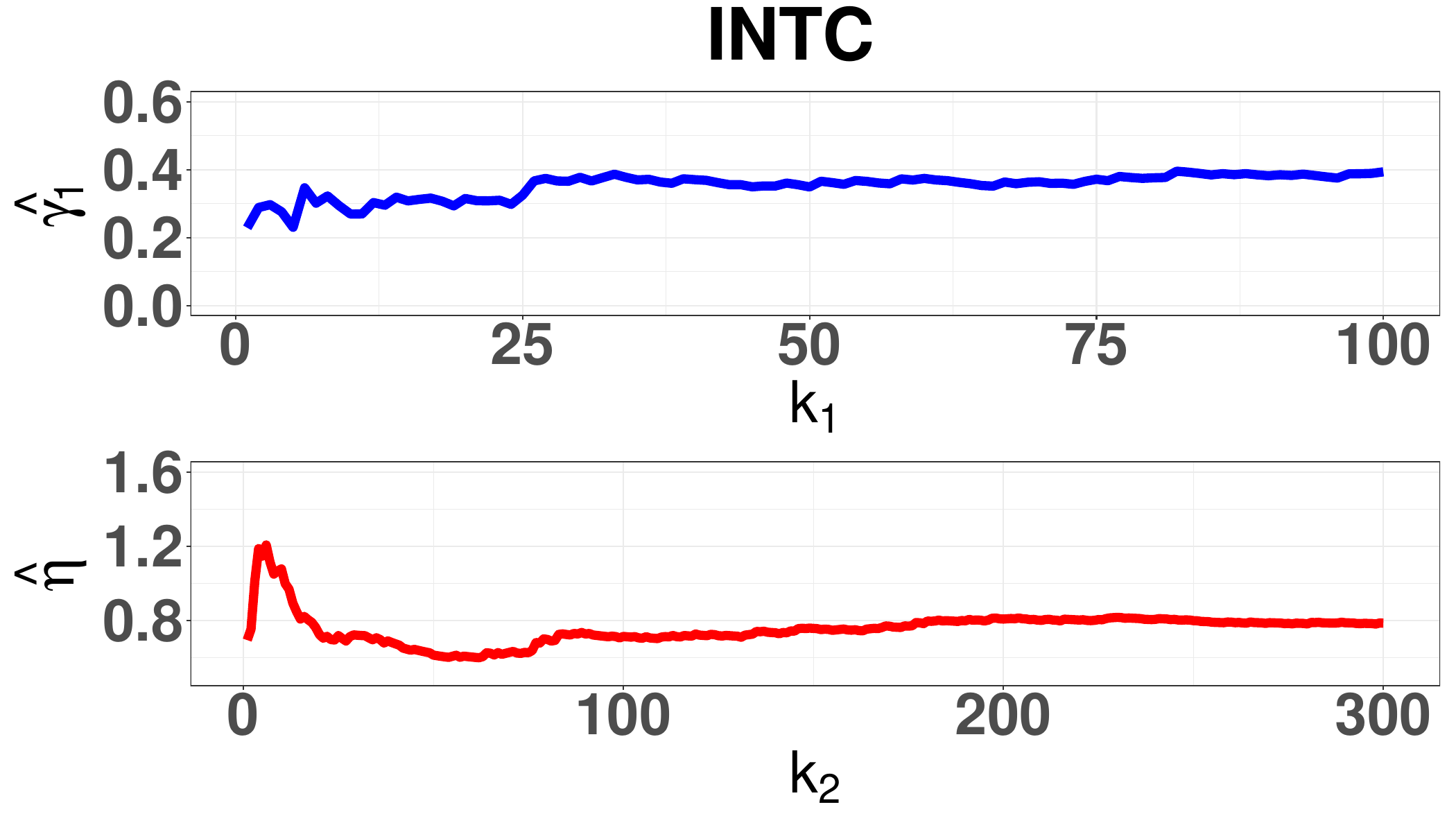}
\end{minipage}
\begin{minipage}[b]{0.32\textwidth}
\includegraphics[width=\textwidth,height = 0.15\textheight]{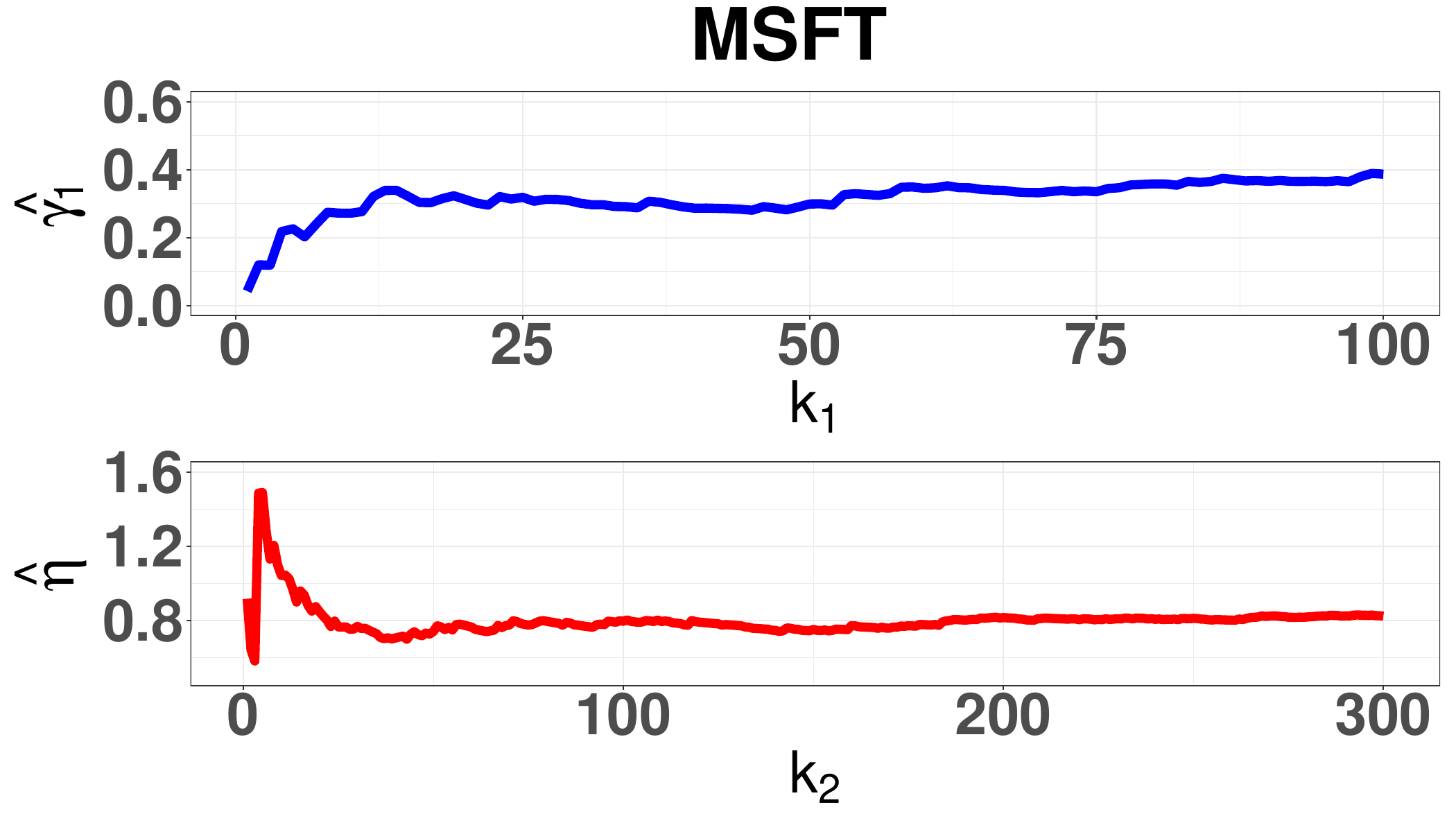}
\end{minipage}
\\[10pt]
\begin{minipage}[b]{0.32\textwidth}
\includegraphics[width=\textwidth,height = 0.15\textheight]{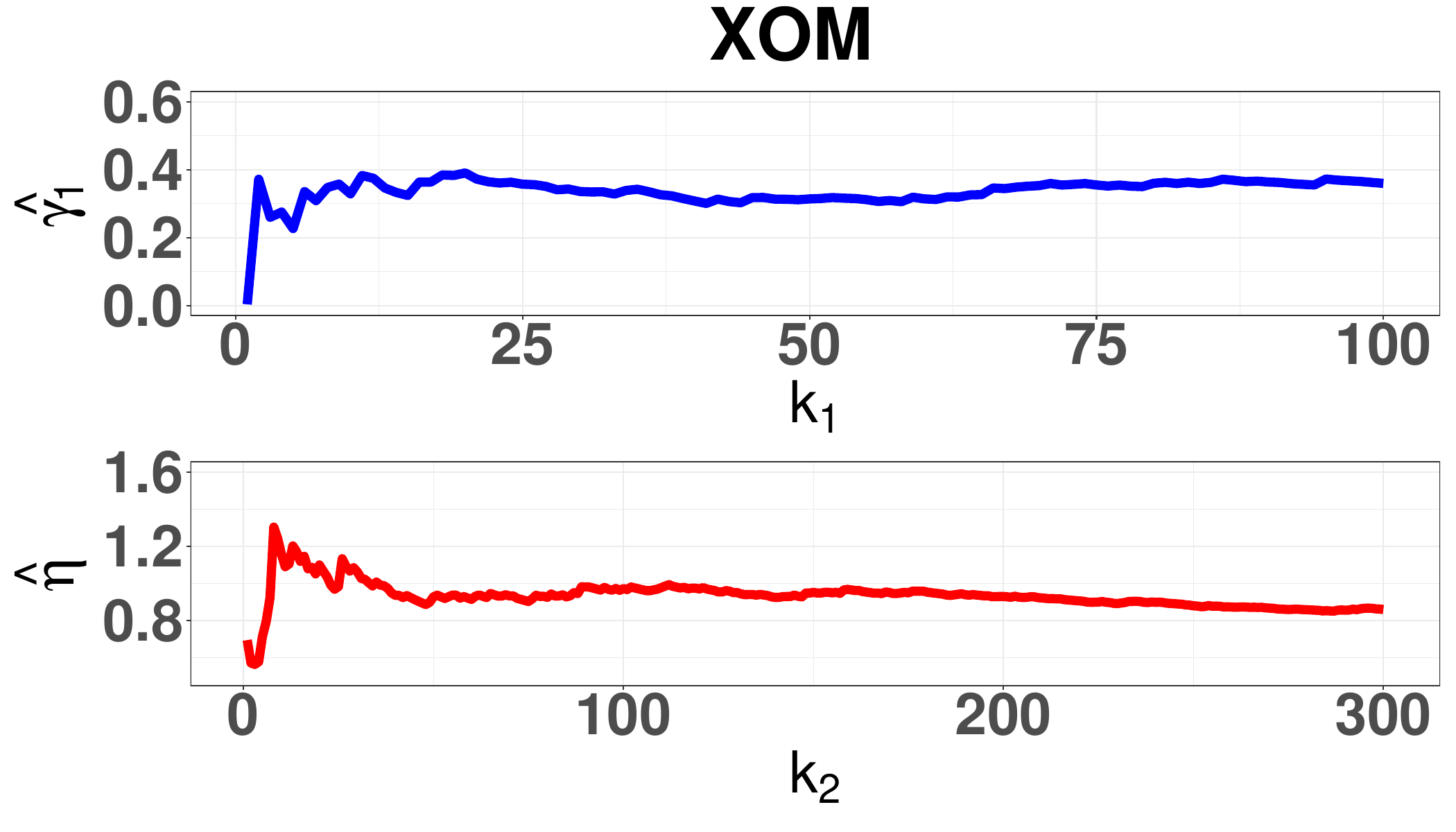}
\end{minipage}
\begin{minipage}[b]{0.32\textwidth}
\includegraphics[width=\textwidth,height = 0.15\textheight]{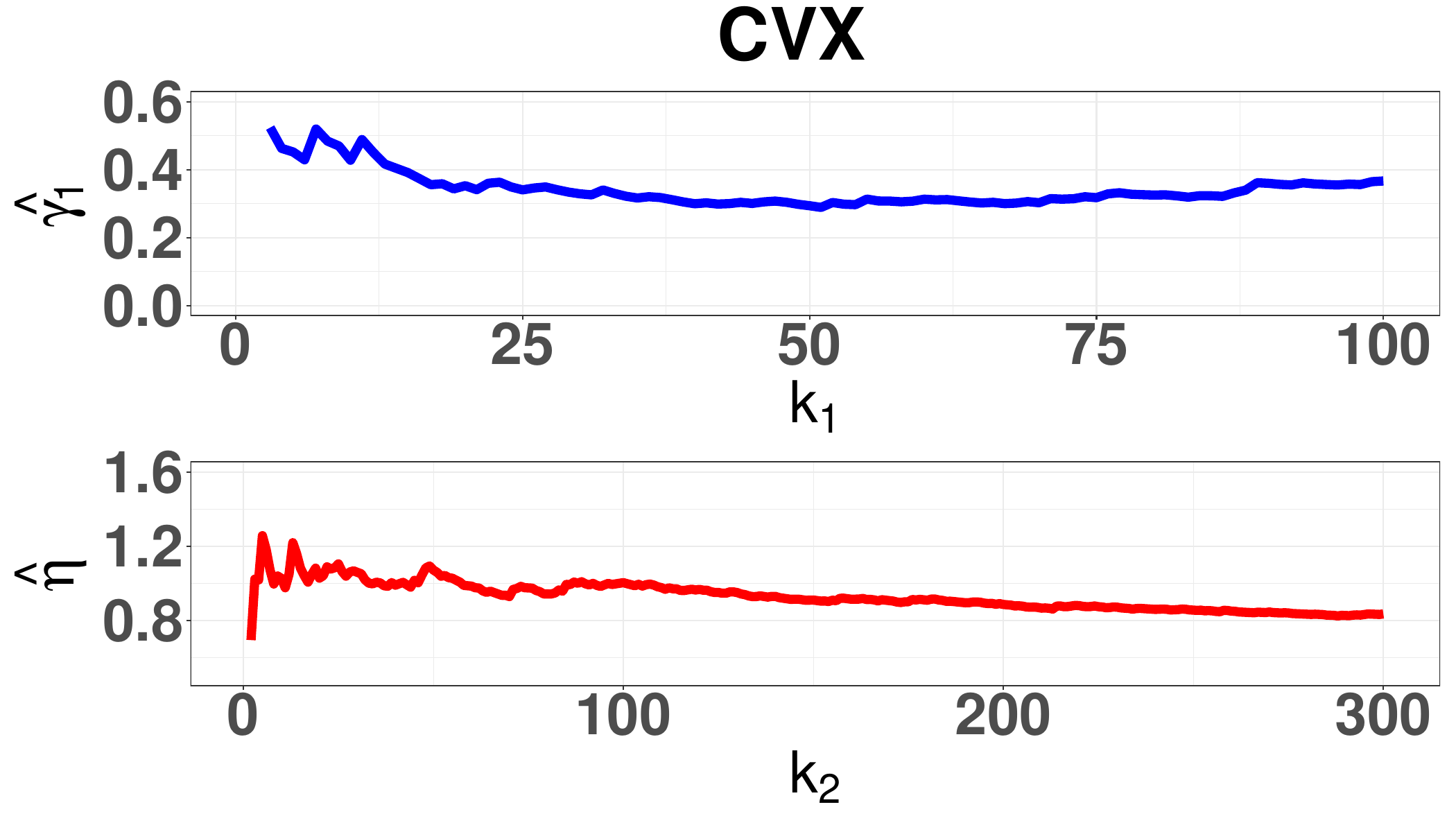}
\end{minipage}
\begin{minipage}[b]{0.32\textwidth}
\includegraphics[width=\textwidth,height = 0.15\textheight]{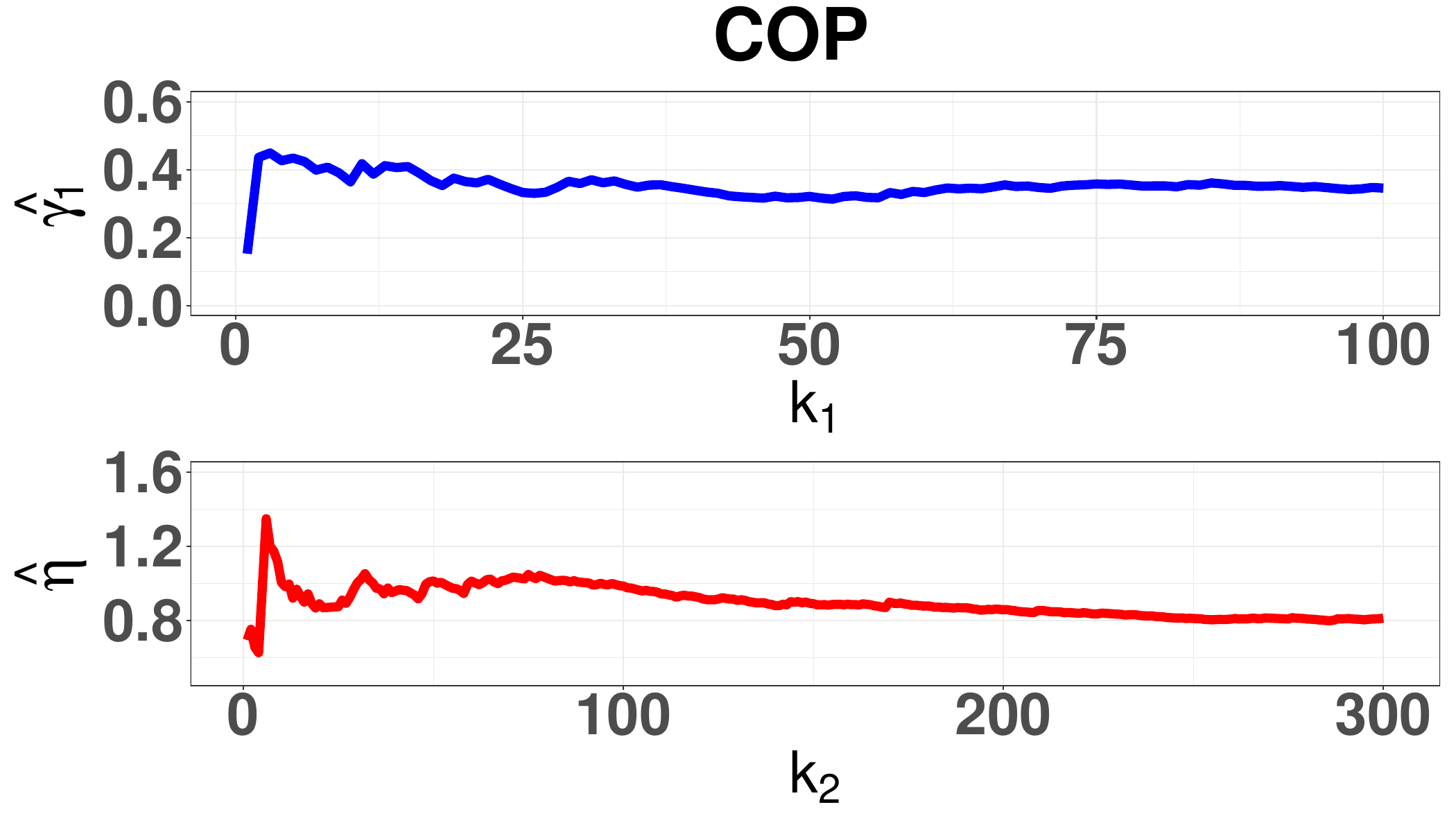}
\end{minipage}
\\[10pt]
\begin{minipage}[b]{0.32\textwidth}
\includegraphics[width=\textwidth,height = 0.15\textheight]{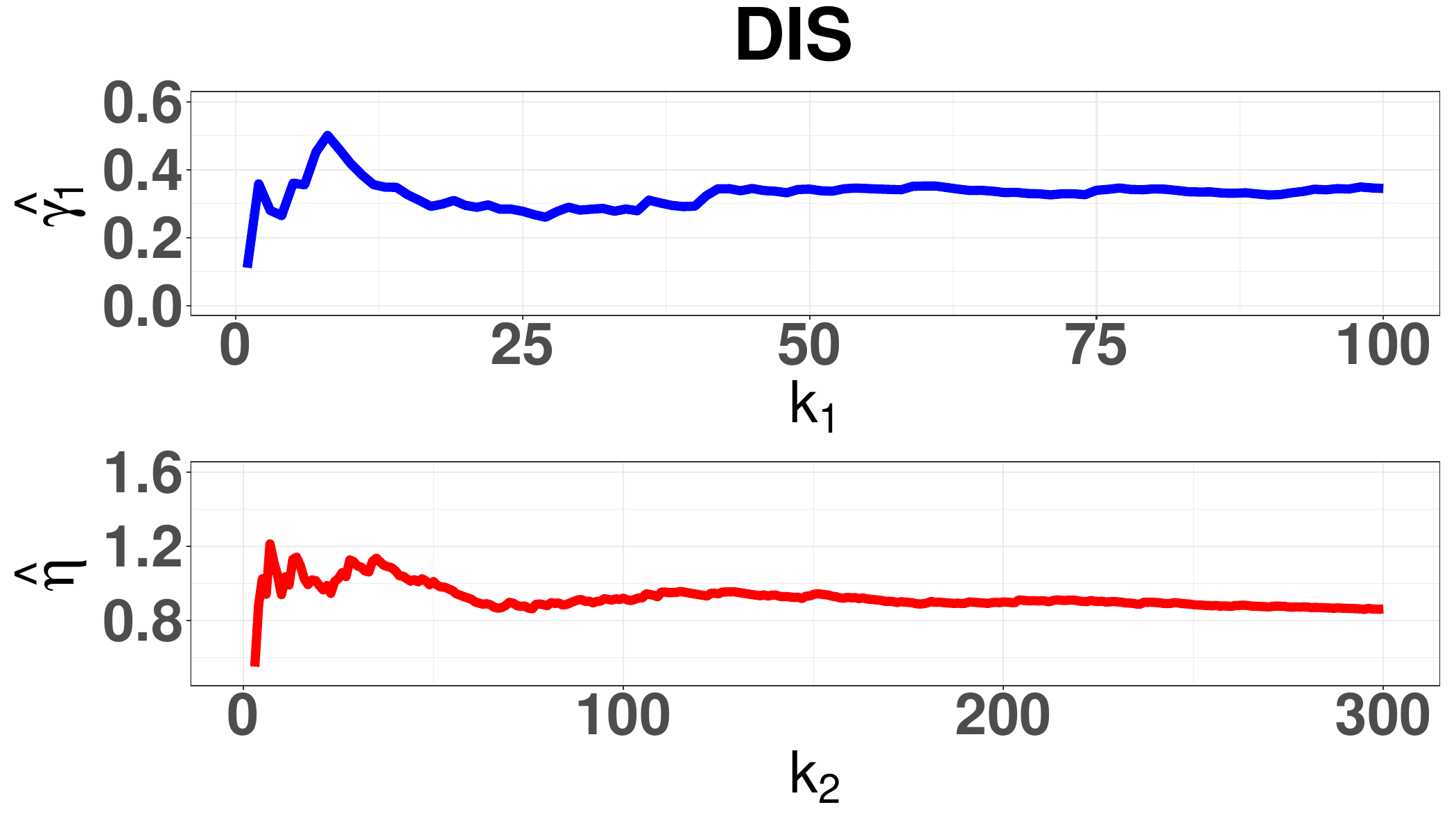}
\end{minipage}
\begin{minipage}[b]{0.32\textwidth}
\includegraphics[width=\textwidth,height = 0.15\textheight]{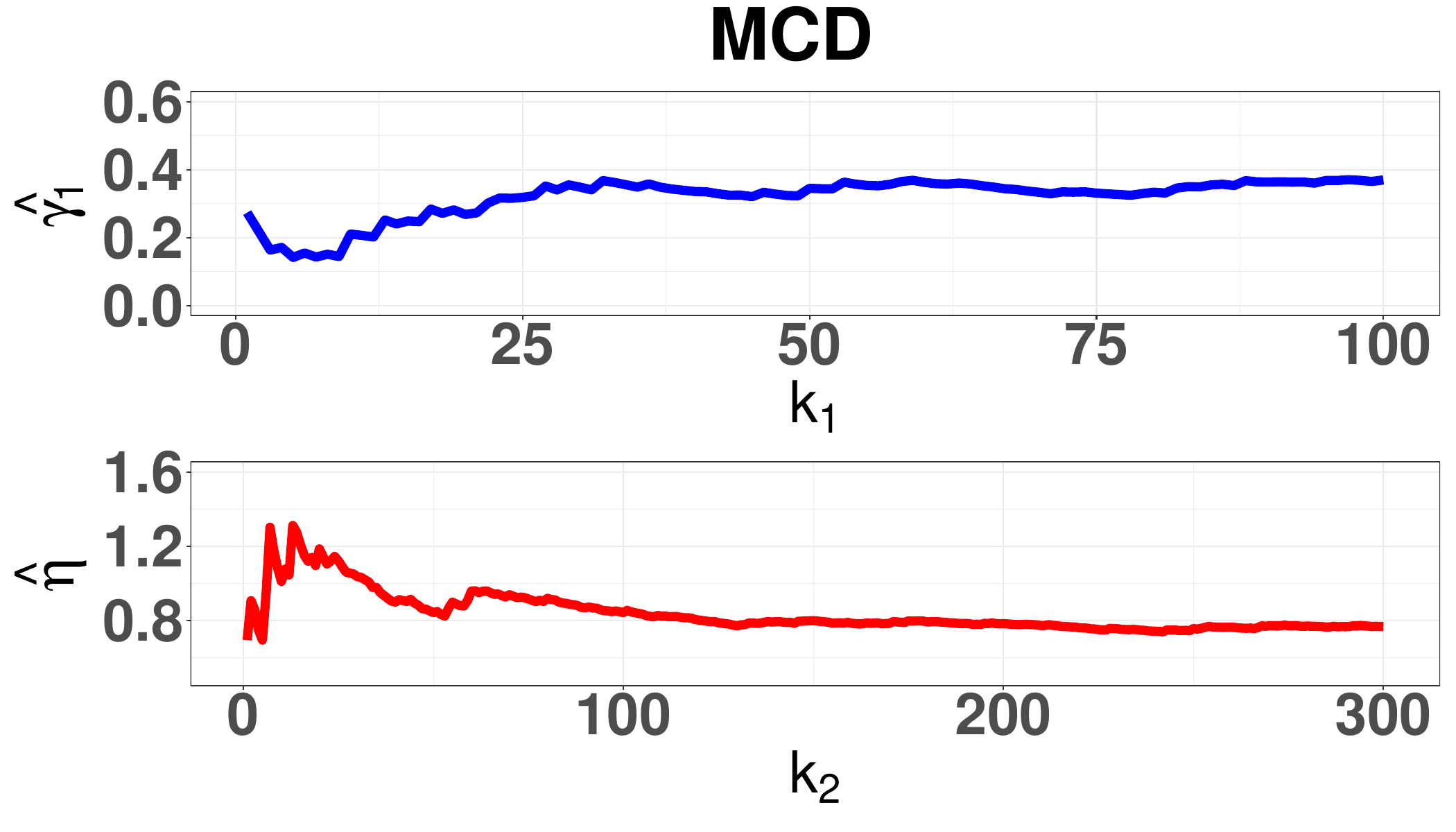}
\end{minipage}
\begin{minipage}[b]{0.32\textwidth}
\includegraphics[width=\textwidth,height = 0.15\textheight]{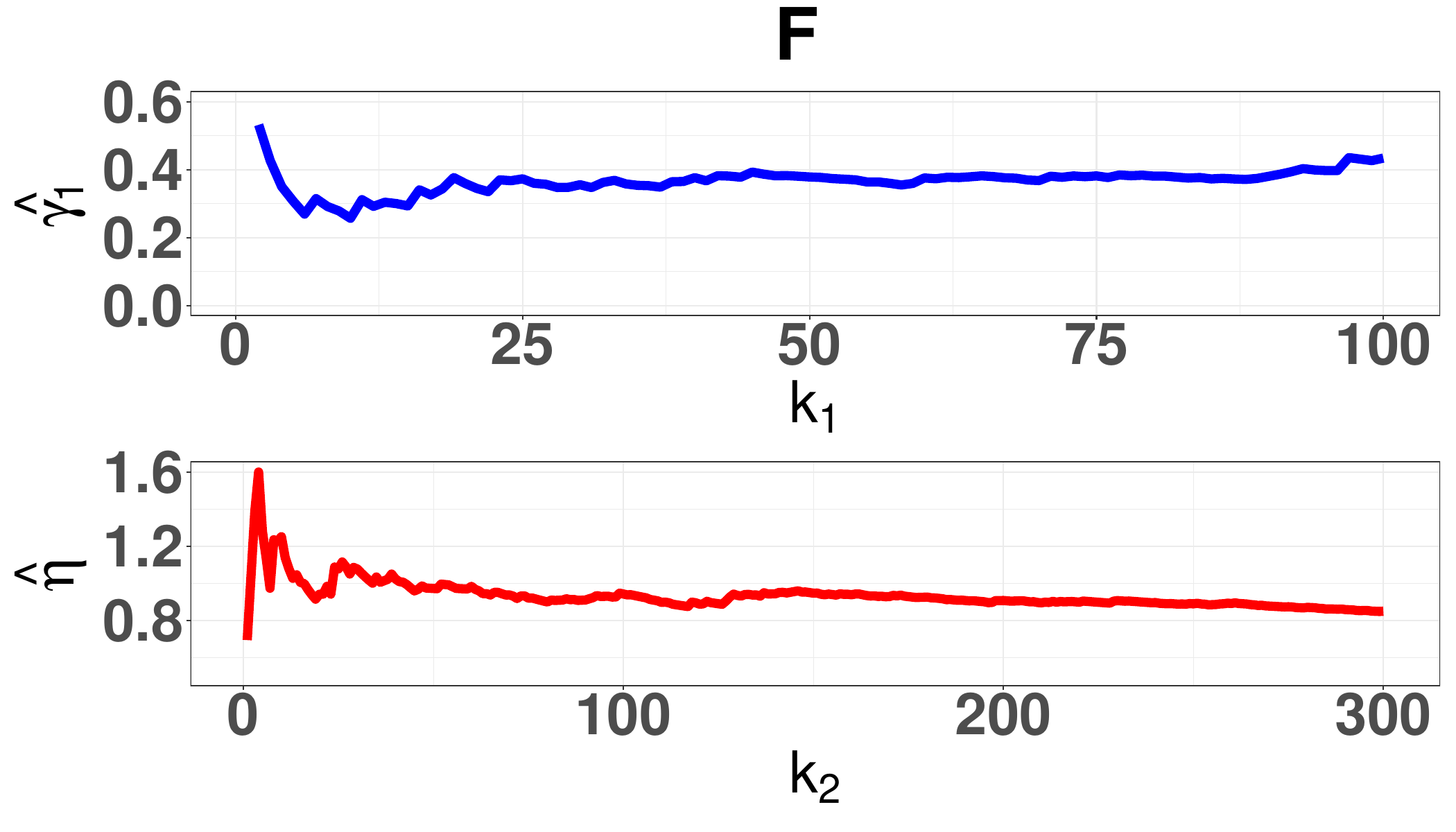}
\end{minipage}
\\[10pt]
\begin{minipage}[b]{0.32\textwidth}
\includegraphics[width=\textwidth,height = 0.15\textheight]{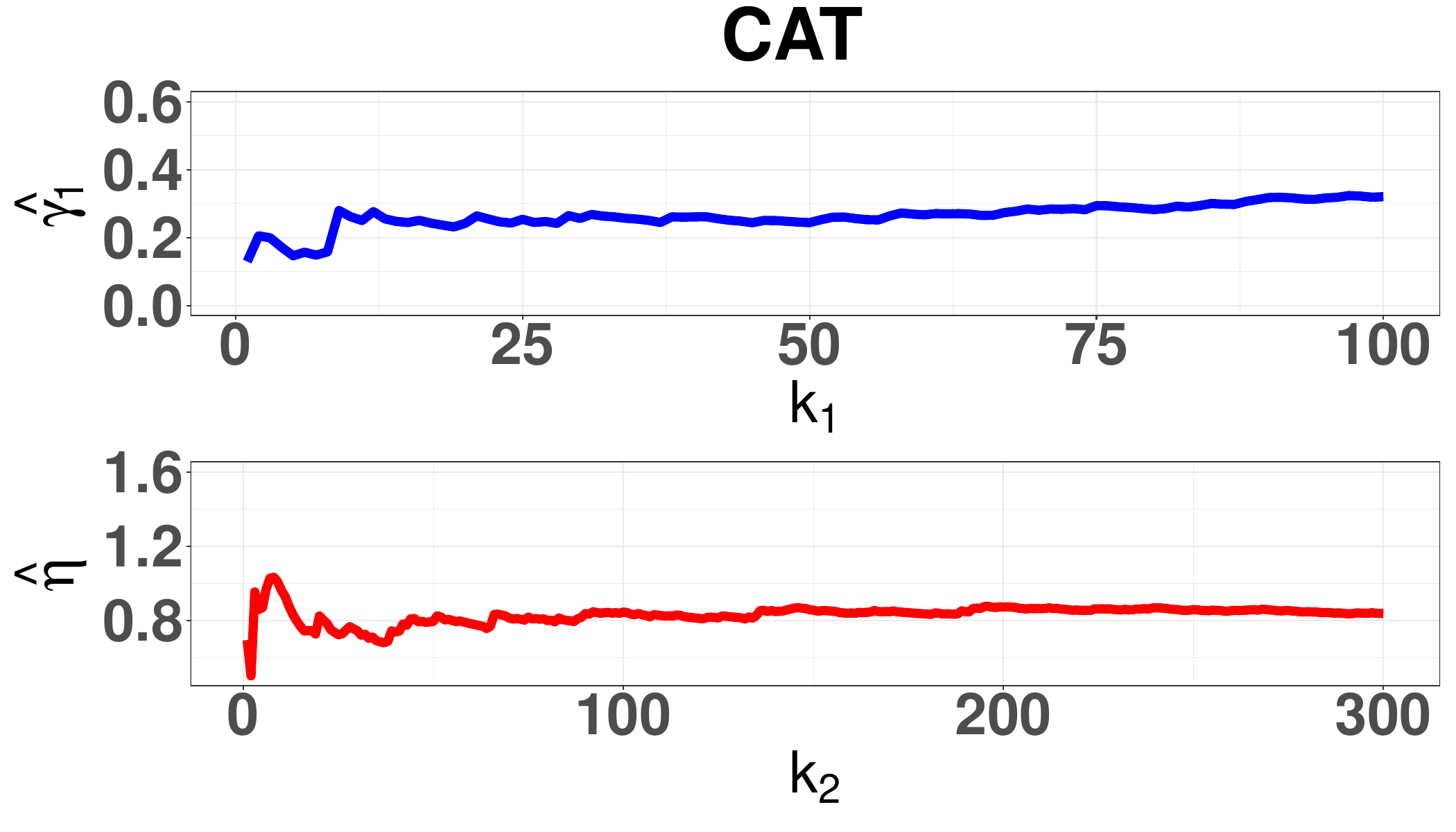}
\end{minipage}
\begin{minipage}[b]{0.32\textwidth}
\includegraphics[width=\textwidth,height = 0.15\textheight]{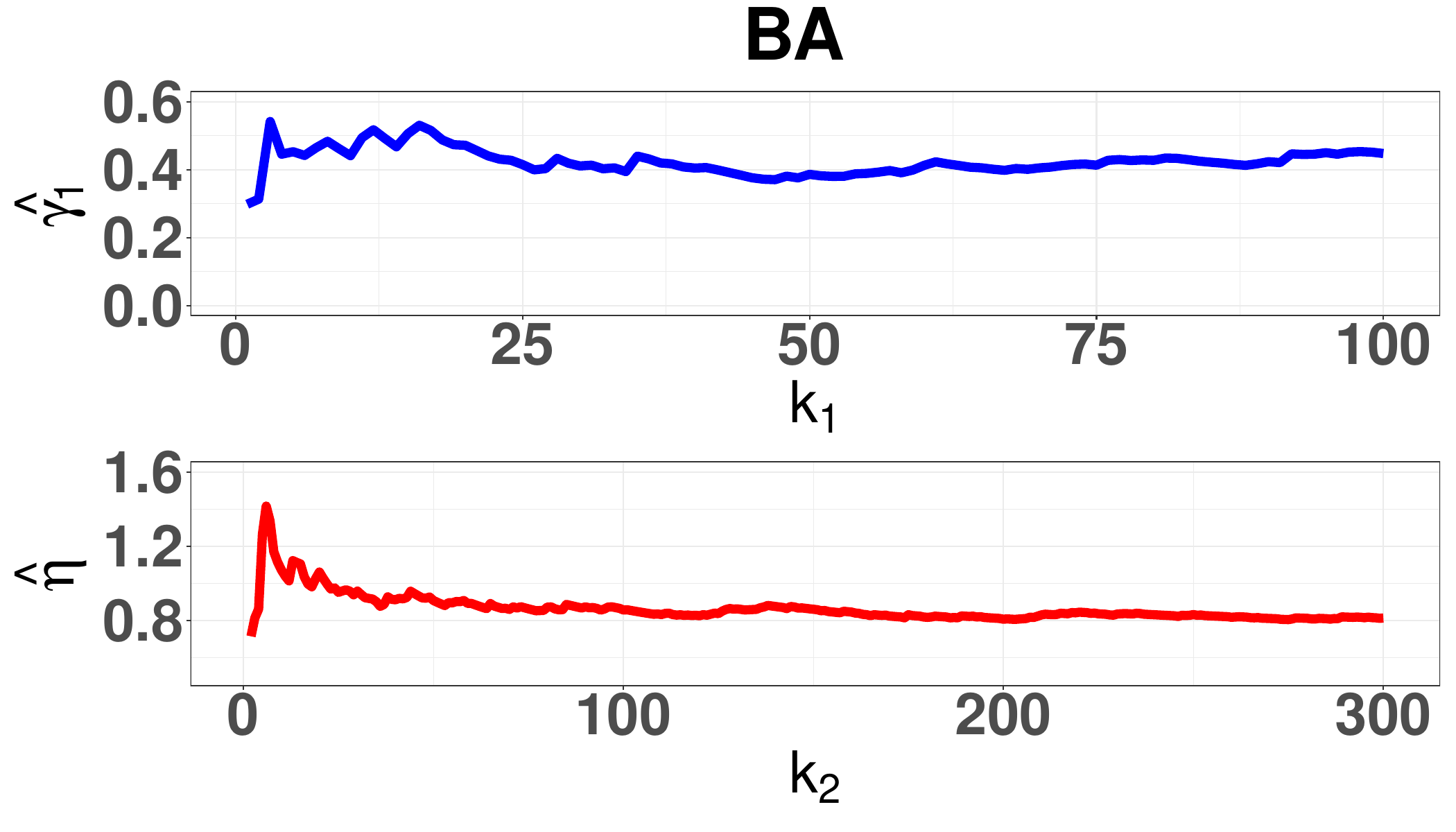}
\end{minipage}
\begin{minipage}[b]{0.32\textwidth}
\includegraphics[width=\textwidth,height = 0.15\textheight]{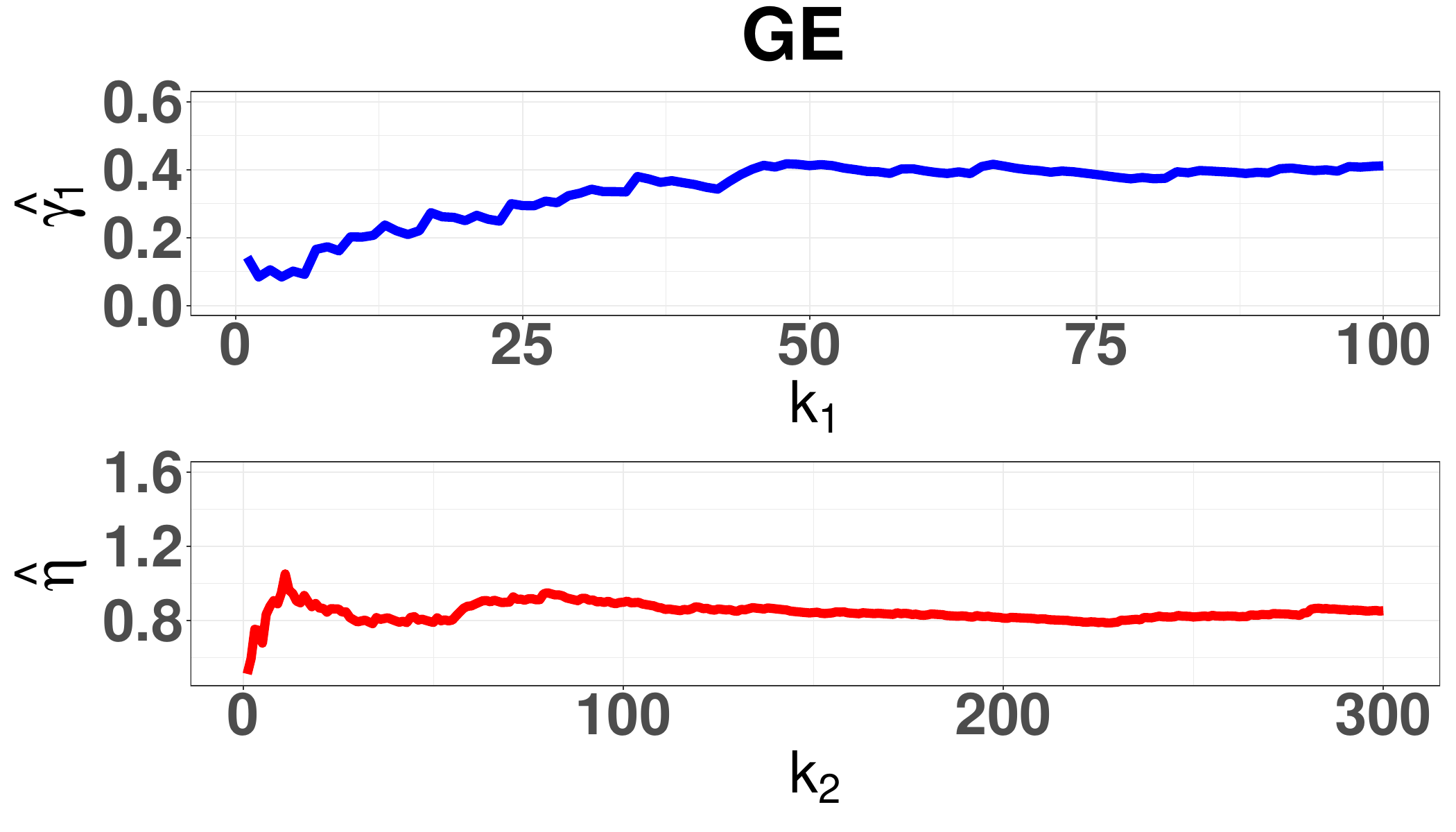}
\end{minipage}
\caption{The estimations for $\gamma_1$ (blue lines) and $\eta$ (red lines) against $k_1$ and $k_2$ for 12 individual losses conditional on S\&P500 Index loss.}
\label{fig:Gam_Eta}
\end{figure}

\begin{figure}[htbp]
\centering
\begin{minipage}[b]{0.25\textwidth}
\includegraphics[width=\textwidth,height = 0.11\textheight]{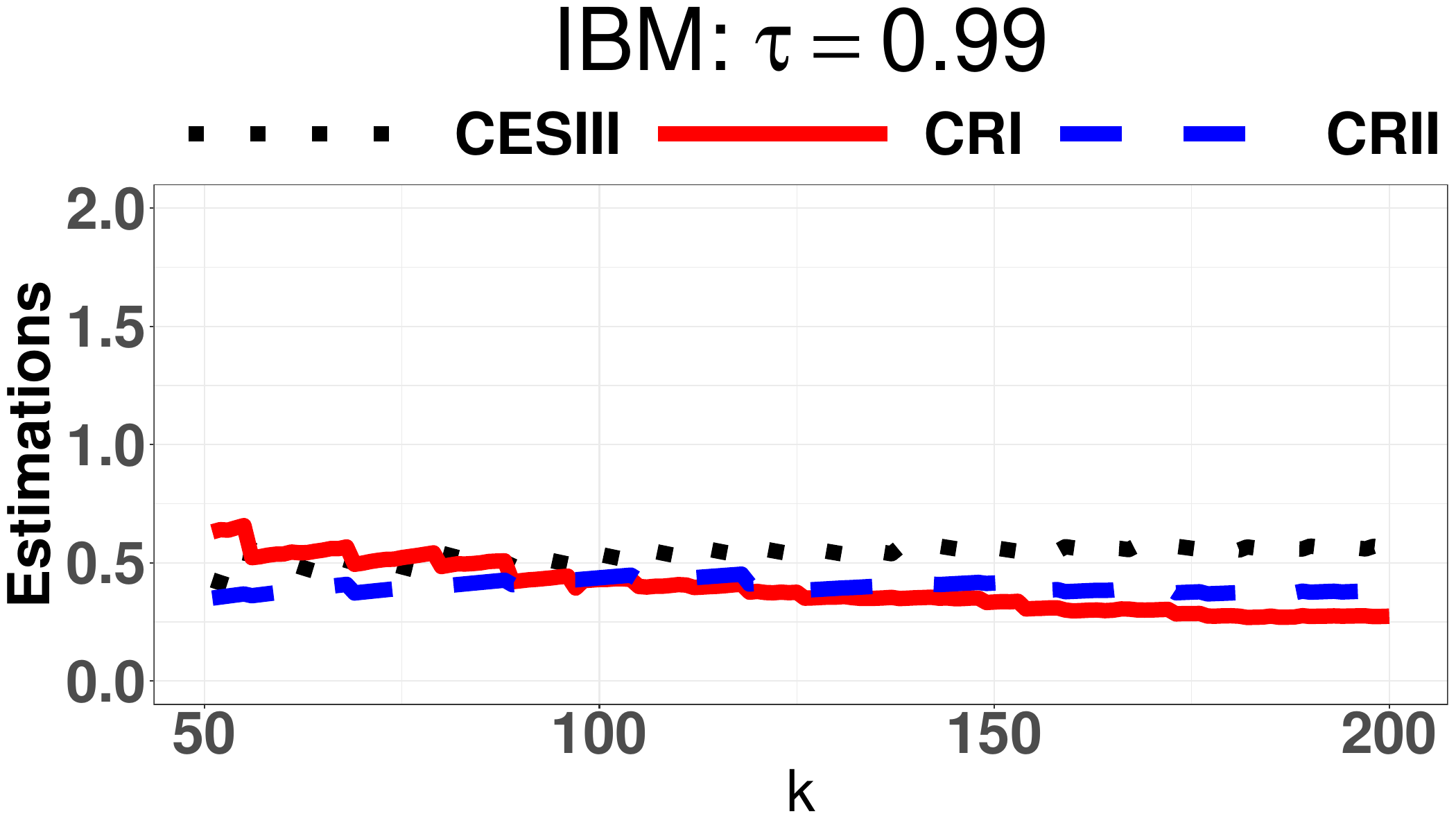}
\end{minipage}
\hspace{0.02\textwidth}
\begin{minipage}[b]{0.25\textwidth}
\includegraphics[width=\textwidth,height = 0.11\textheight]{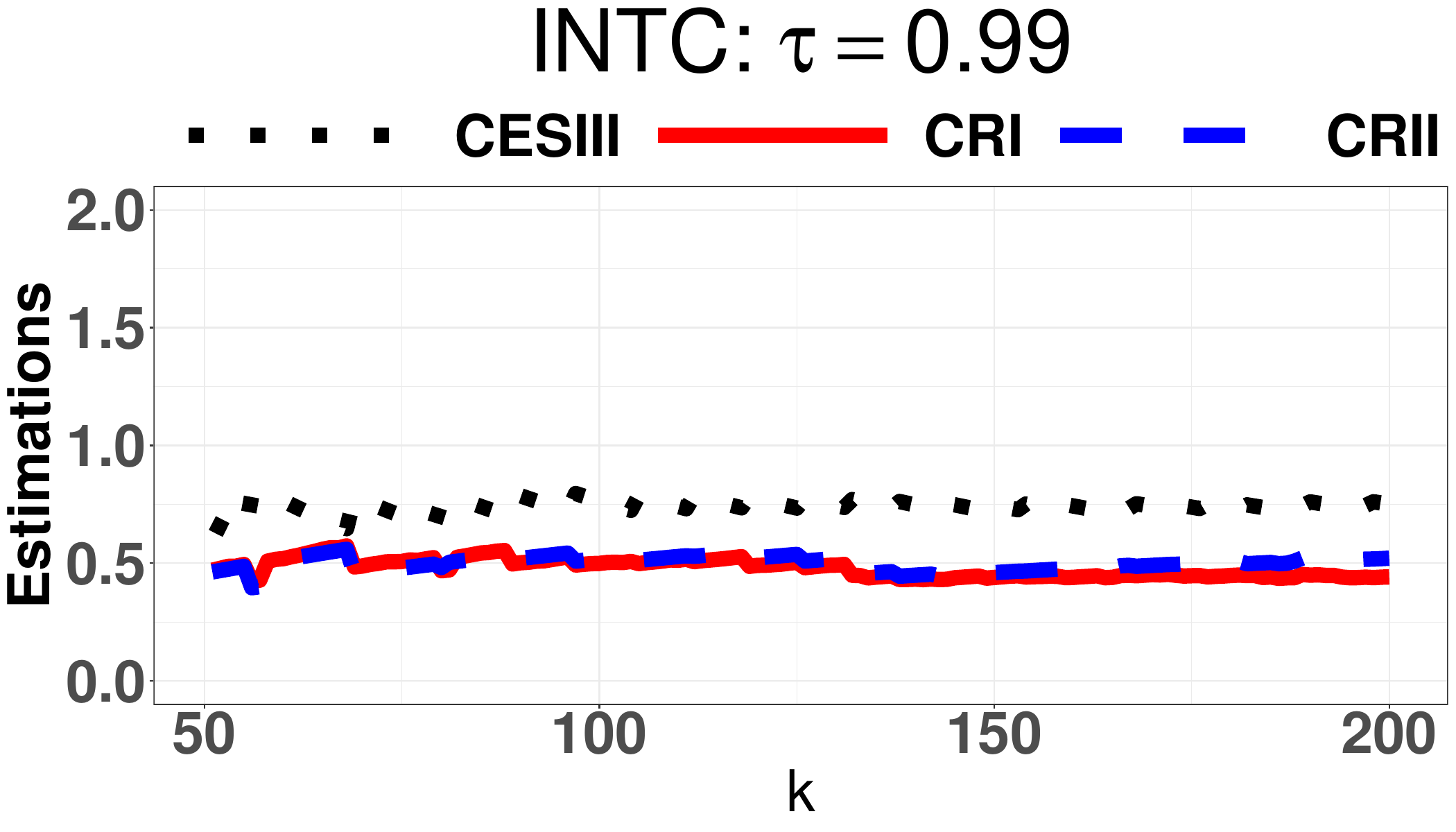}
\end{minipage}
\hspace{0.02\textwidth}
\begin{minipage}[b]{0.25\textwidth}
\includegraphics[width=\textwidth,height = 0.11\textheight]{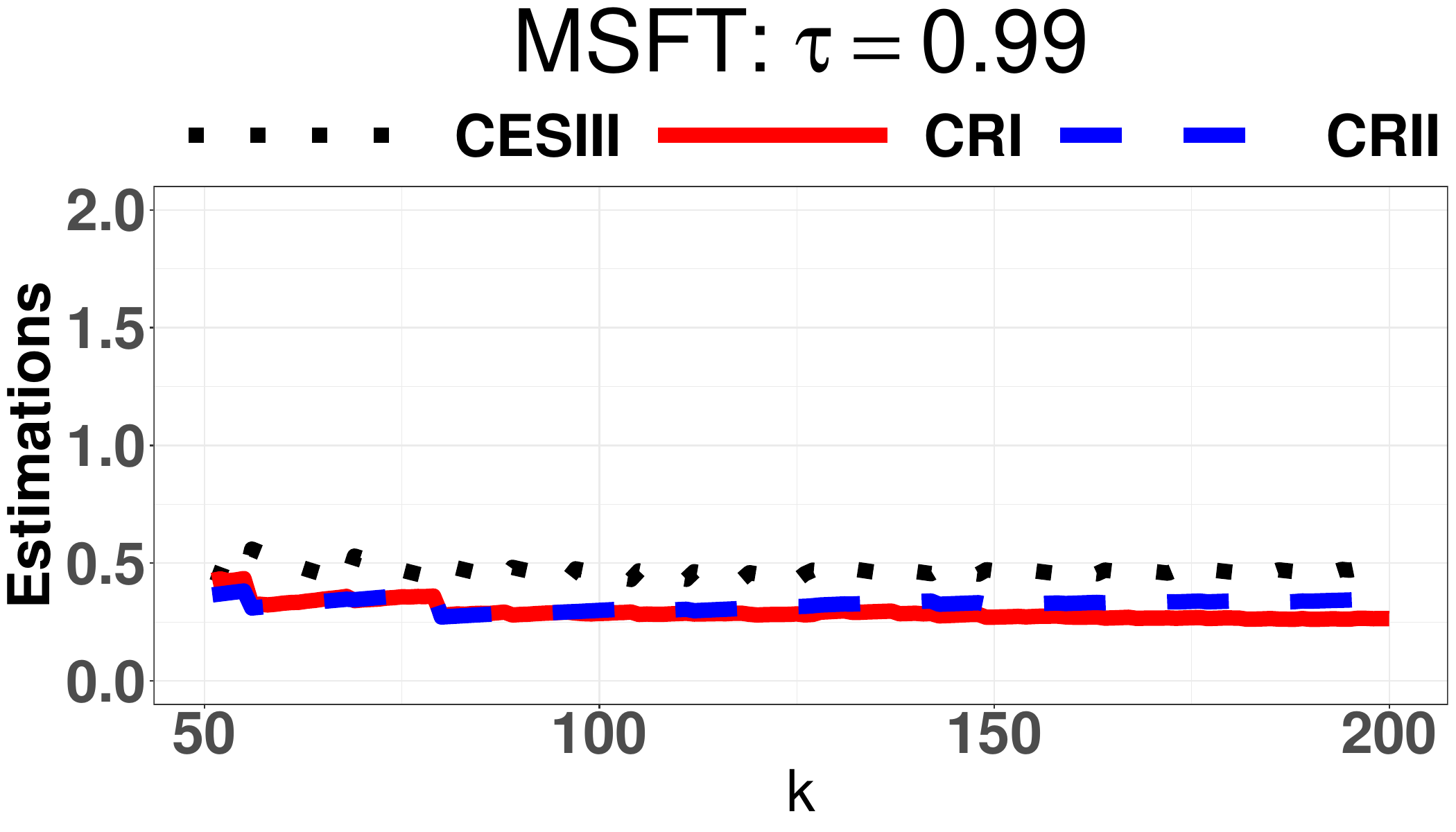}
\end{minipage}
\\
\begin{minipage}[b]{0.25\textwidth}
\includegraphics[width=\textwidth,height = 0.11\textheight]{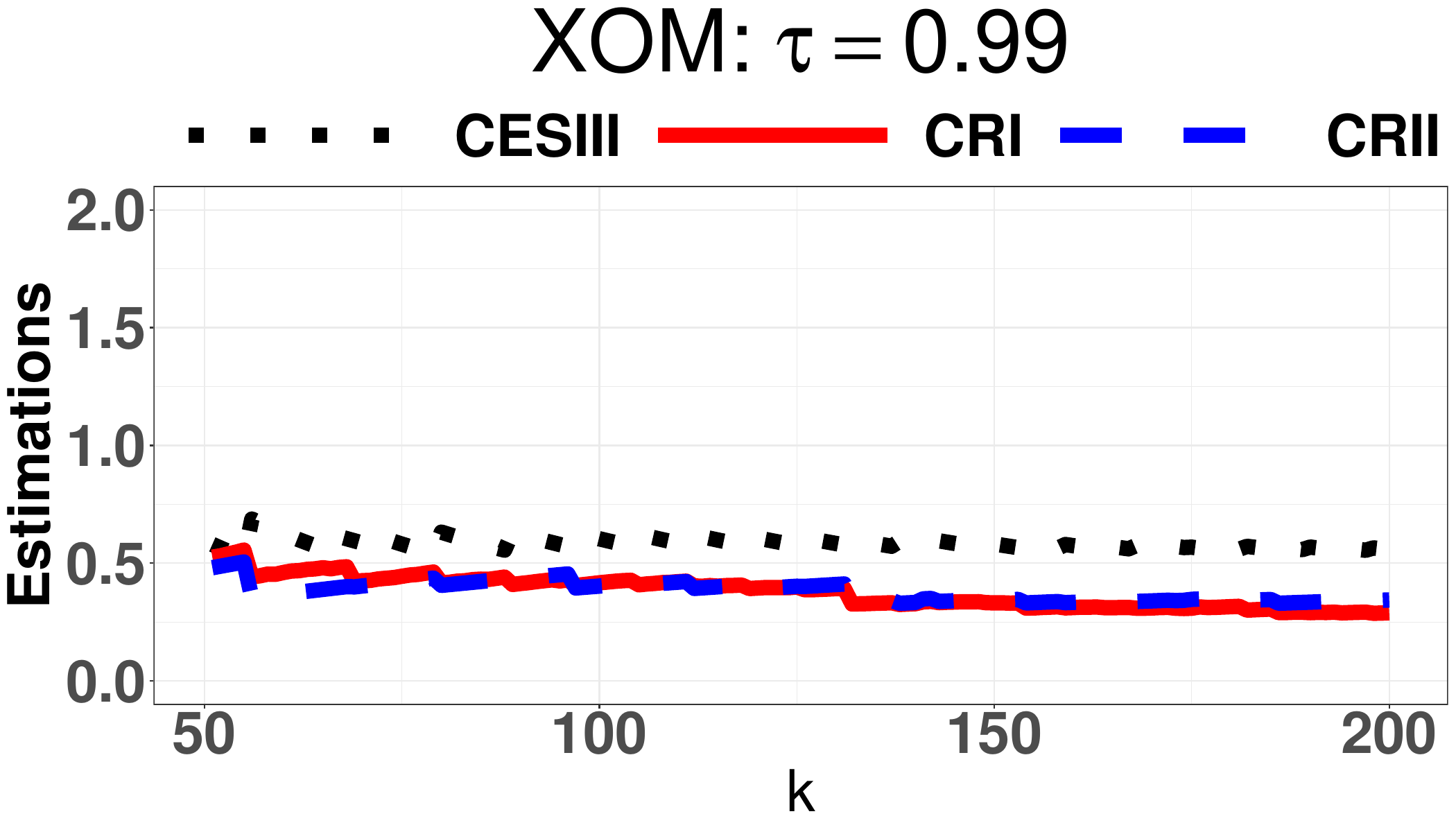}
\end{minipage}
\hspace{0.02\textwidth}
\begin{minipage}[b]{0.25\textwidth}
\includegraphics[width=\textwidth,height = 0.11\textheight]{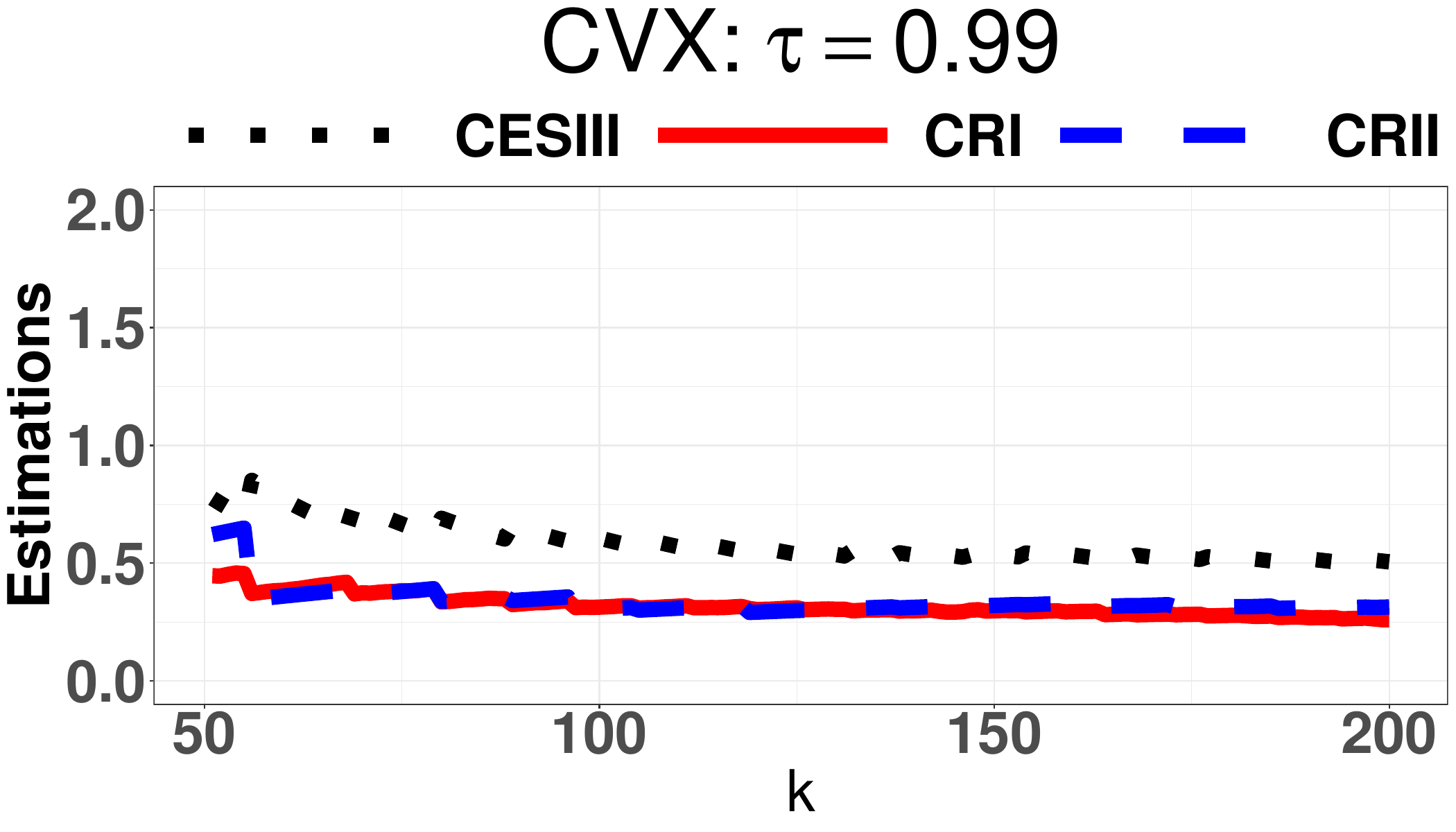}
\end{minipage}
\hspace{0.02\textwidth}
\begin{minipage}[b]{0.25\textwidth}
\includegraphics[width=\textwidth,height = 0.11\textheight]{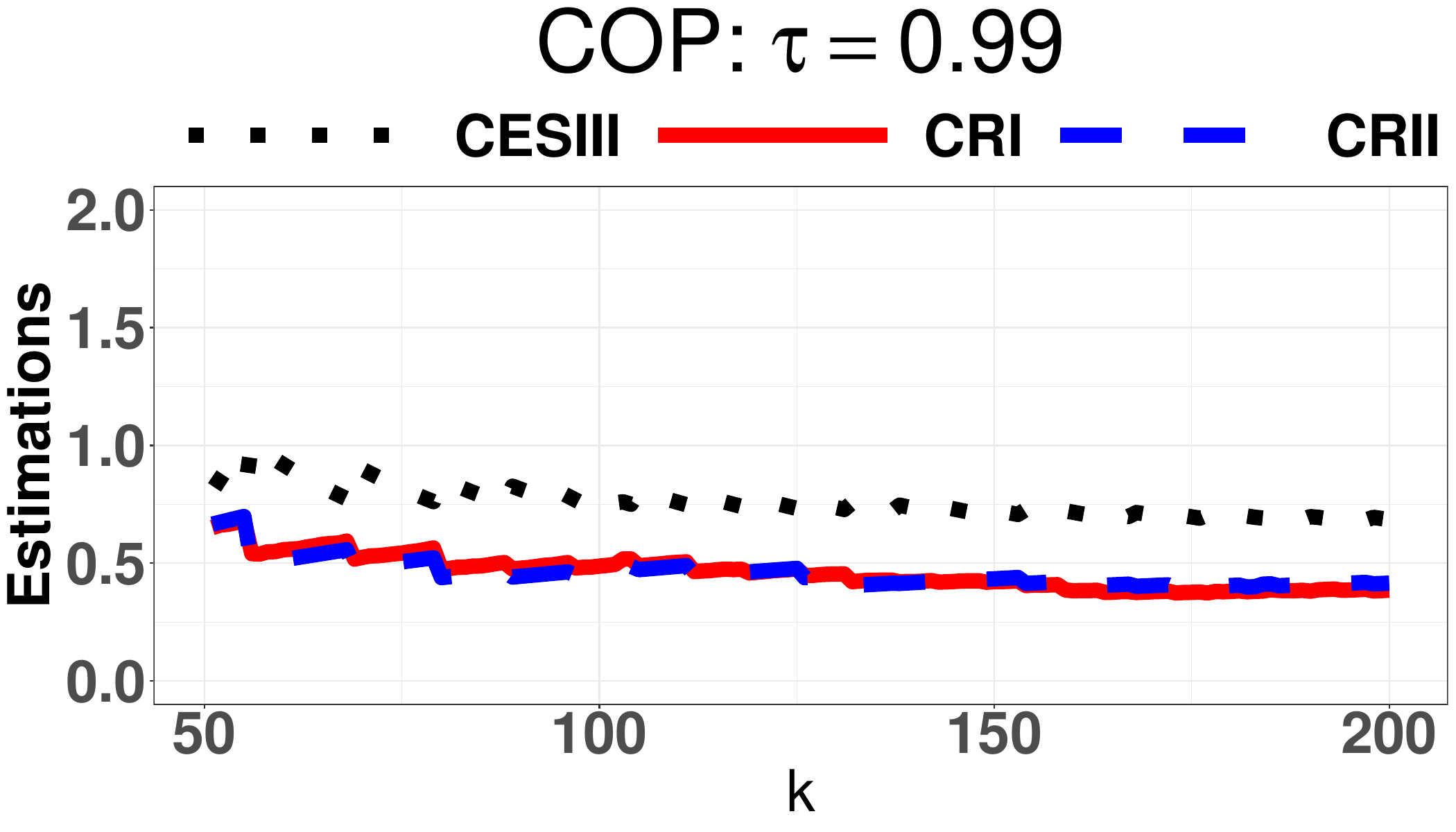}
\end{minipage}
\\
\begin{minipage}[b]{0.25\textwidth}
\includegraphics[width=\textwidth,height = 0.11\textheight]{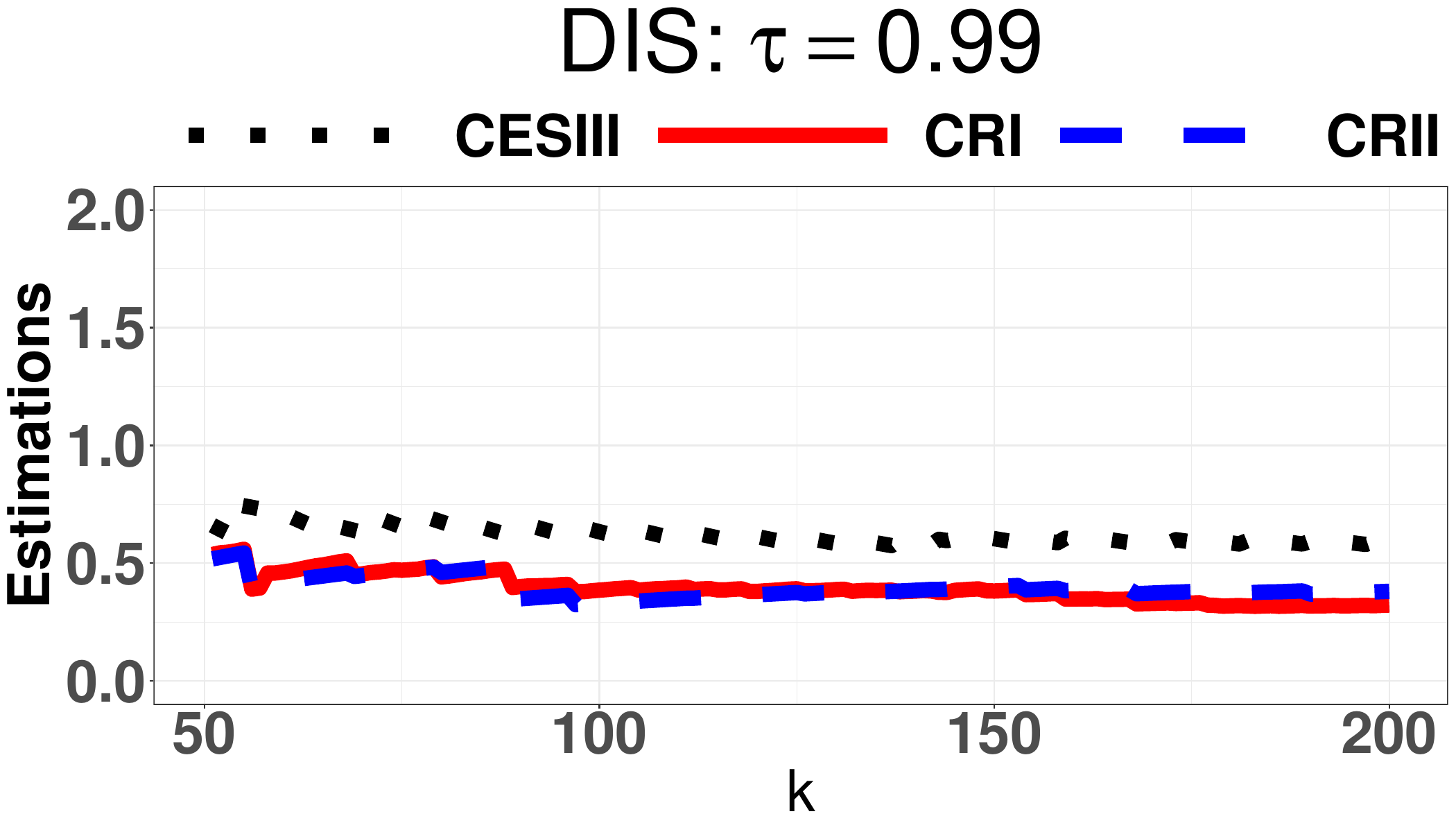}
\end{minipage}
\hspace{0.02\textwidth}
\begin{minipage}[b]{0.25\textwidth}
\includegraphics[width=\textwidth,height = 0.11\textheight]{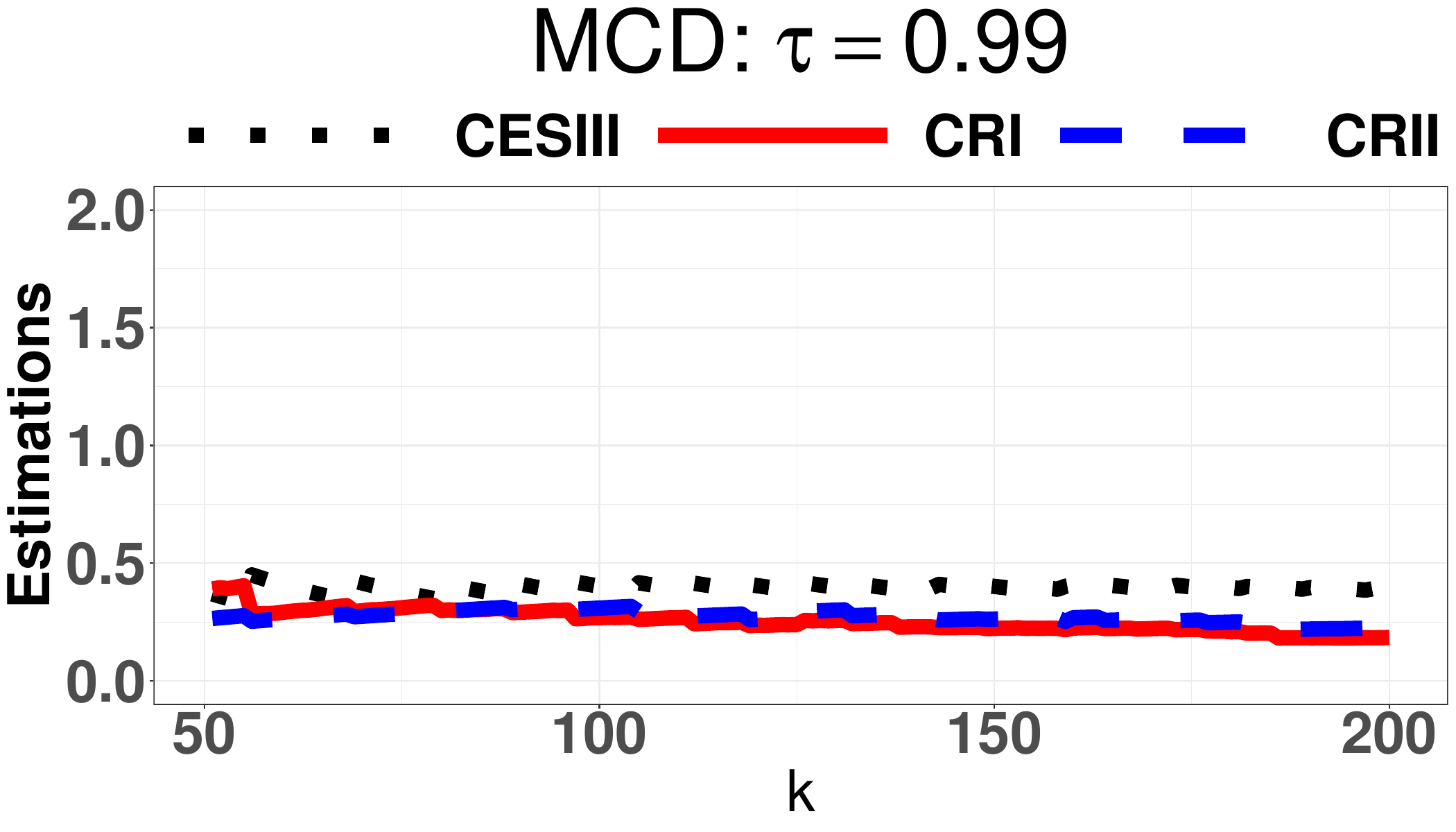}
\end{minipage}
\hspace{0.02\textwidth}
\begin{minipage}[b]{0.25\textwidth}
\includegraphics[width=\textwidth,height = 0.11\textheight]{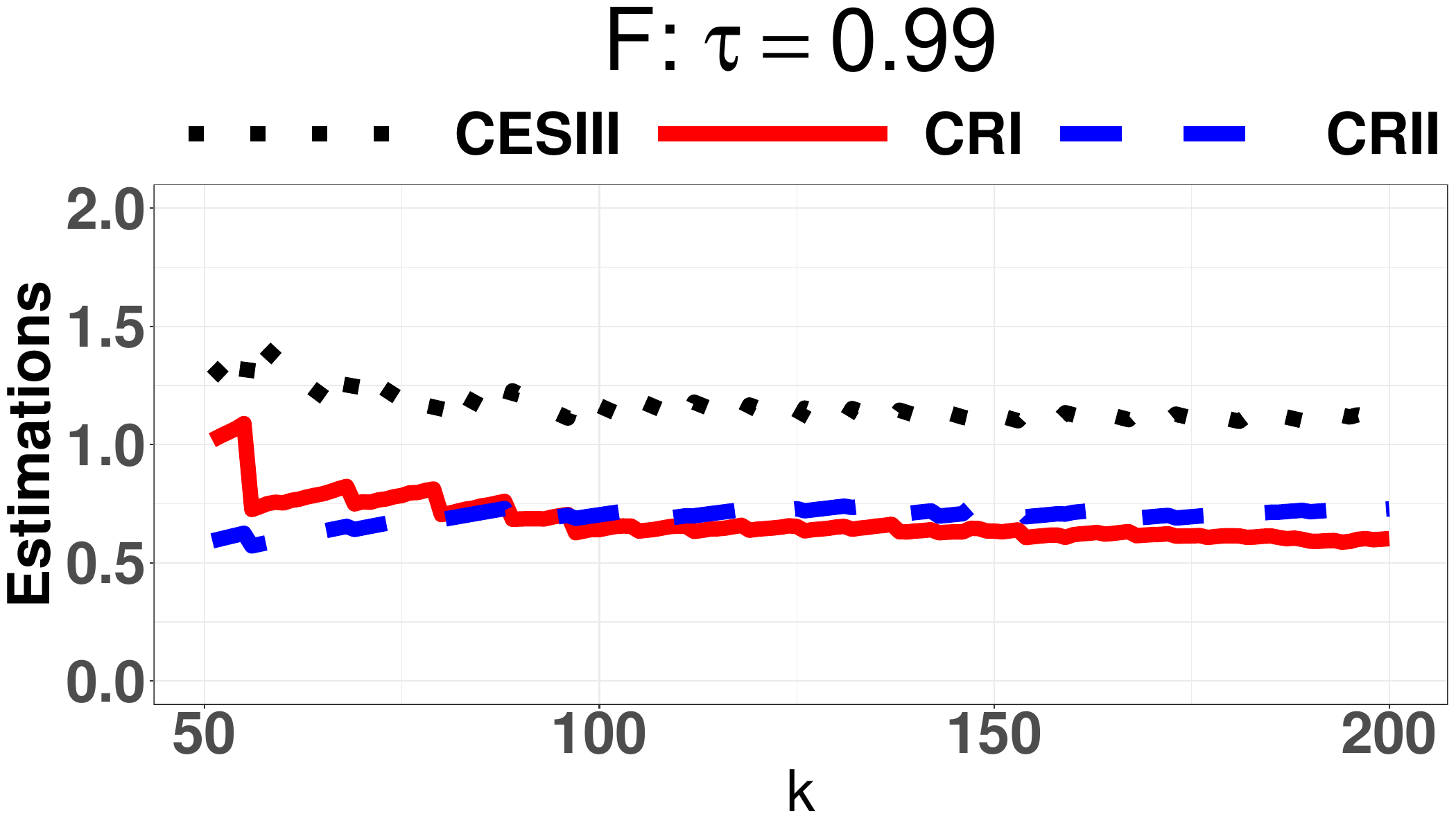}
\end{minipage}
\\
\begin{minipage}[b]{0.25\textwidth}
\includegraphics[width=\textwidth,height = 0.11\textheight]{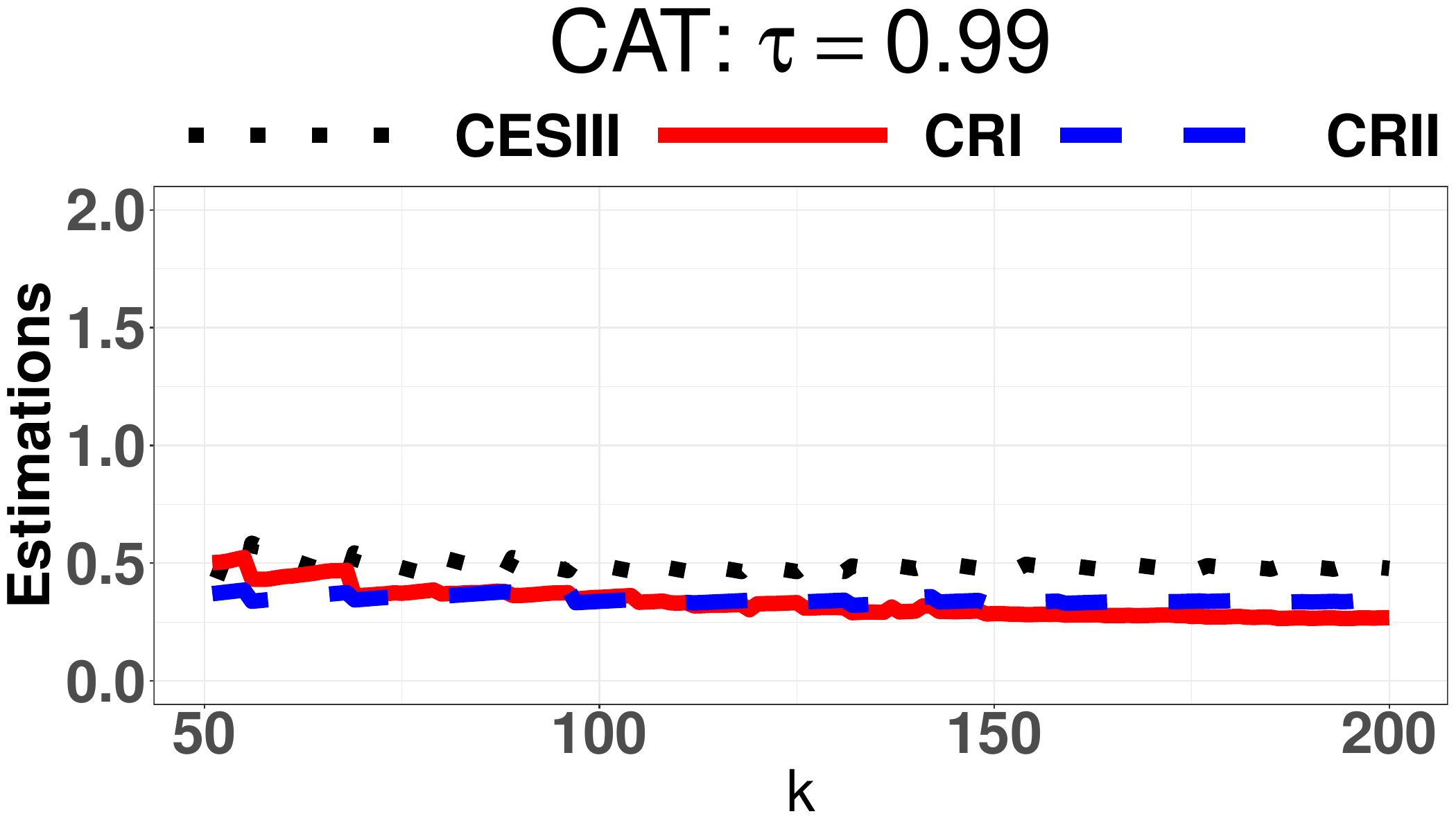}
\end{minipage}
\hspace{0.02\textwidth}
\begin{minipage}[b]{0.25\textwidth}
\includegraphics[width=\textwidth,height = 0.11\textheight]{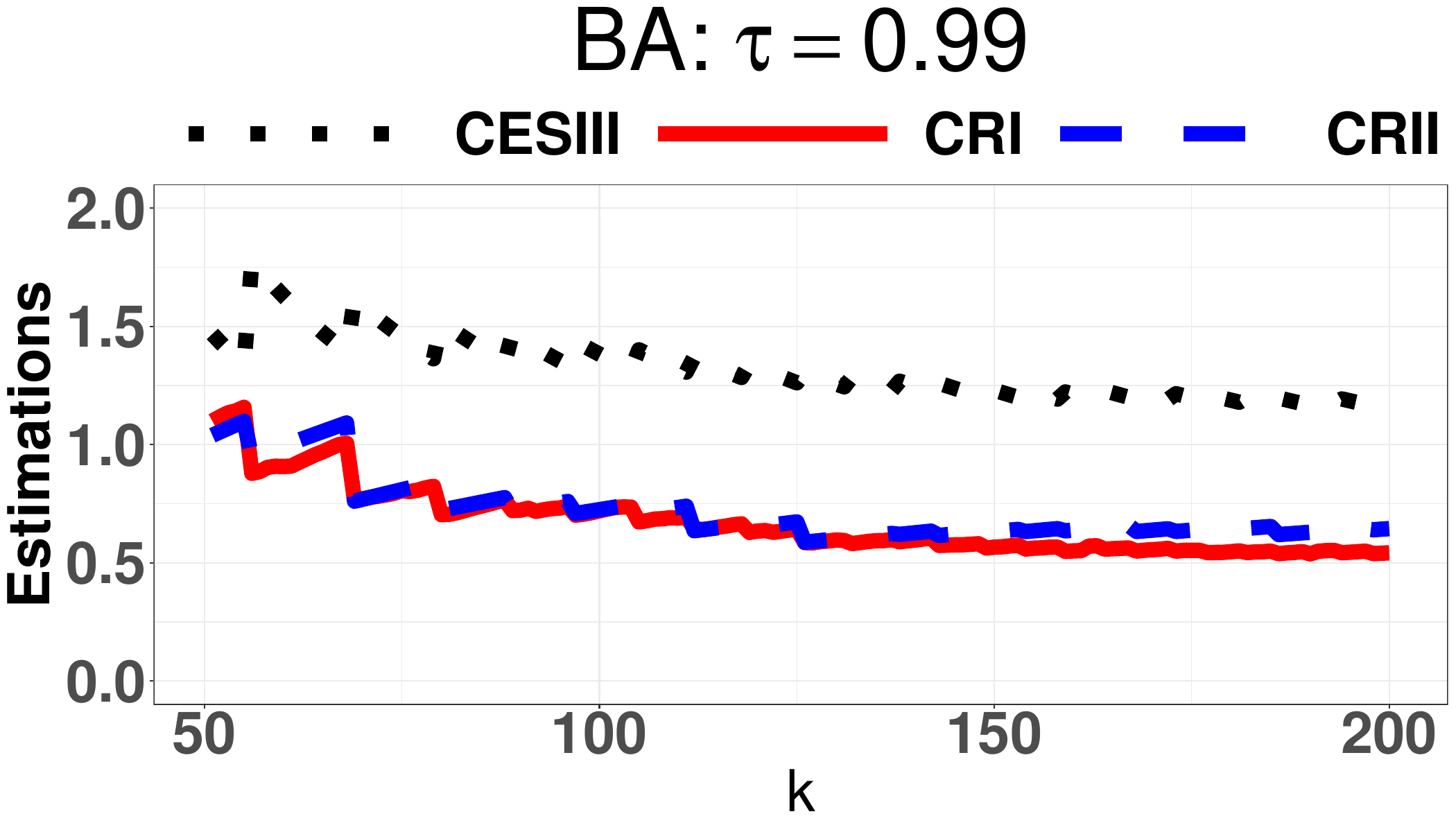}
\end{minipage}
\hspace{0.02\textwidth}
\begin{minipage}[b]{0.25\textwidth}
\includegraphics[width=\textwidth,height = 0.11\textheight]{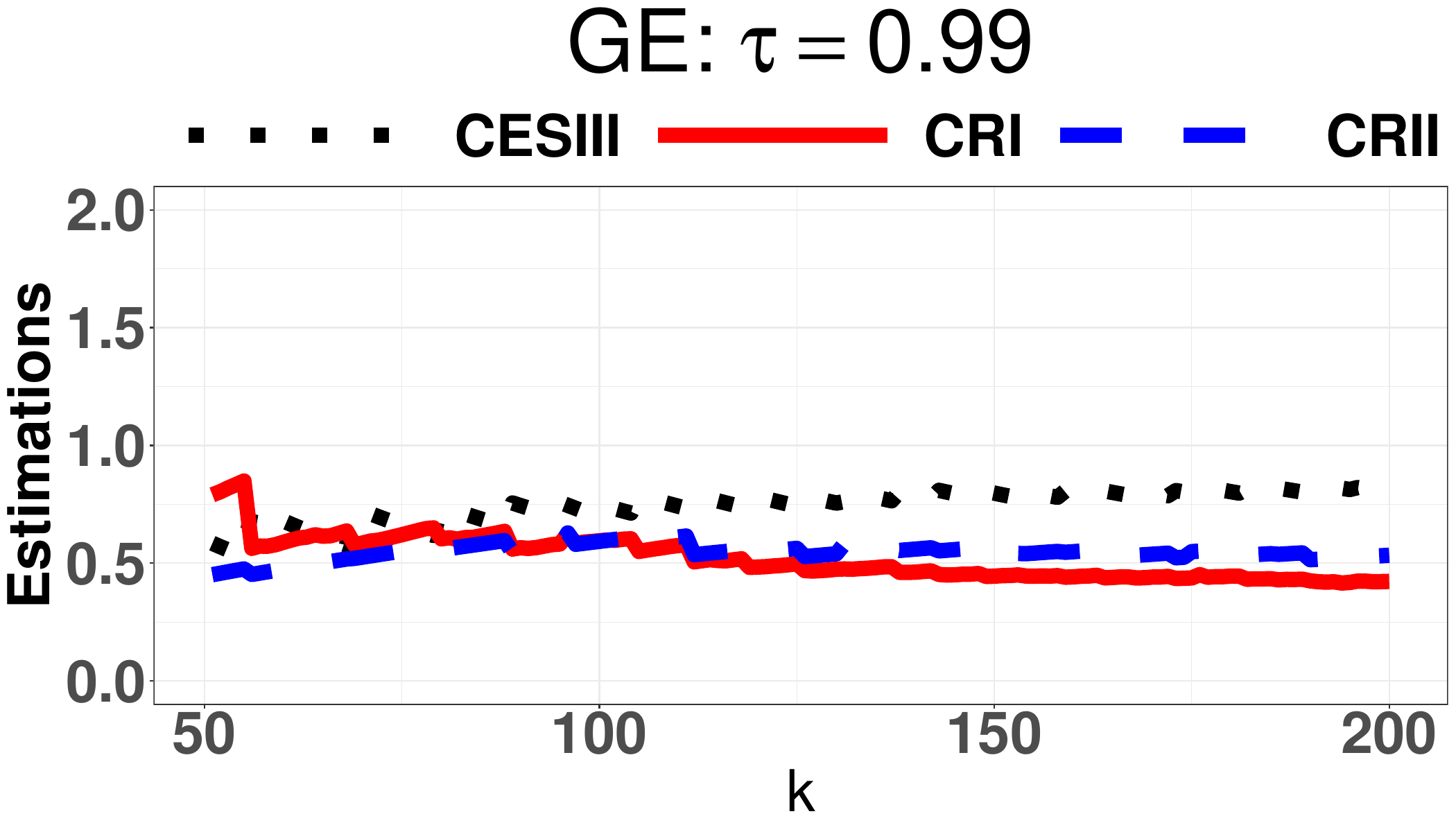}
\end{minipage}
\\
\begin{minipage}[b]{0.25\textwidth}
\includegraphics[width=\textwidth,height = 0.11\textheight]{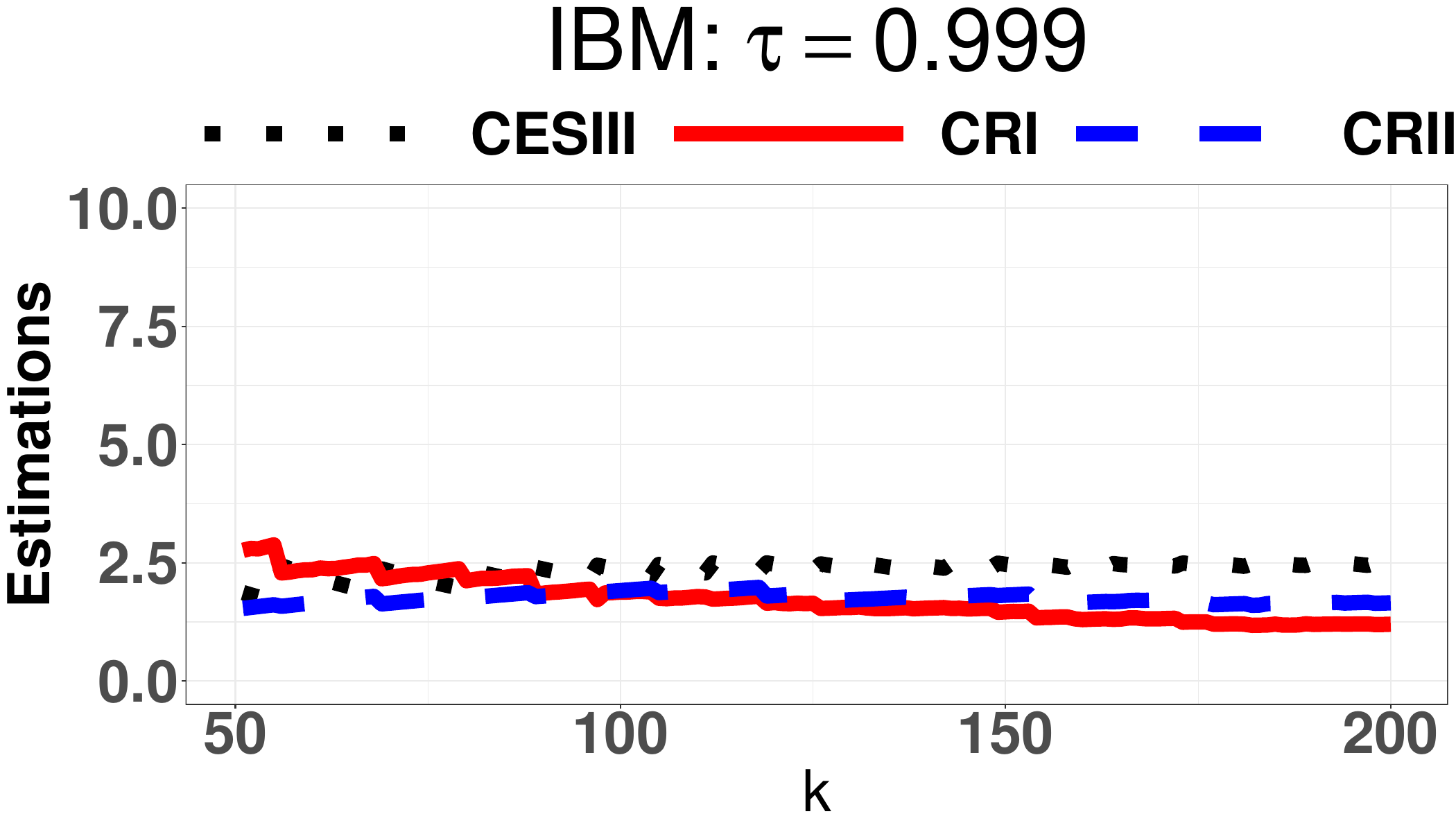}
\end{minipage}
\hspace{0.02\textwidth}
\begin{minipage}[b]{0.25\textwidth}
\includegraphics[width=\textwidth,height = 0.11\textheight]{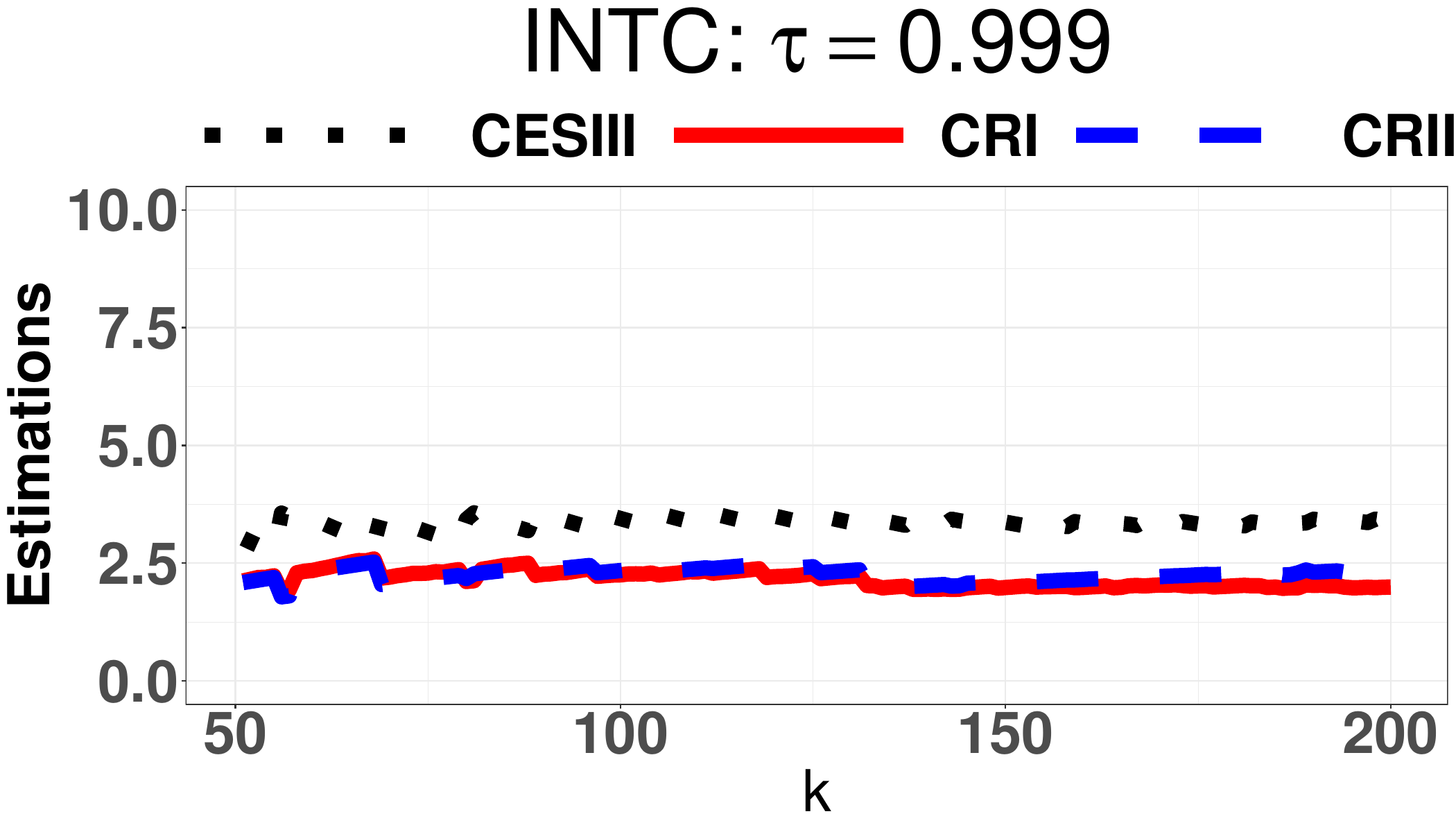}
\end{minipage}
\hspace{0.02\textwidth}
\begin{minipage}[b]{0.25\textwidth}
\includegraphics[width=\textwidth,height = 0.11\textheight]{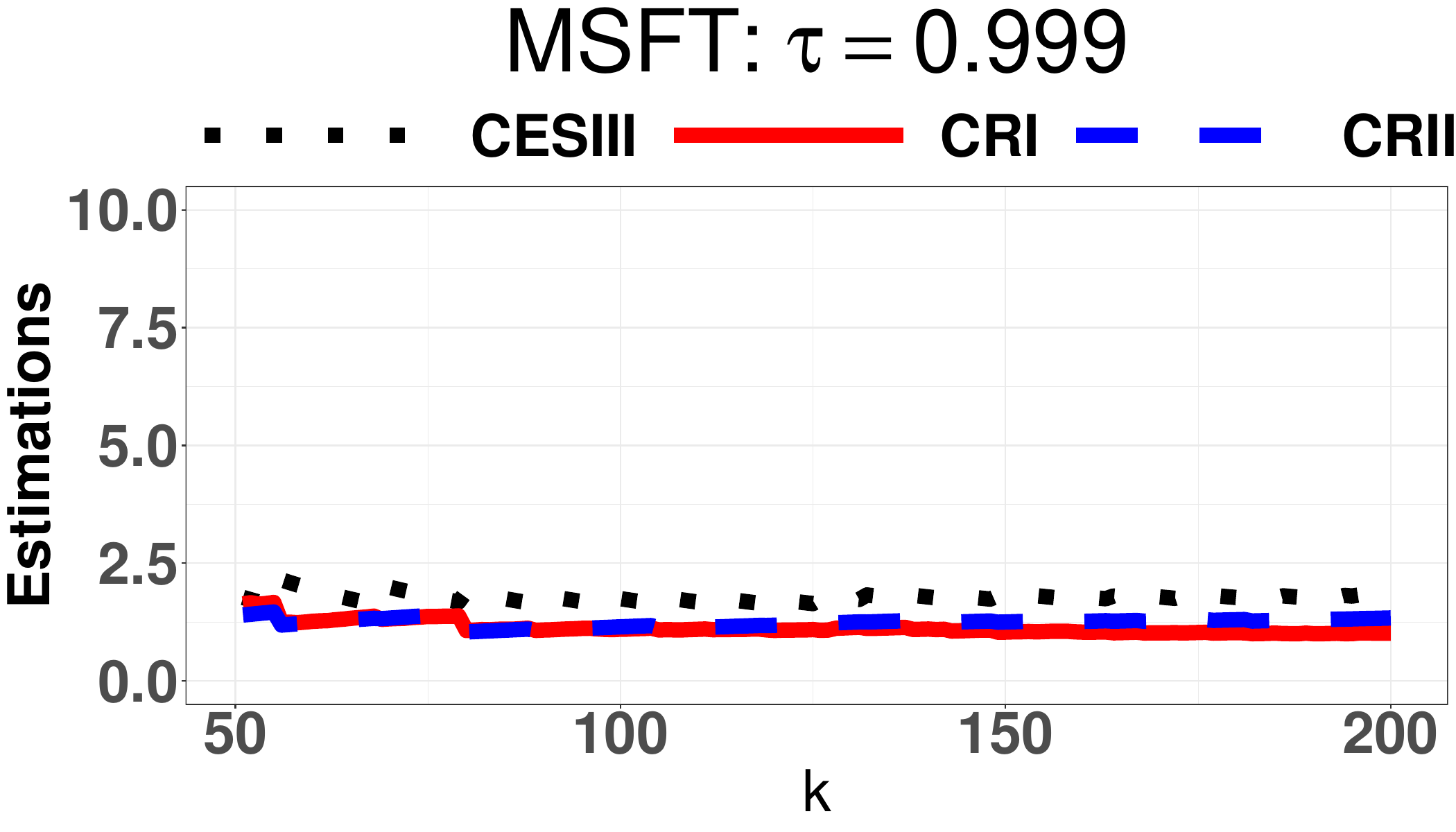}
\end{minipage}
\\
\begin{minipage}[b]{0.25\textwidth}
\includegraphics[width=\textwidth,height = 0.11\textheight]{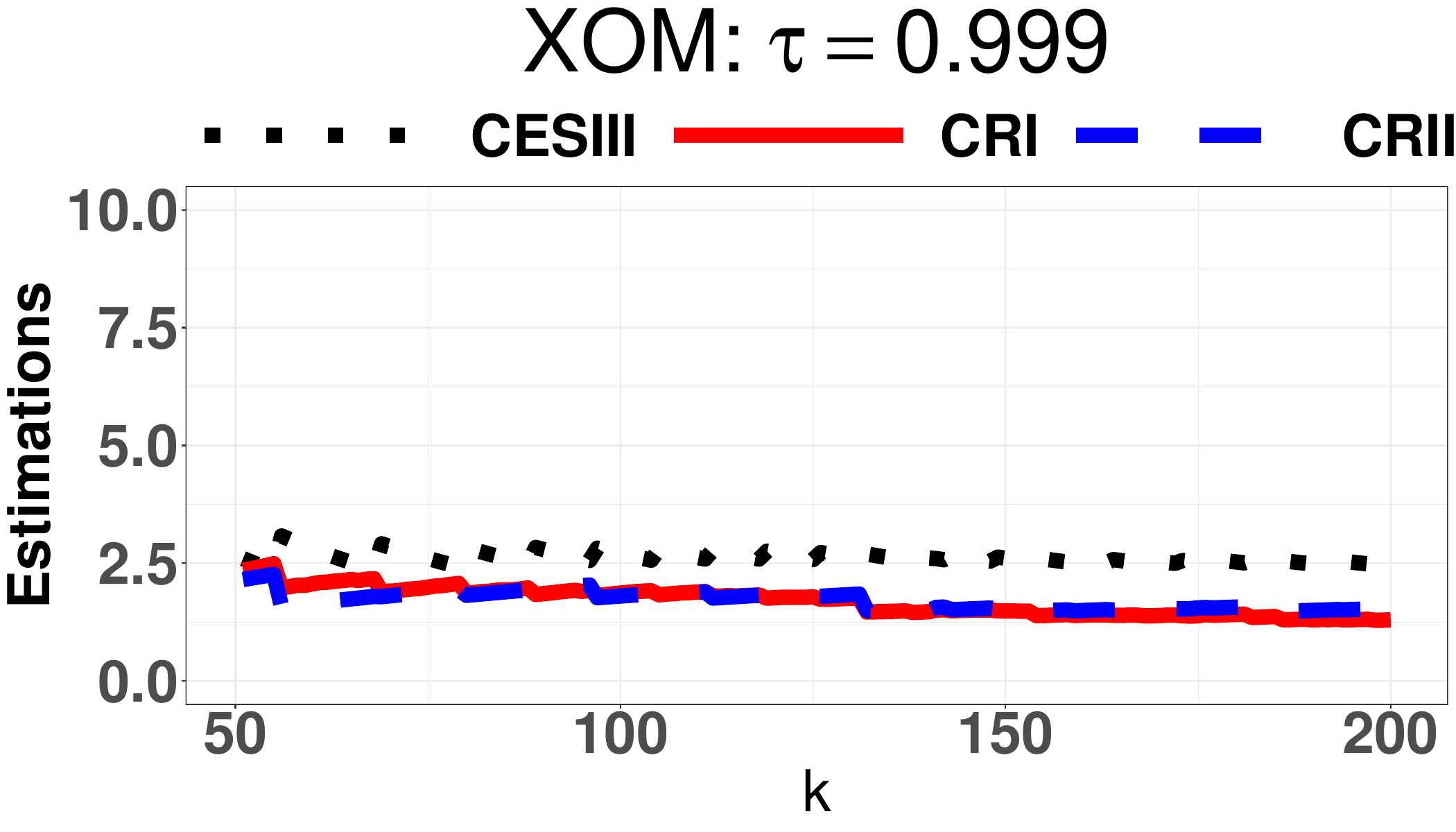}
\end{minipage}
\hspace{0.02\textwidth}
\begin{minipage}[b]{0.25\textwidth}
\includegraphics[width=\textwidth,height = 0.11\textheight]{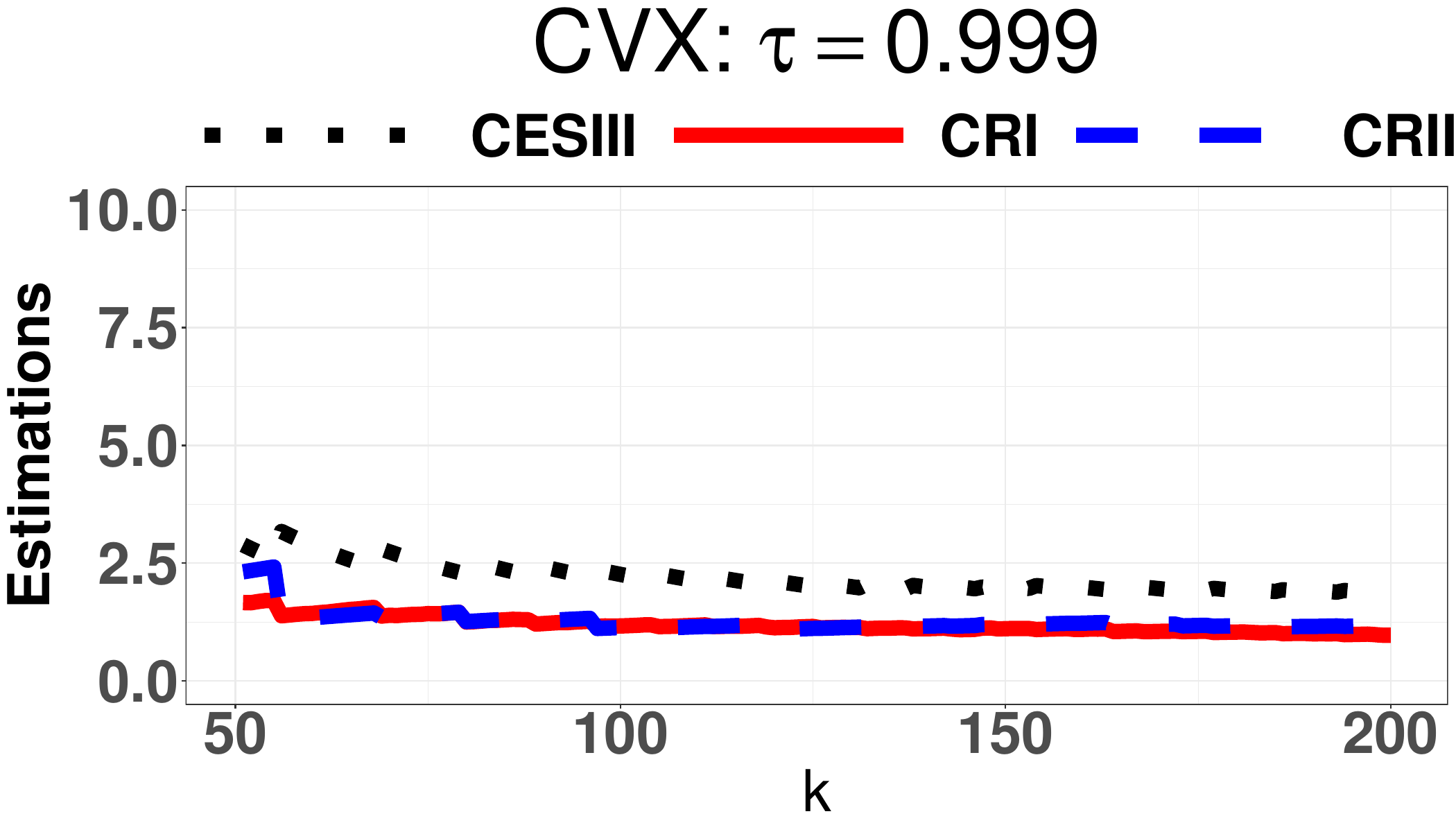}
\end{minipage}
\hspace{0.02\textwidth}
\begin{minipage}[b]{0.25\textwidth}
\includegraphics[width=\textwidth,height = 0.11\textheight]{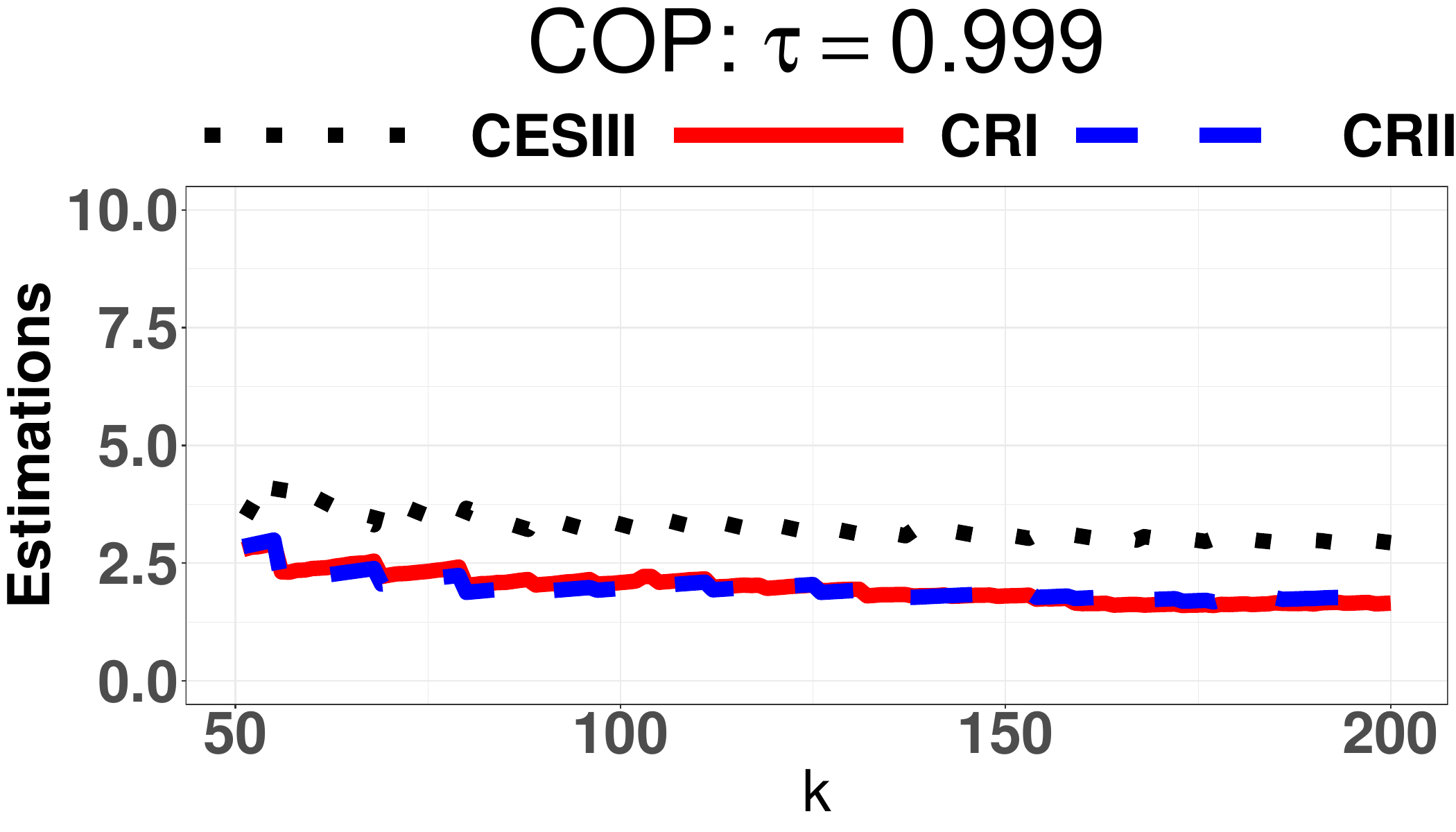}
\end{minipage}
\\
\begin{minipage}[b]{0.25\textwidth}
\includegraphics[width=\textwidth,height = 0.11\textheight]{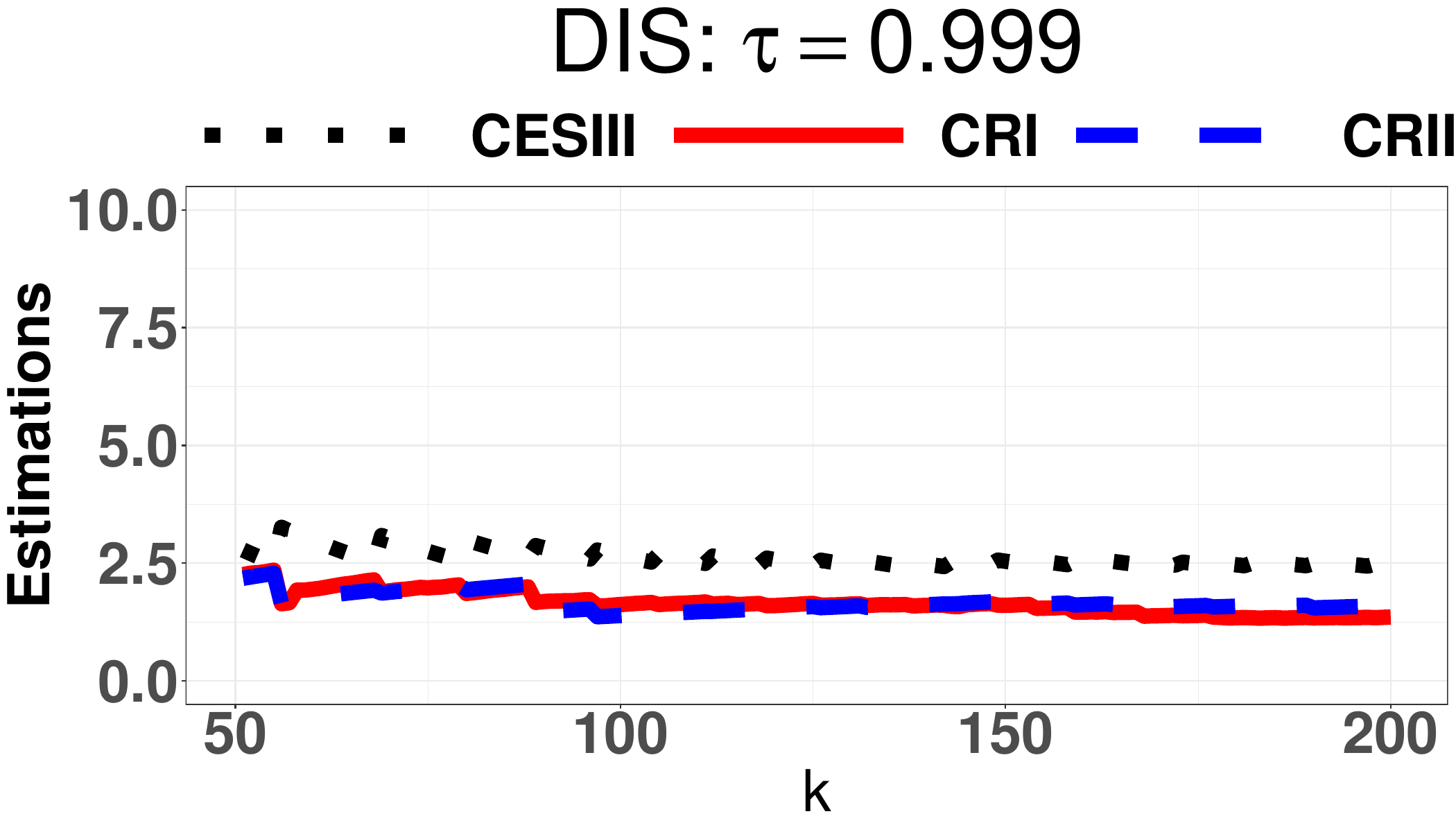}
\end{minipage}
\hspace{0.02\textwidth}
\begin{minipage}[b]{0.25\textwidth}
\includegraphics[width=\textwidth,height = 0.11\textheight]{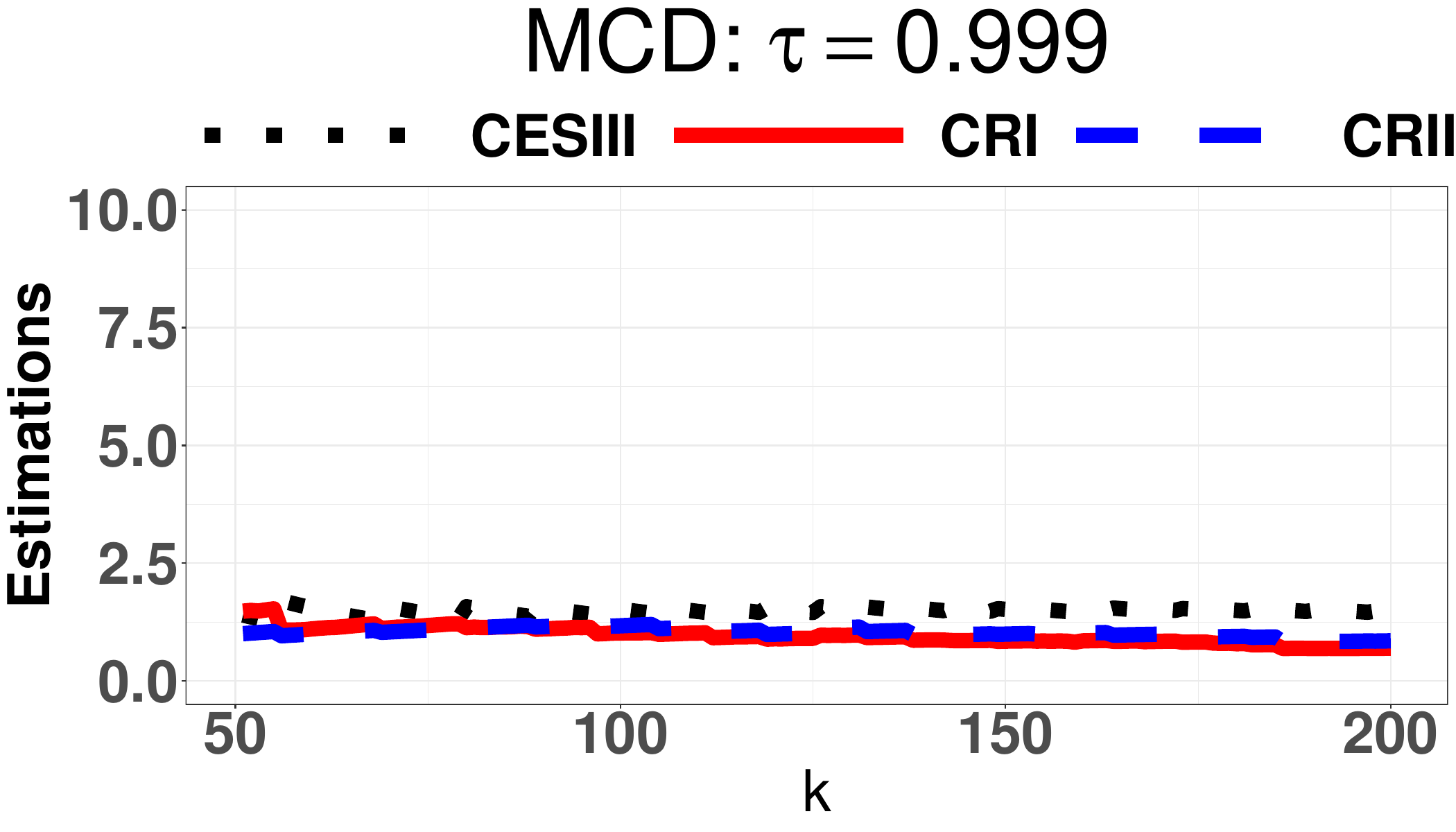}
\end{minipage}
\hspace{0.02\textwidth}
\begin{minipage}[b]{0.25\textwidth}
\includegraphics[width=\textwidth,height = 0.11\textheight]{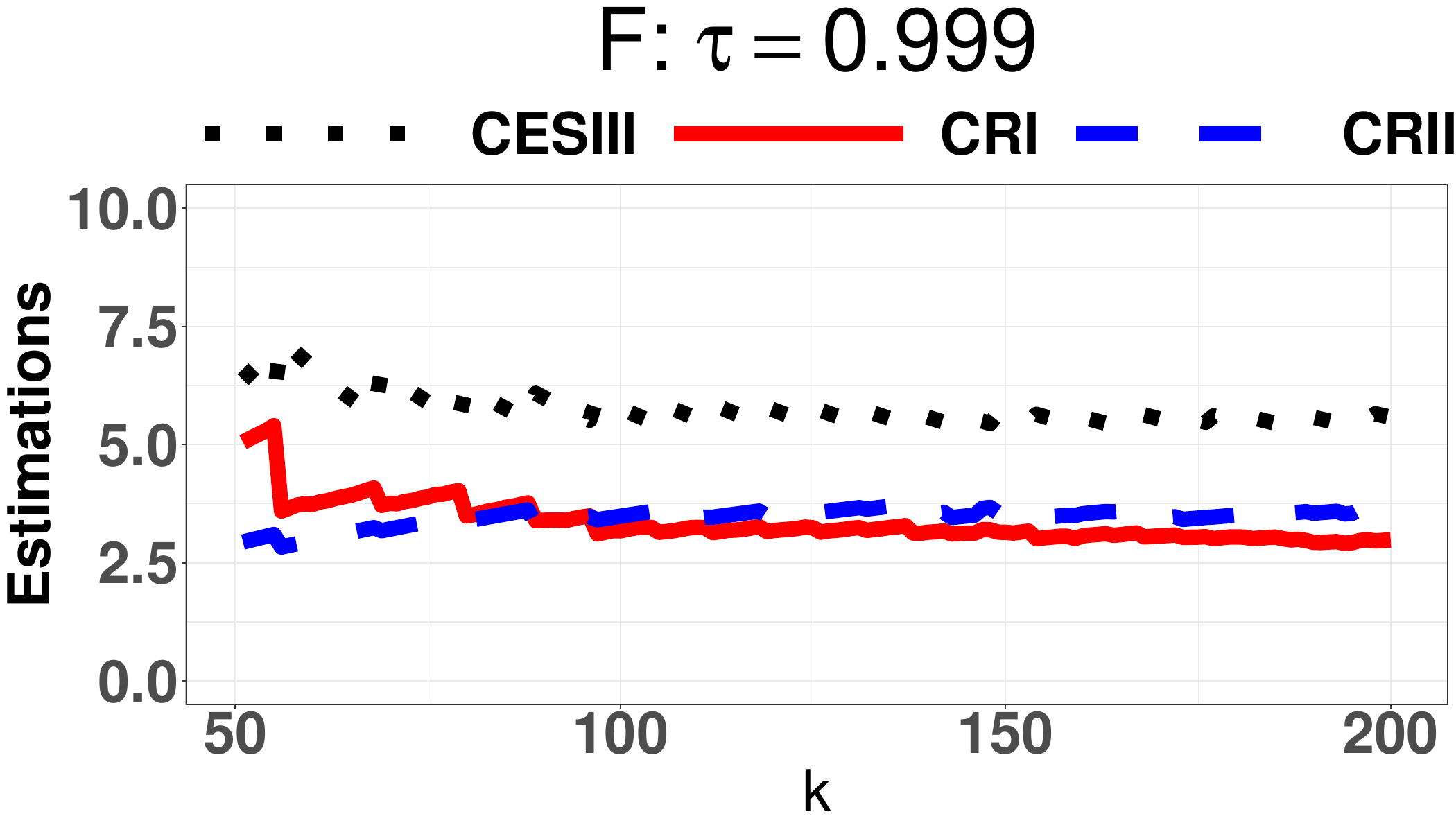}
\end{minipage}
\\
\begin{minipage}[b]{0.25\textwidth}
\includegraphics[width=\textwidth,height = 0.11\textheight]{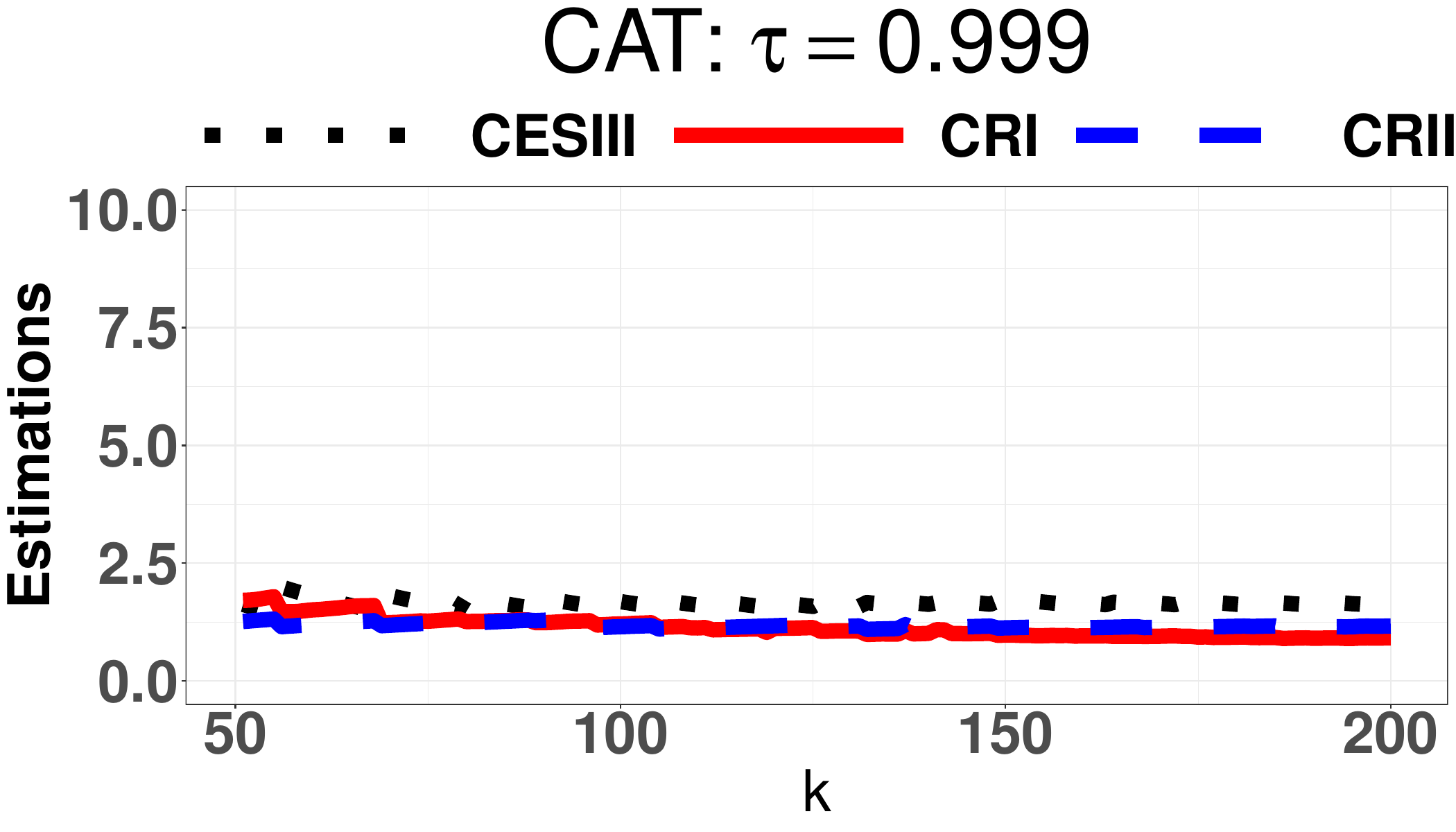}
\end{minipage}
\hspace{0.02\textwidth}
\begin{minipage}[b]{0.25\textwidth}
\includegraphics[width=\textwidth,height = 0.11\textheight]{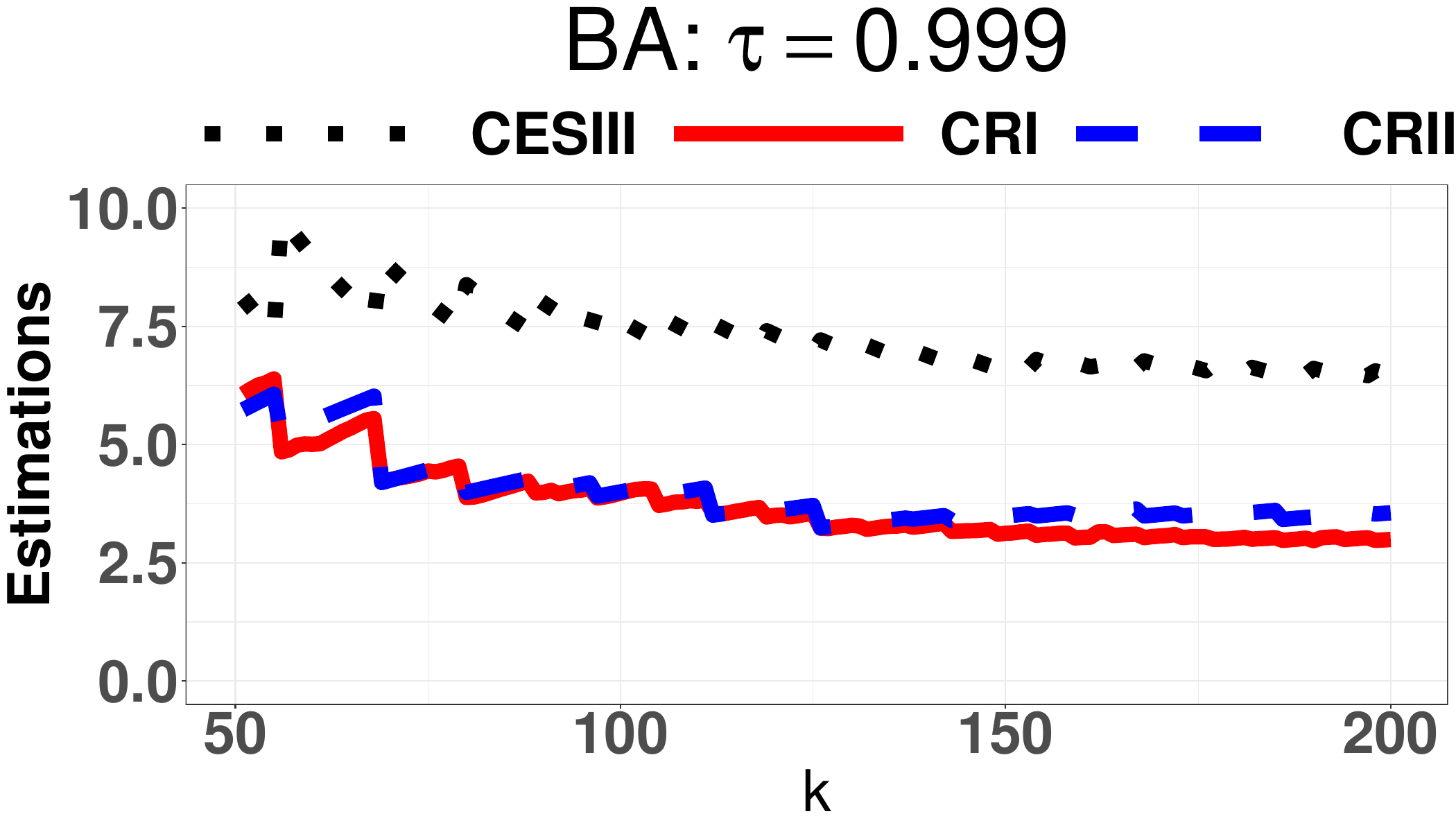}
\end{minipage}
\hspace{0.02\textwidth}
\begin{minipage}[b]{0.25\textwidth}
\includegraphics[width=\textwidth,height = 0.11\textheight]{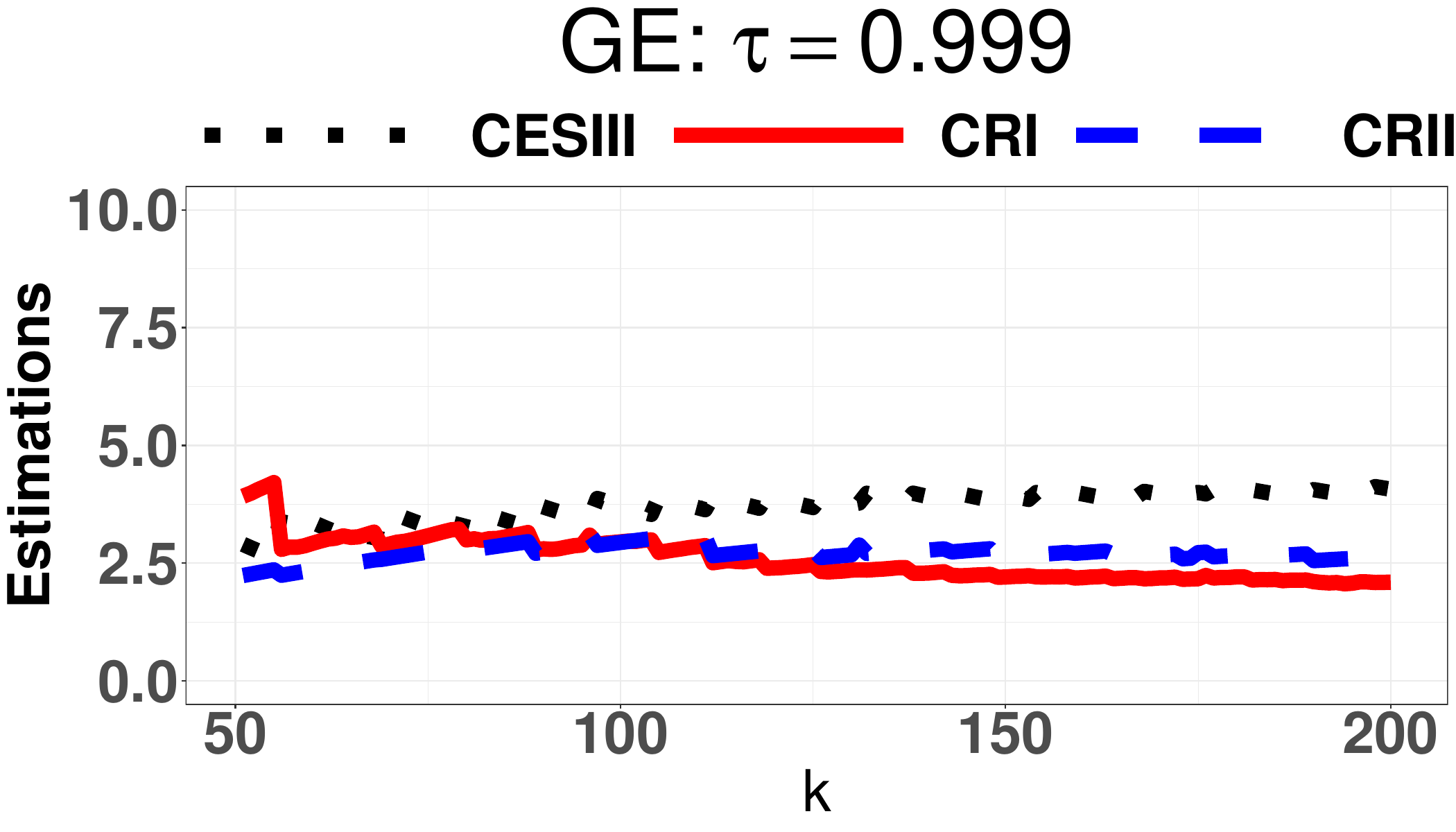}
\end{minipage}
\caption{The estimations $\widetilde{\covar}_{X|Y}^{(1)}(\tau'_n)$ (CRI in red solid lines), $\widetilde{\covar}_{X|Y}^{(2)}(\tau'_n)$ (CRII in blue dashed lines) and $\widetilde{\coes}_{X|Y}^{(3)}(\tau'_n)$ (CESIII in black dotted lines) against $k$ for the 12 individual stocks conditional on S\&P500 Index, with $\tau'_n = 0.99$ (top four panels) and 0.999 (bottom four panels).}
\label{fig:CRES_k}
\end{figure}

Before investigating the extreme losses measured by CoVaR and CoES of individual stocks conditional on S\&P500 Index, we would like to check the asymptotic independence between each pair of $(X_i,Y)$ for $i=1,...,12$ and choose suitable values for three hyperparameters $k, k_1$ and $k_2$. The estimations for $\gamma_1$ and $\eta$ against varying levels of $k_1$ and $k_2$ are represented in Figure \ref{fig:Gam_Eta}. From the Hill plots, it can be observed that the estimations of $\gamma_1$ for all 12 individual losses generally fall within the range of $(0,0.5)$. We also observe that, for most levels of $k_2$, the estimations of $\eta$ lie within the interval $(0.78,0.87)$ and approach around 0.8 as $k_2$ grows. These results provide significant evidence that $X_i$ and $Y$ are asymptotically independent for all $i=1,...,12$, as well as with positive association between their tails. The values of $k_1$ and $k_2$ can be chosen such that the estimations $\hat{\gamma}_1$ and $\hat{\eta}$ stabilize. We summarize the specific values of $k_1$ and $k_2$ and corresponding $\hat{\gamma}_1$ and $\hat{\eta}$ in Table \ref{tab:CRES_values}. For the selection of $k$, as shown in Figure \ref{fig:CRES_k}, we plot the estimations $\widetilde{\covar}_{X|Y}^{(1)}(\tau'_n)$, $\widetilde{\covar}_{X|Y}^{(2)}(\tau'_n)$ and $\widetilde{\coes}_{X|Y}^{(3)}(\tau'_n)$ with $\tau'_n = 0.99$, 0.999 against varying $k$. As one can observe that all the three estimations exhibit consistent patterns across different $\tau'_n$ and tend to stable as $k$ increase; therefore, we select the values of $k$ for which the estimations achieve stability. These chosen values of $k$ are also collected in Table \ref{tab:CRES_values}.

\begin{table}
\centering
\caption{The estimations $\widetilde{\covar}_{X|Y}^{(i)}(\tau'_n)$ ($i = 1, 2$) and $\widetilde{\coes}_{X|Y}^{(i)}(\tau'_n)$ ($i = 1, 2, 3$) of the 12 individual firms conditional on S\&P500 Index, with $\tau'_n = 0.99$ and 0.999.}
\label{tab:CRES_values}
\setlength{\tabcolsep}{4pt}
\begin{tabular}{@{}cccccccccc@{}}
\hline\hline
\addlinespace[1.5pt]
Ticker & $(k, k_1, k_2)$ & $\hat{\gamma}_1$ & $\hat{\eta}$ & $\tau'_n$ & CoVaR-I & CoVaR-II & CoES-I & CoES-II & CoES-III \\[1.5pt]
\hline
& \multicolumn{9}{c}{Information Technology Sector} \\[1.5pt]
\hline
\textbf{IBM} & $(120,75,100)$ & 0.35785 & 0.82820 & 0.99 & 0.37879 & 0.41245 & 0.58989 & 0.64230 & 0.56368 \\[1.5pt]
             &                &         &         & 0.999& 1.65912 & 1.80652 & 2.58371 & 2.81326 & 2.46895 \\[1.5pt]
\textbf{INTC}& $(125,75,200)$ & 0.37186 & 0.80667 & 0.99 & 0.50422 & 0.53746 & 0.80272 & 0.85564 & 0.73419 \\[1.5pt]
             &                &         &         & 0.999& 2.27626 & 2.42634 & 3.62382 & 3.86274 & 3.31445 \\[1.5pt]
\textbf{MSFT}& $(100,75,100)$ & 0.33510 & 0.79462 & 0.99 & 0.28707 & 0.30003 & 0.43175 & 0.45124 & 0.45597 \\[1.5pt]
             &                &         &         & 0.999& 1.10047 & 1.15016 & 1.65509 & 1.72982 & 1.74794 \\[1.5pt]
\hline
& \multicolumn{9}{c}{Energy Sector} \\[1.5pt]
\hline
\textbf{XOM} & $(120,75,275)$ & 0.35499 & 0.85807 & 0.99 & 0.39466 & 0.39174 & 0.61186 & 0.60734 & 0.60774 \\[1.5pt]
             &                &         &         & 0.999& 1.76799 & 1.75493 & 2.74102 & 2.72078 & 2.72254 \\[1.5pt]
\textbf{CVX} & $(125,75,280)$ & 0.31764 & 0.83270 & 0.99 & 0.31093 & 0.29959 & 0.45566 & 0.43905 & 0.53785 \\[1.5pt]
             &                &         &         & 0.999& 1.15907 & 1.11682 & 1.69861 & 1.63669 & 2.00500 \\[1.5pt]
\textbf{COP} & $(120,75,250)$ & 0.35798 & 0.80985 & 0.99 & 0.46190 & 0.46588 & 0.71945 & 0.72564 & 0.77490 \\[1.5pt]
             &                &         &         & 0.999& 1.97902 & 1.99605 & 3.08247 & 3.10900 & 3.32006 \\[1.5pt]
\hline
& \multicolumn{9}{c}{Consumer Discretionary Sector} \\[1.5pt]
\hline
\textbf{DIS} & $(135,75,300)$ & 0.33940 & 0.86086 & 0.99 & 0.38434 & 0.37703 & 0.58180 & 0.57075 & 0.58671 \\[1.5pt]
             &                &         &         & 0.999& 1.61683 & 1.58610 & 2.44753 & 2.40102 & 2.46819 \\[1.5pt]
\textbf{MCD} & $(100,75,150)$ & 0.33096 & 0.79819 & 0.99 & 0.26894 & 0.30968 & 0.40198 & 0.46287 & 0.40241 \\[1.5pt]
             &                &         &         & 0.999& 1.01836 & 1.17262 & 1.52214 & 1.75271 & 1.52381 \\[1.5pt]
\textbf{F}   & $(105,75,300)$ & 0.38167 & 0.84876 & 0.99 & 0.63425 & 0.67251 & 1.02575 & 1.08764 & 1.18816 \\[1.5pt]
             &                &         &         & 0.999& 3.14479 & 3.33453 & 5.08597 & 5.39283 & 5.89127 \\[1.5pt]
\hline
& \multicolumn{9}{c}{Industrials Sector} \\[1.5pt]
\hline
\textbf{CAT} & $(120,75,175)$ & 0.29419 & 0.84092 & 0.99 & 0.32756 & 0.34397 & 0.46409 & 0.48734 & 0.49316 \\[1.5pt]
             &                &         &         & 0.999& 1.11689 & 1.17285 & 1.58243 & 1.66170 & 1.68154 \\[1.5pt]
\textbf{BA}  & $(130,75,175)$ & 0.41321 & 0.83014 & 0.99 & 0.59652 & 0.60089 & 1.01658 & 1.02404 & 1.25659 \\[1.5pt]
             &                &         &         & 0.999& 3.29230 & 3.31645 & 5.61073 & 5.65188 & 6.93541 \\[1.5pt]
\textbf{GE}  & $(100,75,175)$ & 0.38565 & 0.83589 & 0.99 & 0.59528 & 0.59290 & 0.96896 & 0.96508 & 0.75045 \\[1.5pt]
             &                &         &         & 0.999& 2.95330 & 2.94148 & 4.80717 & 4.78793 & 3.72307 \\[1.5pt]
\hline\hline
\end{tabular}
\end{table}

For the given values of $k, k_1$ and $k_2$, we calculate the extrapolative estimations $\widetilde{\covar}^{(i)}_{X|Y}(\tau'_n)$ ($i = 1,2$) and $\widetilde{\coes}^{(i)}_{X|Y}(\tau'_n)$ ($i = 1,2,3$) with extreme levels $\tau'_n = 0.99$ and 0.999, based on 1565 weekly losses of 12 individual stocks conditional on S\&P500 Index losses from January 8, 1995, to December 29, 2024. The results are also reported in Table \ref{tab:CRES_values}, which represent the average weekly losses of the 12 selected firms under a market crisis within a period of nearly 30 years. A straightforward observation from Table \ref{tab:CRES_values} gives that, as risk level becomes more extreme, all extrapolative estimations of CoVaR and CoES grow, which accords with conventional intuition. Moreover, the estimations $\widetilde{\covar}^{(1)}_{X|Y}(\tau'_n)$ and $\widetilde{\covar}^{(2)}_{X|Y}(\tau'_n)$ are generally very close in most cases, though their discrepancy increases with the level $\tau'_n$, implying that the extrapolations based on estimated $\hat{\xi}_{1-k/n}$ and $\widehat{\covar}_{X|Y}(1-k/n)$ exhibit comparable empirical performance. Note that estimations $\widetilde{\coes}^{(i)}_{X|Y}(\tau'_n)$ ($i = 1,2$) follow the same pattern as the corresponding $\widetilde{\covar}^{(i)}_{X|Y}(\tau'_n)$, since $\widetilde{\coes}^{(i)}_{X|Y}(\tau'_n)$ are obtained by multiplying $\widetilde{\covar}^{(i)}_{X|Y}(\tau'_n)$ by a $\hat{\gamma}_1$-dependent term. Except rare cases such as BA and GE, $\widetilde{\coes}^{(3)}_{X|Y}(\tau'_n)$ yields results comparable to those of the first two.

\begin{figure}[htbp]
\centering
\begin{minipage}[b]{0.32\textwidth}
\includegraphics[width=\textwidth,height = 0.14\textheight]{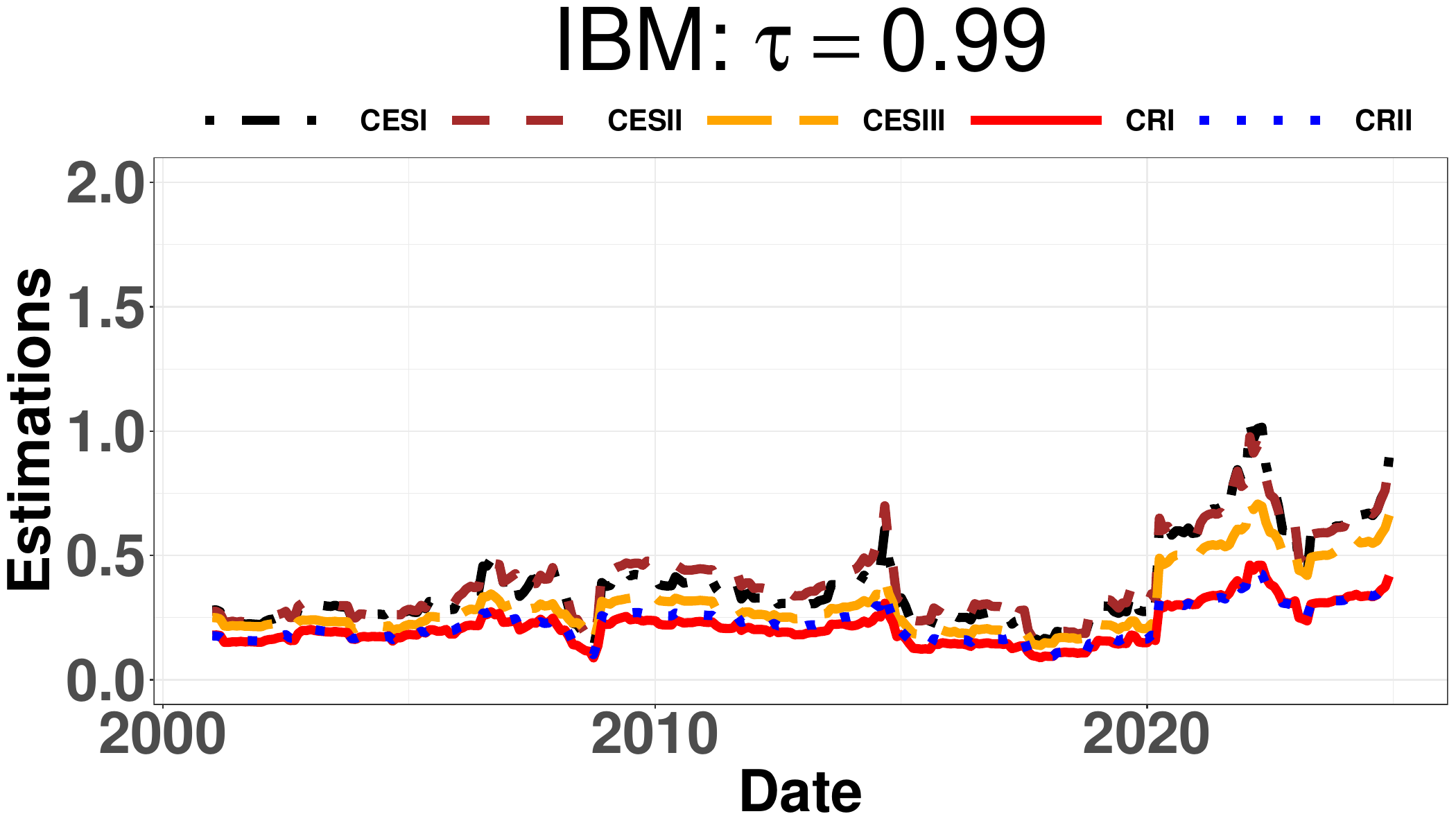}
\end{minipage}
\begin{minipage}[b]{0.32\textwidth}
\includegraphics[width=\textwidth,height = 0.14\textheight]{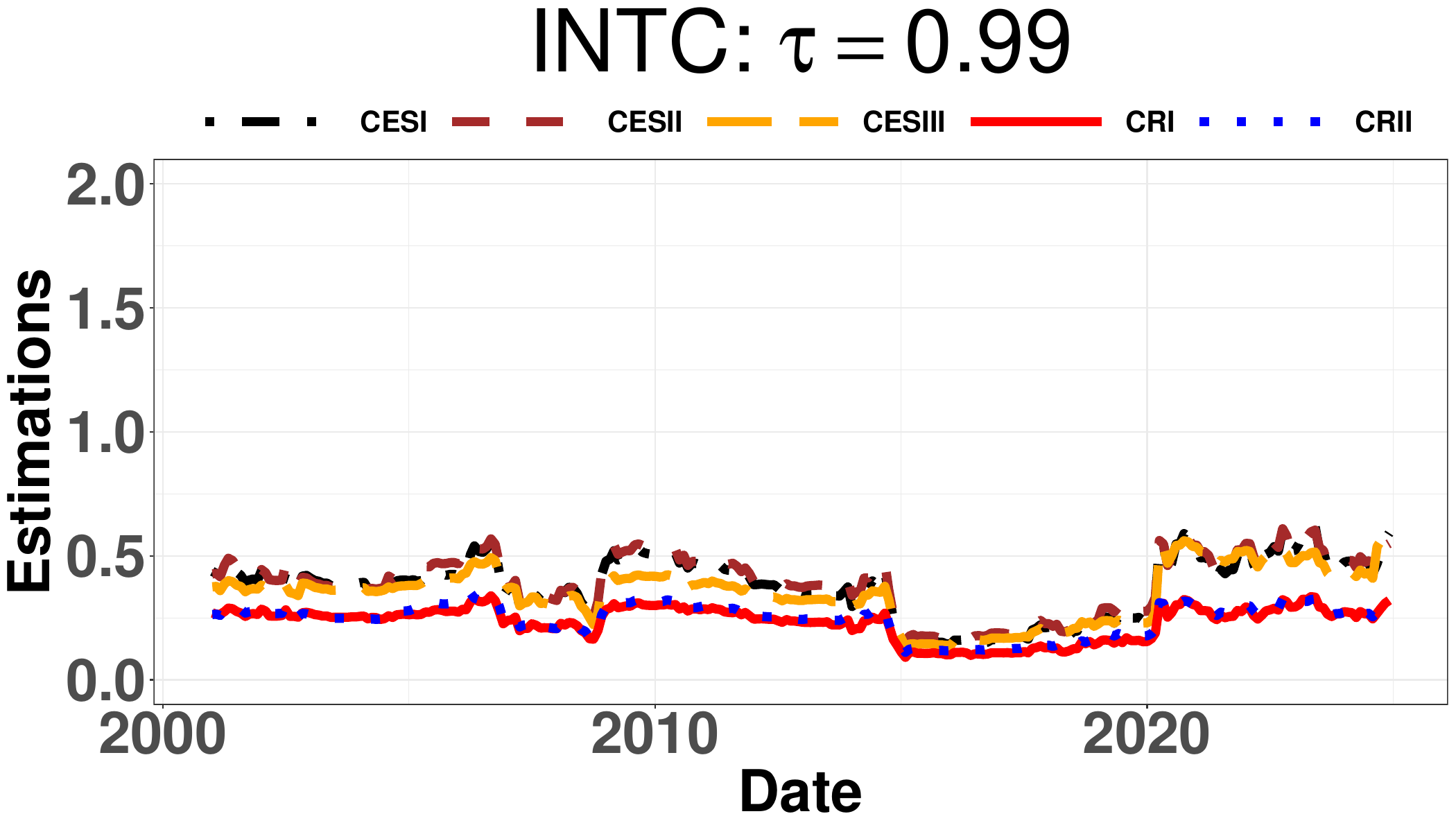}
\end{minipage}
\begin{minipage}[b]{0.32\textwidth}
\includegraphics[width=\textwidth,height = 0.14\textheight]{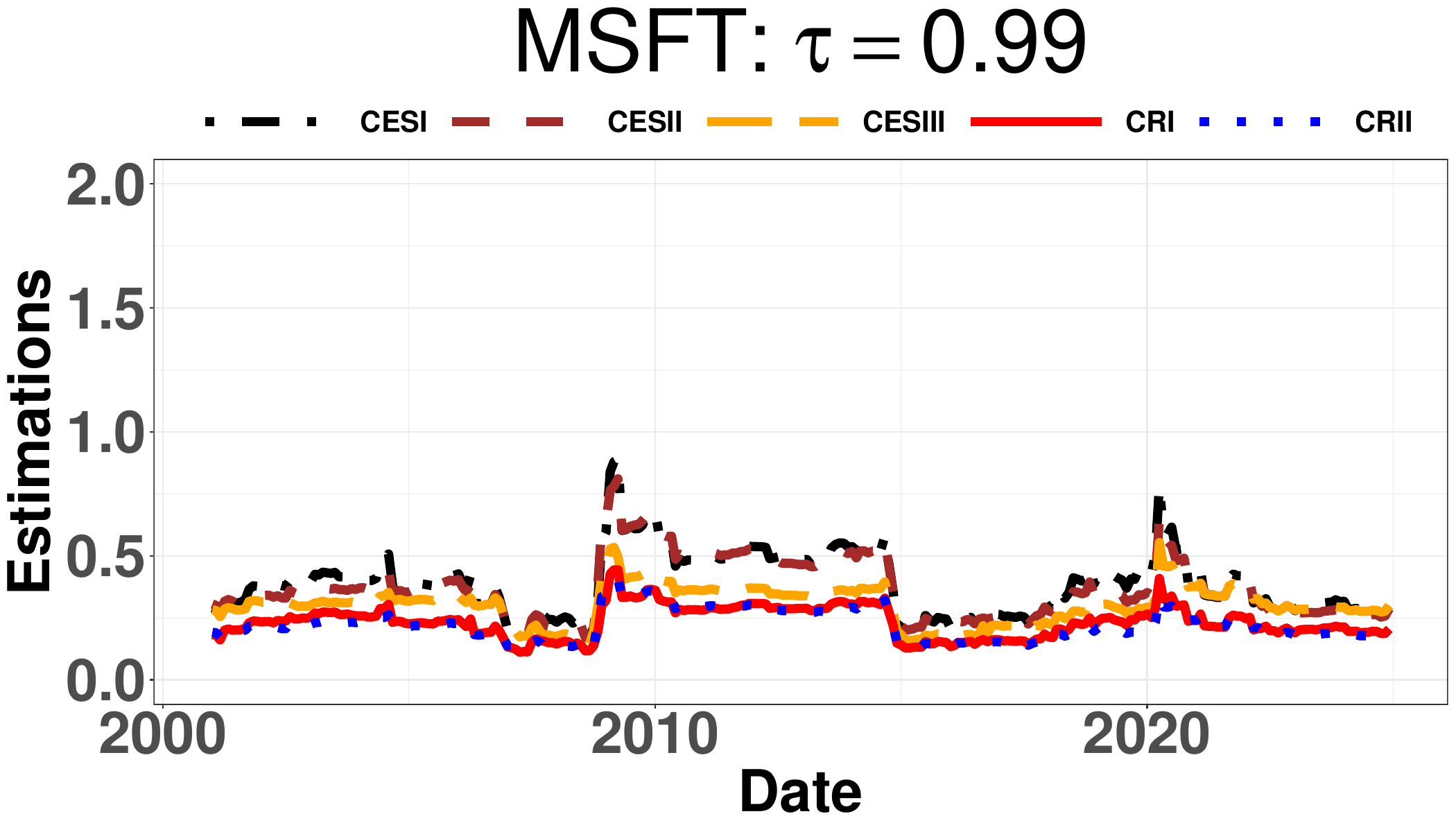}
\end{minipage}
\\[10pt]
\begin{minipage}[b]{0.32\textwidth}
\includegraphics[width=\textwidth,height = 0.14\textheight]{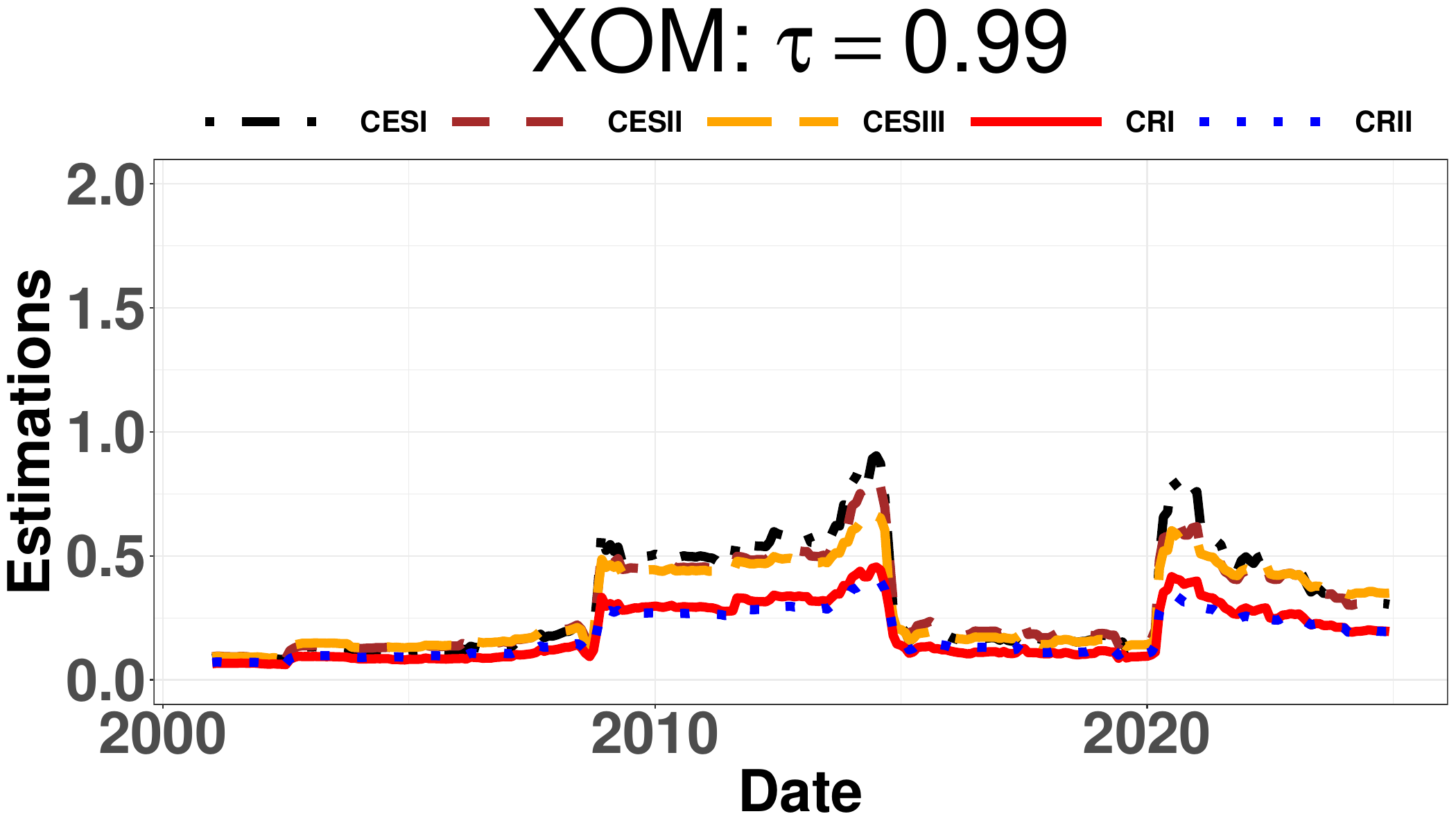}
\end{minipage}
\begin{minipage}[b]{0.32\textwidth}
\includegraphics[width=\textwidth,height = 0.14\textheight]{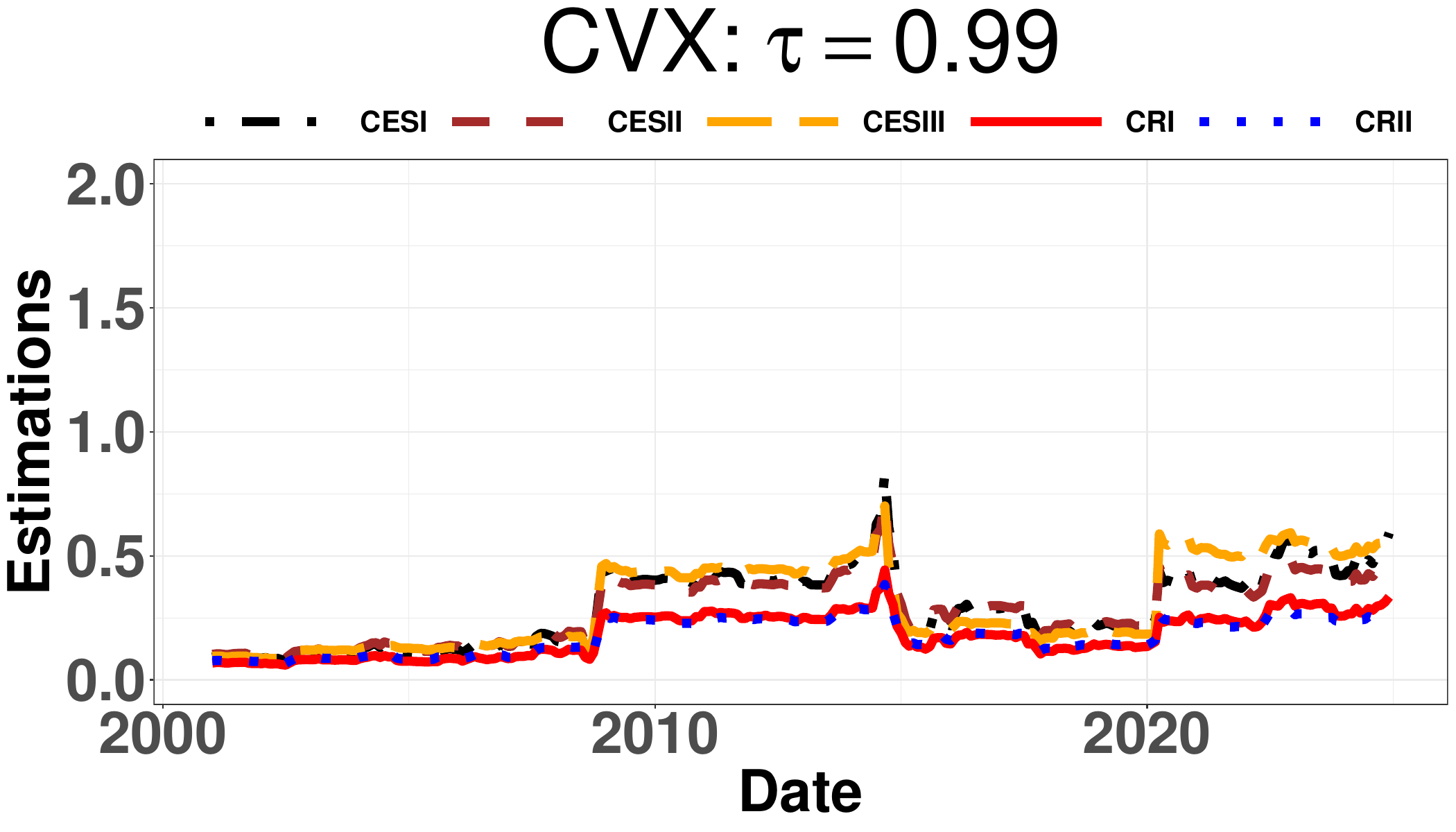}
\end{minipage}
\begin{minipage}[b]{0.32\textwidth}
\includegraphics[width=\textwidth,height = 0.14\textheight]{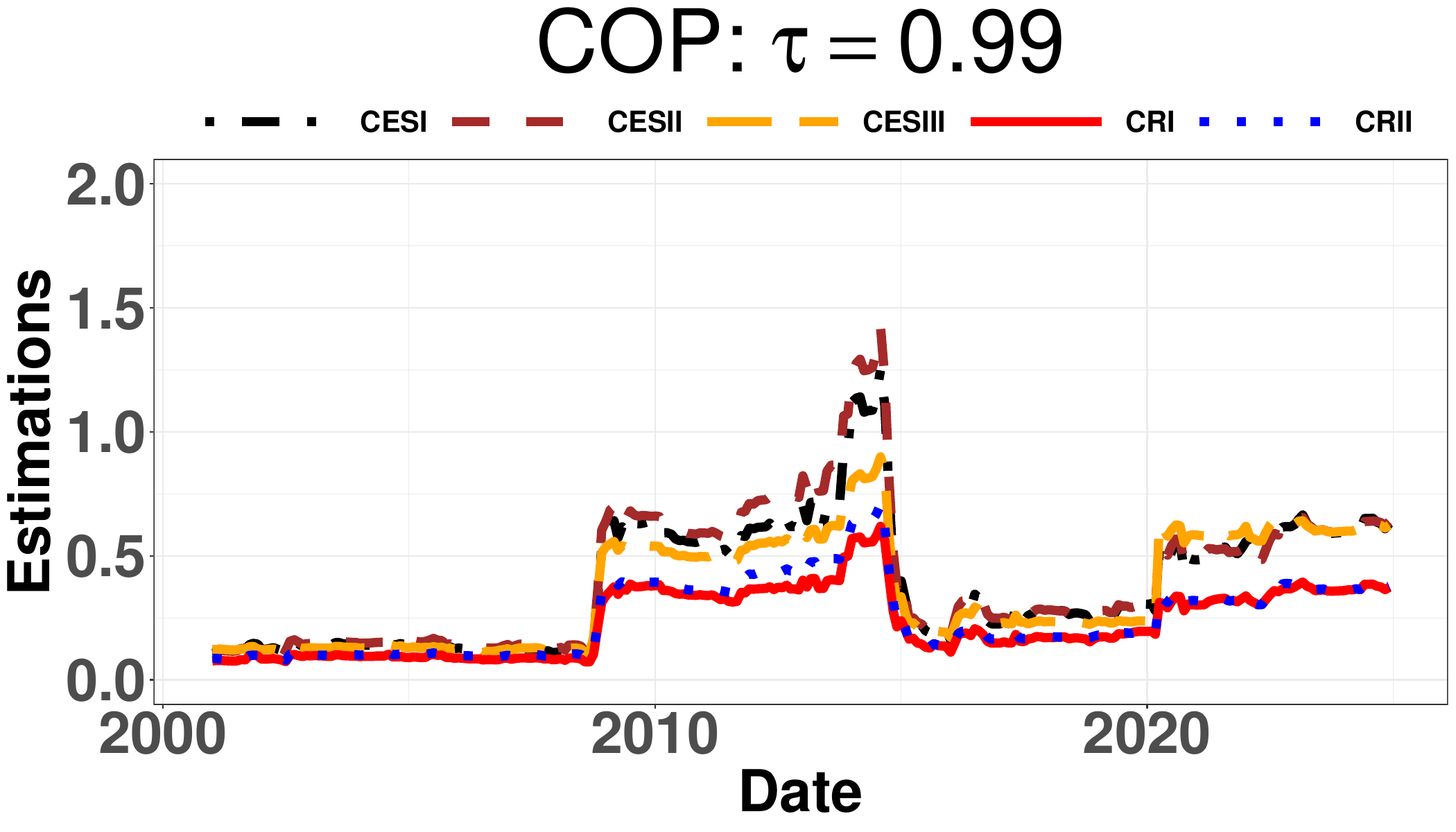}
\end{minipage}
\\[10pt]
\begin{minipage}[b]{0.32\textwidth}
\includegraphics[width=\textwidth,height = 0.14\textheight]{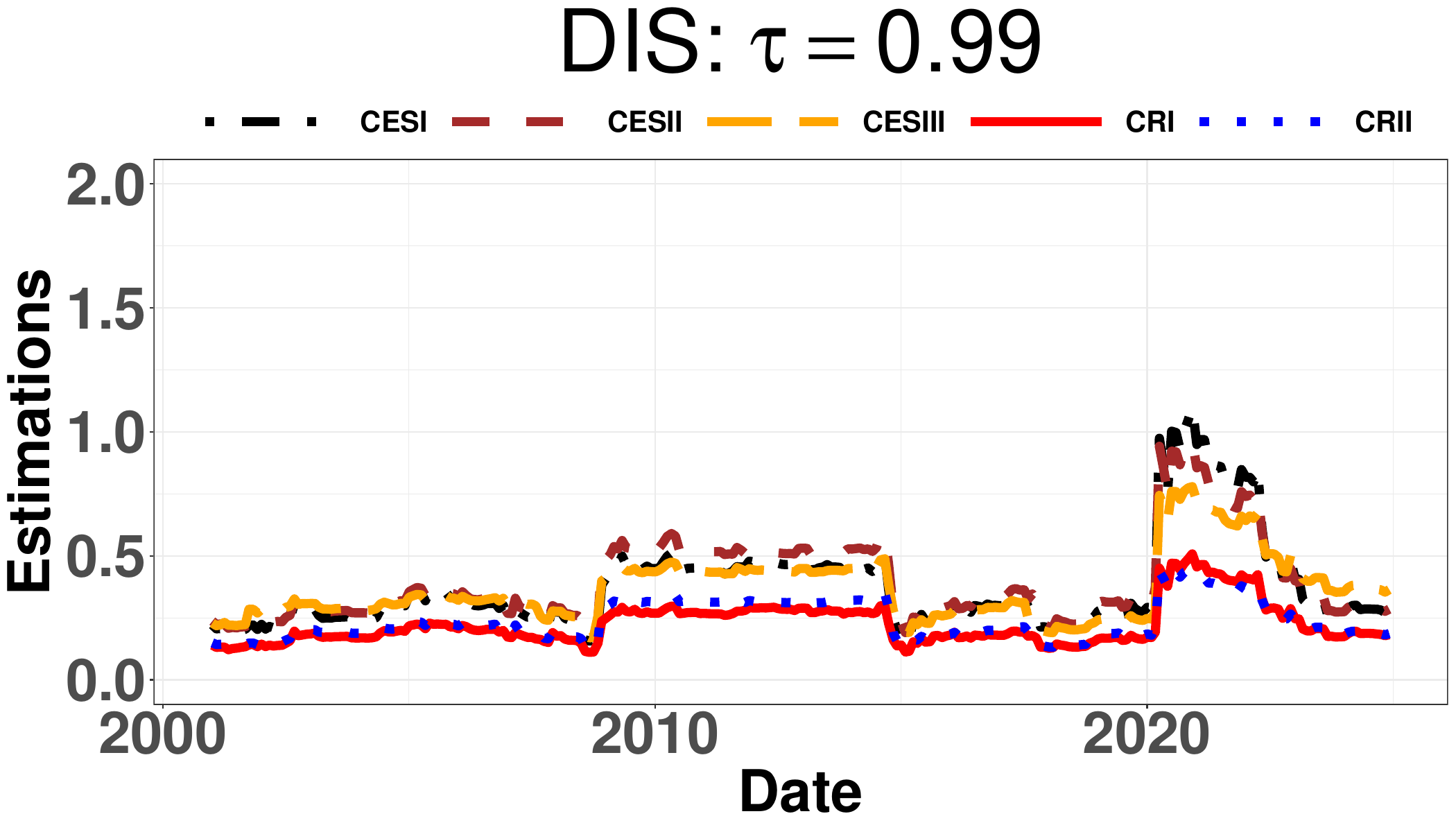}
\end{minipage}
\begin{minipage}[b]{0.32\textwidth}
\includegraphics[width=\textwidth,height = 0.14\textheight]{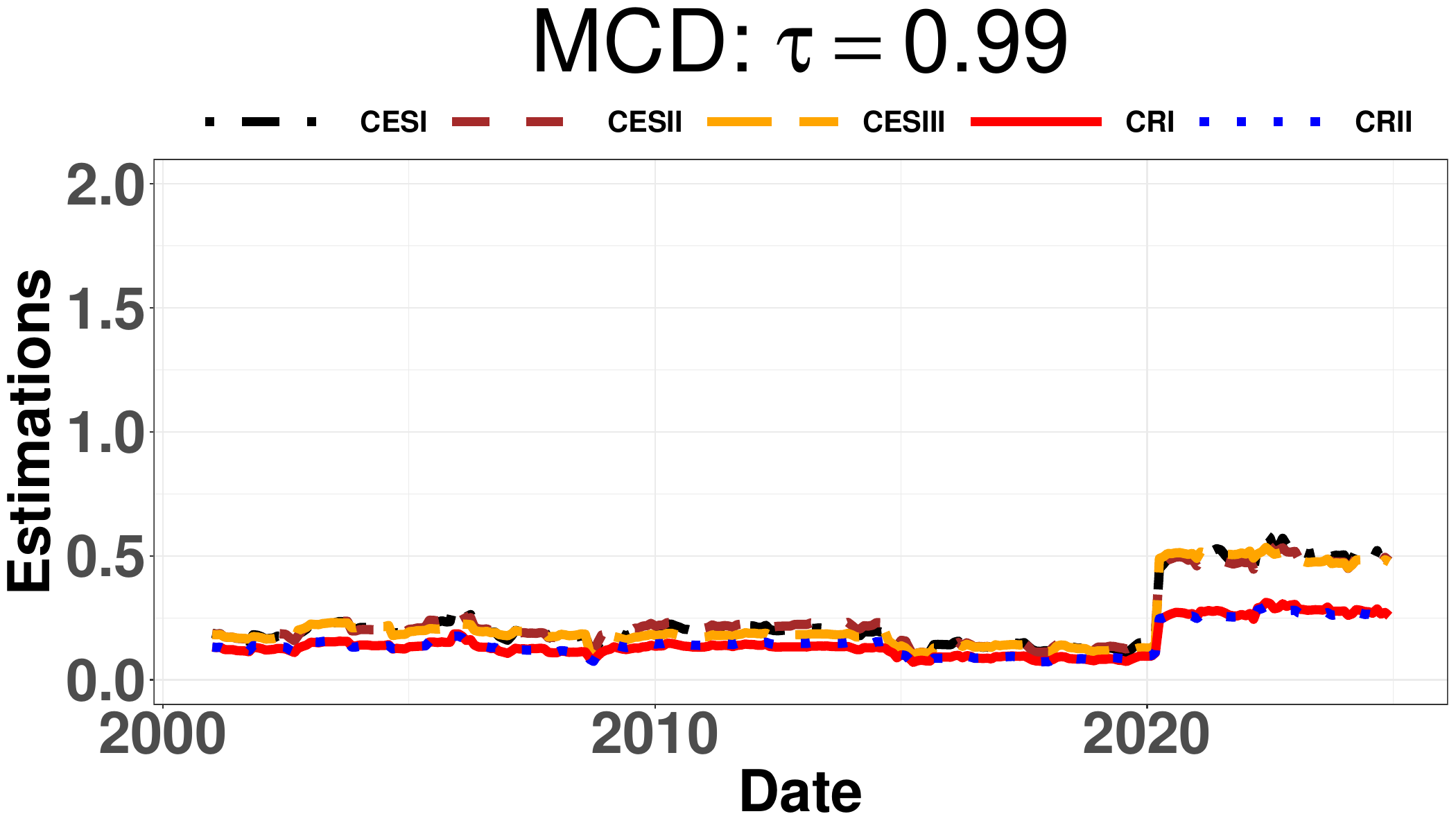}
\end{minipage}
\begin{minipage}[b]{0.32\textwidth}
\includegraphics[width=\textwidth,height = 0.14\textheight]{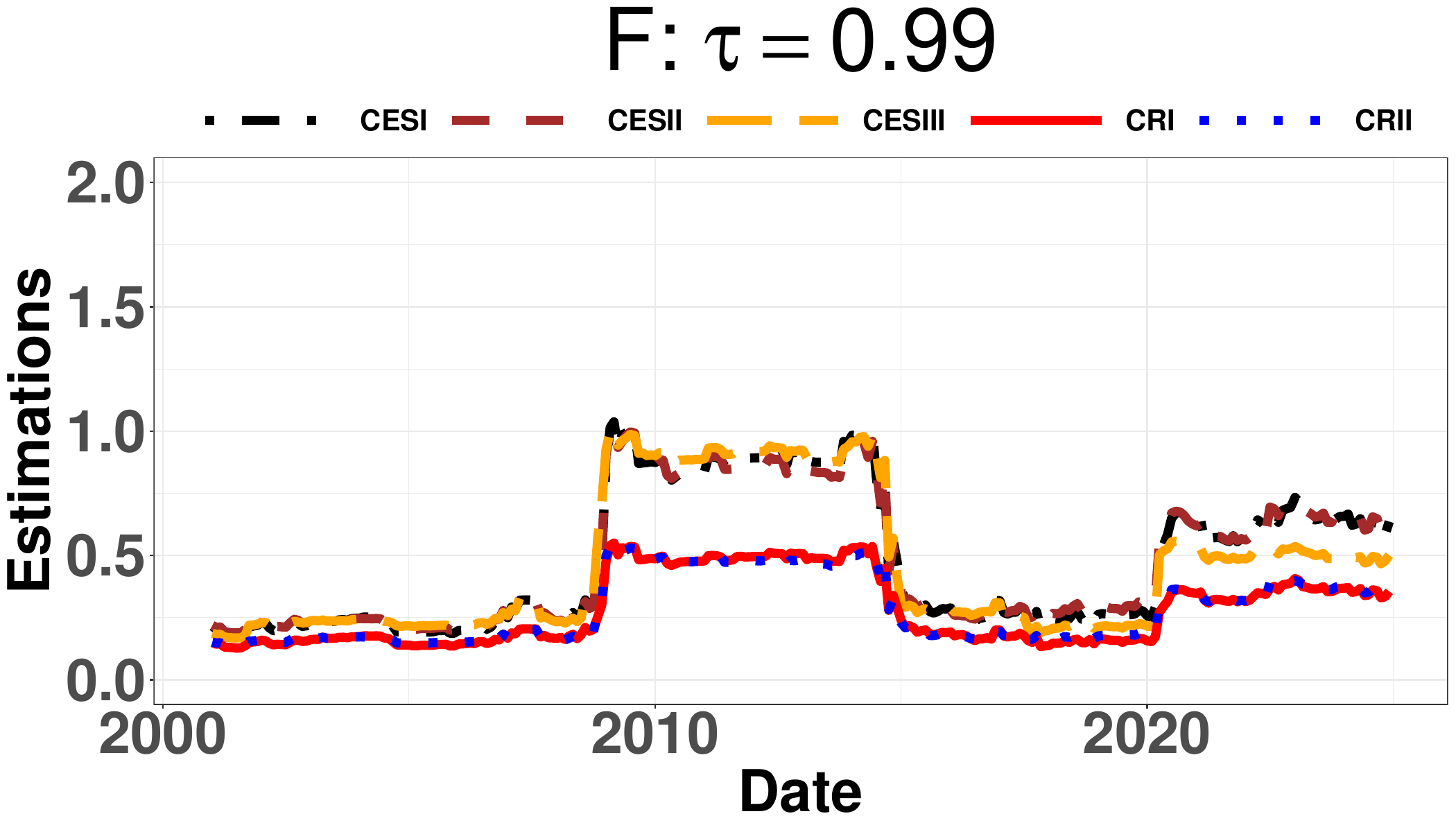}
\end{minipage}
\\[10pt]
\begin{minipage}[b]{0.32\textwidth}
\includegraphics[width=\textwidth,height = 0.14\textheight]{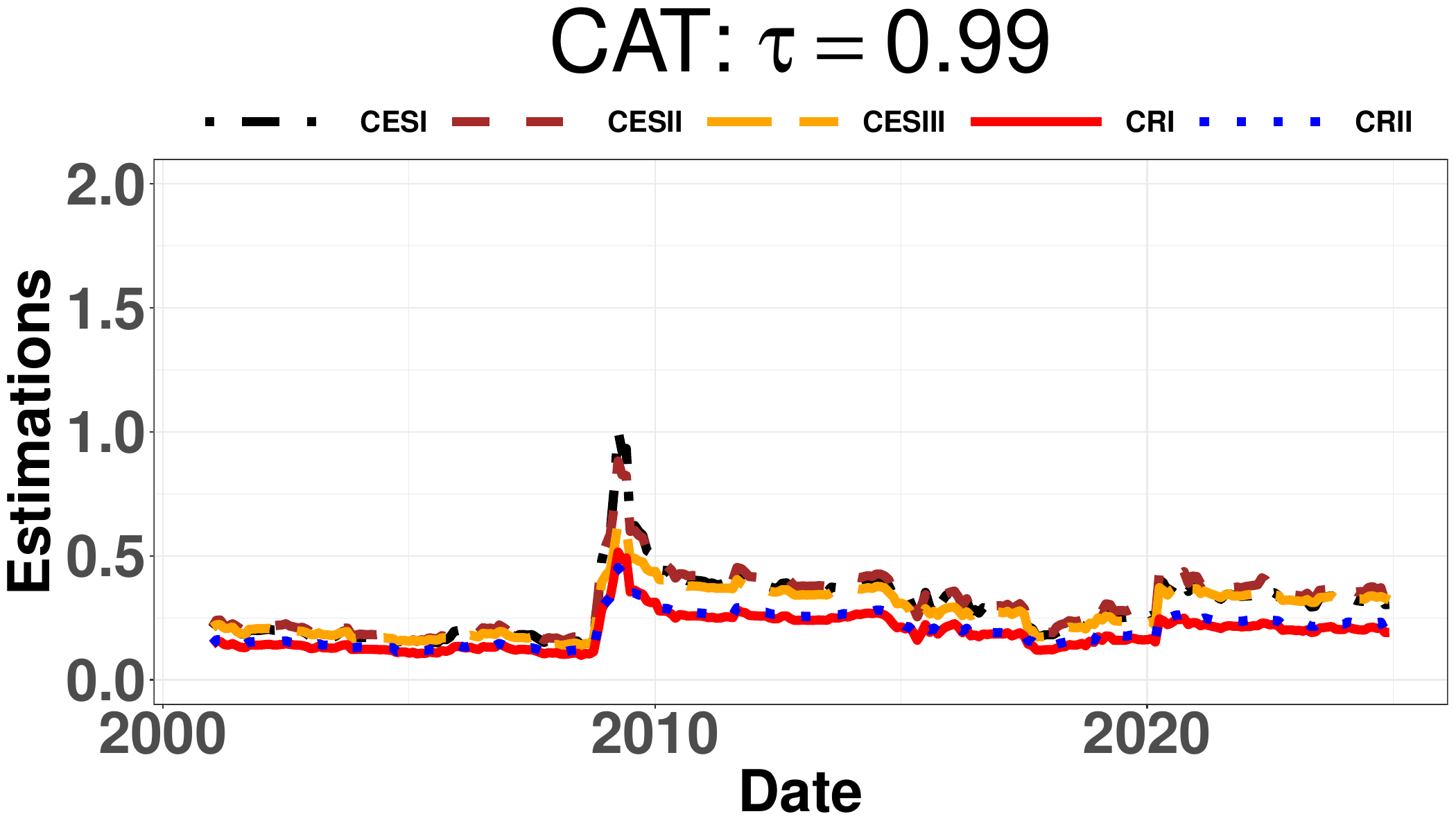}
\end{minipage}
\begin{minipage}[b]{0.32\textwidth}
\includegraphics[width=\textwidth,height = 0.14\textheight]{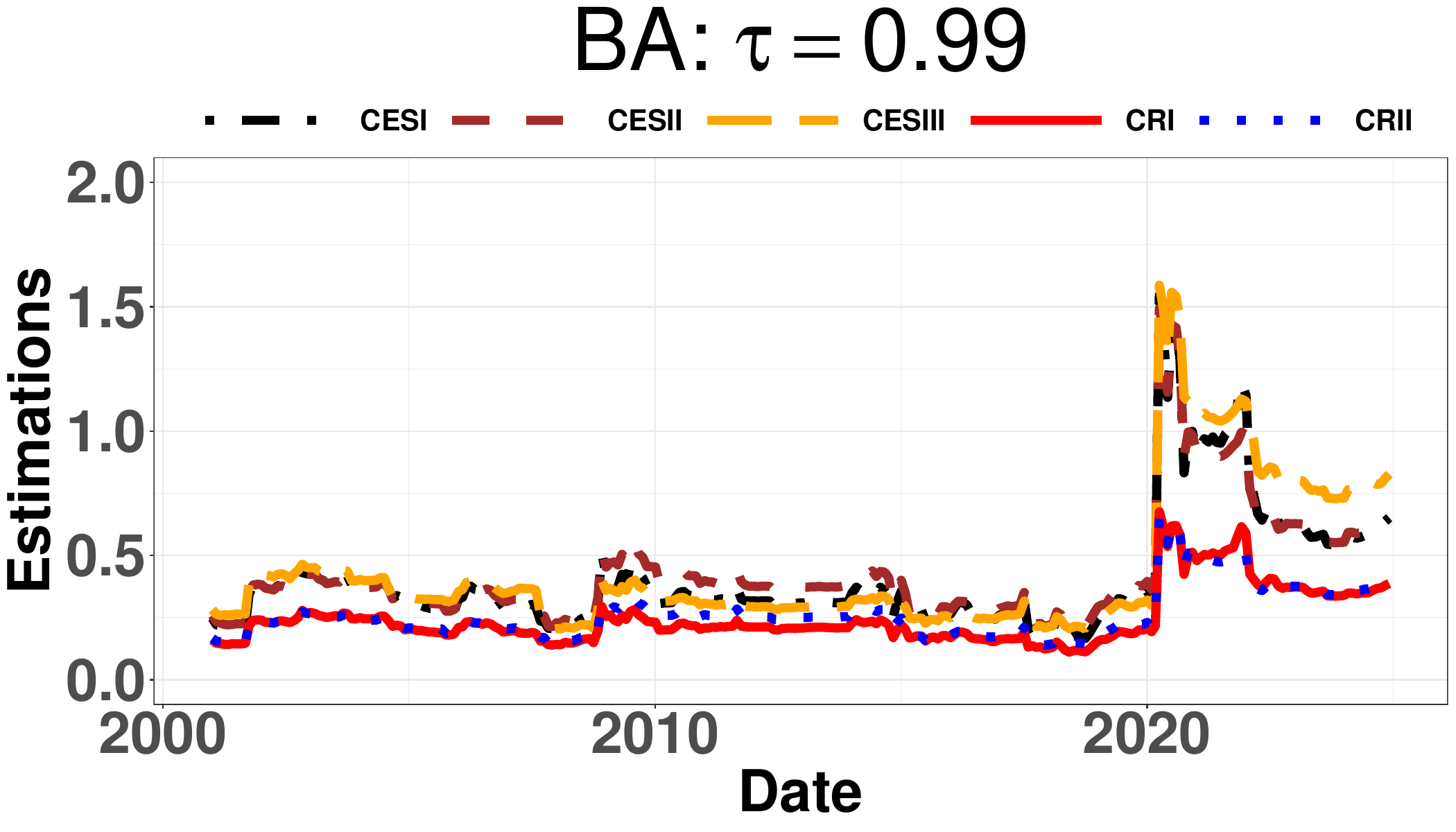}
\end{minipage}
\begin{minipage}[b]{0.32\textwidth}
\includegraphics[width=\textwidth,height = 0.14\textheight]{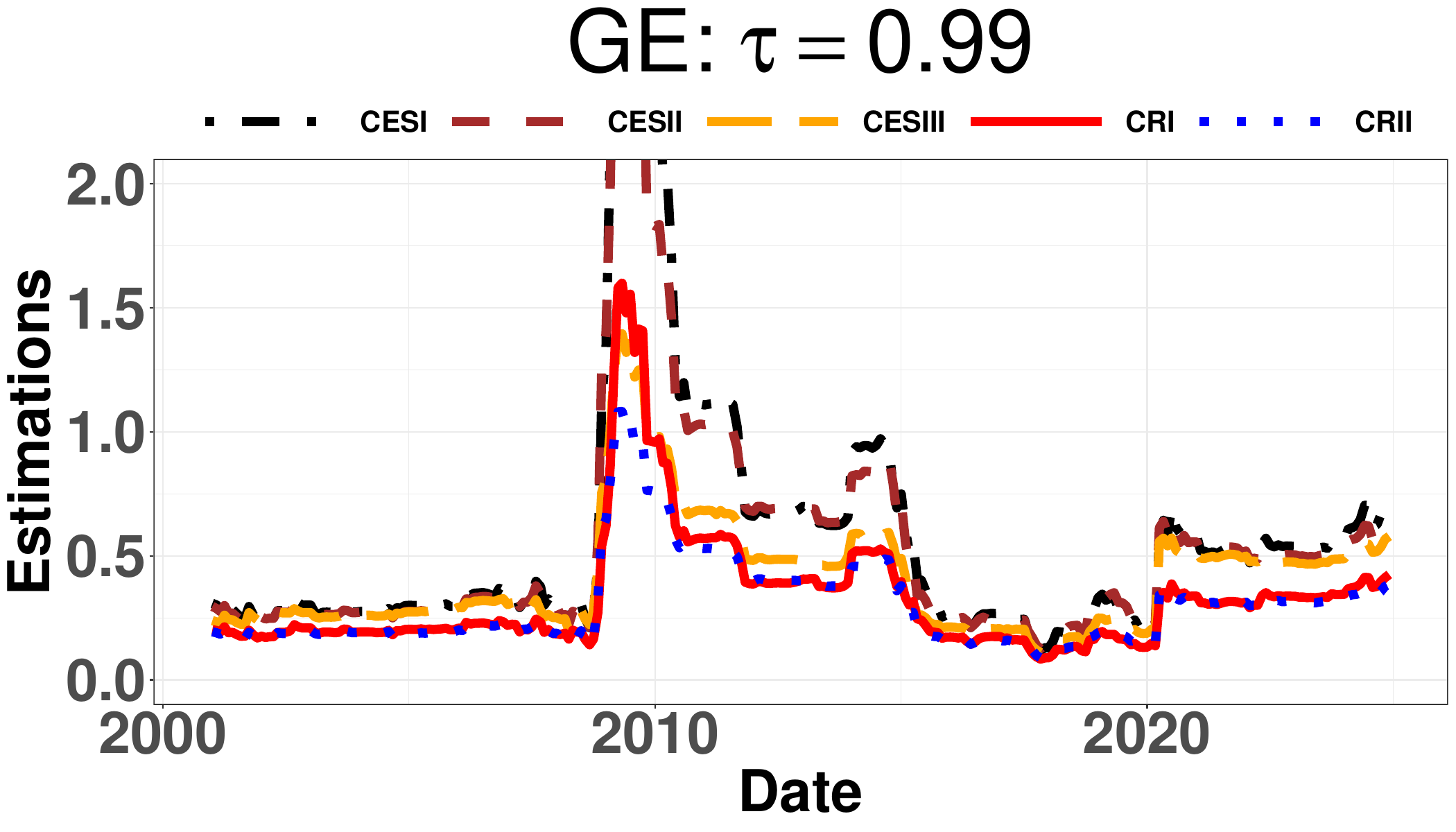}
\end{minipage}
\caption{The rolling estimations $\widetilde{\covar}_{X|Y}^{(1)}(\tau'_n)$ (CRI in red solid lines), $\widetilde{\covar}_{X|Y}^{(2)}(\tau'_n)$ (CRII in blue dotted lines), $\widetilde{\coes}_{X|Y}^{(1)}(\tau'_n)$ (CESI in black dotdash lines), $\widetilde{\coes}_{X|Y}^{(2)}(\tau'_n)$ (CESII in brown dashed lines), and $\widetilde{\coes}_{X|Y}^{(3)}(\tau'_n)$ (CESIII in orange longdash lines) for 12 individual stocks conditional on S\&P500 Index, with $\tau'_n = 0.99$.}
\label{fig:Roll_CRES99}
\end{figure}

\begin{figure}[htbp]
\centering
\begin{minipage}[b]{0.32\textwidth}
\includegraphics[width=\textwidth,height = 0.14\textheight]{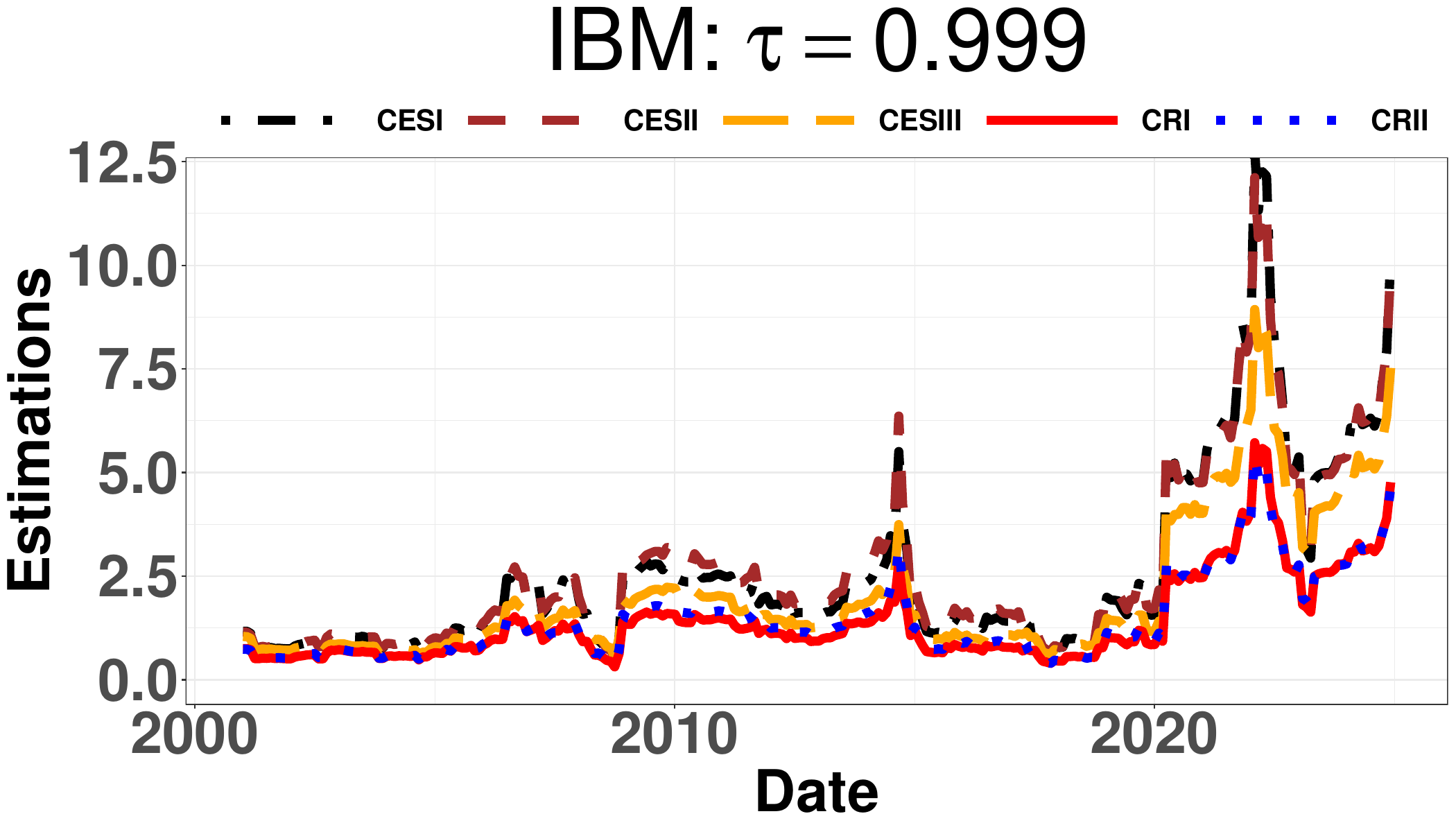}
\end{minipage}
\begin{minipage}[b]{0.32\textwidth}
\includegraphics[width=\textwidth,height = 0.14\textheight]{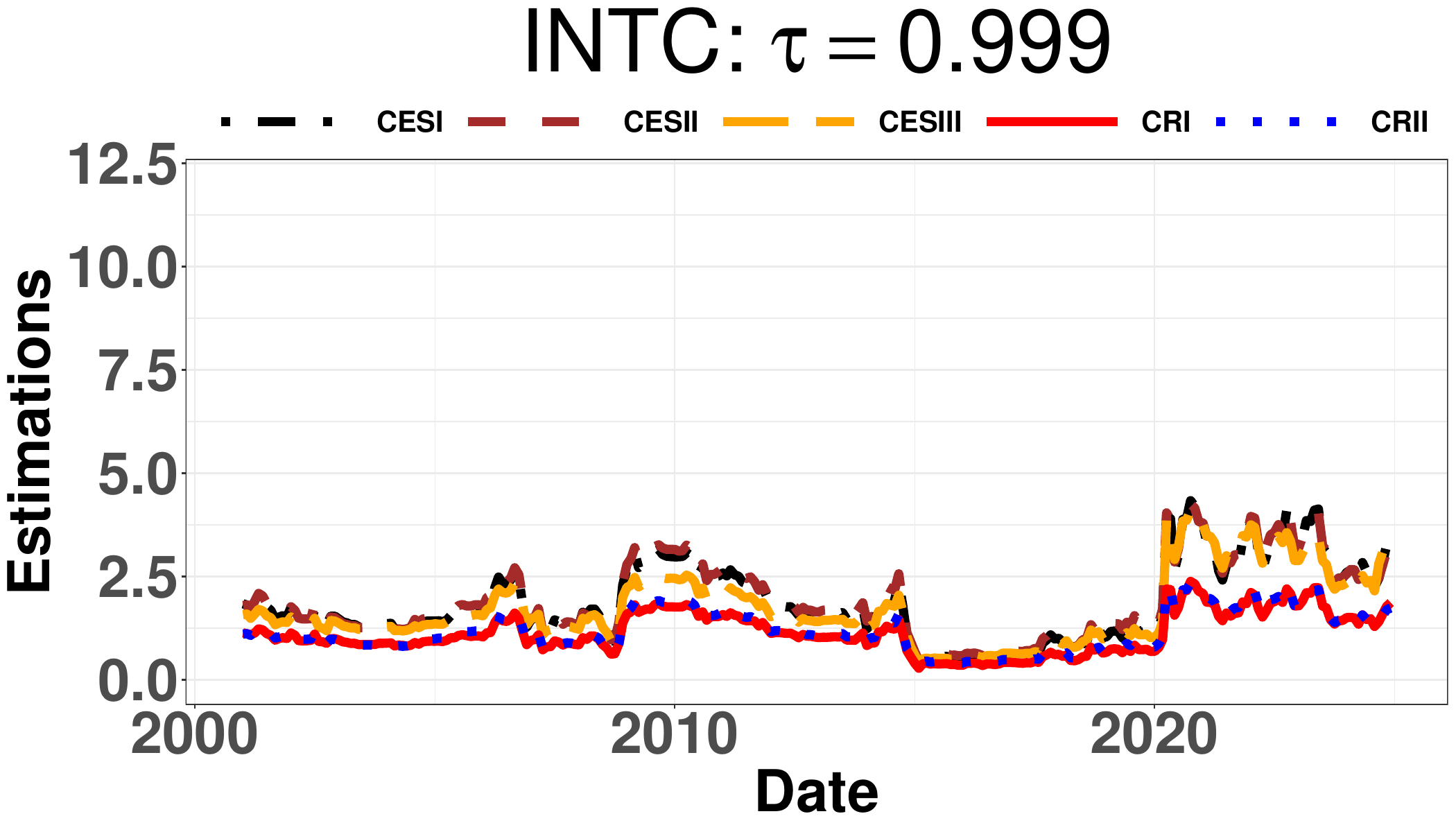}
\end{minipage}
\begin{minipage}[b]{0.32\textwidth}
\includegraphics[width=\textwidth,height = 0.14\textheight]{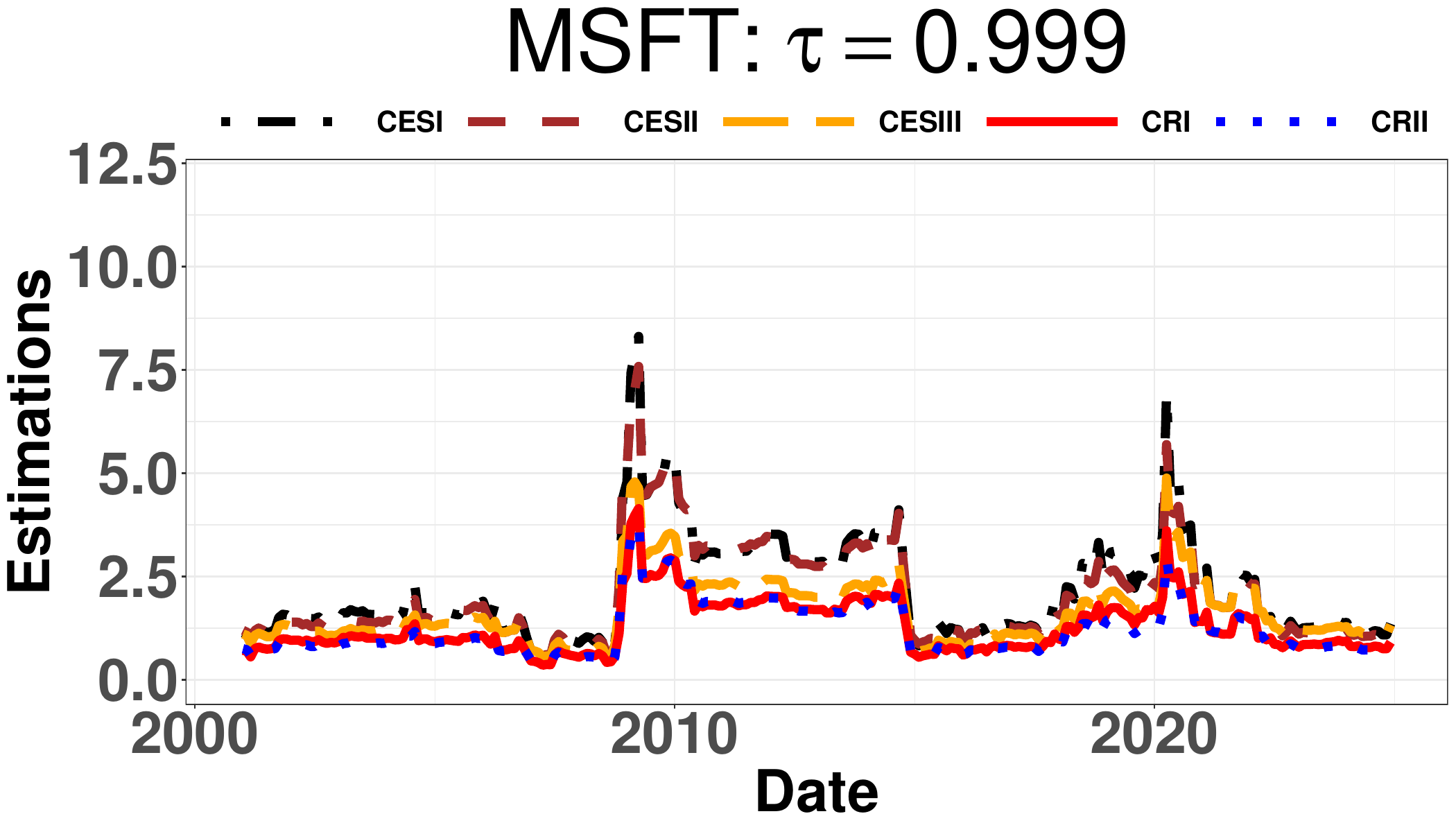}
\end{minipage}
\\[10pt]
\begin{minipage}[b]{0.32\textwidth}
\includegraphics[width=\textwidth,height = 0.14\textheight]{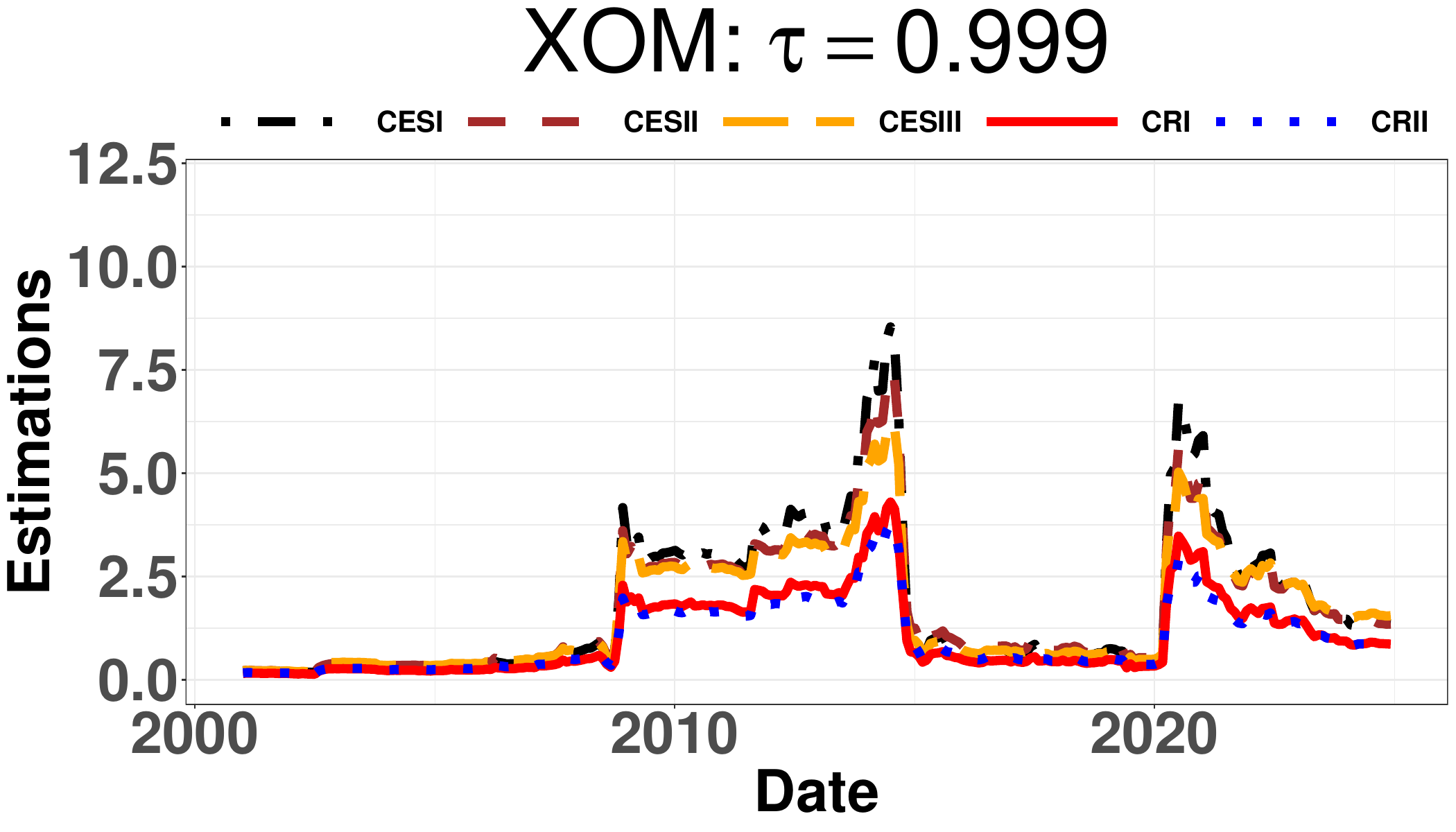}
\end{minipage}
\begin{minipage}[b]{0.32\textwidth}
\includegraphics[width=\textwidth,height = 0.14\textheight]{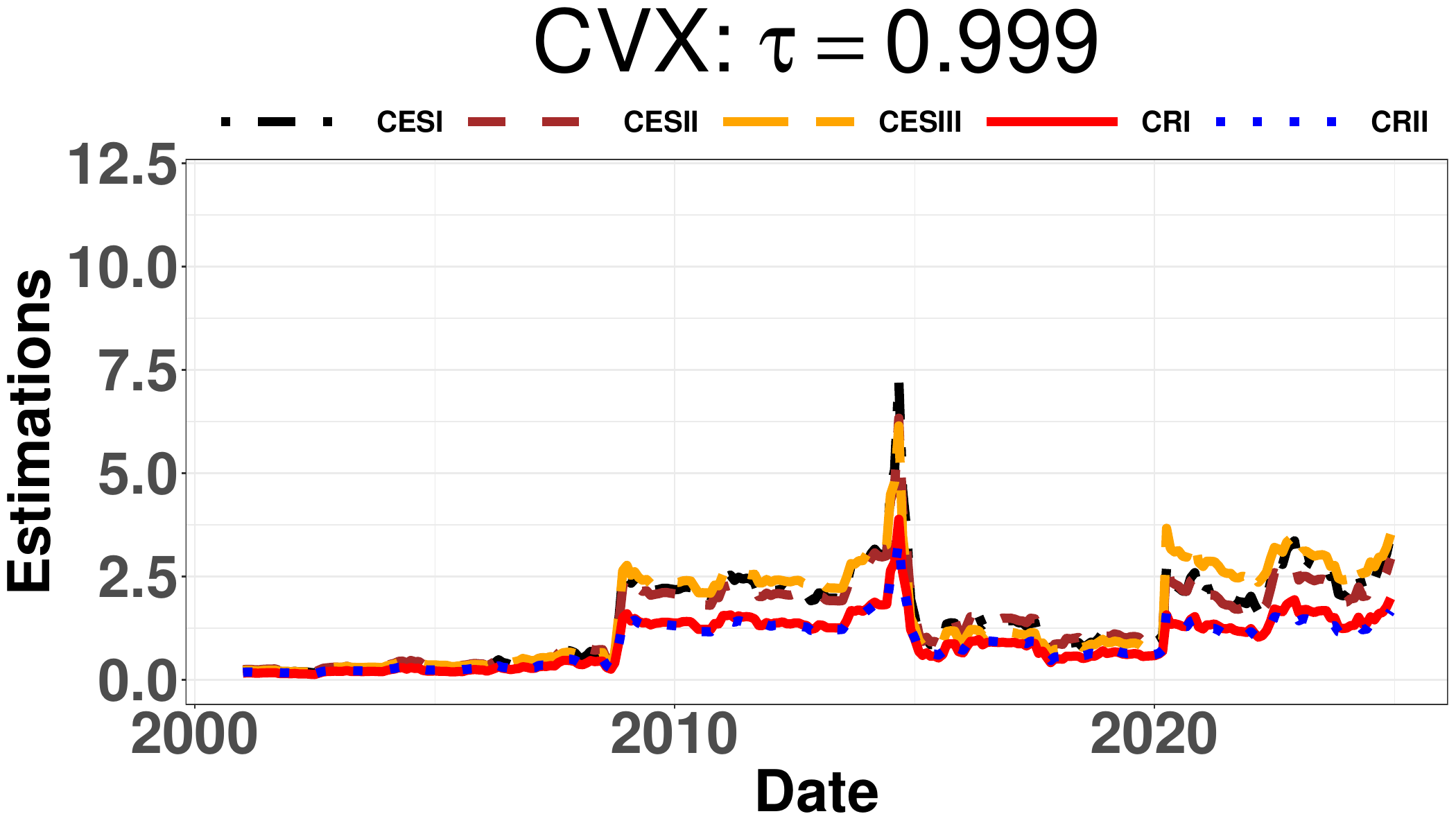}
\end{minipage}
\begin{minipage}[b]{0.32\textwidth}
\includegraphics[width=\textwidth,height = 0.14\textheight]{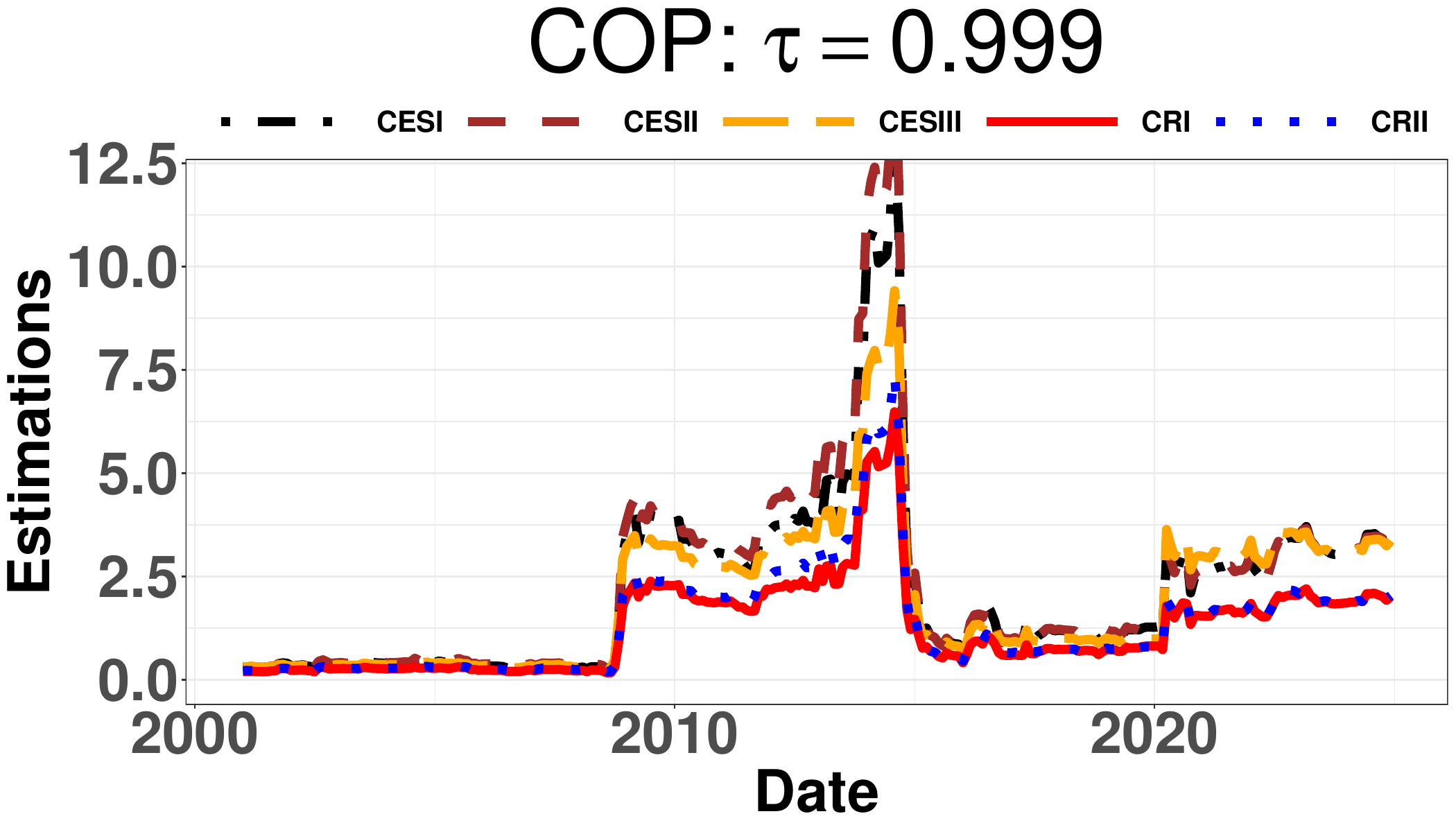}
\end{minipage}
\\[10pt]
\begin{minipage}[b]{0.32\textwidth}
\includegraphics[width=\textwidth,height = 0.14\textheight]{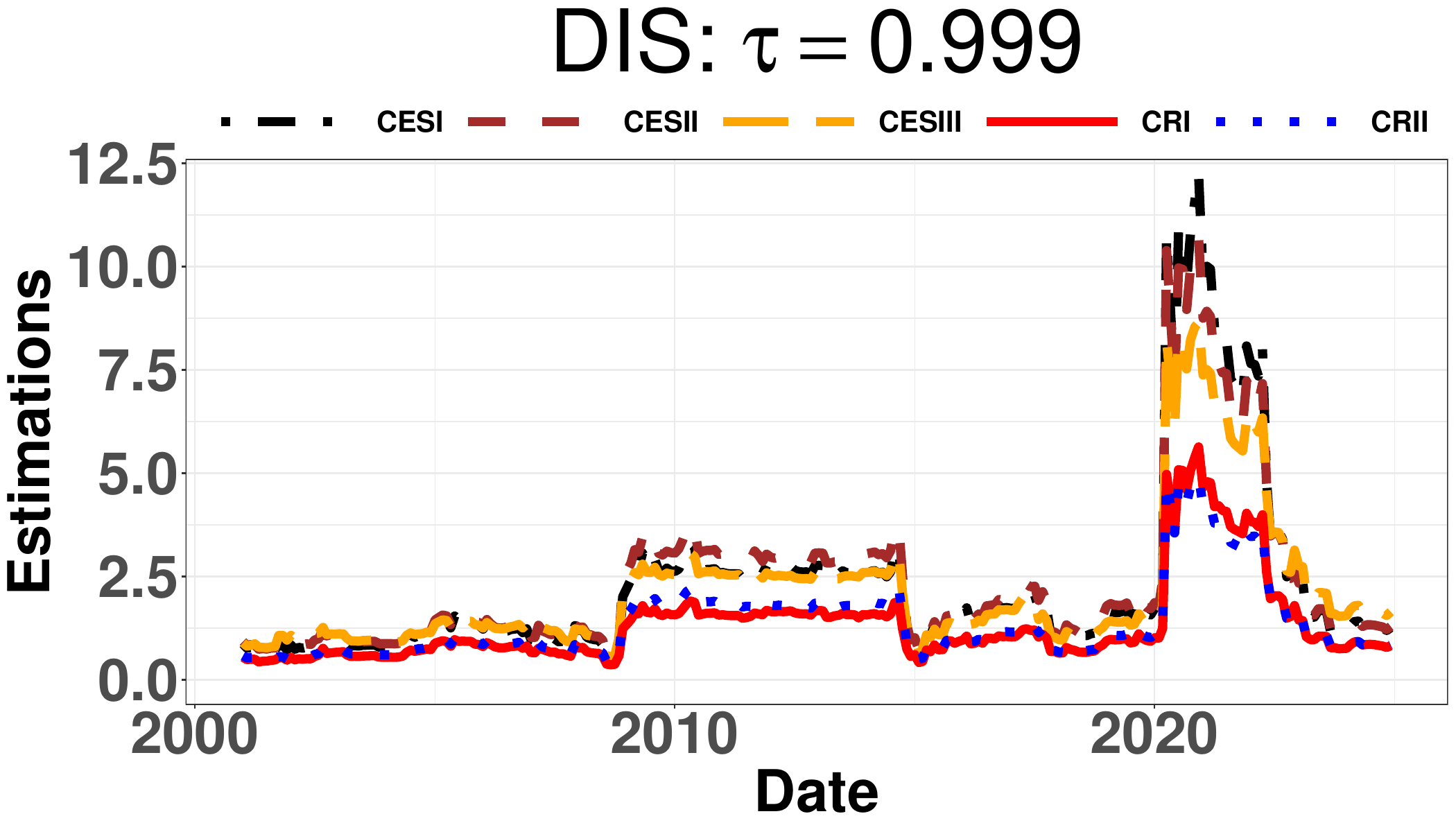}
\end{minipage}
\begin{minipage}[b]{0.32\textwidth}
\includegraphics[width=\textwidth,height = 0.14\textheight]{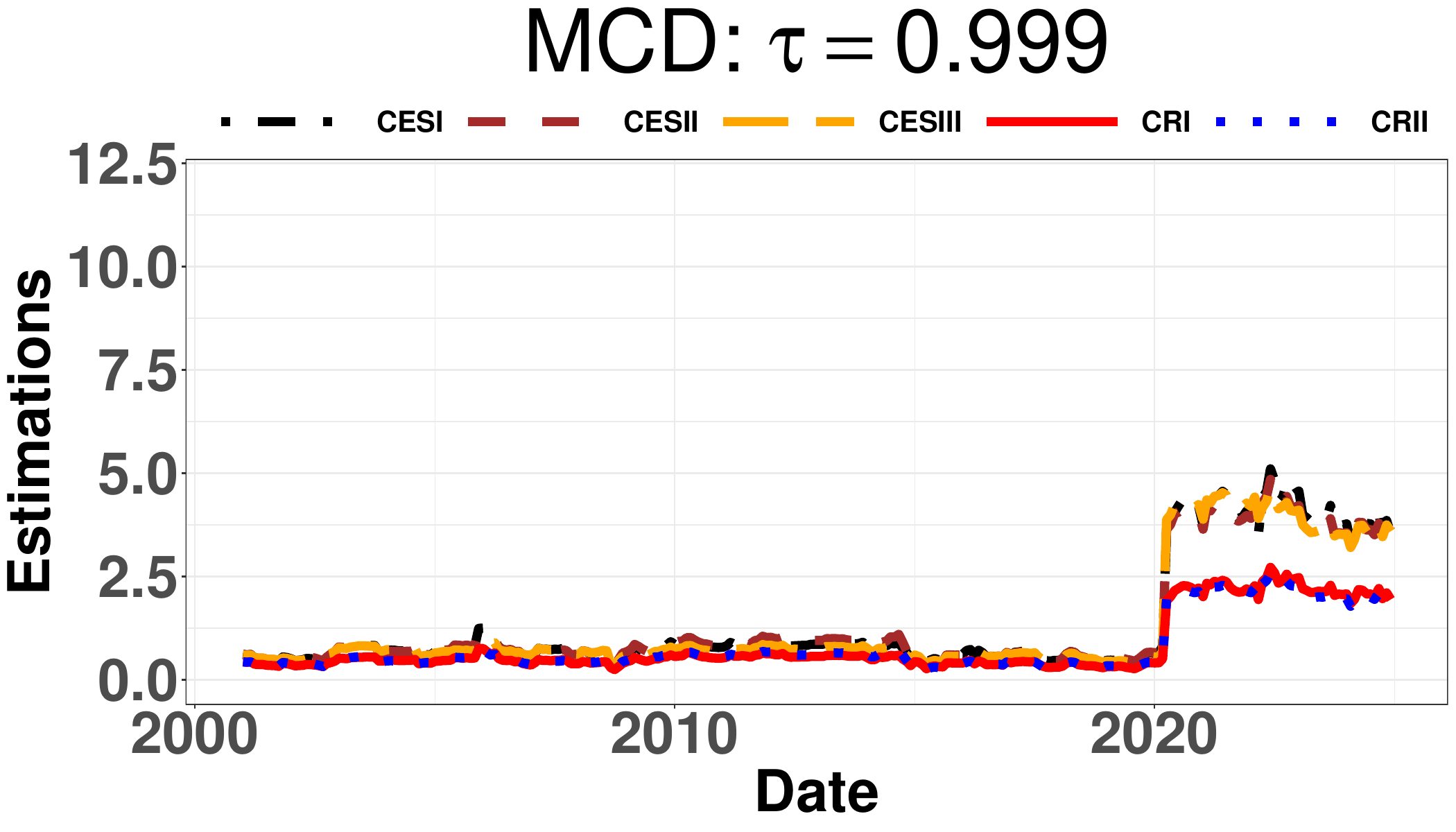}
\end{minipage}
\begin{minipage}[b]{0.32\textwidth}
\includegraphics[width=\textwidth,height = 0.14\textheight]{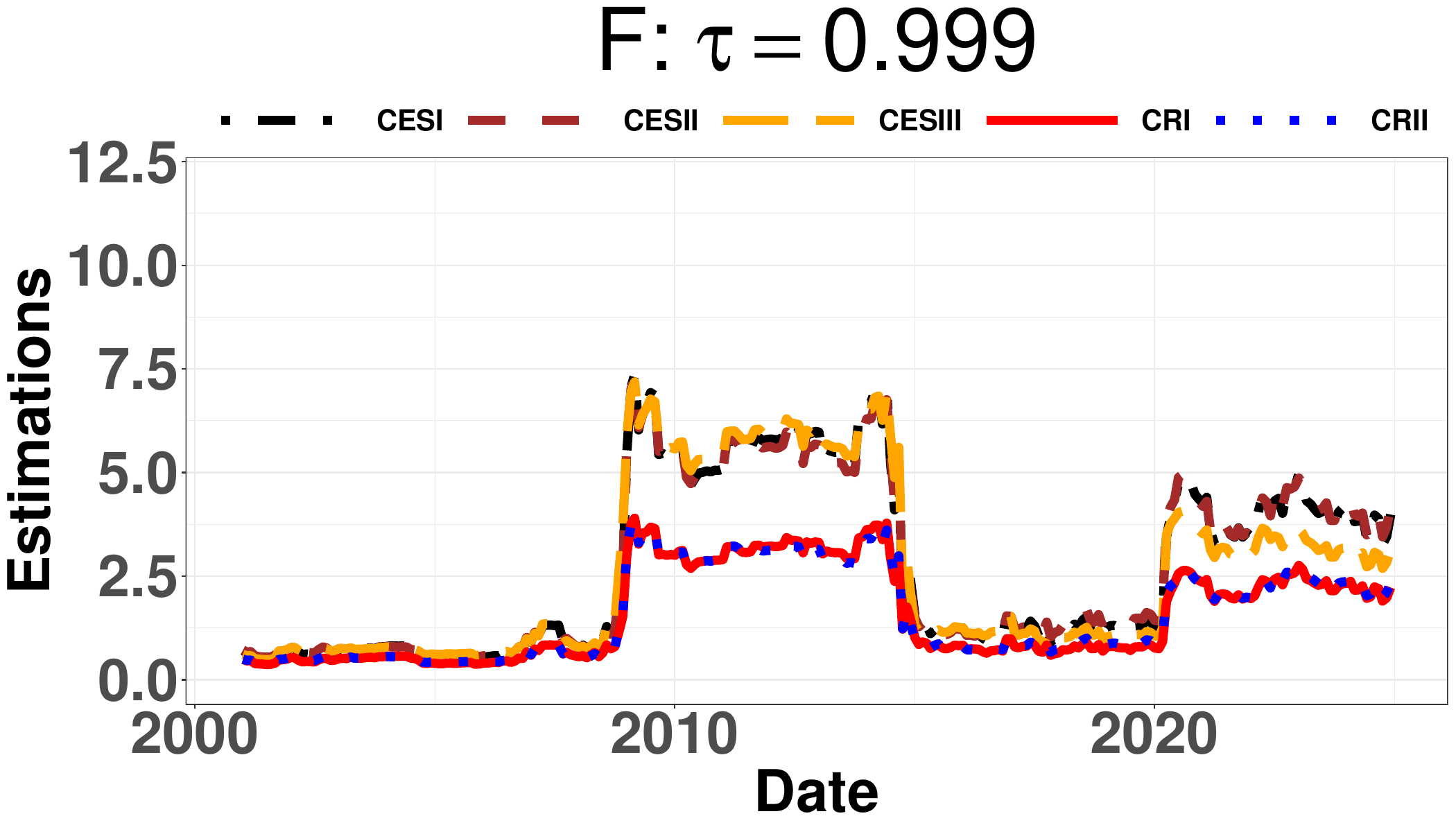}
\end{minipage}
\\[10pt]
\begin{minipage}[b]{0.32\textwidth}
\includegraphics[width=\textwidth,height = 0.14\textheight]{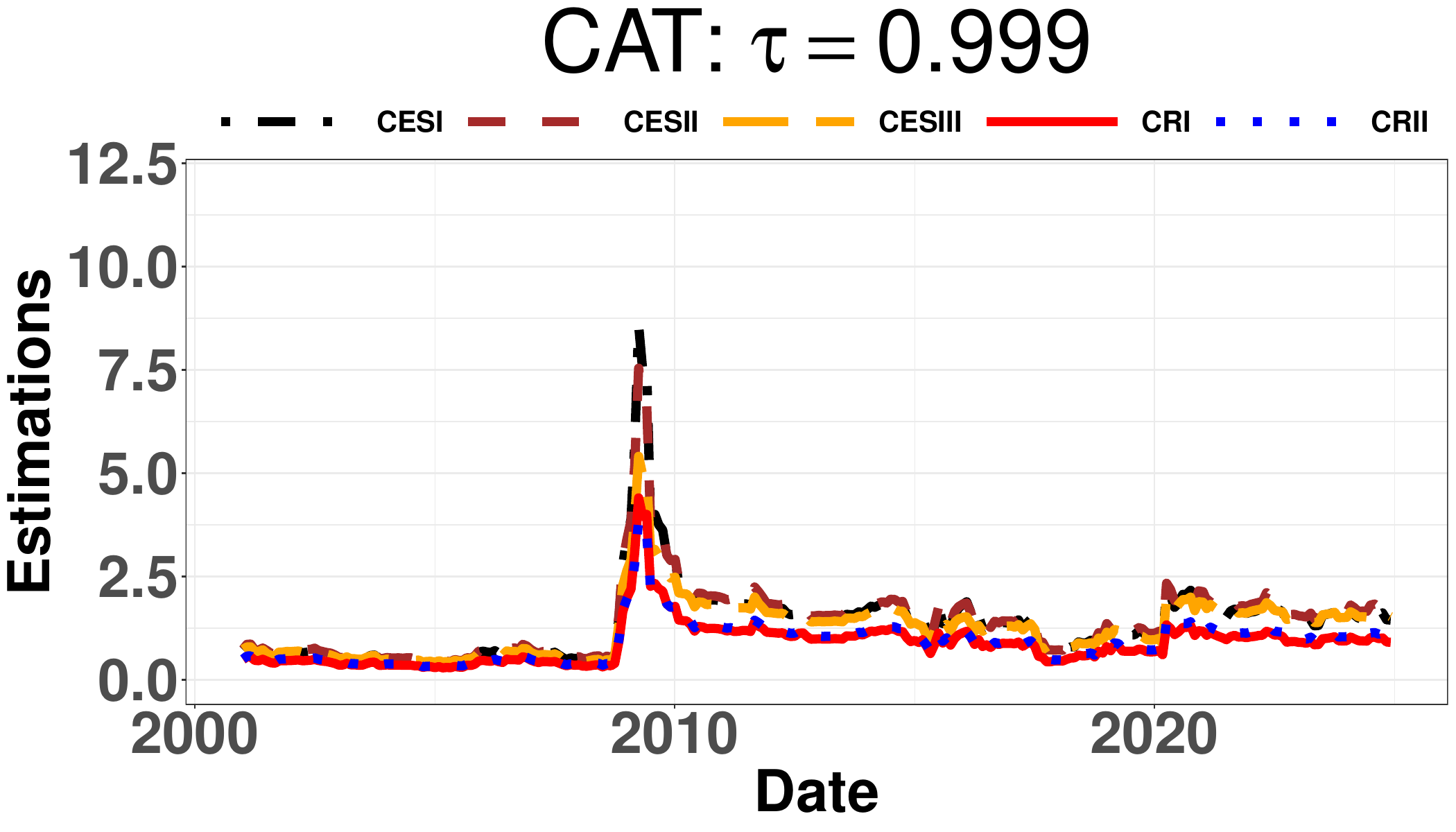}
\end{minipage}
\begin{minipage}[b]{0.32\textwidth}
\includegraphics[width=\textwidth,height = 0.14\textheight]{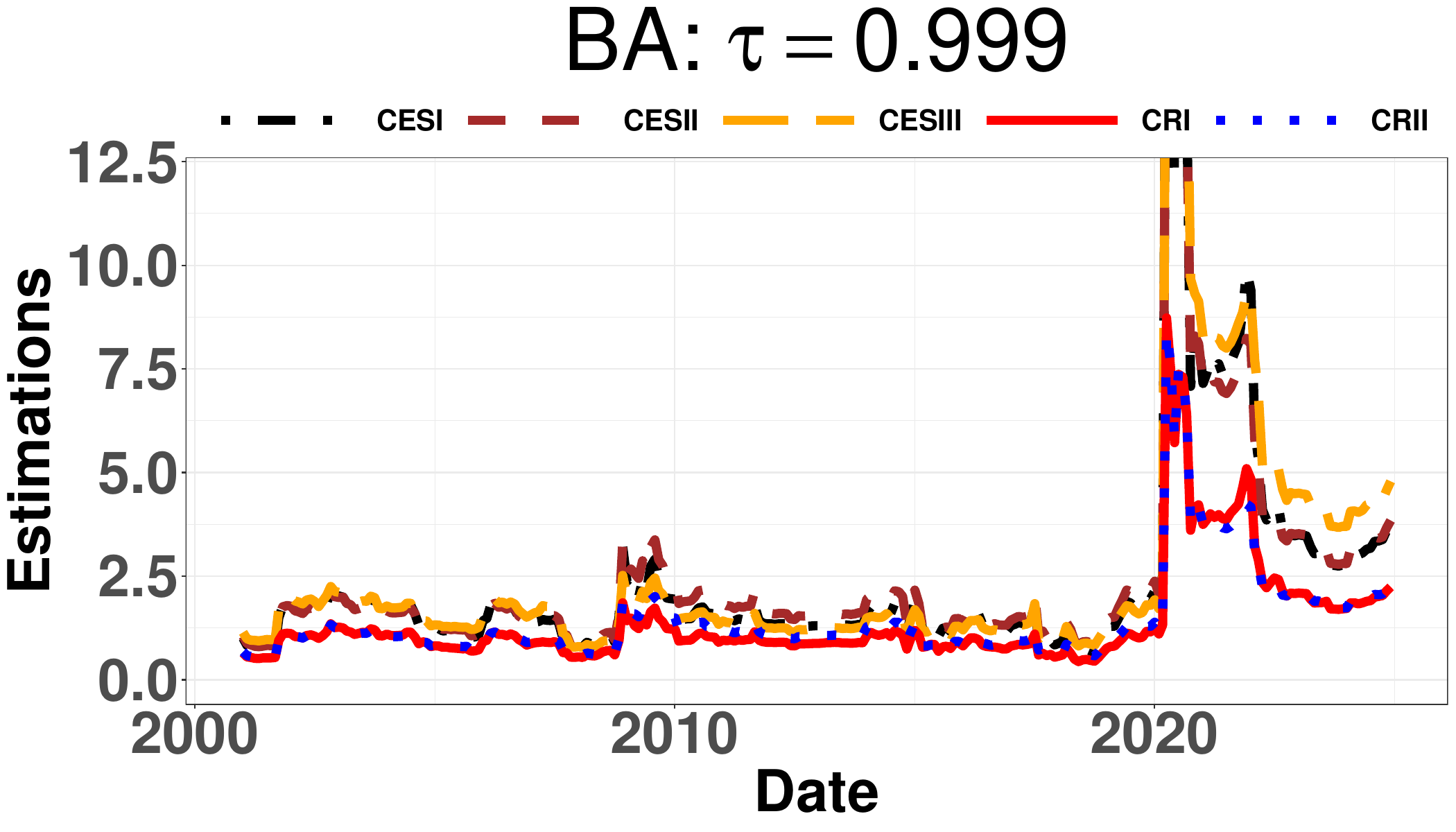}
\end{minipage}
\begin{minipage}[b]{0.32\textwidth}
\includegraphics[width=\textwidth,height = 0.14\textheight]{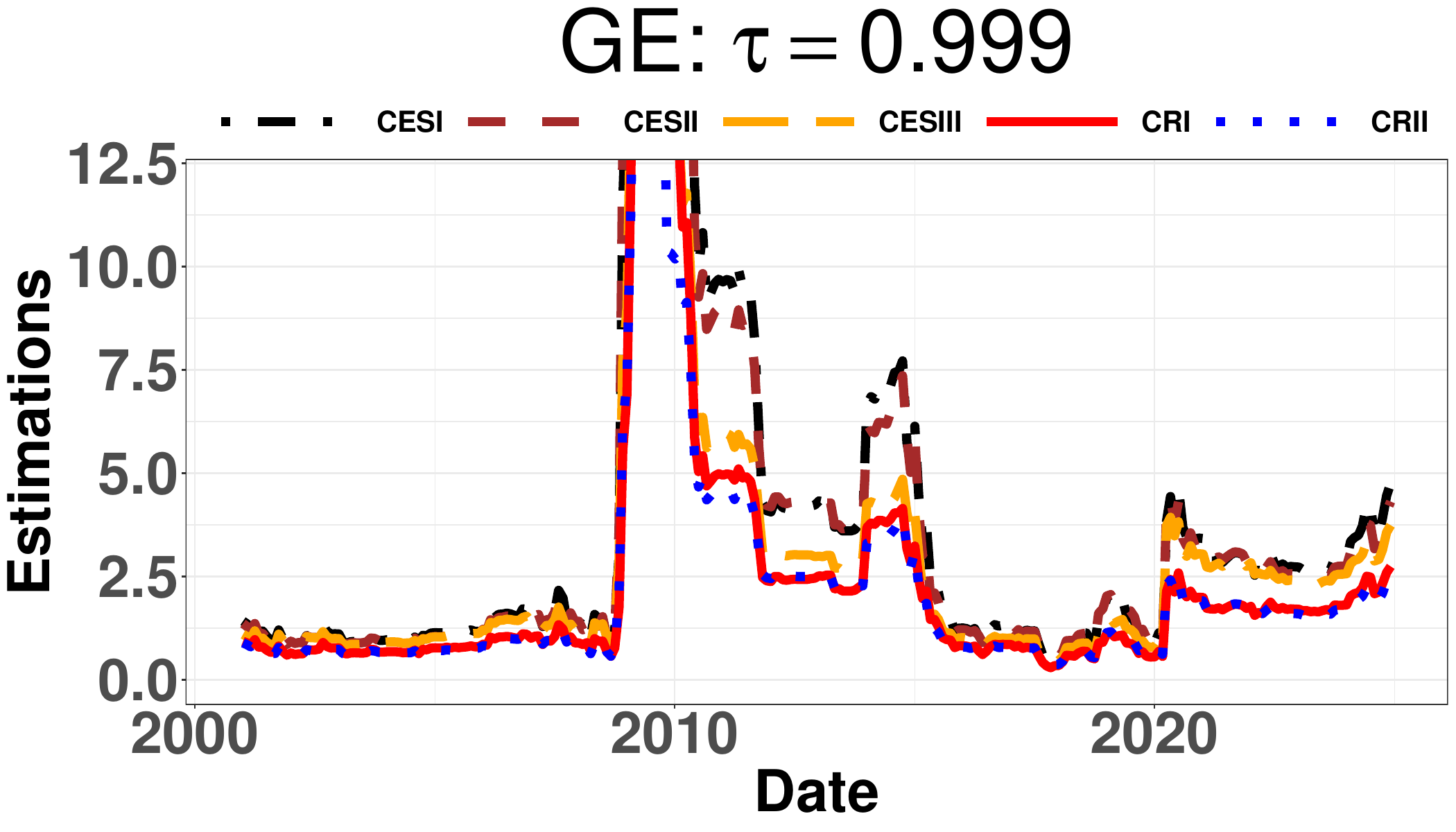}
\end{minipage}
\caption{The rolling estimations $\widetilde{\covar}_{X|Y}^{(1)}(\tau'_n)$ (CRI in red solid lines), $\widetilde{\covar}_{X|Y}^{(2)}(\tau'_n)$ (CRII in blue dotted lines), $\widetilde{\coes}_{X|Y}^{(1)}(\tau'_n)$ (CESI in black dotdash lines), $\widetilde{\coes}_{X|Y}^{(2)}(\tau'_n)$ (CESII in brown dashed lines), and $\widetilde{\coes}_{X|Y}^{(3)}(\tau'_n)$ (CESIII in orange longdash lines) for 12 individual stocks conditional on S\&P500 Index, with $\tau'_n = 0.999$.}
\label{fig:Roll_CRES999}
\end{figure}

In order to effectively capture the time-varying dynamics of extreme risk measured by CoVaR and CoES, we accordingly apply a rolling-window approach to generate monthly estimates of $\covar_{X|Y}(\tau'_n)$ and $\coes_{X|Y}(\tau'_n)$. The sample is enlarged to daily observations from January 3, 1995 to December 31, 2024, yielding 7552 observations, and we implement the procedure with a 1500-day moving window.
For instance, the estimates on October 1st, 2014 are calculated by using 1500 observations from October 15, 2008 to September 30, 2014. This long window captures the evolution of tail-risk interdependence over time. Figures \ref{fig:Roll_CRES99} - \ref{fig:Roll_CRES999} display the dynamics rolling paths of estimations $\widetilde{\covar}^{(i)}_{X|Y}(\tau'_n)$ and $\widetilde{\coes}^{(i)}_{X|Y}(\tau'_n)$ with $\tau'_n = 0.99$ and 0.999, respectively. We observe that all extrapolations move in lockstep, corroborating their similar empirical behaviours. In addition, different industries have been affected by the prevailing economic environment during various periods. For example, almost all firms were affected by the COVID-19 pandemic, with particularly pronounced impacts on IBM, DIS, and BA, all of which exhibit sharp spike-like behaviors. In contrast, the impact of the global financial crisis appears to be much minor: MCD was barely affected by the crisis, and firms in IT sector were also only mildly influenced, whereas GE experienced the most severe impact. Moreover, the Oil excess capacity bubble around 2014–2016 led to a significant increase in tail risk for three energy companies: XOM, CVX, COP.

\section{Conclusion}

In this paper, we investigate the estimation methods of extreme CoVaR and CoES within the framework of asymptotic independence but positively association, namely, the coefficient of tail dependence $\eta \in (1/2,1)$. We develop two extrapolative approaches, where the first relies on intermediate VaR and an adjustment factor, while the second exploits intermediate CoVaR and CoES. Simulation study demonstrates that the proposed estimations are robust and perform well in finite samples. The case of asymptotic independence but negatively association, \emph{i.e.}, $\eta \in (0,1/2)$, is technically more challenging and is therefore left for future research.


\section*{Supplementary Material}

\begin{description}
\item[Title:] {\bf Supplementary Material for ``Estimations of Extreme CoVaR and CoES under Asymptotic Independence"}\newline
This document includes all theoretical proofs. 
\end{description}


\end{document}